\documentclass{book}
\usepackage[mathscr]{euscript}
\usepackage[utf8]{inputenc}
\usepackage{amsmath,amsthm,amsfonts,amssymb}
\usepackage{graphicx}
\usepackage{hyperref}
\usepackage{color}
\definecolor{custom}{rgb}{0.0, 0.0, 0.5}
\hypersetup{
     colorlinks   = true,
     citecolor    = custom
}
\usepackage[backend=biber]{biblatex}
\bibliography{Refbook}

\numberwithin{thm}{chapter}

\newcommand{\beq}{\begin{eqnarray}}
\newcommand{\eeq}{\end{eqnarray}}

\newcommand{\la}{\lambda}

\newcommand{\s}{\sigma}
\newcommand{\qq}{(\s_{y}\otimes\s_{y})}
\newcommand{\uu}{\qq\rho^{*}\qq}

\usepackage[T1]{fontenc}
\usepackage{imakeidx}
\makeindex[columns=3, title=Alphabetical Index]

\begin{document}

\begin{center}

\vspace{0.5cm}
\begin{LARGE}
\textbf{Quantum Information Processing}\\
\textit{A brief overview on Quantum Teleportation}\\
\end{LARGE}
\vspace{0.5cm}

\begin{center}
by\\
{\large\textsc{DR. SOVIK ROY}}\\
\vspace{1.28cm}
\large\textbf{Associate Professor}\\
{\large\textbf{Department of Mathematics}}\\
\textbf{Techno Main Salt Lake (Engg. Colg.)\\ Techno India Group, Kolkata - 91, India}\\
\end{center}

\end{center}
\newpage
\thispagestyle{empty}
\begin{flushleft}
{\large \textsl{Dedicated to my daughter Tulip
\textbf{}}}
\end{flushleft}
\tableofcontents
\newpage 
\listoffigures
\newpage
\pagenumbering{arabic}

\hspace{3.9in}\textsc{Sovik Roy}
\\\\ 
\newpage
\thispagestyle{empty}
\chapter{What Mathematics do we need to know?}
\label{ch:mpp}
\section{Unit - I}\label{sec:unit1}
	\subsection{Born's Rule and quantum interference:}
	To do Quantum Mechanics (QM), the understanding of only classical probability laws won't help; rather, one needs to have a new set of probabilistic rules. This set of rules connects probability amplitudes with probabilities. Probability amplitudes are complex numbers, and the correspondence between probability amplitudes and probabilities is given by what is known as  \textbf{BORN'S RULE}. According to this rule, if the probability amplitude is represented by the complex number $z$, then the associated probability is $|z|^{2}$. The axioms of Born's Rule are as follows:\cite{sakurai1967advanced,feynman1979feynman}\newpage
	\noindent $Rule 1:$ Whenever a change from one particular state (of the system) to another state happens in a sequence of independent steps, we multiply the amplitudes of each step to get the total amplitude.
	\begin{figure}[h!]
		\begin{center}
			\includegraphics[width=5.5cm]{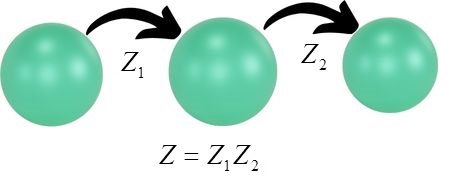}
		\end{center}
        \caption{The figure shows the state change concerning Born's first probability law.}
		\label{f2a}
	\end{figure}\\\\
	$Rule 2:$ Whenever a change from one particular state (of the system) to another state occurs in mutually exclusive steps, we add the amplitudes of each step to obtain the total amplitude.
	\begin{figure}[h!]
		\begin{center}
			\includegraphics[width=5.5cm]{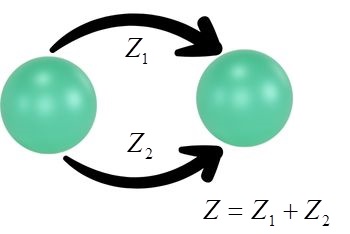}
		\end{center}
        \caption{The figure shows the state change concerning Born's second probability law.}
		\label{f2}
	\end{figure}
	\\\\
	Now suppose a (classical) system evolves from one state $A$ to another state $B$ in mutually exclusive ways, then if $P_{1}$ and $P_{2}$ are the two probabilities associated with two different paths, by \textbf{Kolmogorov's} probability axiom the total probability $P$ will be
	\begin{eqnarray}
		\label{kolmo1}
		P = P_{1} + P_{2}.
	\end{eqnarray}
	In quantum theory if $z_{1}$ and $z_{2}$ are the probability amplitudes for one system changing from one state $A$ to another state $B$, we know that the total probability amplitude is $z = z_{1} + z_{2}$ and hence the total probability $P$ will become
	\begin{eqnarray}
		\label{kolmoborn1}
		P &=& \vert z \vert^{2} = \vert z_{1} + z_{2} \vert^{2}\nonumber\\
		&=& \vert z_{1}\vert^{2} + \vert z_{2}\vert^{2} + z_{1}z_{2}^{*} + z_{1}^{*}z_{2}\nonumber\\
		&=& P_{1} + P_{2} + 2\sqrt{P_{1}P_{2}}\cos(\phi_{1}-\phi_{2}),
	\end{eqnarray}
	where, $P_{i}=|z_{i}|^{2}$, for $i=1,~2$ and $z_{1}=|z_{1}|e^{i~\phi_{1}}$, $z_{2}=|z_{2}|e^{i~\phi_{2}}$ in polar form. In (\ref{kolmoborn1}) the term $2\sqrt{P_{1}P_{2}}\cos(\phi_{1}-\phi_{2})$ is called \textbf{quantum interference} which can both be positive or negative as cosine function can move from $-1$  to $1$ and back. This interference term can enhance probability for the system to go from state $A$ to state $B$ or diminish the probability depending upon the phase difference $\phi_{1}-\phi_{2}$. This extra term is what separates classical probability law from quantum mechanical probability law. \textit{The mystery, however, is in the fact that it is not known how the quantum mechanical system knows the phase setting so that it can react to the difference in the phase setting}. Therefore in QM, the system changes from state $A$ to the state $B$ ,feeling (apparently) the two mutually exclusive paths. This is a statement about a real physical situation that is consistent with many experimental observations \cite{gerlach1989experimentelle}.
	\subsection{What is Quantum Bit (or Qubit)?}
Just as in the heory of classical computation and classical communication \textit{bit} is considered to be the fundamental object, for quantum computation and quantum information, an analogous concept, known as \textit{qubit}, has been developed.	A qubit is a mathematical object having certain properties. It is a vector in a two-dimensional($2D-$) complex vector space in which the inner product is defined. Unlike classical bits, that can be found in either state $0$ or in state $1$, a qubit, apart from being in the state $\vert 0\rangle$ or in $\vert 1\rangle$, can also be found in a superposition of states $\vert 0\rangle$ and $\vert 1\rangle$, which is expressed as
	\begin{eqnarray}
		\label{qubit1}
		\vert \varphi\rangle = \alpha \vert 0\rangle + \beta \vert 1\rangle .
	\end{eqnarray}
	However, the classical state $0 ~(1)$ is different from its quantum mechanical version $\vert 0\rangle ~(\vert 1\rangle)$ in the context of Dirac notation. While $0$ (or $1$) denotes the label, $\vert 0\rangle$ (or $\vert 1\rangle$) defines the state vector corresponding to the label.  Also note that $\alpha$ and $\beta$ are probability amplitudes and $|\alpha |^{2}$ and $|\beta |^{2}$ are respectively the probabilities of finding the system in either state $\vert 0\rangle$ or in state $\vert 1\rangle$ after the state $\vert \varphi\rangle$ of the system is measured. It is also to be noted that the normalization condition $|\alpha |^{2} + |\beta |^{2} = 1$ is preserved in (\ref{qubit1}). If we think of a coin tossing experiment in a classical system, then if the system is unbiased we expect to get outcomes that are either \textit{head} or \textit{tail}. In a biased system, however, a funny case like a coin standing on the edge may happen. In contrast to this a quantum coin, if tossed in an unbiased system, will not only give \textit{quantum head state} or \textit{quantum tail state}, but also a superposition of \textit{quantum head} and \textit{quantum tail} can also be expected. We say in such a scenario that the \textit{quantum coin} is in \textit{continuum} of states between \textit{quantum head} and \textit{quantum tail}. Mathematically we can correspond \textit{quantum state} to $\vert 0\rangle$ and \textit{quantum tail} to $\vert 1\rangle$\cite{nielsen2010quantum}.\\\\
	\textbf{Problem:} For each of the following qubits, if measurement is made, what is the probability that we find the qubit in state $\vert 0\rangle$ and also in state $\vert 1\rangle$? (a)~$\vert \phi\rangle = \dfrac{1}{\sqrt{3}}\vert 0\rangle + \sqrt{\dfrac{2}{3}}\vert 1\rangle$, (b)~$\vert \psi\rangle = \dfrac{i}{2}\vert 0\rangle + \dfrac{\sqrt{3}}{2}\vert 1\rangle$ and (c)~$\vert \chi\rangle = \dfrac{(1+i)}{\sqrt{3}}\vert 0\rangle - \dfrac{i}{\sqrt{3}}\vert 1\rangle$.\\
	\subsubsection{Bloch Sphere representation of qubits:}
	Geometrically we can represent a qubit in the following way:
	\begin{eqnarray}
		\label{blochsphere1}
		\vert \varphi\rangle = e^{i~\gamma}~\Big(\cos~\dfrac{\theta}{2}~\vert 0\rangle + e^{i~\eta}~\sin~\dfrac{\theta}{2}~\vert 1\rangle\Big).
	\end{eqnarray}
	The term $e^{i~\gamma}$, known as the global phase factor, can however be ignored as it has no observable affect. Hence (\ref{blochsphere1}) can also sometimes be expressed as
	\begin{eqnarray}
		\label{blochsphere2}
		\vert \varphi\rangle = \cos~\frac{\theta}{2}~\vert 0\rangle + e^{i~\eta}~\sin~\frac{\theta}{2}~\vert 1\rangle.
	\end{eqnarray}
	\begin{figure}[h!]
		\begin{center}
			\includegraphics[width=5.5cm]{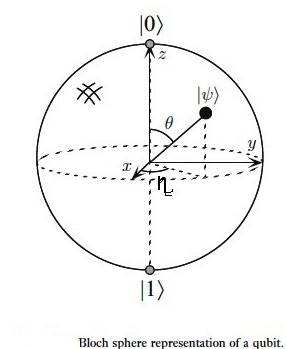}
		\end{center}
		\label{f2}
	\end{figure}\\\\
	The qubit has many other representations, one in which the state $\vert 0\rangle$ can be considered as the lowest energy state of the atom (or grounded state) and $\vert 1\rangle$ as the highest energy state (or excited state) of the atom whereas $\vert \varphi\rangle$ is the superposition of the grounded and the excited state, which we denote by $\vert g\rangle$ and $\vert e\rangle$. Likewise horizontal polarization and vertical polarization ($\vert H\rangle$ and $\vert V\rangle$) can also be equivalently realized for $\vert 0\rangle$ and $\vert 1\rangle$ respectively.
	\begin{figure}[h!]
		\begin{center}
			\includegraphics[width=5.5cm]{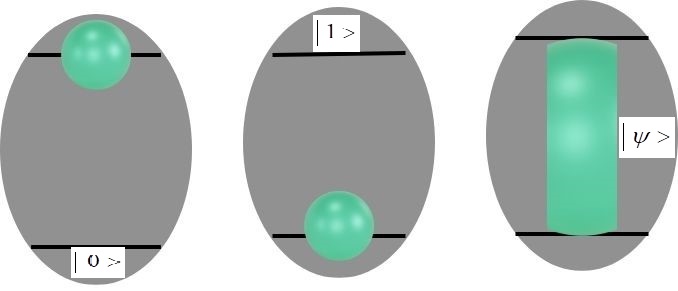}
		\end{center}
		\label{f2}
	\end{figure}\\\\
	Paradoxically, since on the Bloch sphere there are infinitely many points, so in principle one can store all the literary works of Charles Dickens in the infinite binary expansion of $\theta$ (\textit{Think why it is paradox!}) \cite{nielsen2010quantum}.
\subsubsection{Why qubit is represented in the way it is represented?}
We have come across from the previous sessions that a qubit is represented by equation (\ref{qubit1}). A natural question is therefore to ask if such a representation is valid! To put it in other words we may ask whether this mathematical formalism of qubit has its counterpart in physical world. The answer lies in the experiment conducted by Stern and Gerlach in the early 20th century \cite{gerlach1989experimentelle}. Here we are going to describe, keeping deep physics aside, only from the pedagogical point of view, why this experiment is so popular in scientific community.\\\\
Suppose we have a particle (a photon, say, generated from a source) which has three spin components along $x$, $y$ and $z$, respectively denoted by $S_{x}\:, S_{y}\:, S_{z}$.(Remember that the spin is an intrinsic property of the micro particle whose mathematics was formalized to eliminate paradoxes which couldn't be solved by classical physics in the early years of 20th century). Just like the $z-$ axis has two directions (one positive and another negative), the spin component $S_{z}$ can have two phases, one spin-up and another spin-down which we respectively denote here by $S_{z},+$ and $S_{z}, -$. We can similarly consider for spins along $x-$ axis and $y-$ axis. $S_{x},+$ and $S_{x}, -$ are spin up and spin down states with respect to $x-$ axis while $S_{y},+$ and $S_{y}, -$ are spin up and spin down states with respect to $y-$ axis. For $z$ component of the spin we will represent the corresponding state vectors as $\vert S_{z},+\rangle$ and $\vert S_{z},-\rangle$. 
\begin{figure}[h!]
		\begin{center}
			\includegraphics[width=8.5cm]{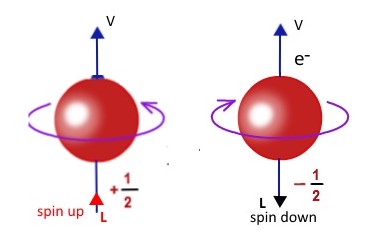}
		\end{center}
        \caption{The figure shows spin-up and spin-down state of a photon. The spin is intrinsic abstract notion introduced by P. A. M. Dirac. The image courtesy is google.}
		\label{sg1}
	\end{figure}\\\\
	After the generation of the photon, it is subjected to the Stern-Gerlach (SG-1) apparatus (which is a magnetic field having the capacity of splitting the particle into two of its spin components along certain axis). The two spin components then pass through a collimator having two slits open. If the particle spins are $S_{z},+$ and $S_{z}, -$ and the orientation of the collimator is aligned with the particle spin then the two $z-$ components will pass through the slits with equal probability (experimentally verified fact). Other spin components $S_{x},+,\:\: S_{x}, -$  and $S_{y},+,\:\: S_{y}, -$ will be (apparently!) blocked. Now instead of putting the two slits of the collimator open, if we keep one slit closed, then either $S_{z},+$  beam (or $S_{z},-$  beam) will come out of the collimator. \\\\
	Let us now put another SG apparatus in front of the collimator slits and the beam $S_{z},+$ or ($S_{z},-$) is subjected to that second SG apparatus (SG-2). But this time the orientation of the SG-2 apparatus is kept along $x-$ direction with its two slits open. Initially it was thought that the $S_{x},+,\:\: S_{x}, -$ spin components were completely blocked, but to the great amazement of the scientific community it was observed that when $S_{z}, +$ beam was subjected to SG-2 apparatus, it splits the beam into two halves $S_{x},+,\:\: S_{x}, -$ (with equal probability). But if we block, as before, one of the slits of the collimator (placed in front of the SG-2 apparatus), it is quite obvious that either $S_{x},+$ (or $S_{x}, -$) will pass through. This time the $S_{z},+$ component is completely missing from the output coming out of the collimator.\\\\
	We again place a third SG apparatus (SG-3) in front of the second collimator oriented along $y-$ axis, with two slits open. It is of great astonishment to the scientific fraternity that this time SG-3 apparatus taking $S_{x},+$ beam as input (or $S_{x}, -$ beam as input) gives $S_{y},+,\:\: S_{y}, -$ beam as two outputs\footnote{For detailed and enlightening discussion on Stern Gerlach experiment one can view https://www.youtube.com/watch?v=AX9769eQV24.}. How can this be explained!\\\\
	After much thought the physicists came to the conclusion that this bizarre phenomenon could only be explained if we consider the following two perspectives.
	\begin{itemize}
	\item \textbf{Physicist's view:} State of the system (here that of the photon particle) cannot be determined beforehand until one measures it.\footnote{Read about Schrodingers Cat problem \cite{nielsen2010quantum}.}. 
	\item \textbf{Mathematician's view:} If $\vert S_{z}, +\rangle$ is the state of the spin component $S_{z}, +$ , then (probably!) it can be written mathematically as
	\begin{eqnarray}
	\label{qubitz}
	\vert S_{z}, +\rangle &=& \frac{1}{\sqrt{2}} \vert S_{x}, +\rangle + \frac{1}{\sqrt{2}} \vert S_{x}, -\rangle \nonumber\\
	\vert S_{x}, +\rangle &=& \frac{1}{\sqrt{2}} \vert S_{y}, +\rangle + \frac{1}{\sqrt{2}} \vert S_{y}, -\rangle,
	\end{eqnarray}
    where $\frac{1}{\sqrt2}$ is the probability amplitude and $(\frac{1}{\sqrt2})^{2}$ is the corresponding probability. 
	\end{itemize}
If we carefully observe at equation (\ref{qubitz}) then we will be able to realize the fact that quantum mechanical state represented as qubit of the form (\ref{qubit1}) is quite justified.
	\subsection{Hilbert Space:}
	Quantum Computation and Quantum Information are studied in the domain of a \textit{finite dimensional complex vector space having inner product defined in it}. This space is called \textbf{Hilbert Space}. We shall now try to understand the mathematics of Hilbert space with Dirac terminologies.\\\\
More formally, \textit{complex vector space} is a set $V$ defined over a set of complex numbers such that given any two vectors $\vert a\rangle$ and $\vert b\rangle$ (that are elements of $V$) and any two complex numbers $\alpha$ and $\beta$ (that are called probability amplitudes), we can form a linear combination $\alpha~\vert a\rangle + \beta~\vert b\rangle$ which is also a vector in $V$. (It is to be noted that the elements of the vector space satisfy a few algebraic properties). However a \textit{subspace} of $V$ is any subset of $V$ which is closed under vector addition and scalar multiplication. The elements of a vector space are called vectors\cite{insel2003linear}. In the table below we show how we write the vectors in Dirac notation.
	\begin{center}
		\begin{tabular}{|c|c|}
			\hline Classical Notation for vectors & Dirac Notation for (Ket) Vectors\\
			\hline $\overrightarrow{a}$ & $\vert a\rangle$\\
			\hline
		\end{tabular}
	\end{center}\
	\newline
	A \textit{Basis} in $V$ is a collection of vectors $\vert e_{1}\rangle,~\vert e_{2}\rangle,~\cdots,\vert e_{n}\rangle$ such that every vector $\vert v\rangle$ in $V$ can be expressed in a unique way as a linear combination of these basis vectors, i.e.
	\begin{eqnarray}
		\label{basis1}
		\vert v\rangle = \lambda_{1}\vert e_{1}\rangle + \lambda_{2}\vert e_{2}\rangle + \cdots + \lambda_{n}\vert e_{n}\rangle
		=  \sum_{i} \lambda_{i} \vert e_{i}\rangle,
	\end{eqnarray}
	where the vectors $\vert e_{i}\rangle$'s are linearly independent (L.I), (here $\lambda_{1}, \lambda_{2},\cdots \lambda_{n}$ are scalars). The number of elements in a basis is called the \textit{dimension} of the vector space. The most common $n-$ dimensional complex vector space is the space of ordered $n-$ tuples of complex numbers usually represented in column vectors. This vector space is denoted by $\mathbb{C}^{n}$. The following algebraic operations are defined on $\mathbb{C}^{n}$ \cite{insel2003linear}.
	\begin{eqnarray}
		\label{C^{n}}
		\vert a\rangle = \left(%
		\begin{array}{c}
			a_{1}\\
			a_{2}\\
			.\\
			.\\
			a_{n}
		\end{array}%
		\right),
		\vert b\rangle = \left(%
		\begin{array}{c}
			b_{1}\\
			b_{2}\\
			.\\
			.\\
			b_{n}
		\end{array}%
		\right), 
		\alpha\vert a\rangle + \beta\vert b\rangle = \left(%
		\begin{array}{c}
			\alpha a_{1}+\beta b_{1}\\
			\alpha a_{2}+\beta b_{2}\\
			.\\
			.\\
			\alpha a_{n}+\beta b_{n}
		\end{array}%
		\right).
	\end{eqnarray}
Here $a_{1}, a_{2}, \cdots ,a_{n}$ are components of $\vert a\rangle$ and $b_{1}, b_{2}, \cdots ,b_{n}$ are components of $\vert b\rangle$.	In $2D-$ system i.e. in $\mathbb{C}^{2}$ space the basis is given by the set $\lbrace \vert 0\rangle,~\vert 1\rangle\rbrace$. Physicists also sometimes like to write $\vert 0\rangle \equiv \vert \uparrow\rangle$ and $\vert 1\rangle \equiv \vert \downarrow\rangle$, the first one is called the spin - up state and the second one is the spin - down state so that $\lbrace \vert \uparrow\rangle,~\vert \downarrow\rangle\rbrace$ represents the basis to work with\footnote{If a set $S$ is basis of a vector space $V$ it satisfies two characteristics, viz. $S$ is L.I and $S$ spans $V$. \cite{insel2003linear}}. Here I would like to make a comment that, as when required, I shall use these various notations of representing qubits in later chapters. The ket vectors $\vert 0\rangle$, $\vert 1\rangle$ and arbitrary vector $\vert \varphi\rangle$ can be represented in terms of the basis elements as follows:
	\begin{eqnarray}\label{vs2}
		\vert 0\rangle = \left(%
		\begin{array}{c}
			1\\
			0
		\end{array}%
		\right),
		\vert 1\rangle =  \left(%
		\begin{array}{c}
			0\\
			1
		\end{array}%
		\right),
		\nonumber\\
		\vert \varphi\rangle = \left(%
		\begin{array}{c}
			\alpha\\
			\beta
		\end{array}%
		\right) = \alpha\left(%
		\begin{array}{c}
			1\\
			0
		\end{array}%
		\right) + \beta \left(%
		\begin{array}{c}
			0\\
			1
		\end{array}%
		\right)
	\end{eqnarray}
\textbf{Problem:} Show that the set $\lbrace \vert a\rangle,~\vert b\rangle\rbrace$ where $\vert a\rangle = \dfrac{1}{\sqrt{2}}\left(%
	\begin{array}{c}
		1\\
		1
	\end{array}%
	\right)$ and $\vert b\rangle = \dfrac{1}{\sqrt{2}}\left(%
	\begin{array}{c}
		1\\
		-1
	\end{array}%
	\right)$ together represent a spanning set\footnote{Let $S$ be a non-empty subset of a vector space $V$. The span of $S$, denoted by $span(S)$, is the set consisting of all linear combinations of the vectors in $S$.} for the vector space $\mathbb{C}^{2}$.\footnote{Remember $\bar{0}$ is different from $\vert 0\rangle$ as $\bar{0} =\left(%
		\begin{array}{c}
			0\\
			0
		\end{array}%
		\right) $while $\vert 0\rangle = \left(%
		\begin{array}{c}
			1\\
			0
		\end{array}%
		\right)$.} \\\\
Now as we have come to know of the facts related to complex vector space, the next part is to understand the inner product which is defined on it. To define inner product between two vectors, a space known as \textit{dual space}, is taken into consideration. So for any vector $\vert v\rangle$ belonging to a Hilbert space we take a vector denoted by $\langle v\vert$ (often called \textit{Bra vector}, $\langle \vert$) from the dual space, known as the dual vector of $\vert v\rangle$. Remember this fact that
	 if $\vert \varphi\rangle = \left(%
	\begin{array}{c}
		\alpha\\
		\beta
	\end{array}%
	\right)$ (a column vector), its corresponding dual $\langle \varphi\vert = \left(%
	\begin{array}{c}
		\alpha^{*}\\
		\beta^{*}
	\end{array}%
	\right)^{T}$ is the row vector. Here $\alpha^{*}$ is the complex conjugate of $\alpha$ and $\beta^{*}$ is the complex conjugate of $\beta$\footnote{$A^{T}$ is the transpose of the matrix $A$.}.\\\\
Let $V$ be the vector space over the field of complex numbers. A function $\langle \cdot, \cdot\rangle$ from $V^{*} \times V$ to $\mathbb{C}$ is said to be the inner product if for vectors $\langle u\vert \in V^{*}$ and $\vert v\rangle \in V$ we have a complex number $\langle \langle u\vert , \vert v\rangle\rangle$ satisfying certain axioms ($V^{*}$ is the dual space). Remember $\langle u\vert$ is the dual vector to the vector $\vert u\rangle$ while dual is a linear operator from the inner product space $V^{*}$ to the field of complex number $\mathbb{C}$. Thus we have for $\vert w\rangle \in V$, the following equivalent terminologies to denote inner product between the two vectors. 
\begin{eqnarray}
\label{innerproduct1}
 \langle v\vert(\vert w\rangle) &\equiv  \langle \langle v\vert,\vert w\rangle\rangle \equiv & \langle v\vert w\rangle\nonumber\\
  .
\end{eqnarray}
Below we discuss the properties of inner product and write them both in classical as well as quantum terminologies.
	\begin{center}
		\begin{tabular}{|c|c|c|}
			\hline Property & Usual terminology & Dirac terminology\\
			\hline $1$ & $\langle \langle u\vert, \vert v\rangle + \vert w\rangle\rangle = \langle \langle u\vert, \vert v\rangle\rangle + \langle \langle u\vert, \vert w\rangle\rangle$ & $\langle u\vert v+w\rangle = \langle u\vert v\rangle + \langle u\vert w\rangle$\\
			\hline $2$ & $\langle \langle u\vert , \alpha\vert v\rangle\rangle = \alpha \langle \langle u\vert, \vert v\rangle\rangle$ & $ \langle u\vert \alpha v\rangle = \alpha \langle u \vert v\rangle$\\
			\hline $3$ & $\langle \langle u\vert , \vert v\rangle\rangle^{*} = \langle \langle v\vert , \vert u\rangle\rangle$ & $\langle u\vert v\rangle^{*}= \langle v\vert u\rangle$\\
			\hline $4$ & $\langle \langle u\vert , \vert u\rangle\rangle > 0, \forall \vert u\rangle$ & $\langle u\vert u\rangle > 0, \forall \vert u\rangle$\\
			\hline $5$ & $\langle \langle u\vert , \vert u\rangle\rangle = 0 \iff \vert u\rangle = \bar{0}$ & $\langle u\vert u\rangle = 0 \iff \vert u\rangle = \bar{0}$\\
			\hline
		\end{tabular}
	\end{center}\
	\newline
The property $4$ in the above table is the postulate of \textbf{positive definite metric}.  From a physicist's perspective this postulate is essential for the probabilistic interpretation of quantum mechanics. Attempt to abandon this postulate, however will lead to physical theories with \textit{indefinite metric} that won't be required for our study of quantum computation and quantum information\footnote{For detailed discussion on this matter the reader is recommended to the reference \cite{sakurai1967advanced}}.\\\\ 
Using properties described above and noting that $*$ is complex conjugation we have the following results.
\begin{eqnarray}
	\label{conjugatelinear2a}
     \Big\langle \sum_{i}\lambda_{i}w_{i} \vert v\Big\rangle &=&  \langle \lambda_{1}w_{1} + \lambda_{2}w_{2} +\cdots \vert v\rangle \nonumber\\
     &=& \langle v \vert \lambda_{1}w_{1} + \lambda_{2}w_{2} +\cdots\rangle^{*}\nonumber\\
     &=& \langle v \vert \lambda_{1}w_{1}\rangle^{*} + \langle v \vert \lambda_{2}w_{2}\rangle^{*} + \cdots\nonumber\\ 
     &=&  \lambda_{1}^{*}   \langle v \vert w_{1}\rangle^{*} + \lambda_{2}^{*}   \langle v \vert w_{2}\rangle^{*} + \cdots\nonumber\\
     &=& \lambda_{1}^{*}   \langle w_{1} \vert v\rangle + \lambda_{2}^{*}   \langle w_{2} \vert v\rangle + \cdots
     \end{eqnarray}
	Also we have
	\begin{eqnarray}
	\label{conjugatelinear2b}
	\Big\langle w\vert \sum_{j}\zeta_{j}v_{j} \Big\rangle &=& \langle w\vert \zeta_{1}v_{1} + \zeta_{2}v_{2} + \cdots \rangle \nonumber\\ 
	&=& \zeta_{1} \langle w\vert v_{1}\rangle + \zeta_{2} \langle w\vert v_{2}\rangle + \cdots \nonumber\\
	&=& \zeta_{1}\langle w\vert v_{1}\rangle + \zeta_{2}\langle w\vert v_{2}\rangle + \cdots
		\end{eqnarray}
Here $\lambda_{i},\:\zeta_{j}$ are all scalars. From eqs.(\ref{conjugatelinear2a}) and (\ref{conjugatelinear2b}) it is observed that the way inner product has been introduced here, the inner product is conjugate linear in its first component and linear in its second component\footnote{This is the Dirac way of representing inner product to validate quantum mechanical physical laws\cite{sakurai1967advanced}.}.\\\\
	\textbf{Problem:} Consider the vector space $\mathbb{C}^{n}$ over $\mathbb{C}$ and show, if we define the operation on $\mathbb{C}^{n}$ in the way shown below
	\begin{eqnarray}
		\label{innerproduct1}
		\langle (x_{1}, x_{2}, \cdots, x_{n}), (y_{1}, y_{2}, \cdots, y_{n})\rangle = \sum_{i} X_{i}^{*} Y_{i} = (x_{1}^{*}~ x_{2}^{*}~ \cdots x_{n}^{*}) \left(%
		\begin{array}{c}
			y_{1}\\
			y_{2}\\
			.\\
			.\\
			.\\
			y_{n}
		\end{array}%
		\right),
	\end{eqnarray}
	the operation $\langle ., .\rangle$ defined in (\ref{innerproduct1}) is an inner product.
	\subsubsection{Norm:}
	For a vector $\vert v\rangle$ the norm is defined as 
	\begin{eqnarray}
		\label{norm1}
		\|\vert v\rangle\| = \sqrt{\langle v\vert v\rangle}.
	\end{eqnarray}
	If the norm of a vector is $1$, that vector is called \textit{unit} vector. We say that a vector $\vert v\rangle$ is normalized if $\|\vert v\rangle\| = 1$. By normalization of a vector $\vert v\rangle$ we conventionally mean dividing the vector $\vert v\rangle$ by its norm. Hence normalized form of the vector $\vert v\rangle$ will be
	\begin{eqnarray}
		\label{norm2}
		\vert v\rangle_{normalized} = \frac{\vert v\rangle}{\|\vert v\rangle\|}.
	\end{eqnarray}
Thus for vector $\vert 0\rangle$ the corresponding norm is $\|\vert 0\rangle\|=\sqrt{\langle 0\vert 0\rangle} = \sqrt{\left(%
		\begin{array}{cc}
			1 & 0\\
		\end{array}%
		\right)\left(%
		\begin{array}{c}
			1\\
			0\\
		\end{array}%
		\right)} = 1$. Similarly $\|\vert 1\rangle\| = 1$. It is clear to see that $\vert 0\rangle_{normalized}= \vert 0\rangle$ and $\vert 1\rangle_{normalized}= \vert 1\rangle$.
	\subsubsection{Orthogonal vectors:}
	Two vectors $\vert v\rangle$ and $\vert w\rangle$ are said to be orthogonal if their inner product is zero i.e. $\langle v\vert w\rangle = 0$. Orthogonality of two vectors geometrically implies that the vectors are perpendicular to one another. To put it in other words, one can say that the orthogonality of the two vectors means that the corresponding states are mutually exclusive. For example the vectors $\vert 0\rangle$ and $\vert 1\rangle$ are orthogonal as $\langle 0\vert 1\rangle = \left(%
		\begin{array}{cc}
			1 & 0\\
		\end{array}%
		\right)\left(%
		\begin{array}{c}
			0\\
			1\\
		\end{array}%
		\right) = 0$. Also $\langle 1\vert 0\rangle = 0$.
\subsubsection{Orthonormal vectors:}
	A set $\mathcal{I}$  of vectors $\vert i\rangle$ (while $i$ is the index and $\mathcal{I}$ is the index set) is orthonormal if $(a)$ each vector $\vert i\rangle$  is a unit vector and $(b)$ distinct vectors $\vert i\rangle$  and $\vert j\rangle$ in the set are orthogonal i.e. $\langle i\vert j\rangle =\delta_{ij}$\footnote{$\delta_{ij}$ is the Dirac-Delta function where $\delta_{ij}= 1$ if $i = j$ and  $\delta_{ij}= 0$ if $i \ne j$.}, both $i$ and $j$ are chosen from the index set $\mathcal{I}$. Therefore the set $\lbrace \vert 0\rangle,\: \vert 1\rangle\rbrace $ is an orthonormal set as each of the vectors of the set are unit vectors and one vector is orthogonal to the other. Moreover the set $\lbrace \vert 0\rangle,\: \vert 1\rangle\rbrace $ is the orthonormal basis of the vector space $\mathbb{C}^{2}$.\\\\
	\textbf{Problem:} Verify that the vectors $\vert v_{1}\rangle = \left(%
		\begin{array}{c}
			1\\
			1\\
		\end{array}%
		\right)$ and  $\vert v_{2}\rangle = \left(%
		\begin{array}{c}
			1\\
			-1\\
		\end{array}%
		\right)$ are orthogonal. What are the normalized forms of these vectors?\\\\
We shall now discuss how one can construct from a set of vectors another set of vectors which is orthonormal.
	\subsubsection{Gram-Schmidt Process:}
	Suppose $\lbrace \vert w_{1}\rangle,\vert w_{2}\rangle,\cdots, \vert w_{d}\rangle\rbrace$ is a basis set for some vector space $V$ (on the vector space an inner product is defined). We want to construct an orthonormal basis set  $\lbrace \vert v_{1}\rangle,\vert v_{2}\rangle,\cdots, \vert v_{d}\rangle\rbrace$ using the elements $\vert w_{1}\rangle,\vert w_{2}\rangle,\cdots, \vert w_{d}\rangle$. Then to fulfill this purpose \textit{Gram-Schmidt Process} is helpful.
	\begin{itemize}
		\item First construct $\vert v_{1}\rangle$ as $\frac{\vert w_{1}\rangle}{\|\vert w_{1}\rangle\|}$.
		\item For $k = 1,2,\cdots, d-1$ define $\vert v_{k+1}\rangle$ inductively by $\vert v_{k+1}\rangle = \frac{\vert w_{k+1}\rangle-\sum_{i = 1}^{k}\langle v_{i} \vert w_{k+1}\rangle\vert v_{i}\rangle}{\|\vert w_{k+1}\rangle-\sum_{i = 1}^{k}\langle v_{i} \vert w_{k+1}\rangle\vert v_{i}\rangle\|}$ .
	\end{itemize}
	To understand the construction in a simpler way, let us take a set of three vectors $\lbrace \vert w_{1}\rangle, \vert w_{2}\rangle, \vert w_{3}\rangle\rbrace$. Using the above methodology of Gram - Schmidt process we will try to construct a set of orthonormal vectors $\lbrace \vert v_{1}\rangle, \vert v_{2}\rangle, \vert v_{3}\rangle\rbrace$ as follows:
	\begin{itemize}
		\item $\vert v_{1}\rangle = \frac{\vert w_{1}\rangle}{\|\vert w_{1}\rangle\|}$.
		\item For $k = 1,\cdots, 3-1$ i.e. for $k = 1,2$, we get
		$\vert v_{1+1}\rangle = \frac{\vert w_{1+1}\rangle-\sum_{i = 1}^{1}\langle v_{i} \vert w_{1+1}\rangle\vert v_{i}\rangle}{\|\vert w_{1+1}\rangle-\sum_{i = 1}^{1}\langle v_{i} \vert w_{1+1}\rangle\vert v_{i}\rangle\|} $ i.e. $\vert v_{2}\rangle = \frac{\vert w_{2}\rangle-\langle v_{1} \vert w_{2}\rangle\vert v_{1}\rangle}{\|\vert w_{2}\rangle-\langle v_{1} \vert w_{2}\rangle\vert v_{1}\rangle\|}$. Also $\vert v_{2+1}\rangle = \frac{\vert w_{2+1}\rangle-\sum_{i = 1}^{2}\langle v_{i} \vert w_{2+1}\rangle\vert v_{i}\rangle}{\|\vert w_{2+1}\rangle-\sum_{i = 1}^{2}\langle v_{i} \vert w_{2+1}\rangle\vert v_{i}\rangle\|} $ i.e. $\vert v_{3}\rangle = \frac{\vert w_{3}\rangle-\lbrace\langle v_{1} \vert w_{3}\rangle\vert v_{1}\rangle + \langle v_{2} \vert w_{3}\rangle\vert v_{2}\rangle\rbrace}{\|\vert w_{3}\rangle-\lbrace \langle v_{1} \vert w_{3}\rangle\vert v_{1}\rangle + \langle v_{2} \vert w_{3}\rangle\vert v_{2}\rangle \rbrace\|}$.
	\end{itemize}
	Final conclusion is $\lbrace \vert v_{1}\rangle, \vert v_{2}\rangle, \vert v_{3}\rangle\rbrace$ is a set of orthonormal vectors\footnote{The proof of the fact that Gram-Schmidt process produces an orthonormal basis for vector space $V$ can be found in\cite{insel2003linear}}.\\\\
	\textbf{Problem:} Construct an orthonormal basis from the non-orthonormal set $\lbrace\vert e_{1}\rangle, \vert e_{2}\rangle; \vert e_{1}\rangle=\left(%
	\begin{array}{c}
		2\\
		1
	\end{array}%
	\right), \vert e_{2}\rangle =\left(%
	\begin{array}{c}
		1\\
		3
	\end{array}%
	\right) \rbrace$.\\\\
\textbf{Important result:} If $\lbrace \vert e_{1}\rangle, \vert e_{2}\rangle\rbrace$ is not linearly independent then the Gram-Schmidt process fails to produce an orthonormal basis $\lbrace \vert e_{1}^{\prime}\rangle, \vert e_{2}^{\prime}\rangle\rbrace$.\\\\
	\textbf{\textit{Proof:}} Suppose the set $\lbrace \vert e_{1}\rangle, \vert e_{2}\rangle\rbrace$ is not linearly independent. Then the vectors $\vert e_{1}\rangle$ and $\vert e_{2}\rangle$ are linearly dependent. Accordingly let, $\vert e_{1}\rangle = k\vert e_{2}\rangle$ where $k$ is constant. Now assume that $\|\vert e_{1}\rangle\|= b$ (say). The first orthonormal vector $\vert e_{1}^{\prime}\rangle$ is then constructed as, $\vert e_{1}^{\prime}\rangle = \frac{\vert e_{1}\rangle}{\|\vert e_{1}\rangle\|}=\frac{\vert e_{1}\rangle}{b}$ and so $\vert e_{2}\rangle = \frac{\vert e_{1}\rangle}{k} \Rightarrow \vert e_{2}\rangle = \frac{b}{k} \vert e_{1}^{\prime}\rangle$. To get the second orthonormal vector $\vert e_{2}^{\prime}\rangle$ we first try to find the term $\|\vert e_{2}\rangle - \langle e_{1}^{\prime} \vert e_{2}\rangle\vert e_{1}^{\prime}\rangle\|$. We see that $\vert e_{2}\rangle - \langle e_{1}^{\prime} \vert e_{2}\rangle\vert e_{1}^{\prime}\rangle = \frac{b}{k}\vert e_{1}^{\prime}\rangle - \frac{k}{b}(\frac{b}{k})^{2}\vert e_{1}^{\prime}\rangle = 0$ and consequently $\|\vert e_{2}\rangle - \langle e_{1}^{\prime} \vert e_{2}\rangle\vert e_{1}^{\prime}\rangle\|=0$. Thus we are unable to construct the vector $\vert e_{2}^{\prime}\rangle$ and hence Gram-Schmidt process fails.~~~~~~ \textbf{Q.E.D}\footnote{Q.E.D stands for \textit{Quod Erat Demonstrandum} meaning which was to be demonstrated.}.
	\section{Unit - II}\label{sec:unit2}
	\subsection{Linear operators:}
A \textit{linear transformation} is a mapping $L$ from a vector space $V$ to another vector space $W$ ($L: V\rightarrow W$) satisfying the following property \cite{insel2003linear}.  
	\begin{eqnarray}
		\label{linearoperator}
		L\Big(\sum_{i}a_{i}\vert v_{i}\rangle\Big) = \sum_{i}a_{i}L(\vert v_{i}\rangle) = \sum_{i}a_{i}\vert w_{i}\rangle.
	\end{eqnarray}
	In practice we like to simply write $L\vert v\rangle$ instead of writing $L(\vert v\rangle)$. In equation (\ref{linearoperator}) the input $\vert v\rangle \in V$ and the output $\vert w\rangle \in W$. However when $V = W$ then the linear transformation is said to be a \textit{linear operator} defined on the vector space $V$. Trivial linear operators are of two types. One is \textit{identity operator} $I_{V}: V\rightarrow V$, that maps the vector in $V$ into itself (such as $I_{V}\vert v\rangle =\vert v\rangle$). Another trivial linear operator is the \textit{zero operator} defined as $\bar{0}: V \rightarrow W$ which maps any vector $\vert v\rangle$ of the vector space $V$ to the zero vector $\vert 0\rangle_{W}$ of the vector space $W$ (such as $\bar{0}\vert v\rangle = \vert 0\rangle_{W}$).\\\\
Suppose $V$, $W$ and $X$ are three vector spaces and $L: V\rightarrow W$ and $K: W\rightarrow X$ are linear operators. Then we use the notation $KL$ to denote the composition of $K$ with $L$ defined as $(KL)\vert v\rangle = K (L\vert v\rangle)$\cite{sakurai1967advanced,insel2003linear}.
	\subsubsection{Matrix representation of linear operators:}
	Suppose $L:V \rightarrow W$ is a linear operator between vector spaces $V$ and $W$. Let $\lbrace\vert v_{1}\rangle,\vert v_{2}\rangle,\cdots ,\vert v_{m}\rangle\rbrace$ be the basis for $V$ and $\lbrace\vert w_{1}\rangle,\vert w_{2}\rangle,\cdots ,\vert w_{n}\rangle\rbrace$ be the basis for $W$. Then for each $j = 1,2,\cdots ,m$ there exist complex numbers $L_{1j},L_{2j},\cdots ,L_{nj}$ such that $L\vert v_{j}\rangle = \sum_{i}L_{ij} \vert w_{i}\rangle$. If we write explicitly then we get
	\begin{eqnarray}
		\label{linearoperator1}
		L\vert v_{1}\rangle &=& L_{11}\vert w_{1}\rangle  + L_{21}\vert w_{2}\rangle + \cdots L_{n1}\vert w_{n}\rangle \nonumber\\
		L\vert v_{2}\rangle &=& L_{12}\vert w_{1}\rangle  + L_{22}\vert w_{2}\rangle + \cdots L_{n2}\vert w_{n}\rangle \nonumber\\
		\cdots \nonumber\\
		L\vert v_{m}\rangle &=& L_{1m}\vert w_{1}\rangle  + L_{2m}\vert w_{2}\rangle + \cdots L_{nm}\vert w_{n}\rangle ,
	\end{eqnarray} 
	so that the corresponding matrix representation is
	\begin{eqnarray}
		\label{linearoperator2}
		L=\left(%
		\begin{array}{cccc}
			L_{11} & L_{21} & \cdots & L_{n1}\\
			L_{12} & L_{22} & \cdots & L_{n2}\\
			\cdots\\
			L_{1m} & L_{2m} & \cdots & L_{nm}\\
		\end{array}%
		\right)^{T}.
	\end{eqnarray} \\
	\subsubsection{A few examples of linear operators and their matrix representations:}
	Let us consider the vector space $V = \mathbb{C}^{2}$ whose basis is $\lbrace \vert 0\rangle, \vert 1\rangle\rbrace$. We define a linear operator $I$ on $\mathbb{C}^{2}$ as $I\vert 0\rangle = \vert 0\rangle$ and $I\vert 1\rangle = \vert 1\rangle$, then we get the following.
	\begin{eqnarray}
		\label{P1}
		I\vert 0\rangle &=& \vert 0\rangle = 1\vert 0\rangle + 0\vert 1\rangle\nonumber\\
		I\vert 1\rangle &=& \vert 1\rangle = 0\vert 0\rangle + 1\vert 1\rangle\nonumber\\\nonumber\\
		I&=&\left(%
		\begin{array}{cc}
			1 & 0 \\
			0 & 1 \\
		\end{array}%
		\right)^{T}=\left(%
		\begin{array}{cc}
			1 & 0 \\
			0 & 1 \\
		\end{array}%
		\right).
	\end{eqnarray} 
	The next operator we define on $V$ is $X$ given as $X\vert 0\rangle = \vert 1\rangle$ and $X\vert 1\rangle = \vert 0\rangle$ such that 
	\begin{eqnarray}
		\label{P2}
		X\vert 0\rangle &=& \vert 1\rangle = 0\vert 0\rangle + 1\vert 1\rangle\nonumber\\
		X\vert 1\rangle &=& \vert 0\rangle = 1\vert 0\rangle + 0\vert 1\rangle\nonumber\\\nonumber\\
		X&=&\sigma_{x}= \left(%
		\begin{array}{cc}
			0 & 1 \\
			1 & 0 \\
		\end{array}%
		\right)^{T}=\left(%
		\begin{array}{cc}
			0 & 1 \\
			1 & 0 \\
		\end{array}%
		\right).
	\end{eqnarray} 
	Then we can define an operator $iY$ on $V$ which is governed by the rule $iY\vert 0\rangle = -\vert 1\rangle$ and $iY\vert 1\rangle = \vert 0\rangle$. Consequently, we observe that
	\begin{eqnarray}
		\label{P3}
		iY\vert 0\rangle &=& -\vert 1\rangle = 0\vert 0\rangle + (-1)\vert 1\rangle\nonumber\\
		iY\vert 1\rangle &=& \vert 0\rangle = 1\vert 0\rangle + 0\vert 1\rangle\nonumber\\\nonumber\\
		iY&=& i\sigma_{y}=\left(%
		\begin{array}{cc}
			0 & -1 \\
			1 & 0 \\
		\end{array}%
		\right)^{T}=\left(%
		\begin{array}{cc}
			0 & 1 \\
			-1 & 0 \\
		\end{array}%
		\right).
	\end{eqnarray} 
	Next we have the operator $Z$ defined on $V$ that takes the vectors in the basis of $V$ by the rule $Z\vert 0\rangle = \vert 0\rangle$ and $Z\vert 1\rangle = -\vert 1\rangle$ so that we have
	\begin{eqnarray}
		\label{P4}
		Z\vert 0\rangle &=& \vert 0\rangle = 1\vert 0\rangle + 0\vert 1\rangle\nonumber\\
		Z\vert 1\rangle &=& -\vert 1\rangle = 0\vert 0\rangle + (-1)\vert 1\rangle\nonumber\\\nonumber\\
		Z&=&\sigma_{z} = \left(%
		\begin{array}{cc}
			1 & 0 \\
			0 & -1 \\
		\end{array}%
		\right)^{T}=\left(%
		\begin{array}{cc}
			1 & 0 \\
			0 & -1 \\
		\end{array}%
		\right).
	\end{eqnarray} 
	Another linear operator that we would like to mention here (for its fundamental importance in quantum computing) is as follows. This operator $H$ defined on $V$ is governed by the rule $H\vert 0\rangle = \frac{\vert 0\rangle + \vert 1\rangle}{\sqrt{2}}$ and $H\vert 1\rangle = \frac{\vert 0\rangle - \vert 1\rangle}{\sqrt{2}}$. Subsequently we get
	\begin{eqnarray}
		\label{P5}
		H\vert 0\rangle &=& \frac{\vert 0\rangle + \vert 1\rangle}{\sqrt{2}} = \frac{1}{\sqrt{2}}\vert 0\rangle + \frac{1}{\sqrt{2}}\vert 1\rangle\nonumber\\
		H\vert 1\rangle &=& \frac{\vert 0\rangle - \vert 1\rangle}{\sqrt{2}} = \frac{1}{\sqrt{2}}\vert 0\rangle - \frac{1}{\sqrt{2}}\vert 1\rangle\nonumber\\\nonumber\\
		H&=&\left(%
		\begin{array}{cc}
			\frac{1}{\sqrt{2}} & \frac{1}{\sqrt{2}} \\
			\frac{1}{\sqrt{2}} & -\frac{1}{\sqrt{2}}\\
		\end{array}%
		\right)^{T}=\left(%
		\begin{array}{cc}
			\frac{1}{\sqrt{2}} & \frac{1}{\sqrt{2}} \\
			\frac{1}{\sqrt{2}} & -\frac{1}{\sqrt{2}}\\
		\end{array}%
		\right).
	\end{eqnarray} 
	The operators defined in equations (\ref{P1}) - (\ref{P5}) are deep-seated in the study of quantum computation and quantum information. The operator of equation (\ref{P1}) is the \textbf{Identity operator} and the operators defined in (\ref{P2}), (\ref{P3}) and (\ref{P4}) are respectively called \textbf{Pauli X operator}, \textbf{Pauli iY operator} and \textbf{Pauli Z operator}. The first four matrices i.e. (\ref{P1})-(\ref{P4}) are also called \textbf{Pauli Spin matrices}. It is to be noted that in some books authors prefer to write \textit{Pauli Spin iY} as simply \textit{Pauli spin Y} \cite{nielsen2010quantum,pathak2013elements}, while simply they ignore the phase factor $i$ so that the corresponding matrix takes the form $Y = \left(%
	\begin{array}{cc}
		0 & -i \\
		i & 0 \\
	\end{array}%
	\right).$ The operator defined in (\ref{P5}) is called \textbf{Hadamard operator}.\\\\
		\textbf{Problem:} Consider a vector space $V$ whose basis is $\Big\lbrace \frac{\vert 0\rangle + \vert 1\rangle}{\sqrt{2}},\frac{\vert 0\rangle - \vert 1\rangle}{\sqrt{2}}\Big\rbrace$. Define a linear operator $A$ on $V$ such that $A\vert 0\rangle = \vert 1\rangle$ and $A\vert 1\rangle = \vert 0\rangle$. With respect to this linear operator transform the basis elements of $V$. (a) Write the matrix representation of $A$ in terms of the output basis which is same as the input basis of $V$. (b) Write the matrix representation of $A$ in terms of the basis $\lbrace \vert 0\rangle, \vert 1\rangle\rbrace$. \\\\
	\textbf{Problem:} Suppose we take an arbitrary single qubit $\alpha \vert 0\rangle + \beta \vert 1\rangle$ with $|\alpha|^{2} + |\beta|^{2} = 1$. This has the matrix representation $\left(%
	\begin{array}{c}
		\alpha\\
		\beta\\
	\end{array}%
	\right)$. How do $X$, $Y$, $Z$ and $H$ transform this single qubit? Write the output qubits in matrix form.\\\\
	The Pauli Operators and Hadamard operator are also known as \textbf{single qubit logic gates}. These are often used to construct multi-qubit quantum logic gates. The Pauli $X$ gate is also called \textbf{Quantum NOT gate} which is analogous to classical NOT gates. The $X$ operator is also called \textit{bit-flip operator} as we see in eq.(\ref{P2}) it changes bit $0$ to bit $1$ and under this operator bit $1$ changes to bit $0$. The operators $iY$ and $Z$ are respectively known  as \textit{bit-phase flip operator} and \textit{phase-flip operator}. From eqs.(\ref{P2}) - (\ref{P5}) we can easily see that under these quantum logic gates a single qubit like $\alpha\vert 0\rangle + \beta\vert 1\rangle$ transforms in the following manner.
\begin{eqnarray}
\label{Qubittransforms}
X(\alpha\vert 0\rangle + \beta\vert 1\rangle) &=& \alpha\vert 1\rangle + \beta\vert 0\rangle \nonumber\\
iY(\alpha\vert 0\rangle + \beta\vert 1\rangle) &=& \beta\vert 0\rangle - \alpha\vert 1\rangle \nonumber\\
Z(\alpha\vert 0\rangle + \beta\vert 1\rangle) &=& \alpha\vert 0\rangle - \beta\vert 1\rangle \nonumber\\
H(\alpha\vert 0\rangle + \beta\vert 1\rangle) &=& \frac{\alpha + \beta}{\sqrt{2}}\vert 0\rangle + \frac{\alpha -\beta}{\sqrt{2}}\vert 1\rangle.
\end{eqnarray}
	\subsubsection{Outer Product:}
	There is a useful way of representing linear operators which makes use of the inner product. This representation is called \textit{Outer Product Representation} \cite{sakurai1967advanced,nielsen2010quantum,pathak2013elements}. Let $\vert v\rangle$ be a vector in an inner product space $V$ and $\vert w\rangle$ be another vector in an inner product space $W$. Define a function from $V$ to $W$ denoted as $\vert w\rangle \langle v\vert : V \rightarrow W$ that is defined as
	\begin{eqnarray}
		\label{outerproduct1}
		(\vert w\rangle \langle v\vert)\vert v^{\prime}\rangle = \vert w\rangle \langle v\vert v^{\prime}\rangle = \langle v\vert v^{\prime}\rangle\vert w\rangle.
	\end{eqnarray}
	The above equation can be interpreted in two different ways. 
	\begin{itemize}
		\item Either we can say the operator $\vert w\rangle \langle v\vert$ acts on $\vert v^{\prime}\rangle$.
		\item Or we can think of the vector $\vert w\rangle$ getting multiplied by the complex number $\langle v\vert v^{\prime}\rangle$.
	\end{itemize}
	The first way however is more often used to understand the intricacies of outer product. 
	\subsubsection{Completeness relation (in outer product formalism):}
The usefulness of the outer product notation can be discerned from an imperative result known as \textit{completeness relation}. Let $\lbrace \vert i\rangle\rbrace_{i \in \mathcal{I}}$ be any orthonormal basis of the vector space $V$ (here $\mathcal{I}$ stands for the index set). Then an arbitrary vector $\vert v\rangle \in V$ can be expressed in terms of the basis elements as $\vert v\rangle = \sum_{i} \alpha_{i}\vert i\rangle$ where $\lbrace \alpha_{i}\rbrace_{i \in \mathcal{I}}$ is some set of complex numbers. Note that if we now take inner product $\langle i\vert v\rangle$ we get
	\begin{eqnarray}
		\label{outerproduct2}
		\langle i\vert v\rangle &=& \langle i\vert \alpha_{1}\vert 1\rangle + \alpha_{2}\vert 2\rangle +\cdots +  \alpha_{i}\vert i\rangle + \cdots\rangle \nonumber\\
		&=& \langle i\vert \alpha_{1}\vert 1\rangle + \langle i\vert \alpha_{2}\vert 2\rangle + \cdots + \langle i\vert \alpha_{i}\vert i\rangle + \cdots \nonumber\\
		&=& \alpha_{1}\langle i\vert 1\rangle + \alpha_{2}\langle i\vert 2\rangle + \cdots + \alpha_{i}\langle i\vert i\rangle + \cdots \nonumber\\
		&=& \alpha_{i}.
	\end{eqnarray}
	Recall $\langle i\vert j\rangle = \delta_{ij}$ ($\delta_{ij}$ is the Dirac-Delta function). Hence we get
	\begin{eqnarray}
		\label{outerproduct3}
		\Big(\sum_{i}\vert i\rangle \langle i\vert\Big)\vert v\rangle &=&  \sum_{i}\vert i\rangle \langle i\vert v\rangle \nonumber\\ &=& \sum_{i}\vert i\rangle \alpha_{i} = \vert v\rangle
		.\end{eqnarray}
	Since (\ref{outerproduct3}) is true for all $\vert v\rangle$ it follows that 
	\begin{eqnarray}
		\label{completeness}
		\sum_{i}\vert i\rangle \langle i\vert  = I.
	\end{eqnarray}
	$I$ is the identity operator. The eq.(\ref{completeness}) is known as \textit{completeness relation}.

		\subsubsection{Representation of an operator in the outer product notation:}
	Suppose $A: V \rightarrow W$ is a linear mapping, $\lbrace \vert i\rangle\rbrace_{i \in \mathcal{I}}$ is an orthonormal basis for $V$ and $\lbrace \vert j\rangle\rbrace_{j \in \mathcal{J}}$ is the orthonormal basis of $W$ (here $\mathcal{I}$ and $\mathcal{J}$ are respectively the two index sets). Applying identity operators $I_{W}$ and $I_{V}$ of the vector spaces $W$ and $V$ on both sides of the operator $A$ and then using the completeness relation (\ref{completeness}) we get
	\begin{eqnarray}
		\label{outerproduct4}
		A &=& I_{W}AI_{V} = \sum_{i,j} \vert j\rangle \langle j \vert A \vert i\rangle \langle i \vert\nonumber\\
		&=& \sum_{i,j}\langle j\vert A\vert i\rangle\vert j\rangle \langle i\vert.
	\end{eqnarray}
	This is an \textit{outer product representation} for the operator $A$. This also implies that $A$ has matrix representation with matrix elements in the $i^{th}$ column and $j^{th}$ row as $\langle j\vert A\vert i\rangle$ concerning the input basis elements $\vert i\rangle$ and the output basis elements $\vert j\rangle$. The matrix form of the operator $A$ is thus given by
\begin{eqnarray}
\label{outerproductmatrix}
A = \left(%
	\begin{array}{ccccc}
		\langle 1\vert A\vert 1\rangle & \langle 1\vert A\vert 2\rangle & \cdots & \langle 1\vert A\vert i\rangle & \cdots\\
		\langle 2\vert A\vert 1\rangle & \langle 2\vert A\vert 2\rangle & \cdots & \langle 2\vert A\vert i\rangle & \cdots\\
		\cdots & \cdots & \cdots & \cdots & \cdots\\
		\langle j\vert A\vert 1\rangle & \langle j\vert A\vert 2\rangle & \cdots & \langle j\vert A\vert i\rangle & \cdots\\
		\cdots & \cdots & \cdots & \cdots & \cdots\\
	\end{array}%
	\right)
\end{eqnarray}	
 Let us try to understand this by considering a simple situation. Let $I_{V}:V\rightarrow V$ be the identity operator whose outer product version we want to find. The basis is however same for both domain and codomain which is in fact $\lbrace \vert 0\rangle, \vert 1\rangle\rbrace$. Using (\ref{outerproduct4}) we thus obtain
	\begin{eqnarray}
		\label{outerproduct5}
		I_{V} &=& \vert 0\rangle \langle 0\vert I_{V}\vert 0\rangle \langle 0\vert + \vert 0\rangle \langle 1\vert I_{V}\vert 0\rangle \langle 1\vert + \vert 1\rangle \langle 0\vert I_{V}\vert 1\rangle \langle 0\vert +\vert 1\rangle \langle 1\vert I_{V}\vert 1\rangle \langle 1\vert  \nonumber\\
		&=& \langle 0\vert I_{V}\vert 0\rangle \vert 0\rangle \langle 0\vert + \langle 0\vert I_{V}\vert 1\rangle \vert 0\rangle \langle 1\vert  +\langle 1\vert I_{V}\vert 0\rangle \vert 1\rangle \langle 0\vert + \langle 1\vert I_{V}\vert 1\rangle \vert 1\rangle \langle 1\vert \nonumber\\
		&=& \left(%
	\begin{array}{cc}
	\langle 0\vert I_{V}\vert 0\rangle & \langle 0\vert I_{V}\vert 1\rangle\\
	\langle 1\vert I_{V}\vert 0\rangle & \langle 1\vert I_{V}\vert 1\rangle
	\end{array}%
	\right)\nonumber\\
		&=& \left(%
	\begin{array}{cc}
	\langle 0\vert 0\rangle & \langle 0\vert 1\rangle\\
	\langle 1\vert  0\rangle & \langle 1\vert 1\rangle
	\end{array}%
	\right)\nonumber\\
		&=& \left(%
	\begin{array}{cc}
		1 & 0 \\
		0 & 1\\
	\end{array}%
	\right)\nonumber\\
	&=& \vert 0\rangle\langle 0\vert + \vert 1\rangle\langle 1\vert.
	\end{eqnarray}
	\textbf{Problem:} Find the outer product representations of the Pauli operators and Hadamard operator. \\\\
	\textbf{Problem:} The \textit{Cauchy-Schwarz's inequality} is an important geometric fact about Hilbert spaces. It states that for any two vectors $\vert v\rangle$ and $\vert w\rangle$, we get $\vert \langle v\vert w\rangle \vert \leq \| \vert v\rangle\|\:\|\vert w\rangle\|$. Prove this inequality! 
	\subsection{Eigenvalues and eigenvectors:}
	Suppose a linear operator $A$ is defined on a vector space $V$. By \textit{eigenvector} of $A$ we mean a non-zero vector $\vert v\rangle$ such that 
	\begin{eqnarray}
		\label{eigenvector}
		A\vert v\rangle = \lambda\vert v\rangle,
	\end{eqnarray} 
where $\lambda$ is the eigenvalue and $\vert v\rangle$ is the corresponding eigenvector. To find the eigenvalue/s of an operator we equate the characteristic function, which is represented as $\det(A-\lambda\:I)$,  to zero.\footnote{$\det$ is the determinant and $I$ is the identity matrix.} Solving this characteristic equation we get solutions which are called characteristic values or \textit{eigenvalues}. \textit{Fundamental theorem of algebra} states that every polynomial has at least one complex root\cite{fine1997fundamental}, so consequently every operator $A$ has at least one eigenvalue and a corresponding eigenvector. It is to be noted that corresponding to an eigenvalue there may exist infinitely many eigenvectors.
	Suppose for an eigenvalue $\lambda$ we have some eigenvectors. The \textit{eigenspace} is the set of all eigenvectors along with the zero vector of the vector space. Thus corresponding to the eigenvalue $\lambda$, the set $E_{\lambda} = \lbrace v \in V: A\vert v\rangle = \lambda\vert v\rangle\rbrace$ is called its eigenspace.\footnote{Remember that not all the vectors of eigenspace are eigenvectors as because zero vector is present in the eigenspace and zero vector is not an eigenvector.} Clearly the eigenspace $E_{\lambda}$ consisting of zero vector and the eigenvectors of $A$ is a subspace of $V$. The maximum number of linearly independent eigenvectors of $A$ corresponding to the eigenvalue $\lambda$ is therefore the dimension of $E_{\lambda}$. Take for example the Pauli spin operator $X$. Its characteristic equation gives $\lambda = \pm 1$ (\textbf{Solve!}). Using the equation (\ref{eigenvector}) we obtain that the eigenvector $\left(%
	\begin{array}{c}
		1\\
		1\\
	\end{array}%
	\right)$ corresponding to the eigenvalue $1$ and $\left(%
	\begin{array}{c}
		-1\\
		1\\
	\end{array}%
	\right)$ corresponding to the eigenvalue $-1$ (\textbf{Solve!}). At this point, I would like to make a comment. It is often a misconception that the two eigenvectors are the only eigenvectors corresponding to the eigenvalues $\pm 1$. A little mathematical analysis using eq.(\ref{eigenvector}) will reveal that, corresponding to the eigenvalue $1$, the eigenvectors are $\lbrace t (1,1): t\in \mathbb{C}-\lbrace 0\rbrace$ while with respect to the eigenvalue $-1$, the eigenvectors are $\lbrace t^{\prime} (-1,1): t^{\prime}\in \mathbb{C}-\lbrace 0\rbrace$, all in the context of the Pauli operator $X$. It is to be noted that mathematical softwares like Mathematica, Maple, et.al will not reveal all the eigenvectors, only the special ones. In other words, corresponding to the two eigenvalues $1$ and $-1$, there exist infinitely many eigenvectors.
When an eigenspace is more than one dimensional we say that it is \textit{degenerate}. In other words if $A$ is an operator with two (or more) linearly independent eigenvectors having the same eigenvalue  then the eigenvalues of these eigenvectors are said to be degenerate. For example if we take $\left(%
	\begin{array}{ccc}
		2 & 0 & 0\\
		0 & 2 & 0\\
		0 & 0 & 0\\
	\end{array}%
	\right)$ it can be easily seen that the eigenvalues are $0,\:2,\: 2$ and the corresponding eigenvectors are $\left(%
	\begin{array}{c}
		0\\
		0\\
		1\\
	\end{array}%
	\right)$, $\left(%
	\begin{array}{c}
		0\\
		1\\
		0\\
	\end{array}%
	\right)$ and $\left(%
	\begin{array}{c}
		1\\
		0\\
		0\\
	\end{array}%
	\right)$. This clearly implies that $2$ is the degenerate eigenvalue but $0$ is not.\\\\
\textbf{Problem:} Find eigenvalues and eigenvectors of the Pauli gates $iY$ and $Z$ and also of the Hadamard gate $H$.\\\\
Remember that $E_{\lambda_{1}=1} = \lbrace t \left(%
	\begin{array}{c}
		1\\
		1\\
	\end{array}%
	\right) : t \in \mathbb{C} \rbrace$ and  $E_{\lambda_{2}=1} = \lbrace t^{\prime} \left(%
	\begin{array}{c}
		-1\\
		1\\
	\end{array}%
	\right) : t^{\prime} \in \mathbb{C} \rbrace$ are the eigenspaces corresponding to the eigenvalues $1$ and $-1$ respectively.
	One of the most important theorem in this area goes as follows:\\\\
	\textbf{Theorem:} Let $A \in M_{n\times n}(F)$\footnote{$M_{n\times n}(F)$ is the set of all square matrices of order $n\times n$ whose elements have been chosen from the field $F$.}, then (a) the characteristic polynomial of $A$ is a polynomial of degree $n$ with leading coefficient $(-1)^{n}$ and (b) $A$ has at most $n$ distinct eigenvalues \cite{insel2003linear}.\\\\ Thus we see that for the Pauli spin matrix $X \in M_{2\times 2}(\mathbb{C})$, its characteristic equation is $\lambda^{2}-1=0$ and so the leading coefficient of $\det(X-\lambda\:I)$ is $(-1)^{2}$. Consequently $X$ has at most two (here in this case exactly two) distinct eigenvalues. The theorem can also be verified for the matrices of \textit{Problem $10$} (\textbf{verify!}). \\\\
Geometrically, what we mean by a linear operator $A$ acting on an eigenvector $\vert v\rangle$ in the context of a vector space $V$ over the field $\mathbb{C}$ is the following. One can think of a subspace $W$ of $V$ which is spanned by the vector $\vert v\rangle$. Such a subspace may be thought of containing lines in $V$ that passes through the zero vector (origin) and the $\vert v\rangle$.   The operator $A$ acting on a vector in $W$ can now imply multiplying each such vector by $\lambda$ \cite{insel2003linear}.
\begin{itemize}
\item If $\lambda > 1$, then $A$ moves vectors in $W$ farther from origin by a factor $\lambda$.
\item If $\lambda  = 1$, then $A$ acts as the identity operator on $W$.
\item If $0 <\lambda < 1$, then $A$ moves vectors in $W$ closer to origin by a factor $\lambda$.
\item If $\lambda  = 0$, then $A$ acts as zero transformation on $W$.
\item If $\lambda < 0$, then $A$ reverses the orientation of vectors in $W$; that is $A$ moves vectors in $W$ from one side of origin to the other.
\end{itemize}
It is to be noted that a linear operator $A$ on a finite-dimensional vector space $V$ is diagonalizable if and only if there exists an ordered basis for $V$ consisting of eigenvectors of $A$.
	\subsection{Diagonalizability:}
	A diagonal representation for an operator $A$ on a vector space $V$ is a representation $A=\sum_{i}\lambda_{i}\vert i\rangle\langle i\vert$ where the vectors $\vert i\rangle$ form an orthonormal set of eigenvectors for $A$ with corresponding eigenvalues $\lambda_{i}$. An operator is said to be \textit{diagonalizable} if it has a diagonal representation. For instance we consider the the Pauli spin matrix $Z$ whose eigenvalues are $1$ and $-1$ and the corresponding eigenvectors are $\left(%
	\begin{array}{c}
		1\\
		0\\
	\end{array}%
	\right)=\vert 0\rangle$ and $\left(%
	\begin{array}{c}
		0\\
		1\\
	\end{array}%
	\right)=\vert 1\rangle$ respectively. These vectors constitute an orthonormal set so that the required diagonal representation is $Z = 1\vert 0\rangle\langle 0\vert -1\vert 1\rangle\langle 1\vert = \vert 0\rangle\langle 0\vert -\vert 1\rangle\langle 1\vert $.\\\\
	\textbf{Working criteria:} A working criterion in this regard becomes very helpful to diagonalize a given operator $A$ \cite{insel2003linear}. 
	\begin{itemize}
		\item First, find the eigenvalues of the operator $A$. For simplicity, we assume that there are two eigenvalues $\lambda_{1}$ and $\lambda_{2}$. 
		\item Suppose the corresponding eigenvectors are $\left(%
		\begin{array}{c}
			a\\
			b\\
		\end{array}%
		\right)$ for $\lambda_{1}$ and $\left(%
		\begin{array}{c}
			c\\
			d\\
		\end{array}%
		\right)$ for $\lambda_{2}$. 
		\item Construct the matrix $Q = \left(%
		\begin{array}{cc}
			a & c\\
			b & d\\
		\end{array}%
		\right)$ and find its inverse which we denote by $Q^{-1}$.
		\item Then construct $D = Q^{-1}\:A\:Q$. The matrix $D$ is diagonal.\footnote{The working criteria have been written here in a very simplistic way. For a detailed discussion, the author of this article recommends reference \cite{insel2003linear}}.
	\end{itemize}
	\textbf{Problem:} Using the above working criteria diagonalize the Pauli matrices $X$, $Y$ and $Z$. This is known as \textbf{eigen-decomposition} of the Pauli matrices.\\\\
	\textbf{Problem:} Verify whether the matrix $A=\left(%
	\begin{array}{cc}
	1 & 1\\
	4 & 1
	\end{array}%
	\right)$ is diagonalizable! \\\\
	\textit{Solution:} We shall use the above working criteria to check the diagonalizability of the given matrix.
	\begin{itemize}
	\item Considering the equation $\det(A-\lambda\:I) = 0$, we get $\det\:\left(%
	\begin{array}{cc}
	1-\lambda & 1\\
	4 & 1-\lambda
	\end{array}%
	\right) = 0$ which gives $(1-\lambda)^{2} - 4 = 0$, whose roots in return give two eigenvalues respectively, $\lambda_{1} = 3$ and $\lambda_{2} = -1$. \\
	\item Using eq. (\ref{eigenvector}) we will now find out the eigenvectors corresponding to each of the two eigenvalues calculated in the previous step.\\
	(i) For the eigenvalue $\lambda_{1} = 3$ let the non-zero eigenvector be $\left(%
		\begin{array}{c}
			v_{1}\\
			v_{2}\\
		\end{array}%
		\right)$ such that eq. (\ref{eigenvector}) gives $\left(%
		\begin{array}{cc}
			-2 & 1\\
			4 & -2\\
		\end{array}%
		\right)\left(%
		\begin{array}{c}
			v_{1}\\
			v_{2}\\
		\end{array}%
		\right)=\left(%
		\begin{array}{c}
			0\\
			0\\
		\end{array}%
		\right)$. The \textit{reduced row echelon form} of the matrix $\left(%
		\begin{array}{cc}
			-2 & 1\\
			4 & -2\\
		\end{array}%
		\right)$ gives that the matrix is equivalent to $\left(%
		\begin{array}{cc}
			1 & -\frac{1}{2}\\
			0 & 0\\
		\end{array}%
		\right)$ and in this way we get  $\left(%
		\begin{array}{cc}
			1 & -\frac{1}{2}\\
			0 & 0\\
		\end{array}%
		\right)\left(%
		\begin{array}{c}
			v_{1}\\
			v_{2}\\
		\end{array}%
		\right)=\left(%
		\begin{array}{c}
			0\\
			0\\
		\end{array}%
		\right)$. Thus $v_{2} = 2\:v_{1}$ so that the eigenvector is $\left(%
		\begin{array}{c}
			v_{1}\\
			2v_{1}\\
		\end{array}%
		\right)$. Remember that the set $\Big\lbrace v_{1}\left(%
		\begin{array}{c}
			1\\
			2\\
		\end{array}%
		\right) : v_{1} \in R\Big\rbrace$ is the set of all eigenvectors corresponding to the eigenvalue $\lambda_{1} = 3$. Similarly if we assume that $\left(%
		\begin{array}{c}
			u_{1}\\
			u_{2}\\
		\end{array}%
		\right)$ is the eigenvector corresponding to the eigenvalue $\lambda_{2} = -1$ then eq.(\ref{eigenvector}) gives $\left(%
		\begin{array}{cc}
			2 & 1\\
			4 & 2\\
		\end{array}%
		\right)\left(%
		\begin{array}{c}
			u_{1}\\
			u_{2}\\
		\end{array}%
		\right)=\left(%
		\begin{array}{c}
			0\\
			0\\
		\end{array}%
		\right)$. The \textit{reduced row echelon form} of the coefficient matrix transforms the given eigenvalue-eigenvector equation into the following equivalent form $\left(%
		\begin{array}{cc}
			2 & 1\\
			0 & 0\\
		\end{array}%
		\right)\left(%
		\begin{array}{c}
			u_{1}\\
			u_{2}\\
		\end{array}%
		\right)=\left(%
		\begin{array}{c}
			0\\
			0\\
		\end{array}%
		\right)$. This gives $u_{2} = -2u_{1}$ so that the eigenvector is $\left(%
		\begin{array}{c}
			u_{1}\\
			-2u_{1}\\
		\end{array}%
		\right)$. The set $\Big\lbrace u_{1}\left(%
		\begin{array}{c}
			1\\
			-2\\
		\end{array}%
		\right) : u_{1} \in R\Big\rbrace$ is the set of all eigenvectors corresponding to the eigenvalue $\lambda_{2} = -1$.
		\item To diagonalize the matrix (or a linear operator) is to find a basis of eigenvectors and the corresponding eigenvalues\cite{insel2003linear}. We now consider the set of eigenvectors $\beta = \Big\lbrace \left(%
		\begin{array}{c}
			1\\
			2\\
		\end{array}%
		\right), \left(%
		\begin{array}{c}
			1\\
			-2\\
		\end{array}%
		\right)\Big\rbrace$ and it can easily be verified that this set is linearly independent and spans the Hilbert space $\mathbb{C}^{2}$. The matrix $Q$ of order $2\times 2$ is now constructed whose columns are constituted by the vectors of $\beta$ and so $Q = \left(%
		\begin{array}{cc}
			1 & 1\\
			2 & -2\\
		\end{array}%
		\right)$. 
		\item We now find the inverse of the matrix $Q$ that is $Q^{-1} = \left(%
		\begin{array}{cc}
			\frac{1}{2} & \frac{1}{4}\\\\
			\frac{1}{2} & -\frac{1}{4}\\\\
		\end{array}%
		\right)$. We compute the matrix $D = Q^{-1}\:A\:Q$ which is $D=\left(%
		\begin{array}{cc}
			3 & 0\\
			0 & -1\\
		\end{array}%
		\right)$. This shows that $D$ is a diagonal matrix.
	\end{itemize}
	\textbf{A brief comment:} A matrix is said to be in its \textbf{reduced row echelon} form if the following three conditions are satisfied \cite{insel2003linear}.
\begin{itemize}
\item Any row containing a non-zero entry precedes any row in which all the entries are zero (if any).
\item The first non-zero entry in each row is the only non-zero entry in its column.
\item The first non-zero entry in each row is $1$ and it occurs in a column to the right of the first non-zero entry in the preceding row.
\end{itemize}	
However, it is to be noted that we must be a little bit thoughtful about finding this basis of eigenvectors. The diagonalizability of a linear operator (or of a matrix which corresponds to that linear operator) defined on a vector space is guaranteed by the existence of an ordered basis  of that vector space which will consist of eigenvectors of that linear operator. It is a known fact from linear algebra that if $T$ be a linear operator defined on a vector space $V$ and $\lambda_{1},\lambda_{2},\cdots ,\lambda_{n}$ are eigenvalues of $T$ while $v_{1},v_{2},\cdots ,v_{n}$ are eigenvectors which correspond to the eigenvalues  $\lambda_{1},\lambda_{2},\cdots ,\lambda_{n}$ respectively then the set $\lbrace v_{1},v_{2},\cdots ,v_{n}\rbrace$ of these eigenvectors are linearly independent. Also we know that any linearly independent subset of the vector space $V$ that contains exactly $n$ vectors is a basis for $V$. Combining these two facts one immediately can conclude that $\lbrace v_{1},v_{2},\cdots v_{n}\rbrace$ is a basis of $T$ containing eigenvectors of $T$. We can thus summarize that if $T$ is a linear operator defined on an $n-$ dimensional vector space $V$ and $T$ has $n$ distinct eigenvalues, then $T$ is diagonalizable \cite{insel2003linear}, which is happening in case of the above problem. But the converse statement that if linear operator $T$ is diagonalizable then it has distinct eigenvalues is not true. To justify this fact one can immediately cite the example of Pauli spin matrix $\left(%
	\begin{array}{cc}
		1 & 0\\
		0 & 1\\
	\end{array}%
	\right)$ which is diagonalizable but it has only one eigenvalue $1$, not distinct.\\\\
	\textbf{Problem $1.13$:} Verify whether the matrix $\left(%
	\begin{array}{cc}
		1 & 0\\
		1 & 1\\
	\end{array}%
	\right)$ is diagonalizable!
[Hints: To test the diagonalizability, one can calculate \textit{algebraic} and \textit{geometric} multiplicity of eigenvalues. If $\lambda$ be an eigenvalue of a linear operator or matrix with characteristic polynomial (say) $f(t)$, then the \textbf{algebraic multiplicity} of $\lambda$ is the largest positive integer $k$ for which $(t-\lambda)^{k}$ is a factor of $f(t)$ \cite{insel2003linear}. Show that the algebraic multiplicity of the eigenvalue $1$ in problem $13$ is $2$! On the other hand, the \textbf{geometric multiplicity} of an eigenvalue $\lambda$ of a linear operator is the dimension of the eigenspace corresponding to that eigenvalue. Show that the dimension of the eigenspace $E_{\lambda = 1}$ of the given matrix is $1$! An operator is not diagonalizable if there exists at least one eigenvalue whose algebraic and geometric multiplicity are not the same. Verify!]\\\\
In \textit{problem:} the algebraic and geometric multiplicity of both eigenvalues are the same, which justifies that the matrix is diagonalizable.
	\subsection{Adjoint and Hermitian Operator:}
	Let $A$ be any operator defined on a Hilbert space $V$. Then there always exists a unique linear operator $A^{\dagger}$ on $V$ corresponding to $A$, such that for all vectors $\vert v\rangle,\:\vert w\rangle\:\in V$ we have
	\begin{eqnarray}
		\label{adjoint1}
		\langle \langle v\vert,\: A\vert w\rangle\rangle &=& \langle A^{\dagger}\langle v\vert, \vert w\rangle\rangle\nonumber\\
		\Rightarrow \langle v\vert A\:w\rangle &=& \langle A^{\dagger}\: v \vert w\rangle .
	\end{eqnarray} 
	This linear operator $A^{\dagger}$ is called \textbf{adjoint} or \textbf{Hermitian conjugate} of the operator $A$. Remember that the adjoint defined here is not the classical adjoint; rather, it is best known as \textit{Hermitian adjoint}\footnote{Physicists use the term adjoint instead of Hermitian adjoint often.}.\\\\
	To illustrate this with an example we take the operator $A$ whose matrix representation is, say, $A = \left(%
	\begin{array}{cc}
		a_{11} & a_{12}\\
		a_{21} & a_{22}\\
	\end{array}%
	\right)$. Then $A^{\dagger} = \left(%
	\begin{array}{cc}
		a_{11}^{*} & a_{12}^{*}\\
		a_{21}^{*} & a_{22}^{*}\\
	\end{array}%
	\right)^{T}$ which implies $A^{\dagger} = \left(%
	\begin{array}{cc}
		a_{11}^{*} & a_{21}^{*}\\
		a_{12}^{*} & a_{22}^{*}\\
	\end{array}%
	\right)$. So suppose our matrix is $\left(%
	\begin{array}{cc}
		1+i & 2\\
		4 & 2-i\\
	\end{array}%
	\right)$, then its Hermitian conjugate will be $\left(%
	\begin{array}{cc}
		1-i & 4\\
		2 & 2+i\\
	\end{array}%
	\right)$.\\\\
	To cite the difference between the classical adjoint and Hermitian adjoint, we can consider the following example. Suppose $A = \left(%
	\begin{array}{cc}
		1 & 1\\
		0 & 0\\
	\end{array}%
	\right)$. If we calculate the classical adjoint of the matrix $A$ (which is of course the transpose of the matrix of cofactors of each element of $A$), then this will yield $adj(A)_{classical} = \left(%
	\begin{array}{cc}
		0 & -1\\
		0 & 1\\
	\end{array}%
	\right)$ while the Hermitian adjoint of the matrix $A$ will yield $adj(A)_{Hermitian}=\left(%
	\begin{array}{cc}
		1 & 0\\
		1 & 0\\
	\end{array}%
	\right)$. The following properties need to be remembered:
\begin{itemize}
\item For $\vert v\rangle\:\in V$, we denote the following, $\vert v\rangle^{\dagger}\equiv \langle v\vert$. So if $\vert v\rangle = \left(%
\begin{array}{c}
a_{1}\\
a_{2}\\
\cdots\\
a_{n}\\  
\end{array}%
\right)$ is in $\mathbb{C}^{n}$ then 
$\langle v\vert = \left(%
\begin{array}{cccc}
a_{1}^{*} & a_{2}^{*} & \cdots & a_{n}^{*}\\
\end{array}%
\right)$ 
belongs to  $\mathbb{C^{*}}^{n}$.
\item For two linear operators $A$ and $B$ defined on $V$ we have $(A\:B)^{\dagger} = B^{\dagger}\:A^{\dagger}$.\:(\textbf{Prove!})\\\\
	\textit{Hints:} To prove this result we assume a vector $\vert v\rangle = \left(%
	\begin{array}{c}
		\alpha\\
		\beta\\
	\end{array}%
	\right)$. Also, assume the operators to be of the forms $A = \left(%
	\begin{array}{cc}
		a_{11} & a_{12}\\
		a_{21} & a_{22}\\
	\end{array}%
	\right)$ and $B=\left(%
	\begin{array}{cc}
		b_{11} & b_{12}\\
		b_{21} & b_{22}\\
	\end{array}%
	\right)$. Then show that $(A\:B)^{\dagger}\vert v\rangle = (B^{\dagger}\:A^{\dagger})\vert v\rangle$, $\forall$ $\vert v\rangle \in V$. Generalize the proof by taking $V=\mathbb{C}^{n}$.
\item For a vector $\vert v\rangle$, $(A\:\vert v\rangle)^{\dagger} = \vert v\rangle^{\dagger}\:A^{\dagger} = \langle v\vert\: A^{\dagger}$. 
\end{itemize}
	\textbf{Problem:} If $\vert w\rangle$ and $\vert v\rangle$ are any two vectors, show that $(\vert w\rangle\langle v\vert)^{\dagger} = \vert v\rangle\langle w\vert$.\\\
	\textit{Proof:} Let us consider the two vectors $\vert v\rangle = \left(%
\begin{array}{c}
a_{1}\\
a_{2}\\
\cdots\\
a_{n}\\  
\end{array}%
\right),\: \vert v\rangle = \left(%
\begin{array}{c}
b_{1}\\
b_{2}\\
\cdots\\
b_{n}\\  
\end{array}%
\right) \in \mathbb{C}^{n}(\mathbb{C})$. Then $\vert w\rangle\langle v\vert = \left(%
\begin{array}{cccc}
a_{1}b_{1} & a_{1}b_{2} & \cdots & a_{1}b_{n}\\
a_{2}b_{1} & a_{2}b_{2} & \cdots & a_{2}b_{n}\\
\cdots\\
a_{n}b_{1} & a_{n}b_{2} & \cdots & a_{n}b_{n}\\  
\end{array}%
\right)$.\\\\ Now $(\vert w\rangle\langle v\vert)^{\dagger} = \left(%
\begin{array}{cccc}
b_{1}^{*}a_{1}^{*} & b_{1}^{*}a_{2}^{*} & \cdots & b_{1}^{*}a_{n}^{*}\\
b_{2}^{*}a_{1}^{*} & b_{2}^{*}a_{2}^{*} & \cdots & b_{2}^{*}a_{n}^{*}\\
\cdots\\
b_{n}^{*}a_{1}^{*} & b_{n}^{*}b_{2}^{*} & \cdots & b_{n}^{*}a_{n}^{*}\\  
\end{array}%
\right)$. This implies that $(\vert w\rangle\langle v\vert)^{\dagger} = \left(%
\begin{array}{c}
b_{1}^{*}\\
b_{2}^{*}\\
\cdots\\
b_{n}^{*}\\  
\end{array}%
\right)\left(%
\begin{array}{cccc}
a_{1}^{*} & a_{2}^{*} & \cdots & a_{n}^{*}\\
\end{array}%
\right)=\vert v\rangle\langle w\vert$. ~~~~~~\textbf{Q.E.D}\\\\ 
	\textbf{Problem:} \textbf{Anti-linearity of adjoint:} Show that the adjoint operation is anti-linear i.e.
	\begin{eqnarray}
		\label{adjoint2}
		(\sum_{i}a_{i}\:A_{i})^{\dagger} = \sum_{i}a_{i}^{*}\:A^{\dagger}_{i}.
	\end{eqnarray} \\\\
	\textbf{Problem:} Show that $(A^{\dagger})^{\dagger} = A$.\\\\
	\textit{Hints:} Let us take a vector $\vert v\rangle\: \in V$. Then we get \\
    $(A^{\dagger})^{\dagger}\vert v\rangle = \left(%
	\begin{array}{cc}
		a_{11}^{*} & a_{21}^{*}\\
		a_{12}^{*} & a_{22}^{*}\\
	\end{array}%
	\right)^{\dagger}\:\left(%
	\begin{array}{c}
		\alpha\\
		\beta\\
	\end{array}%
	\right) = \left(%
	\begin{array}{cc}
		a_{11} & a_{12}\\
		a_{21} & a_{22}\\
	\end{array}%
	\right)\:\left(%
	\begin{array}{c}
		\alpha\\
		\beta\\
	\end{array}%
	\right)=A\:\vert v\rangle, \: \forall\: v\:\\ \Rightarrow (A^{\dagger})^{\dagger} = A$.\\\\
	\textbf{Problem:} Find $\left(%
	\begin{array}{cc}
		1+3\:i & 2\:i\\
		1+\:i & 1-4\:i\\
	\end{array}%
	\right)^{\dagger}$.\\\\
It is to be noted that $\sigma_{x}^{\dagger}=X^{\dagger}=X$, $\sigma_{y}^{\dagger}=Y^{\dagger}=Y$ and $\sigma_{z}^{\dagger}=Z^{\dagger}=Z$.\: (\textbf{Verify!})\\\\
See that $(i\:\sigma_{y})=i\:\left(%
	\begin{array}{cc}
		0 & -i\\
		i & 0\\
	\end{array}%
	\right)=\left(%
	\begin{array}{cc}
		0 & 1\\
		-1 & 0\\
	\end{array}%
	\right)\Rightarrow (i\:\sigma_{y})^{\dagger}=\left(%
	\begin{array}{cc}
		0 & -1\\
		1 & 0\\
	\end{array}%
	\right)\neq (i\:\sigma_{y})$. Therefore $iY$ is not hermitian.\\\\
	An operator $A$ whose adjoint is $A$ (i.e. $A^{\dagger}=A$) is known as a \textbf{Hermitian} or \textbf{Self adjoint operator}. It is easily seen from above discussions that $(i\:\sigma_{y})$ is not Hermitian while $X,\:Y\:,Z$ are Hermitian.\\\\
	\textbf{Problem:} Is Hadamard operator Hermitian? Verify!
	\subsection{Normal and Unitary Operator:}
An operator $A$ is said to \textbf{Normal} if $A^{\dagger}\:A = A\:A^{\dagger}$. It is easy to verify the fact that if an operator is Hermitian then it is always normal. \\\\
\textit{Proof:}	To show this we consider an operator $A$ which is Hermitian. Then $A^{\dagger} = A$. It is easy to see now that $A^{\dagger}A=AA=A^{2}=AA^{\dagger}$. This proves that Hermitian operator $A$ is necessarily normal. ~~~~~\textbf{Q.E.D}\\\\
	\textbf{Theorem:} A normal matrix (or operator) is Hermitian if and only if it has real eigenvalues.\\\\
	\textit{Proof:} In this proof we need to show the necessary and sufficient condition, the necessary condition to prove is to take a normal operator $A$ and to show if $A$ is Hermitian then it has real eigenvalues whereas for sufficiency we need to show that if the normal operator $A$ has real eigenvalues then it is Hermitian. (The theorem is very important from a Physicist's point of view). Given $A$ to be a normal operator for which we have $A^{\dagger}\:A = A\:A^{\dagger}$. \\\\
	\textit{Necessity:} Suppose that $A$ is Hermitian. Then $A^{\dagger}=A$. Let $\lambda$ be the eigenvalue of the operator $A$ and $\vert v\rangle$ be the corresponding eigenvector such that we have $A\:\vert v\rangle = \lambda\vert v\rangle$. Now 
\begin{eqnarray}
\label{proof1}
(A\:\vert v\rangle)^{\dagger} &=&(\lambda\:\vert v\rangle)^{\dagger}\nonumber\\
\Rightarrow \vert v\rangle^{\dagger}\:A^{\dagger}&=&\vert v\rangle^{\dagger}\:\lambda^{\dagger}\nonumber\\ \Rightarrow \langle v\vert\: A^{\dagger} &=& \langle v\vert\: \lambda^{*}\nonumber\\ \Rightarrow \langle v\vert\: A = \langle v\vert \:\lambda^{*}.
\end{eqnarray} 
Then from eq.(\ref{proof1}) we get
\begin{eqnarray}
\label{proof2}
\langle v\vert \:A\:\vert v\rangle &=& \langle v\vert \:\lambda^{*}\:\vert v\rangle \nonumber\\ 
&=& \lambda^{*}\langle v\vert v\rangle =\lambda^{*}\nonumber\\
 \Rightarrow \langle v\vert \:\lambda\:\vert v\rangle &=& \lambda^{*}\nonumber\\
 \Rightarrow \lambda\:\langle v\vert v\rangle &=& \lambda^{*}\nonumber\\
  \Rightarrow \lambda &=& \lambda^{*}. 
\end{eqnarray}
This proves that the eigenvalue $\lambda$ is real. \\\\
	\textit{Sufficiency:} Conversely let the operator $A$ have real eigenvalues. Let $\lambda$ be one such eigenvalue which is real and corresponding to this the eigenvector be $\vert v\rangle$. Then from eigenvalue-eigenvector equation we get $A\:\vert v\rangle = \lambda\vert v\rangle$. Now as $A$ is normal operator then if $\vert v\rangle$ is the eigenvector of $A$ corresponding to the eigenvalue $\lambda$, it implies that $\vert v\rangle$ is also the eigenvector of $A^{\dagger}$ corresponding to the eigenvalue $\lambda^{*}$\cite{insel2003linear}. Therefore we have $A^{\dagger}\:\vert v\rangle = \lambda^{*}\vert v\rangle$.  Also since $\lambda$ is real, $\lambda^{*}=\lambda$. Now 
\begin{eqnarray}
\label{proof3a}
A\:\vert v\rangle &=& \lambda\vert v\rangle\nonumber\\
\Rightarrow A^{\dagger}A\: \vert v\rangle  &=& A^{\dagger}\:\lambda\vert v\rangle =  \lambda\:A^{\dagger}\vert v\rangle.
\end{eqnarray}
Similarly
\begin{eqnarray}
\label{proof3b}
A^{\dagger}\:\vert v\rangle &=& \lambda^{*}\vert v\rangle\nonumber\\
\Rightarrow AA^{\dagger}\: \vert v\rangle  &=& A\:\lambda^{*}\vert v\rangle =  \lambda^{*}\:A\vert v\rangle.
\end{eqnarray}
Since $A$ is normal operator then $AA^{\dagger} = A^{\dagger}A$ ~~i.e.~~$AA^{\dagger}\vert v\rangle = A^{\dagger}A\vert v\rangle$ for all $\vert v\rangle$. Hence from eqs. (\ref{proof3a}) and (\ref{proof3b}) we get 
\begin{eqnarray}
\label{proof3c}
\lambda\:A^{\dagger}\vert v\rangle = \lambda^{*}\:A\vert v\rangle. 
\end{eqnarray}
Eq.(\ref{proof3c}) is true for all $\vert v\rangle$ and as $\lambda^{*}=\lambda$, consequently we have $A^{\dagger}\vert v\rangle = A\vert v\rangle$ for all $\vert v\rangle$. This shows that $A = A^{\dagger}$. Hence $A$ is Hermitian.  ~~~~~\textbf{Q.E.D}
\\\\
A very important fact to be noted is, any operator having real eigenvalues does not imply that the operator is Hermitian. Suppose again we consider the matrix $A = \left(%
	\begin{array}{cc}
		1 & 1\\
		0 & 0\\
	\end{array}%
	\right)$ whose eigenvalues are $0$ and $1$ (which are real). However $A^{\dagger} \ne A$ implying that the matrix $A$ is not Hermitian. So we have to work with normal operators for the above theorem to hold.\\\\
	A matrix (or operator) $U$ is \textbf{Unitary} if  $U^{\dagger}\:U = U\:U^{\dagger} = I$ where $I$ is the identity operator. (By definition $U$ is also a kind of normal operator). \\\\
	If we consider the operator $X$, we can show that $X^{\dagger}\:X =\left(%
	\begin{array}{cc}
		0 & 1\\
		1 & 0\\
	\end{array}%
	\right)\:\left(%
	\begin{array}{cc}
		0 & 1\\
		1 & 0\\
	\end{array}%
	\right)=X^{2}=\left(%
	\begin{array}{cc}
		1 & 0\\
		0 & 1\\
	\end{array}%
	\right)=X\:X^{\dagger}$. Hence $X$ is a unitary operator. Similarly we can show that the other two Pauli operators $Y$ and $Z$ are also unitary. This is also to be noted that Pauli operators $X$, $Y$ and $Z$ have real eigenvalues.\\\\
Not all unitary operators are hermitian. (Verify with the operator $i\:Y$!)\\\\
	\textbf{Problem $1.19$:} Prove that Hadamard operator is also a unitary operator.\\\\
	An important class of Hermitian operators is the \textbf{Projectors}. Suppose $W$ is a $k-$ dimensional vector subspace of the $d-$ dimensional vector space $V$. Using Gram-Schmidt process we can construct an orthonormal basis $\lbrace \vert 1\rangle,\cdots, \vert d\rangle\rbrace$ for $V$ such that $\lbrace \vert 1\rangle,\cdots, \vert k\rangle\rbrace$ becomes an orthonormal basis for $W$. By definition
	\begin{eqnarray}
		\label{projector}
		P = \sum_{i\:= 1}^{k}\vert i\rangle\langle i\vert 
	\end{eqnarray}
	is the projector onto the subspace of $W$. The \textbf{orthogonal complement} of $P$ is again an operator which is defined as $Q=\:I-\:P$. $Q$ is a projector on to the vector space spanned by $\lbrace \vert k+1\rangle,\cdots, \vert d\rangle\rbrace$. \\\\
	\textbf{Problem:} Show that any projector $P$ satisfies the equation $P^{2} = P$.\\\\
	\textit{Proof:} Let $P$ be a projector having the form as defined in eq. (\ref{projector}). Now $P^{2} = PP=(\sum_{i}^{k}\vert i\rangle\langle i\vert)(\sum_{i}^{k}\vert i\rangle\langle i\vert)\\=\sum_{i}^{k}\vert i\rangle\langle i\vert i\rangle \langle i\vert = \sum_{i}^{k}\vert i\rangle \langle i \vert =P$. This is because $\langle i\vert i\rangle = 1$ as $\lbrace \vert i\rangle\rbrace$ is considered to be an orthonormal basis. ~~~~~\textbf{Q.E.D}\\\\
	\textbf{Problem:}  Prove that two eigenvectors of a Hermitian operator with different eigenvalues are necessarily orthogonal.\\\\
	\textit{Proof:} Let $A$ be a Hermitian operator such that $A^{\dagger}=A$. Then if $\lambda_{1}$ and $\lambda_{2}$ be two distinct eigenvalues and the corresponding eigenvectors are respectively $\vert v_{1}\rangle$ and $\vert v_{2}\rangle$. Then
\begin{eqnarray}
\label{proof4}
A\vert v_{1}\rangle &=& \lambda_{1}\vert v_{1}\rangle\nonumber\\
A\vert v_{2}\rangle &=& \lambda_{2}\vert v_{2}\rangle
\end{eqnarray}
Thus
\begin{eqnarray}
\label{proof5}
\langle v_{2}\vert A\vert v_{1}\rangle = \langle v_{2}\vert \lambda_{1}\vert v_{1}\rangle = \lambda_{1}\langle v_{2}\vert v_{1}\rangle.
\end{eqnarray}
From eq.(\ref{proof4}) (and noting the fact that $A$ is Hermitian with real eigenvalues) we have
\begin{eqnarray}
\label{proof6}
\langle v_{2}\vert A^{\dagger} &=& \langle v_{2}\vert \lambda_{2}^{*}\nonumber\\
\Rightarrow \langle v_{2}\vert A &=& \langle v_{2}\vert \lambda_{2}
\end{eqnarray}
From eq.(\ref{proof6}) 
\begin{eqnarray}
\label{proof7}
\langle v_{2}\vert A\vert v_{1}\rangle = \langle v_{2}\vert \lambda_{2}\vert v_{1}\rangle = \lambda_{2}\langle v_{2}\vert v_{1}\rangle.
\end{eqnarray}
From eqs.(\ref{proof5}) and (\ref{proof7}) we then obtain
\begin{eqnarray}
\label{proof8}
(\lambda_{1} - \lambda_{2})\langle v_{2}\vert v_{1}\rangle = 0.
\end{eqnarray}
Since $\lambda_{1} \ne \lambda_{2}$, therefore $\langle v_{2}\vert v_{1}\rangle = 0$ implying that the distinct eigenvectors $\vert v_{1}\rangle$ and $\vert v_{2}\rangle$ are orthogonal. ~~~~~\textbf{Q.E.D}\\\\
\textbf{Geometrical interpretation} of the above result can be put in the following way. If we have two vectors $\vert v_{1}\rangle$ and $\vert v_{2}\rangle$, then the span of these two vectors can be another vector $\vert v_{3}\rangle$, different from both  $\vert v_{1}\rangle$ and $\vert v_{2}\rangle$ and falling on a different straight line. But if $\vert v_{1}\rangle$ and $\vert v_{2}\rangle$ are eigenvectors then these vectors will coincide with some spans of these vectors (for suitable choice of the scalars). Now by orthogonality of the two vectors $\vert v_{1}\rangle$ and $\vert v_{2}\rangle$  we understand that the vectors are coinciding with two different straight lines which are orthogonal.\\\\
	There is a remarkable representation theorem for normal operators known as \textbf{spectral decomposition}, which states that an operator is a normal operator if and only if it is diagonalizable\cite{nielsen2010quantum,insel2003linear}. Formally we can place this in the form of a theorem as follows\\\\
\textbf{Theorem:} Any normal operator $M$ (defined on a vector space $V$) is diagonal with respect to some orthonormal basis for $V$. Conversely, any diagonalizable operator is normal.\footnote{We are skipping the proof of this result. Any interested reader can refer \cite{nielsen2010quantum} for the detailed proof of the theorem.}\\\\
If $A$ be a linear operator on a finite dimensional complex inner product space $V$, then $A$ is normal if and only if there exists an orthonormal basis for $V$ consisting of eigenvectors of $A$\cite{insel2003linear}. We know that the existence of an ordered basis of a linear operator containing the eigenvectors authenticates the diagonalizability of that linear operator. So in other words $A$ being the normal operator is diagonalizable with respect to some basis with eigenvectors as elements. Moreover that basis is also an orthonormal basis. The Pauli spin operators, for example, are all diagonalizable as these are normal operators.\\\\  
\textbf{Problem:} Prove that eigenvalues of a projector $P$ are all either zero or $1$.\\\\
\textit{Proof:} Let $P$ be the projector and remember any projector satisfies the equation $P^{2}=P$. Let $\lambda$ be the eigenvalue of the projector $P$ and its corresponding non-zero eigenvector is $\vert v\rangle$. Then we have $P\:\vert v\rangle = \lambda \vert v\rangle$ and  $P^{2}\:\vert v\rangle = \lambda^{2} \vert v\rangle$.\footnote{Let $T$ be a linear operator on a vector space $V$ and let $x$ be an eigenvector of $T$ corresponding to the eigenvalue $\lambda$. For any positive integer $m$ it can be shown that $x$ is an eigenvector of $T^{m}$ corresponding to the eigenvalue $\lambda^{m}$.\cite{insel2003linear}} Now as $P^{2} =P$  then $\lambda \vert v\rangle = \lambda^{2}\vert v\rangle$ which implies $(\lambda - \lambda^{2})\vert v\rangle = 0$ i.e. $\lambda = 0,\: 1$.	
\subsection{Trace of an operator:}
Let $A$ be a square matrix of order $n$. The trace of $A$ (which we shall denote by $Tr(A)$) is defined to be the sum of its diagonal elements.
\begin{eqnarray}
     \label{t1}
             Tr(A) = \sum_{i = 1}^{n} A_{ii}.
\end{eqnarray}
The trace of Pauli operators $X,\:Y\:,Z$ are all zero. The trace of Hadamard operator is also zero while the trace of identity operator is $2$.\\\\
\textbf{Properties of Trace:}
\begin{itemize}
\item If $A$ be any $m \times n$ matrix and $B$ is any $n \times m$ matrix then $Tr(A\:B)=Tr(B\:A)$. This property is called the \textit{cyclic property} of trace.
\item The trace is a linear map i.e. $Tr(A+B) = Tr(A) + Tr(B)$ and $Tr(k\:A)=k\:Tr(A)$. This is \textit{linear property} of trace.
\item Suppose we give a unitary similarity transformation on $A$ which is given as $A \rightarrow U\:A\:U^{\dagger}$, it follows that trace of matrix (or of operator) is invariant under such transformation which means mathematically that $Tr(U\:A\:U^{\dagger})=Tr(U^{\dagger}\:U\:A)=Tr(A)$. \\
\end{itemize}
\textbf{Problem $1.23$:} Suppose a $2\times 2$ matrix $M$ (not necessarily Hermitian, nor unitary) is written as $X=a_{0} + \bar{\sigma}\bullet \bar{a}$, where $a_{0}$, $a_{1}$, $a_{2}$ and $a_{3}$ are numbers. How are $a_{0}$ and $a_{k}$, $k = 1\:,2\:,3$ related to $Tr(X)$ and $Tr(\sigma_{k}X)$?( $\sigma_{i}$' are Pauli spin matrices, $\bar{\sigma}$ and $\bar{a}$ are the vectors with three components each.)\cite{sakurai1967advanced}

	\subsection{Expectation and variance of an operator:}
Suppose $\vert \psi\rangle$ is a unit vector and $A$ is an arbitrary operator. To evaluate $Tr(A\:\vert \psi\rangle\langle \psi\vert)$ we use Gram-Schmidt procedure to extend $\vert \psi\rangle$ to an orthonormal basis $\vert i\rangle$ which includes $\vert \psi\rangle$ as the first element. Then we have 
\begin{eqnarray}
\label{exp}
Tr(A\:\vert \psi\rangle\langle \psi\vert)=\sum_{i} \langle i\vert A\vert \psi\rangle\langle \psi\vert i\rangle = \langle \psi\vert A\vert \psi\rangle.
\end{eqnarray}\\	
The quantity $\langle \psi\vert A\vert \psi\rangle$ of equation (\ref{exp}) is called \textbf{expectation} of the operator $A$ and is denoted by $\langle A\rangle$. Likewise the \textbf{variance} (which is a measure of deviation) is given by the formula:
\begin{eqnarray}
\label{variance}
\langle (\Delta A)^{2}\rangle = \langle A^{2}\rangle - \langle A\rangle^{2}
\end{eqnarray}
\textbf{Problem:} If the state $\vert \psi\rangle = \vert S_{x};+\rangle = \frac{1}{\sqrt{2}}\vert +\rangle + \frac{1}{\sqrt{2}}e^{i\:\delta_{1}}\vert -\rangle$ and the corresponding operator is $\hat{S_{x}}=\frac{\hbar}{2}\lbrace e^{-i\:\delta_{1}}(\vert +\rangle\langle -\vert) + e^{i\:\delta_{1}}(\vert -\rangle\langle +\vert)\rbrace$ then show that $\langle \hat{S_{x}}\rangle$ = $\frac{\hbar}{2}$ while $\langle (\Delta \hat{S_{x}})^{2}\rangle = 0$\footnote{Here $\hbar$ is the Planck's constant and $\vert +\rangle \equiv \vert 0\rangle$ and $\vert -\rangle \equiv \vert 1\rangle$.} (where $e^{\pm i \delta_{1}}$ is the phase factor).\\\\
\textbf{Problem $1.25$:} If the state $\vert S_{y}; +\rangle = \frac{1}{\sqrt{2}}\vert +\rangle + \frac{1}{\sqrt{2}}e^{i\:\delta_{2}}\vert -\rangle$ and the chosen operator is $\hat{S_{x}}=\frac{\hbar}{2}\lbrace e^{-i\:\delta_{1}}(\vert +\rangle\langle -\vert) + e^{i\:\delta_{1}}(\vert -\rangle\langle +\vert)\rbrace$, then show that $\langle \hat{S_{x}}\rangle$ = $0$ where $\hbar$ is the Planck's constant. Also $e^{\pm i \delta}$ is the phase factor and the phase difference $\delta_{1}+\delta_{2}=\frac{\pi}{2}$.\\\\
\textit{Solution:} Given $\vert S_{x};+\rangle = \frac{1}{\sqrt{2}}(\vert +\rangle + e^{i\:\delta_{1}}\vert -\rangle)$ and the corresponding operator $\hat{S_{x}}=\frac{\hbar}{2}\lbrace e^{-i\:\delta_{1}}(\vert +\rangle\langle -\vert)+e^{i\:\delta_{1}}(\vert -\rangle\langle +\vert)\rbrace$. Now $\hat{S_{x}^{2}}=\frac{\hbar^{2}}{4}\lbrace \vert +\rangle\langle +\vert +\vert -\rangle\langle -\vert\rbrace$.We know that for an operator $A$ and with respect to the state $\vert \psi\rangle$, the expectation of the operator $A$ is given by $\langle \psi \vert A\vert \psi\rangle$. Using this, the expectation of the operator $\hat{S_{x}^{2}}$ is given as\\
\begin{eqnarray}
	\label{Sx}
	\langle \hat{S_{x}^{2}}\rangle_{\vert S_{x};+\rangle} &=& \lbrace \frac{1}{\sqrt{2}}\langle +\vert +\frac{1}{\sqrt{2}}e^{-i\:\delta_{1}}\langle -\vert\rbrace \frac{\hbar^{2}}{4}(\vert +\rangle\langle +\vert +\vert -\rangle\langle -\vert)\lbrace\frac{1}{\sqrt{2}}\vert +\rangle +\frac{1}{\sqrt{2}}e^{-i\:\delta_{1}}\vert -\rangle\rbrace \nonumber\\
	&=& \frac{\hbar^{2}}{4}
\end{eqnarray}
Now we know that the expectation value of the operator $S_{x}$ is denoted by $\langle S_{x}\rangle$ and is calculated as
\begin{eqnarray}
	\label{Sx1}
	\langle S_{x}\rangle_{\vert S_{x};+\rangle} &=&  \lbrace \frac{1}{\sqrt{2}}\langle +\vert +\frac{1}{\sqrt{2}}e^{-i\:\delta_{1}}\langle -\vert\rbrace \frac{\hbar}{2}(e^{-i\:\delta_{1}}\vert +\rangle\langle -\vert +e^{i\:\delta_{1}}\vert -\rangle\langle +\vert)\lbrace\frac{1}{\sqrt{2}}\vert +\rangle +\frac{1}{\sqrt{2}}e^{-i\:\delta_{1}}\vert -\rangle\rbrace \nonumber\\
	&=& \frac{\hbar}{2}
\end{eqnarray}
Equation (\ref{Sx1}) gives the expectation value of the operator and to calculate the variance of the operator we use the formula given in eq.(\ref{variance}) which is defined for any operator $A$ and thus we have
\begin{eqnarray}
	\label{Sxvariance}
	\langle (\Delta \hat{S_{x}})^{2}\rangle = \langle S_{x}^{2}\rangle -\langle S_{x}\rangle^{2} = 0.
\end{eqnarray}
An alternative approach to prove the above result is to take the outer product of the state $\vert S_{x};+\rangle$ and to calculate $Tr(\hat{S_{x}}\vert S_{x};+\rangle\langle S_{x};+\vert)$ that will give us the expectation value and accordingly we can proceed to obtain subsequent results of the above problem.

\subsection{Positive Operator and its Hermiticity:}
A special sub-class of Hermitian operator is extremely important. These are \textit{Positive operators}. A \textit{positive operator} $A$ is defined to be an operator such that for any vector $\vert v\rangle$, $\langle v\vert A\: v\rangle$ is real non-negative number. However if for all $\vert v\rangle \ne 0\:\:,$  we have $\langle v\vert A\: v\rangle > 0$ then we say that $A$ is \textit{Positive definite}.\\\\
By \textbf{Hermiticity} of Positive operators we mean that \textit{Positive operator is necessarily Hermitian}. To prove it we proceed as follows. We will show that any operator $A$ can be expressed as $A = B + i\: C$ where $B$ and $C$ are Hermitian. Choose $B = \frac{A + A^{\dagger}}{2}$ and $C = \frac{A - A^{\dagger}}{2\:i}$. Then it is obvious that $A = B + i\: C$. Again it is easy to show that $B^{\dagger} = B$ as well as $C^{\dagger} = C$ implying that both $B$ and $C$ are Hermitian.\\\\
Now assume that $A$ is positive and $B$ and $C$ are as defined above. For a positive operator $A$ we know that $\langle v\vert A\:v\rangle \ge 0$, $\forall \vert v\rangle$. Since $C$ is Hermitian, it is normal and hence is diagonalizable (by Spectral decomposition theorem). Then $C$ has the following representation such that for an orthonormal basis $\lbrace \vert i\rangle \rbrace$ we have\\
\begin{eqnarray}
	\label{hermiticity1}
	C = \sum_{i}\:\lambda_{i}\vert i\rangle \langle i\vert .
\end{eqnarray}
Since $\lbrace \vert i\rangle \rbrace$ is orthonormal basis for the vector space under consideration, so any vector $\vert v\rangle$ can be written as
\begin{eqnarray}
	\label{hermiticity2}
	\vert v\rangle = \sum_{j}\:\alpha_{j}\vert j\rangle.
\end{eqnarray}
Then we get
	\begin{eqnarray}
		\label{hermiticity3}
		C\:\vert v\rangle &=& \sum_{i}\:\lambda_{i}\vert i\rangle \langle i\vert\:\sum_{j}\alpha_{j}\vert j\rangle\nonumber\\
		&=& \sum_{i,\:j} \lambda_{i}\alpha_{j} \vert i\rangle \langle i\vert j\rangle\nonumber\\
		&=& \sum_{i}\lambda_{i}\:\alpha_{i} \vert i\rangle.
	\end{eqnarray}
Also the dual of $\vert v\rangle$ is denoted by $\langle v\vert$ and is given by the expression $\sum_{j}\:\alpha_{j}^{*}\langle j\vert$. Thus we get,
\begin{eqnarray}
	\label{hermiticity4}
	\langle v\vert C v\rangle &=& \sum_{j}\:\alpha_{j}^{*}\langle j\vert \sum_{i}\lambda_{i}\alpha_{i}\vert i\rangle \nonumber\\ 
	&=& \sum_{j,\:i}\lambda_{i}\alpha_{j}^{*}\alpha_{i}\langle j\vert i\rangle \nonumber\\
	&=& \sum_{i}\lambda_{i}\vert \alpha_{i}\vert^{2}.
\end{eqnarray}
Since $C$ is Hermitian, its eigenvalues are real and hence $\langle v\vert C v\rangle$ is real as well as it is non-negative from equation (\ref{hermiticity4}) while by the same argument $\langle v\vert B v\rangle$ is also real and can be shown to be non-negative. Thus, $\langle v\vert A\:v\rangle = \langle v\vert (B+i\:C)\:v\rangle = \langle v\vert B\:v\rangle + i\: \langle v\vert C\:v\rangle \ge 0$. ~~~~\textbf{Q.E.D}
\\\\
\textbf{Problem:} Show that for any operator $A$, $A^{\dagger}\:A$ is positive.\\\\
\textit{Proof:} From spectral decomposition theorem, we know that any normal operator is diagonalizable and vice-versa. To put it in other words, we can say that any operator $B$ can be written as $B = U\: D\: U^{\dagger}$, where $D$ is diagonal and $U$ is unitary if and only if $B\:B^{\dagger} = B^{\dagger}\:B$. Let us now consider $B=A^{\dagger}A$. It is observed that $(A^{\dagger}A)^{\dagger}(A^{\dagger}A) = (A^{\dagger}(A^{\dagger})^{\dagger})(A^{\dagger}A)=(A^{\dagger}A)(A^{\dagger}A)$. Again $(A^{\dagger}A)(A^{\dagger}A)^{\dagger} = (A^{\dagger}A)(A^{\dagger}(A^{\dagger})^{\dagger})=(A^{\dagger}A)(A^{\dagger}A)$. This shows that $A^{\dagger}A$ is a normal operator. Then it has spectral decomposition. Hence there exists an orthonormal basis $\lbrace \vert i\rangle\rbrace$ such that
\\
\begin{eqnarray}
	\label{AdaggerA}
	A^{\dagger}A = \sum_{i}\lambda_{i}\vert i\rangle \langle i\vert.  
\end{eqnarray}
Now for any arbitrary vector $\vert j\rangle$, we get $(A^{\dagger}A)\vert j\rangle = \sum_{i}\lambda_{i}\vert i\rangle \langle i\vert j\rangle = \lambda_{j}\vert j\rangle$. This in turn shows that $\langle j\vert A^{\dagger}A\vert j\rangle = \lambda_{j} \Rightarrow \|\vert A\vert j\rangle\rangle\|^{2} = \lambda_{j} \ge 0$. This shows that $A^{\dagger}A$ is positive. \textbf{Q.E.D}\\
\section{Unit - III}\label{sec:unit3}
\subsection{Commutator and Anti-commutator:}
Let $A$ and $B$ be two operators. By \textit{commutator} of $A$ and $B$ we mean
\begin{eqnarray}
	\label{commutator1}
	[A,\: B] = AB - BA.
\end{eqnarray}
Thus operators $A$ and $B$ commute with one another if $[A,\: B] = 0$ i.e.if $AB=BA$. In this case we can also say that the observables measured by $A$ and $B$ are \textit{compatible}. The operator $[\bullet,\bullet]$ is known as \textit{commutator}. Likewise we can also define anti-commutator between two operators $A$ and $B$, denoted by $\lbrace A, \: B\rbrace$ as follows,
\begin{eqnarray}
	\label{anticommutator1}
	\lbrace A,\: B\rbrace = AB + BA.
\end{eqnarray}
The operator $\lbrace \bullet, \bullet \rbrace$ is the anti-commutator.
Thus $A$ and $B$ anti-commutes with one another if $\lbrace A,\: B\rbrace = 0$.\\\\
\textbf{Problem:} Prove that $[AB,\:CD] = -AC\lbrace D,B\rbrace\:+\:A\lbrace C,B\rbrace D\:-\:C \lbrace D,A \rbrace B\:+\: \lbrace C,A \rbrace DB$\cite{sakurai1967advanced}.\\\\
\textbf{Problem:} Prove that $[A,\: B] = -[B,\:A]$.\\\\
\textbf{Problem:} Verify that for any two operators $A$ and $B$, $AB = \frac{[A,\: B] + \lbrace A, \: B\rbrace}{2}$.\\\\
\textbf{Problem:} Show that $[A,\:B]^{\dagger} = [B^{\dagger},\:A^{\dagger}]$.\\\\
\textbf{Problem:} Show that if the operators $A$ and $B$ are Hermitian then $i\:[A,B]$ is also Hermitian.\\\\
\textit{Proof:} Given $A$ and $B$ are Hermitian we have $A^{\dagger} = A$ and $B^{\dagger}=B$. To show that $i\:[A,B]$ is Hermitian we need to prove $(i\:[A,B])^{\dagger} = i\:[A,B]$. Now $(i\:[A,B])^{\dagger} =(i(AB-BA))^{\dagger}=(iAB)^{\dagger}-(iBA)^{\dagger} = -i(B^{\dagger}A^{\dagger}-A^{\dagger}B^{\dagger}) = -i\:[B^{\dagger}, A^{\dagger}]$. This implies that $(i\:[A,B])^{\dagger} = -i\:[B,\:A]$ as $A$ and $B$ are Hermitian. This in turn gives us that $(i\:[A,B])^{\dagger} = i\:[A, B]$, thus proving the fact that $(i\:[A,B])$ is Hermitian. ~~~~~~~~~~~~~~~~~~~\textbf{Q.E.D}\\\\
Some other properties that one can summarize are as follows:
\begin{itemize}
\item $[A, BC] = [A, B]C + B[A, C]$.\\\\
\textit{Proof:} 
\begin{eqnarray}
\label{poissonbracket1}
[A, BC] = A(BC)-(BC)A &=& (AB)C-(BA)C + (BA)C - (BC)A \nonumber\\
&=& (AB-BA)C + B(AC) - B(CA)\nonumber\\
&=& (AB-BA)C + B(AC-CA) \nonumber\\
&=& [A,B]C + B[A,C].
\end{eqnarray}
~~~~~~~~~~~~~~~~~~~~~~~~~~~~~~~~~~\textbf{Q.E.D}
\item $[A+B, C] = [A, C] + [B, C]$\\\\
\textit{Proof:}
\begin{eqnarray}
\label{poissonbracket1a}
[A+B, C] =(A+B)C - C(A+B) &=& AC+BC-CA-CB\nonumber\\
&=& AC-CA+BC-CB\nonumber\\
&=& [A,C] + [B,C].
\end{eqnarray}
~~~~~~~~~~~~~~~~~~~~~~~~~~~~~~~~~~~~~~\textbf{Q.E.D}
\item $[A, c]=0$ where $c$ is a constant.
\item $[A, [B,C]] + [B, [C,A]] + [C, [A,B]] = 0$. This identity is called \textbf{Jacobi's identity}.
\item Let $X$ and $P$ be the position and momentum operators and $X_{i},\:X_{j}$ are $i^{th}$ and $j^{th}$ component of the position operator while $P_{i},\:P_{j}$ are $i^{th}$ and $j^{th}$ component of the momentum operator. Then we have $[X_{i},X_{j}] = 0$, $[P_{i}, P_{j}]=0$ and $[X_{i},P_{j}]=i\hbar\delta_{ij}$. These relations are called \textbf{canonical commutation relations}\cite{sakurai1967advanced}.
\end{itemize}
\subsection{Commutation and Anti-commutation relation of Pauli Operators:}
The Pauli operators $I,\: X,\: Y,\: Z$ have already been defined in equations (\ref{P1}) - (\ref{P4}). Authors sometimes use different notations to denote Pauli operators such as $X = \sigma_{x} = \sigma_{1},\:Y = \sigma_{y} = \sigma_{2},\: Z = \sigma_{z} = \sigma_{3}$ while $I=\sigma_{0}$. Let us now calculate $[\sigma_{1},\sigma_{2}]$, $[\sigma_{2},\sigma_{3}]$ and $[\sigma_{3},\sigma_{1}]$ by using the definition (\ref{commutator1}). By putting the matrix forms of these Pauli operators, it is easy to see that 
\begin{eqnarray}
	\label{paulicommu1}
[\sigma_{1},\sigma_{2}] = 2\:i\:\sigma_{3},\:\:\:
[\sigma_{2},\sigma_{3}] = 2\:i\:\sigma_{1},\:\:\:
[\sigma_{3},\sigma_{1}] = 2\:i\:\sigma_{2}	.
\end{eqnarray}
The relations defined above in equation (\ref{paulicommu1}) can be elegantly expressed in a unique form using an anti-symmetric tensor. The anti-symmetric tensor on three indices is denoted by $\epsilon_{j\: k\: l}$. The property of $\epsilon_{j\: k\: l}$ is as follows
\begin{eqnarray}
	\label{epsilon1}
	 \epsilon_{j\: k\: l} = 0,
\end{eqnarray}
except
\begin{eqnarray}
	\label{epsilon2}
	\epsilon_{1\: 2\: 3} = \epsilon_{2\: 3\: 1} = \epsilon_{3\: 1\: 2} = 1,
\end{eqnarray}
and
\begin{eqnarray}
	\label{epsilon3}
	\epsilon_{3\: 2\: 1} = \epsilon_{2\: 1\: 3} = \epsilon_{1\: 3\: 2} = -1.
\end{eqnarray}
Using equation (\ref{epsilon1})-(\ref{epsilon3}) the relations (\ref{paulicommu1}) can be re-written as
\begin{eqnarray}
	\label{paulicommu2}
	[\sigma_{j},\sigma_{k}] = 2\: i\: \sum_{l=1}^{3}\epsilon_{j\:k\:l}\:\sigma_{l}.
\end{eqnarray}
It is also to be noted that for indices $j,\: k=1,2,3$ we have,
\begin{eqnarray}
\label{epsilon4}
\sigma_{1}\sigma_{2} = \left(%
	\begin{array}{cc}
		0 & 1\\
		1 & 0\\
	\end{array}%
	\right)\left(%
	\begin{array}{cc}
		0 & -i\\
		i & 0\\
	\end{array}%
	\right)=\left(%
	\begin{array}{cc}
		i & 0\\
		0 & -i\\
	\end{array}%
	\right), ~~~ j=1,k=2.
\end{eqnarray}
Also
\begin{eqnarray}
\label{epsilon5}
\delta_{jk}I + i\sum_{l = 1}^{3}\epsilon_{jkl}\sigma_{l} &=& 0 + i[\epsilon_{121}+\epsilon_{122}\sigma_{2}+\epsilon_{123}\sigma_{3}]\nonumber\\
&=& 0 + i\sigma_{3} = i\left(%
	\begin{array}{cc}
		1 & 0\\
		0 & -1\\
	\end{array}%
	\right) = \left(%
	\begin{array}{cc}
		i & 0\\
		0 & -i\\
	\end{array}%
	\right).
\end{eqnarray}
From eqs.(\ref{epsilon4}) and (\ref{epsilon5}) and considering other possible pairs of values for the indices $j,\:k$ one can deduce that
\begin{eqnarray}
\label{epsilon6}
\sigma_{j}\sigma_{k} = \delta_{jk}I + i\sum_{l = 1}^{3}\epsilon_{jkl}\sigma_{l}, ~~~j,k=1,2,3.
\end{eqnarray}
\textbf{Problem:} Expand the relation (\ref{paulicommu2}) and verify using properties of anti-symmetric tensor defined in (\ref{epsilon1})-(\ref{epsilon3}) to deduce the relations (\ref{paulicommu1}).\\\\
\textbf{Simultaneous Diagonalization Theorem:} Suppose $A$ and $B$ are Hermitian operators. Then $[A, B]=0$ if and only if there exists an orthonormal basis such that both $A$ and $B$ are diagonal with respect to that basis. We say that $A$ and $B$ are \textit{simultaneously diagonalizable} in that case\footnote{For proof the author is recommending the references \cite{sakurai1967advanced} and \cite{nielsen2010quantum} for the reader.}.\\\\
With the help of the knowledge acquired so far in the previous sections and to make the reader feel the utilities of commutation and anti-commutation relation, we deduce below what is known as \textit{uncertainty relation} in quantum mechanics. 
\subsection{Uncertainty Principle:} 
Suppose we want to measure an observable $A$ and the uncertainty in the measurement is denoted by $\Delta\:A$. Likewise suppose the uncertainty in the measurement of another observable $B$ is $\Delta\:B$. If $\langle A\rangle$ and $\langle B\rangle$ respectively denote the expectations of the operators $A$  and $B$ i.e. the average value of the measurements of the observables $A$ and $B$, then the quantities $\Delta\:A \equiv A-\langle A\rangle$ and $\Delta\:B \equiv B-\langle B\rangle$ denote the \textit{dispersion}\footnote{Dispersion is also termed as deviation.} of $A$ and $B$ in their measurements respectively. We shall now first deduce the Uncertainty relation by Robertson\cite{sakurai1967advanced,robertson20141}. The uncertainty relation is given by\\
\begin{eqnarray}
	\label{robertson1}
	\langle (\Delta A)^{2}\rangle\langle (\Delta B)^{2}\rangle \geq \frac{1}{4}\vert \langle [A, B]\rangle\vert^{2}.
\end{eqnarray}
This deduction requires initially, proofs of three lemmas. The lemma $1$ considered here is the well known Cauchy-Schwartz's inequality. The inequality was first introduced in \textit{problem} 1.9 of this chapter. We restate this inequality below. This time we have only changed the notation. The Cauchy - Schwartz's inequality is thus defined as
\begin{eqnarray}
	\label{cauchy1}
	\| \vert \alpha\rangle\| \:\:\| \vert \beta\rangle\| \geq \vert \langle \alpha \vert \beta\rangle \vert.
\end{eqnarray}
\textit{Proof of Cauchy Schwartz's Lemma:} Let $\vert \psi\rangle = \vert\alpha\rangle + \lambda \vert \beta\rangle$ be any state. Then its corresponding dual state is $\langle \psi\vert = \langle\alpha \vert + \lambda^{*}\langle \beta\vert$. The norm of $\vert \psi\rangle$ is given by (\ref{norm1}) and we get as $\|\vert  \psi\rangle\|^{2} \geq 0$
\begin{eqnarray}
	\label{cauchy2}
( \langle\alpha \vert + \lambda^{*}\langle \beta\vert)(\vert\alpha\rangle + \lambda \vert \beta\rangle)&=&\langle \alpha\vert \alpha\rangle + \lambda\langle \alpha\vert \beta\rangle + \lambda^{*}\langle \beta \vert \alpha\rangle + \vert \lambda\vert^{2} \langle \beta\vert \beta\rangle \geq 0.\nonumber\\ 
\end{eqnarray}
The scalar $\lambda$ can now be chosen as $-\frac{\langle \beta\vert \alpha\rangle}{\langle \beta\vert \beta\rangle}$. Substituting the value of $\lambda$ in equation (\ref{cauchy2}) we get,
\begin{eqnarray}
	\label{cauchy3}
	\langle \alpha\vert \alpha\rangle + \Big(-\frac{\langle \beta\vert \alpha\rangle}{\langle \beta\vert \beta\rangle}\Big)\langle \alpha\vert \beta\rangle + \Big(-\frac{\langle \beta\vert \alpha\rangle}{\langle \beta\vert \beta\rangle}\Big)^{*}\langle \beta \vert \alpha\rangle + \Big\vert -\frac{\langle \beta\vert \alpha\rangle}{\langle \beta\vert \beta\rangle}\Big\vert^{2} \langle \beta\vert \beta\rangle \geq 0.
\end{eqnarray}
The simplification of the above equation (\ref{cauchy3}) reveals Cauchy-Schwartz's inequality (Check!). \textbf{Q.E.D}\\\\
The next lemma that we would like to prove is that \textit{the expectation value of Hermitian operator is purely real.} We know that for an operator $X$ the matrix representation is in general given by $\langle \beta \vert X\vert \alpha\rangle$. Now $\langle \beta \vert X\vert \alpha\rangle = \langle \beta \vert (X\:\vert \alpha\rangle)=\langle \alpha \vert X^{\dagger}\vert \beta\rangle^{\dagger}$. If $X$ is Hermitian, then $X^{\dagger}=X$. Therefore, $\langle \beta \vert X\vert \alpha\rangle = \langle \alpha \vert X \vert \beta\rangle^{\dagger}$\footnote{If we take $\beta =\alpha$, $\langle \alpha \vert X\vert \alpha\rangle = \langle \alpha \vert X \vert \alpha\rangle^{\dagger}=\langle X\rangle_{\vert \alpha\rangle}$.}. This implies that expectation value of Hermitian operator $X$ is real. This proves the desired result. \textbf{Q.E.D}\\\\
Just like the above, \textit{for anti-Hermitian operator, the expectation value is purely imaginary}. For an anti-Hermitian operator $Y$, $Y^{\dagger}=-Y$. As before, $\langle \beta \vert Y\vert \alpha\rangle = \langle \beta\vert(Y\:\vert \alpha\rangle)=\langle \alpha\vert Y^{\dagger}\vert \beta\rangle^{\dagger} = \langle \alpha\vert (-Y)\vert \beta \rangle^{\dagger}=-\langle \alpha\vert Y\vert \beta \rangle^{\dagger}$ \footnote{For $\beta = \alpha$, $\langle \alpha \vert Y\vert \alpha\rangle^{\dagger}=-\langle \alpha \vert Y\vert \alpha\rangle$.}. This proves the desired result.\textbf{Q.E.D}\\\\
With these lemmas at hand as special weapons we are now in a position to derive the uncertainty relation in Robertson form. We choose $\vert \alpha\rangle = \Delta A\vert \bullet\rangle$ and $\vert \beta\rangle = \Delta B\vert \bullet\rangle$. Here $\vert \bullet\rangle$ is any arbitrary state. Substituting these in Cauchy-Schwartz inequality (\ref{cauchy1}) we get
\begin{eqnarray}
	\label{cauchy4}
	(\Delta A\langle \bullet\vert)(\Delta A \vert \bullet\rangle)(\Delta B\langle \bullet\vert)(\Delta B \vert \bullet\rangle) \geq \vert (\Delta A\langle \bullet\vert)(\Delta B\vert \bullet\rangle)\vert^{2},
\end{eqnarray}
which in turn implies 
\begin{eqnarray}
	\label{cauchy5}
	\langle \bullet\vert (\Delta A)^{2}\vert \bullet\rangle\langle \bullet\vert (\Delta B)^{2}\vert \bullet\rangle \geq \vert \langle \bullet\vert (\Delta A\Delta B\vert \bullet\rangle)\vert^{2}.
\end{eqnarray}
From the fact that $\langle \alpha \vert A\vert \alpha\rangle = \langle A\rangle_{\vert \alpha\rangle}$ we see from equation (\ref{cauchy5}) that
\begin{eqnarray}
	\label{cauchy6}
	\langle (\Delta A)^{2}\rangle \langle (\Delta B)^{2}\rangle \geq \vert \langle \Delta A\Delta B\rangle\vert^{2}.
\end{eqnarray}
We see from eq. (\ref{cauchy6}) that the left hand side of Robertson's inequality has taken shape. We will now work with right hand side of the inequality only.\\\\
Using the relation that for any two operators $A$ and $B$, the operator $AB$ can be expressed as $AB = \frac{[A,B]+\lbrace A, B\rbrace}{2}$, we get $\Delta A\Delta B = \frac{[\Delta A,\Delta B]+\lbrace \Delta A, \Delta B\rbrace}{2}$ and consequently we have $\langle\Delta A\Delta B\rangle = \frac{\langle[\Delta A,\Delta B]\rangle+\langle \lbrace \Delta A, \Delta B\rbrace\rangle}{2}$. Substituting this in equation (\ref{cauchy6}), we find
\begin{eqnarray}
	\label{cauchy7}
		\langle (\Delta A)^{2}\rangle \langle (\Delta B)^{2}\rangle \geq \frac{1}{4} \vert \langle [\Delta A, \Delta B]\rangle + \langle \lbrace \Delta A, \Delta B\rbrace\rangle \vert^{2}.
\end{eqnarray}
We also know that for an operator $A$, $\Delta A = A - \langle A\rangle$. With the help of this fact it is easy to verify that $[\Delta A, \Delta B]=[A, B]$ (\textbf{Check!}). Again for any two real numbers $\dot{a}$ and $\dot{b}$, $\vert \dot{a}+\dot{b}\vert^{2} \geq \vert \dot{a}\vert^{2}+\vert \dot{b}\vert^{2}$. Applying these in equation (\ref{cauchy7}) we get,
\begin{eqnarray}
	\label{cauchy8}
		\langle (\Delta A)^{2}\rangle \langle (\Delta B)^{2}\rangle \geq \frac{1}{4} \vert \langle [A,B]\rangle \vert^{2} + \frac{1}{4}\vert \langle \lbrace \Delta A,\Delta B\rbrace\rangle \vert^{2},
\end{eqnarray}
where $\langle [A,B]\rangle$ is purely real and $\langle \lbrace \Delta A, \Delta B\rbrace\rangle$ is purely imaginary (by the lemmas).  Omission of the second term on the right hand side of the inequality can only make the inequality relation stronger\cite{sakurai1967advanced}. Thus from (\ref{cauchy8}) we have $\langle (\Delta A)^{2}\rangle\langle (\Delta B)^{2}\rangle \geq \frac{1}{4}\vert \langle [A, B]\rangle\vert^{2}$.\\\\
For a particle let the operator $A$ be the \textit{position operator} denoted by $\hat{X}$ and operator $B$ be the \textit{momentum operator} denoted by $\hat{P}$, such that $\Delta A = \Delta \hat{X}$ and $\Delta B = \Delta \hat{P}$. Also $[\hat{X},\hat{P}] = i\:\hbar\:\delta_{i\:j}$\cite{sakurai1967advanced}. Then using these expressions in (\ref{robertson1}) we get\footnote{$\hbar$ is Planck's constant whose value is $6.5822 \times 10^{-16}$ eV -s.}
\begin{eqnarray}
	\label{heisenberg1}
	\langle (\Delta \hat{X})^{2}\rangle\langle (\Delta \hat{P})^{2}\rangle \geq \frac{1}{4} \vert \langle i\:\hbar\:\delta_{ij}\rangle\vert^{2} = \frac{\hbar^{2}}{4}.
\end{eqnarray}
Another representation of relation (\ref{heisenberg1}) is as follows:
\begin{eqnarray}
	\label{heisenberg2}
	\langle \Delta \hat{X}\rangle\langle \Delta \hat{P}\rangle \geq  \frac{\hbar}{2}.
\end{eqnarray}
The relations (\ref{heisenberg1}) and (\ref{heisenberg2}) are known as \textbf{position-momentum uncertainty relation}\footnote{This uncertainty relation is due to Heisenberg. The principle applies to other related (conjugate) pairs of observables, such as energy and time.}. The interpretation of Heisenberg's uncertainty relation is that, the uncertainty in the measurement of the position of the particle is inversely proportional to the uncertainty in the measurement of the momentum of the particle i.e if uncertainty in the position measurement increases the corresponding uncertainty in the momentum measurement will decrease and vice-versa.\\\\
If we want to speak a little more about the uncertainty relation (\ref{robertson1}), then we must say that this uncertainty relation was criticized on the grounds that the right hand side of eq.(\ref{robertson1}) was not a fixed lower bound, but dependent on the state under consideration. To improve the situation, later, \textit{entropic uncertainty} relations were proposed. The first such relations relied upon Shannon entropy\footnote{Shannon entropy is a classical measure of uncertainty.}.  For a general probability distribution $p=(p_{1},p_{2},\cdots, p_{n})$, $p_{i}\geq 0$ and $\sum_{i}p_{i} = 1$, on a set of $n$ possible outcomes, the \textbf{Shannon entropy} is defined as\cite{shannon1993claude}
\begin{eqnarray}
\label{shannonentropy}
H(p) = - \sum_{i}p_{i}\ln(p_{i}).
\end{eqnarray}
Applying this notion to the probability distributions $p$ and $q$, Deutsch devised the entropic uncertainty relation as\cite{deutsch1983uncertainty}
\begin{eqnarray}
\label{deutschuncertainty}
H(p) + H(q)\geq -2 \ln\Bigg(\frac{1+c}{2}\Bigg),
\end{eqnarray}
where $c=\max_{j,k}| \langle a_{j}\vert b_{k}\rangle|$. Here $\lbrace \vert a_{j}\rangle\rbrace$ and $\lbrace \vert b_{j}\rangle\rbrace$ with $j=1,2,\cdots, n$ are the complete sets of normalized eigenvectors of the physical observables $A$ and $B$ belonging to the $n-$ dimensional Hilbert space. Kraus conjectured an uncertainty relation which is given by\cite{kraus1987complementary}
\begin{eqnarray}
\label{krausuncertainty}
H(p)+ H(q) \geq -2 \ln(c).
\end{eqnarray}
The relation (\ref{krausuncertainty}) was later proved by Maassen and Uffink\cite{maassen1988generalized}.
\subsection{Operator Function:}
To understand the concepts of this section, we shall first discuss what we mean by \textit{operator function}. Suppose we have a normal operator $A$ which can be represented by spectral decomposition as $A=\sum_{i}i\:\vert i\rangle\langle i\vert$. We define $f(A)=\sum_{i}f(i)\vert i\rangle\langle i\vert$. In this manner, $f(A)$ can be uniquely defined. This helps us to find the square root of a positive operator, the logarithm of a positive definite operator, or the exponential of a normal operator, and like.\\\\
\textbf{Problem $1.33$:} Find the exponential of the matrix $\sigma_{x} = \left(%
\begin{array}{cc}
	0 & 1\\
	1 & 0\\
\end{array}%
\right)$.\\\\
\textit{Solution:} 
\begin{itemize}
\item It is easy to verify (using \textit{section $1.2.2$}) that the eigenvalues of $\sigma_{x}$ are $1$ and $-1$ and the corresponding eigenvectors are $\Bigg\lbrace t\left(%
\begin{array}{c}
	1 \\
	1 \\
\end{array}%
\right)\: :\: t \in \mathbb{C}; t\ne 0\Bigg\rbrace$ and $\Bigg\lbrace t\left(%
\begin{array}{c}
	-1 \\
	1 \\
\end{array}%
\right)\: :\: t^{\prime} \in \mathbb{C}; t^{\prime} \ne 0\Bigg\rbrace$. 
\item We construct the basis $\beta_{\sigma_{x}} = \Bigg\lbrace \left(%
\begin{array}{c}
	1\\
	1\\
\end{array}%
\right),\left(%
\begin{array}{c}
	1\\
	-1\\
\end{array}%
\right)\bigg\rbrace$ which is linearly independent. Let $Q=\left(%
\begin{array}{cc}
	1 & 1\\
	1 & -1\\
\end{array}%
\right)$. 
\item $Q^{-1} = \left(%
\begin{array}{cc}
	\frac{1}{2} & \frac{1}{2}\\
	-\frac{1}{2} & \frac{1}{2}\\
\end{array}%
\right)$.
\item Let $D_{\sigma_{x}}$ be the diagonal matrix of $\sigma_{x}$ with respect to the basis $\beta_{\sigma_{x}}$. Then $D_{\sigma_{x}} = Q^{-1}\sigma_{x}Q$. We have thus $D_{\sigma_{x}} =  \left(%
\begin{array}{cc}
	1 & 0\\
	0 & -1\\
\end{array}%
\right)$.
\item Now we have $\sigma_{x} = QD_{\sigma_{x}}Q^{-1}$.
\item Then $\exp(\sigma_{x})=I + \sigma_{x} + \frac{\sigma_{x}^{2}}{2!} + \cdots = Q\exp(D_{\sigma_{x}})Q^{-1}$.
\item $\exp(\sigma_{x}) =  \left(%
\begin{array}{cc}
	\exp(1) & 0\\
	0 & \exp(-1)\\
\end{array}%
\right)$.
\item Also it is observed that as $\sigma_{x} = 1\vert 0\rangle\langle 0\vert + (-1)\vert 1\rangle\langle 0\vert$ (diagonalized in this case) in outer product notation, so $\exp(\sigma_{x}) = \exp(1)\vert 0\rangle\langle 0\vert + \exp(-1)\vert 1\rangle\langle 0\vert$\footnote{Remember that $\sigma_{x}$ being the Pauli operator is normal.}.
\end{itemize}
\textbf{Problem:} Find the exponential of the matrix $\sigma_{y} = \left(%
\begin{array}{cc}
	0 & -i\\
	i & 0\\
\end{array}%
\right)$ and $\sigma_{z} = \left(%
\begin{array}{cc}
	1 & 0\\
	0 & -1\\
\end{array}%
\right)$\footnote{$\sigma_{y}$ and $\sigma_{z}$ are also Pauli spin operators and are normal.}.\\\\
\textbf{Problem $1.35$:} Let $\bar{v}$ be any real, three dimensional unit vector and $\theta$ a real number. Prove that $\exp(i\:\theta\:\:\bar{v}\cdot \bar{\sigma})=\cos(\theta)I+i\sin(\theta)\bar{v}\cdot \bar{\sigma}$ where $\bar{v}\cdot \bar{\sigma} = \sum_{i=1}^{3}v_{i}\sigma_{i}$\cite{nielsen2010quantum}.\\\\
\textit{Solution:} The Pauli matrices $\sigma_{i},\:\:i = 1,2,3$ are defined by eqs. (\ref{P1})-(\ref{P4}). Now 
\begin{eqnarray}
\label{34c1}
\bar{v}\cdot\bar{\sigma} &=& v_{1}\left(%
\begin{array}{cc}
	0 & 1\\
	1 & 0\\
\end{array}%
\right) +v_{2}\left(%
\begin{array}{cc}
	0 & -i\\
	i & 0\\
\end{array}%
\right)+v_{3}\left(%
\begin{array}{cc}
	1 & 0\\
	0 & -1\\
\end{array}%
\right)\nonumber\\
&=& \left(%
\begin{array}{cc}
	v_{3} & v_{1}-iv_{2}\\
	v_{1}+iv_{2} & -v_{3}\\
\end{array}%
\right).
\end{eqnarray}
If we let $H=\bar{v}\cdot\bar{\sigma}$ then it is easy to verify that $H$ is Hermitian as $H^{\dagger}=H$. Consequently $H$ is a normal operator and then by spectral decomposition theorem we see that $H$ is diagonalizable.\\\\
Again $H^{2}=I$, $I$ is the Pauli matrix given by eq.(\ref{P1}).[Remember that as $\bar{v}$ is the unit vector so $v_{1}^{2} + v_{2}^{2} + v_{3}^{2} = 1$ or in other words $\|v\|=1$.\\\\
Also $H^{2k} = (H^{2})^{k} = I^{k}=I,\:k \in\mathbb{Z^{+}}$\footnote{$\mathbb{Z^{+}}$ is the set of positive integers.}. This gives $H^{2k+1}=H^{2k}H=I\:H=H,\: k \in \mathbb{Z^{+}}$. Now as $H=\bar{v}\cdot \bar{\sigma}$ then
\begin{eqnarray}
\label{34c2}
\exp(i\theta\bar{v}\cdot\bar{\sigma}) &=& \exp(i\theta H)\nonumber\\
&=& I + (i\theta H) + \frac{(i\theta H)^{2}}{2!}+ \frac{(i\theta H)^{3}}{3!}+\cdots.
\end{eqnarray}
It is easy to verify by the induction process that $(i\theta H)^{k} = i^{k}\theta^{k}H^{k}$ where $k \in \mathbb{Z^{+}}$. Hence from eq.(\ref{34c2}) we get
\begin{eqnarray}
\label{34c3}
\exp(i\theta\bar{v}\cdot\bar{\sigma}) &=& I + (i\theta H) + \frac{i^{2}\theta^{2}H^{2}}{2!} + \cdots +\frac{i^{k}\theta^{k}H^{k}}{k!}+\cdots\nonumber\\
&=& (1 - \frac{\theta^{2}}{2!} + \frac{\theta^{4}}{4!} -\cdots)I + i(\theta -\frac{\theta^{3}}{3!} + \frac{\theta^{5}}{5!} +\cdots)H\nonumber\\
&=& \cos(\theta) + i\sin(\theta)H.
\end{eqnarray}
\textbf{Q.E.D}
\subsection{Polar and singular value decomposition:}
The understanding of structures of general linear operators are sometimes difficult and so they are decomposed into simpler structures such as unitary operators and positive operators. One such decomposition is known as \textbf{Polar decomposition} which is defined as follows.
Let $A$ be any linear operator on a vector space $V$. Then there exist unitary operator $U$ and positive operators $J$ and $K$ such that $A = U\:J = K\:U$ where unique positive operators $J$ and $K$ satisfying these equations are defined by $J \equiv \sqrt{A^{\dagger}\:A}$ and $K \equiv \sqrt{A\:A^{\dagger}}$. Moreover if $A$ is invertible then $U$ is unique.  We call $A=U\:J$ as \textbf{left polar decomposition} of $A$ and $A=K\:U$ is the \textbf{right polar decomposition} of $A$.\footnote{The operators $J$ and $K$ are clearly positive as in \textit{problem} 1.26 we know that for any operator $A$, $A^{\dagger}A$ or $AA^{\dagger}$ is always positive and hence its square root.}\\\\
A few special cases can however be considered.\\\\
\textit{Case $I$:} Suppose the operator $A$ is Hermitian. Then $A^{\dagger}=A$. This implies that $J =\sqrt{A^{\dagger}A}=\sqrt{A^{2}}=A$. Similarly $K=\sqrt{AA^{\dagger}}=\sqrt{A^{2}}=A$. Now if $U$ be any unitary operator then by left polar decomposition of $A$ we have $A = UJ = UA$. If $A$ is invertible, then $U=I$. Similar argument can also be placed with respect to right polar decomposition of $A$. Thus we conclude that for Hermitian operator $A$ (which is non-invertible), the left and right polar decompostion of $A$ will be $A=UA$ and $A=AU$ respectively whereas if $A$ is invertible the left and right polar decomposition of $A$ will be $A=IA$ and $A=AI$ respectively. Here $I$ is the identity operator.\\\\
\textit{Case $II$:} Suppose operator $A$ is unitary. Then $A^{\dagger}A = AA^{\dagger} = I$. Also $\sqrt{A^{\dagger}A} = \sqrt{AA^{\dagger}} = \sqrt{I} = I$. Consequently from both left and right polar decomposition we can say that the unitary matrix $U$ is the operator $A$ itself.\\\\
\textit{Case $III$:} If $A$ is a positive operator and every positive operator is necessarily Hermitian, the case is similar to case $I$.\\\\
\textit{Case $IV$:} If the operator $A$ is normal then $A^{\dagger}A = AA^{\dagger}$ and consequently $J = K$ such that $A = KU = UK$ or $A = UJ = JU$ is the left and right polar decomposition of the operator $A$.\\\\
The another decomposition which is of practical significance is the \textbf{Singular value decomposition} which is defined as follows:
Let $A$ be any square matrix. Then there exist unitary matrices $U$ and $V$ and a diagonal matrix $D$ with non-negative entries such that $A=U\:D\:V$. The diagonal elements of $D$ are called singular values of $A$.
\subsection{Postulates of Quantum Mechanics:}
\begin{itemize}
\item \textit{Postulate $1$:} Associated to any isolated physical system is a complex vector space with inner product (that is, a Hilbert space) known as the \textit{state space} of the system. The system is completely described by its state vector, which is a unit vector in the system's state space\cite{nielsen2010quantum}.
\item \textit{Postulate $2$:} The evolution of the closed quantum system is described by unitary transformation. That is, the state $\vert \psi\rangle$ of the system at time $t_{1}$ is related to the state $\vert \psi^{\prime}\rangle$ of the system at time $t_{2}$ by a unitary operator $U$ which depends only on the times $t_{1}$ and $t_{2}$\cite{nielsen2010quantum},
\begin{eqnarray}
	\label{postulate2}
	\vert \psi^{\prime}\rangle = U\:\vert \psi\rangle .
\end{eqnarray}
\item \textit{Postulate $3$:} The time evolution of the state of a closed quantum system is described by the Schr$\dot{o}$dinger's equation
\begin{eqnarray}
	\label{schrodinger}
	i\:\hbar\:\frac{d\:\vert \psi\rangle}{d\:t} = H\:\vert \psi\rangle,
\end{eqnarray}
where $\hbar$ is the Planck's constant and $H$ is the Hamiltonian of the system\cite{nielsen2010quantum}.
\item \textit{Postulate $4$:} Quantum measurements are described by a collection $\lbrace M_{m}\rbrace$ of measurement operators. These are operators acting on the state space of the system being measured. The index $m$ refers to the measurement outcomes that may occur in the experiment. If the state of the quantum system is $\vert \psi\rangle$ immediately before the measurement then the probability that result $m$ occurs is given by
\begin{eqnarray}
	\label{measurement1}
	p(m)=\langle \psi \vert M_{m}^{\dagger}M_{m}\vert \psi\rangle, 
\end{eqnarray}
and the state of the system after measurement is
\begin{eqnarray}
	\label{measurement2}
	\frac{M_{m}\vert \psi\rangle}{\sqrt{\langle \psi\vert M_{m}^{\dagger}M_{m}\vert \psi\rangle}}.
\end{eqnarray}
The measurement operators satisfy the completeness equation\cite{nielsen2010quantum}
\begin{eqnarray}
	\label{measurement3}
	\sum_{m}M_{m}^{\dagger}M_{m} = I.
\end{eqnarray}\\
To illustrate this postulate by an example we consider a qubit $\vert\phi\rangle = \alpha \vert 0\rangle + \beta \vert 1\rangle$. We consider two measurement operators $M_{0} = \vert 0\rangle\langle 0\vert$ and $M_{1}=\vert 1\rangle\langle 1\vert$. If we want to measure the state $\vert\phi\rangle$ then the probability of obtaining the measurement outcome $0$ is given by $p(0)=\langle \phi \vert M_{0}^{\dagger}M_{0}\vert \phi\rangle$ i.e. $\vert \alpha\vert^{2}$ while the probability of obtaining the measurement outcome $1$ is given by $p(1)=\langle \phi \vert M_{1}^{\dagger}M_{1}\vert \phi\rangle$ i.e. $\vert \beta\vert^{2}$. Remember, according to our choices, both the measurement operators $M_{0}$ and $M_{1}$ are Hermitian. The state of the system $\vert\phi\rangle$ after applying the measurement operator $M_{0}$, by using equation (\ref{measurement2}), is thus obtained as $\frac{M_{0}\vert \phi\rangle}{p(0)} = \frac{a\vert 0\rangle}{\vert a\vert}$ whereas after applying the measurement operator $M_{1}$, again by using (\ref{measurement2}), the post measurement state of the state $\vert\phi\rangle$ is $\frac{M_{1}\vert \phi\rangle}{p(1)} = \frac{b\vert 1\rangle}{\vert b\vert}$. It is also to be noted that the measurement operators $M_{0}$ and $M_{1}$ satisfy the completeness relation (\ref{measurement3}).\\\\
Among all the postulates described above, the postulate $4$ is the most fascinating as it gives us the idea that the measurement changes the state of the system, that is after the measurement is performed on a given state of the system, the state is projected into a new state is always associated with some probability. This distinguishes the quantum system from its classical counterpart.\footnote{This did not go well with Einstein as He was always having the notion that whether we observed any system or not, the system had always an existence. According to him, Moon is out there whether we look at it or don't. This is described in a famous paper by Mermin\cite{mermin1985moon}.} Also one of the  important applications of postulate $3$ is that of \textit{distiguishig quatum states}. Let us discuss below what we mean by quantum distinguishability.\\\\
Suppose we have a set $\lbrace \vert \psi_{i}\rangle: i=1,2,\cdots,n\rbrace$ and let the existence of such a set be in the knowledge of both Alice and Bob. Again we assume that the set to be orthonormal. Suppose from this set, Alice picks one state arbitrarily, (say, $\vert \psi_{i}\rangle$) and sends this state to Bob. Bob's task is now to identify the index $i$ of the state that Alice has given to him. Measurement operators $M_{i}=\vert \psi_{i}\rangle\langle \psi_{i}\vert$ are now defined for each possible index $i$. An additional measurement operator $M_{0}$ is also defined, which is the positive square root of the positive operator $\mathbb{I}-\sum_{i\ne 0}\vert \psi_{i}\rangle\langle \psi_{i}\vert$\footnote{$\mathbb{I}$ is the identity operator}. These operators satisfy the completeness relation (\ref{completeness}). Now if the state $\vert \psi_{i}\rangle$ is prepared then by eq.(\ref{measurement1}), the probability of obtaining the measurement outcome $i$ would be $p(i)= \langle \psi_{i}\vert M_{i}\vert \psi_{i}\rangle = \langle \psi_{i}\vert \psi_{i}\rangle\langle \psi_{i}\vert \psi_{i}\rangle = 1$, since $\vert \psi_{i}\rangle$'s are orthonormal. This shows that the outcome $i$ occurs with certainty. Thus it is possible to reliably distinguish the orthonormal states $\vert \psi_{i}\rangle$. On the other hand, one asks the question, what happens when the set $\lbrace \vert \psi_{i}\rangle: i=1,2,\cdots,n\rbrace$ is not orthonormal? Now if the set is not orthonormal that means the states present in this set are not orthonormal. In other words, either the states are not unit vectors or they are not orthogonal to one another or both. In such a scenario, our objective is to prove that there is no quantum measurement capable of distinguishing such states. Without any loss of generality, assume that the states are non-orthogonal (or not orthogonal) and hence are non-orthonormal. To understand this, we consider two arbitrary non-orthogonal states $\vert \psi_{1}\rangle$ and $\vert \psi_{2}\rangle$ and will show that there exists no measurement that can distinguish between these two non-orthogonal states. Define a function $f(\bullet)$ by
\begin{eqnarray}
\label{fn1}
f(j) = i,
\end{eqnarray}
where this function means when Bob observes $j$ as measurement outcome he guesses that state to be $\vert \psi_{i}\rangle$. If possible let such a measurement exists which can reliably distinguish between the non-orthogonal states $\vert \psi_{1}\rangle$ and $\vert \psi_{2}\rangle$. Then if the state $\vert \psi_{1}\rangle$ is prepared then from eq.(\ref{fn1}) we have $f(j)=1$ which will have probability $1$. (Or, if the state $\vert \psi_{2}\rangle$ is prepared then from eq.(\ref{fn1}) we have $f(j)=2$ which will have probability $1$). Define 
\begin{eqnarray}
\label{fn2}
E_{i}\equiv \sum_{j; f(j)=i}M_{j}^{\dagger}M_{j}.
\end{eqnarray}
Let us say when $j=1$ Bob observes the state $\vert \psi_{1}\rangle$ i.e. $i=1$. Then $\langle \psi_{1}\vert E_{1}\vert \psi_{1}\rangle = \langle \psi_{1}\vert M_{1}^{\dagger}M_{1}\vert \psi_{1}\rangle =1$. Similarly we can argue that $\langle \psi_{2}\vert E_{2}\vert \psi_{2}\rangle = 1$. Since $E_{i}$ of eq.(\ref{fn2}) satisfies completeness relation (\ref{completeness}) i.e. $\sum_{i}E_{i}=\mathbb{I}$, this implies that $\sum_{i}\langle \psi_{1}\vert E_{i}\vert \psi_{1}\rangle=1$ . Again since we have $\langle \psi_{1}\vert E_{1}\vert \psi_{1}\rangle = 1$ one must have $\langle \psi_{1}\vert E_{2}\vert \psi_{1}\rangle = 0$. This implies $\sqrt{E_{2}}\vert \psi\rangle = 0$. According to our assumption, $\vert \psi_{1}\rangle$ and $\vert \psi_{2}\rangle$ are non-orthogonal and we decompose $\vert \psi_{2}\rangle$ into a (non-zero) component parallel to $\vert \psi_{1}\rangle$ and a component orthogonal to $\vert \psi_{1}\rangle$. Let the decomposition be $\vert \psi_{2}\rangle = \alpha\vert \psi_{1}\rangle + \beta \vert \varphi_{1}\rangle$, where $\vert \varphi_{1}\rangle$ is orthonormal to $\vert \psi_{1}\rangle$. Also in this decomposition $\vert \alpha\vert^{2}+\vert \beta\vert^{2}=1$ and $\vert \beta\vert < 1$. Now as $\vert \psi_{1}\rangle$ and $\vert \psi_{2}\rangle$ are non-orthogonal and as $\vert \varphi_{1}\rangle$ is orthonormal to $\vert \psi_{1}\rangle$ (and hence is orthogonal to $\vert \psi_{1}\rangle$), therefore $\vert \varphi_{1}\rangle$ is non-orthogonal to $\vert \psi_{2}\rangle$. Let us take $\vert \psi_{2}\rangle = \beta\vert \varphi_{1}\rangle \Rightarrow \sqrt{E_{2}}\vert \psi_{2}\rangle = \beta \sqrt{E_{2}}\vert \varphi_{1}\rangle$. Then we observe that $\langle \psi_{2}\vert E_{2}\vert\psi_{2}\rangle = \vert \beta\vert^{2}\langle \varphi_{1}\vert E_{2}\vert\varphi_{1}\rangle$\footnote{$\sqrt{E_{2}}\langle\psi_{2}\vert = \beta^{*}\sqrt{E_{2}}\langle \varphi\vert$}. Now $\langle \varphi_{1}\vert E_{2}\vert \varphi_{1}\rangle \leq \sum_{i}\langle \varphi_{1}\vert E_{i}\vert \varphi_{1}\rangle = \langle \varphi_{1} \vert \sum_{i}E_{i}\vert \varphi_{1}\rangle = \langle \varphi_{1} \vert \mathbb{I}\vert \varphi_{1}\rangle = \langle \varphi_{1}\vert \varphi_{1}\rangle=1$. Then $\langle \psi_{2}\vert E_{2}\vert\psi_{2}\rangle \leq \vert \beta\vert^{2} < 1$ which contradicts the fact that $\langle \psi_{2}\vert E_{2}\vert\psi_{2}\rangle =1$. Therefore we conclude that non-orthogonal quantum states can't reliably be distinguished.
\item \textit{Postulate $5$:} The state space of a composite physical system is the tensor product of the state spaces of the component physical systems. Moreover if we have systems numbered $1$ through $n$ and system numbered $i$ is prepared in the state $\vert \psi_{i}\rangle$, then the joint state of the  composite quantum system  is $\vert \psi_{1}\rangle \otimes \vert \psi_{2}\rangle \otimes\cdots \vert \psi_{n}\rangle$ \cite{nielsen2010quantum}.\\\\
Postulate $5$ also enables us to define one of the most interesting and puzzling ideas associated with composite quantum systems- known as \textbf{entanglement}. The notation $\otimes$ is known as \textit{tensor product} or sometimes also known as \textit{Kronecker product}. In the next unit, we shall give a detailed discussion on the concept of tensor product.
\end{itemize}
\subsection{General measurement:} The measurement scheme discussed in postulate $3$ above is the general description of measurements. We below prove an important result based on this general measurement formalism.\\\\
\textbf{Result:} Suppose $\lbrace L_{l}\rbrace$ and $\lbrace M_{m}\rbrace$ are two sets of measurement operators. Then a measurement defined by the measurement operators $\lbrace L_{l}\rbrace$ followed by  a measurement defined by the measurement operators $\lbrace M_{m}\rbrace$ is physically equivalent to a single measurement defined by measurement operators $\lbrace N_{lm}\rbrace$ with the representation $N_{lm}=M_{m}L_{l}$. The sequential applications of measurement operators one after another is called \textit{cascaded measurements}.\\\\
\textit{Proof:} Let $\vert\psi\rangle$ be the quantum state on which we are going to apply cascaded measurements ($\lbrace L_{l}\rbrace$ and $\lbrace M_{m}\rbrace$ sequentially). Once the operators $\lbrace L_{l}\rbrace$ are applied, the post measurement state of $\vert\psi\rangle$ is denoted by $\vert\psi^{\prime}\rangle$ and is given by (using postulate $3$) 
\begin{eqnarray}
\label{cascaded1}
|\psi^{\prime}\rangle = \frac{L_{l}|\psi\rangle}{\sqrt{p(l)}},
\end{eqnarray}
where $p(l) = \langle \psi |L_{l}^{\dagger}L_{l}|\psi\rangle$ and $\sum_{l}{L_{l}^{\dagger}L_{l}}= I$.  On this state $|\psi^{\prime}\rangle$ we again apply the measurement operators $\lbrace M_{m}\rbrace$. Consequently we get new post measurement state as
\begin{eqnarray}
\label{cascaded2}
|\psi^{\prime\prime}\rangle = \frac{M_{m}|\psi^{\prime}\rangle}{\sqrt{p(m)}},
\end{eqnarray}
where $p(m) = \langle \psi^{\prime} |M_{m}^{\dagger}M_{m}|\psi^{\prime}\rangle$ and $\sum_{m}{M_{m}^{\dagger}M_{m}}= I$. Now using (\ref{cascaded1}), we can rewrite $p(m)$ as 
\begin{eqnarray}
\label{cascaded3}
p(m) &=& \langle \psi^{\prime} |M_{m}^{\dagger}M_{m}|\psi^{\prime}\rangle\nonumber\\
&=& \frac{\langle \psi|L_{l}^{\dagger}}{\sqrt{p(l)}} \Big|M_{m}^{\dagger}M_{m}\Big|\frac{L_{l}|\psi\rangle}{\sqrt{p(l)}}\nonumber\\
&=& \frac{\langle \psi | L_{l}^{\dagger} M_{m}^{\dagger} M_{m}L_{l}|\psi\rangle}{p(l)}.
\end{eqnarray}
Now using (\ref{cascaded1}) and (\ref{cascaded3}) in (\ref{cascaded2}) we get the post measurement state $|\psi^{\prime\prime}\rangle$ as
\begin{eqnarray}
\label{cascaded4}
|\psi^{\prime\prime}\rangle = \frac{M_{m}L_{l}|\psi\rangle}{\langle \psi | L_{l}^{\dagger} M_{m}^{\dagger} M_{m}L_{l}|\psi\rangle}.
\end{eqnarray}
Now if we consider a measurement operator of the form $\lbrace N_{lm}\rbrace$, where $N_{lm} = M_{m}L_{l}$, and if we apply these operators on the state $|\psi\rangle$, the post measurement state is given by (say $|\psi^{(1)}\rangle$)
\begin{eqnarray}
\label{cascaded5}
|\psi^{(1)}\rangle = \frac{M_{m}L_{l}|\psi\rangle}{\langle \psi | L_{l}^{\dagger} M_{m}^{\dagger} M_{m}L_{l}|\psi\rangle}.
\end{eqnarray}
We see that eqs.(\ref{cascaded4}) are eqs.(\ref{cascaded5}) are having same expressions. Hence proved. ~~~~~~\textbf{Q.E.D}

\section{Unit - IV}\label{sec:unit4}

\subsection{Tensor Product:}
The pivotal role that the mathematical language of tensor product plays is to capture the essence of joining two or several quantum mechanical systems together. In other words if we have several Hilbert spaces, the tensor product among them help to construct a larger Hilbert space. With this language of tensor product it is possible to develop the physics of multi-qubits or to manage multi-particle quantum gates. Suppose $V$ and $W$ are two Hilbert spaces with dimensions $m$ and $n$ respectively. Then the \textit{Tensor Product} of $V$ and $W$, which is denoted as $V\otimes W$, is a Hilbert space with dimension $mn$. If $\vert v\rangle\in V$ and $\vert w\rangle \in W$, the element of $V\otimes W$ is denoted by $\vert v\rangle \otimes \vert w\rangle$ or by $\vert vw\rangle$ or by $\vert v,\:w\rangle$ or by $\vert v\rangle \vert w\rangle$. Also it is to be noted that if $\lbrace \vert i\rangle\rbrace$ is the orthonormal basis of $V$ and $\lbrace \vert j\rangle\rbrace$ is the orthonormal basis of $W$, then $\lbrace \vert i\rangle \otimes  \vert j\rangle\rbrace$ is the orthonormal basis of $V \otimes W$.  If $\lbrace \vert 0\rangle, \vert 1\rangle\rbrace$ is the orthonormal basis of the Hilbert space $\mathbb{C}^{2}$, then the basis of the Hilbert space $\mathbb{C}^{2}\otimes \mathbb{C}^{2}$ of the joint quantum mechanical system is $\lbrace \vert 0\rangle\otimes \vert 0\rangle, \vert 0\rangle\otimes \vert 1\rangle, \vert 1\rangle\otimes \vert 0\rangle, \vert 1\rangle \otimes \vert 1\rangle\rbrace$\footnote{In this article we will stick to the notation $\vert v\rangle\vert w\rangle$ rather than to $\vert v\rangle \otimes \vert w\rangle$ as when deem fit.}. Some of the properties of tensor product have been summarized below now \cite{nielsen2010quantum}.
\\\\
\textit{Property $1$:} For $\vert v\rangle \in V$ and $\vert w\rangle \in W$ and for arbitrary scalar (which is a complex number) $c$ we have 
\begin{eqnarray}
	\label{tensorproperty1}
	c\:(\vert v\rangle \otimes \vert w\rangle) = (c\vert v\rangle) \otimes \vert w\rangle = \vert v\rangle \otimes (c\vert w\rangle)
\end{eqnarray}\\
\textit{Property $2$:} For $\vert v_{1}\rangle,\:\vert v_{2}\rangle \in V$ and for $\vert w\rangle \in W$ we have
\begin{eqnarray}
	\label{property2}
	(\vert v_{1}\rangle +\vert v_{2}\rangle)\otimes \vert w\rangle = \vert v_{1}\rangle \otimes \vert w\rangle + \vert v_{2}\rangle \otimes \vert w\rangle.
\end{eqnarray}\\
\textit{Property $3$:} For arbitrary vector $\vert v\rangle \in V$ and for $\vert w_{1}\rangle,\:\vert w_{2}\rangle \in W$ we have
\begin{eqnarray}
	\label{property3}
	\vert v\rangle \otimes (\vert w_{1}\rangle + \vert w_{2}\rangle) = \vert v\rangle \otimes \vert w_{1}\rangle + \vert v\rangle \otimes \vert w_{2}\rangle.
\end{eqnarray}\\
\textit{Property $4$:} Let $A$ be the linear operator defined on the Hilbert space $V$ and $B$ be the linear operator defined on the Hilbert space $W$. Then we can define a linear operator $A\otimes B$ on $V\otimes W$ by
\begin{eqnarray}
	\label{property4a}
	(A\otimes B)(\vert v\rangle \otimes \vert w\rangle) = (A\vert v\rangle) \otimes (B \vert w\rangle),
\end{eqnarray} 
where $\vert v\rangle \in V$ and $\vert w\rangle \in W$. The definition of $A\otimes B$ can then be extended to all elements of $V\otimes W$ in the natural way to ensure linearity of $A\otimes B$ that is
\begin{eqnarray}
	\label{property4b}
	(A\otimes B)(\sum_{i}a_{i}\vert v_{i}\rangle\otimes \vert w_{i}\rangle) =\sum_{i}a_{i}A\vert v_{i}\rangle \otimes B\vert w_{i}\rangle.
\end{eqnarray}\\
A convenient representation of the tensor product of two operators is given by \textit{Kronecker Product} between the two operators where the matrix forms of the operators are considered. If $A$ is the matrix of order $p\times q$ and $B$ is the matrix of order $s\times t$ then $A\otimes B$ is the matrix of order $ps\times qt$.\\\\
\textbf{Problem $1.36$:} Let $\vert \psi\rangle =\frac{\vert 0\rangle + \vert 1\rangle}{\sqrt{2}}$. Write out $\vert \psi\rangle^{\otimes 2}$ and $\vert \psi\rangle^{\otimes 3}$\footnote{$\vert \psi\rangle^{\otimes 2}$ means $\vert \psi\rangle \otimes \vert \psi\rangle$ and so on.} explicitly in terms of the tensor product of $\vert 0\rangle$ and $\vert 1\rangle$. Also using matrix form of Kronecker Product find the same.\\\\ 
\textit{Solution:} In tensor product notation we get $\vert \psi\rangle^{\otimes 2} = \frac{\vert 0\rangle + \vert 1\rangle}{\sqrt{2}} \otimes \frac{\vert 0\rangle + \vert 1\rangle}{\sqrt{2}} = \frac{\vert 0\rangle \vert 0\rangle + \vert 0\rangle \vert 1\rangle + \vert 1\rangle \vert 0\rangle + \vert 1\rangle \vert 1\rangle}{2}$. (Similarly find $\vert \psi\rangle^{\otimes 3}$!)  Let us concentrate on the matrix form of the Kronecker product representation. The state $\vert \psi\rangle = \frac{\vert 0\rangle + \vert 1\rangle}{\sqrt{2}}$ is actually written in the matrix form as $\left(%
	\begin{array}{c}
		\frac{1}{\sqrt{2}}\\\\
		\frac{1}{\sqrt{2}}
	\end{array}%
	\right)$. Then $\vert \psi\rangle^{\otimes 2} = \left(%
	\begin{array}{c}
		\frac{1}{\sqrt{2}}\\\\
		\frac{1}{\sqrt{2}}
	\end{array}%
	\right)_{2\times 1}\otimes \left(%
	\begin{array}{c}
		\frac{1}{\sqrt{2}}\\\\
		\frac{1}{\sqrt{2}}
	\end{array}%
	\right)_{2\times 1} = \left(%
	\begin{array}{c}
		\frac{1}{\sqrt{2}}\left(%
	\begin{array}{c}
		\frac{1}{\sqrt{2}}\\\\
		\frac{1}{\sqrt{2}}
	\end{array}%
	\right)\\\\
		\frac{1}{\sqrt{2}}\left(%
	\begin{array}{c}
		\frac{1}{\sqrt{2}}\\\\
		\frac{1}{\sqrt{2}}
	\end{array}%
	\right)
	\end{array}%
	\right)_{4\times 1}=\left(%
	\begin{array}{c}
		\frac{1}{2}\\\\
		\frac{1}{2}\\\\
		\frac{1}{2}\\\\
		\frac{1}{2}
	\end{array}%
	\right)_{4\times 1} =\frac{\vert 0\rangle \vert 0\rangle + \vert 0\rangle \vert 1\rangle + \vert 1\rangle \vert 0\rangle + \vert 1\rangle \vert 1\rangle}{2}$. Similarly we can find $\vert \psi\rangle^{\otimes 3}$. \\\\ It is to be noted from the above calculation that when we consider two matrices $A$ and $B$ where $A = \left(%
	\begin{array}{cccc}
		A_{11} & A_{12} & \cdots & A_{1n}\\\\
		A_{21} & A_{22} & \cdots & A_{2n}\\\\
		\cdots & \cdots & \cdots & \cdots\\\\
		A_{m1} & A_{m2} & \cdots & A_{mn}
	\end{array}%
	\right)$ then 
	\begin{eqnarray}
	\label{tensorproduct}
	A\otimes B = \left(%
	\begin{array}{cccc}
		A_{11}B & A_{12}B & \cdots & A_{1n}B\\\\
		A_{21}B & A_{22}B & \cdots & A_{2n}B\\\\
		\cdots & \cdots & \cdots & \cdots\\\\
		A_{m1}B & A_{m2}B & \cdots & A_{mn}B
	\end{array}%
	\right).
	\end{eqnarray}
	\textbf{Problem $1.37$:} Calculate the matrix representation of the tensor products of the Pauli operators (a) $\sigma_{x}$ and $\sigma_{z}$, (b) $I$ and $\sigma_{x}$ and (c) $\sigma_{x}$ and $I$. Is tensor product commutative?\\\\
	\textbf{Note $16$:} It is to be remembered that (a) the tensor product of two unitary operators is a unitary operator, (b) the tensor product of two Hermitian operators is a Hermitian operator, (c) the tensor product of two positive operators is a positive operator and (d) the tensor product of two projectors is a projector.\\\\
	\textbf{Note $17$:} The following results are also important concerning the tensor product of two linear operators $A$ and $B$. 
	\begin{eqnarray}
	\label{tensorproduct1}
	(A\otimes B)^{*} = A^{*}\otimes B^{*}\\
	(A\otimes B)^{T} = A^{T}\otimes B^{T}\\
	(A\otimes B)^{\dagger} = A^{\dagger}\otimes B^{\dagger}
	\end{eqnarray}
	Tensor product often helps to describe the joint physical system in the following way. Suppose Alice has a qubit $\frac{\vert 0_{A}\rangle +\vert 1_{A}\rangle}{\sqrt{2}}$ in her possession and Bob has another qubit $\frac{\vert 0_{B}\rangle +\vert 1_{B}\rangle}{\sqrt{2}}$ in his possession and that Alice and Bob are spatially separated from one another\footnote{By spatial separation we actually mean that if Alice is situated in City A then Bob is situated in City B where City A and City B are some distance apart.}. Now if the qubits of Alice and Bob are combined together somehow, then the joint quantum mechanical state shared by Alice and Bob will be given by $\lbrace\frac{\vert 0_{A}\rangle +\vert 1_{A}\rangle}{\sqrt{2}}\rbrace\otimes \lbrace\frac{\vert 0_{B}\rangle +\vert 1_{B}\rangle}{\sqrt{2}}\rbrace$ which can also be written as $\frac{\vert 0_{A}\rangle\vert 0_{B}\rangle+\vert 0_{A}\rangle\vert 1_{B}\rangle+\vert 1_{A}\rangle \vert 0_{B}\rangle +\vert 1_{A}\rangle \vert 1_{B}\rangle}{2}$. This is a two qubit state. There are some joint quantum mechanical two qubit states which are of special importance. These states are given as
	\begin{eqnarray}
	\label{bellstates}
	\vert \phi^{+}\rangle = \frac{1}{\sqrt{2}}\Big\lbrace \vert 0_{A}\rangle \vert 0_{B}\rangle + \vert 1_{A}\rangle \vert 1_{B}\rangle\Big\rbrace.\nonumber\\
	\vert \phi^{-}\rangle = \frac{1}{\sqrt{2}}\Big\lbrace \vert 0_{A}\rangle \vert 0_{B}\rangle - \vert 1_{A}\rangle \vert 1_{B}\rangle\Big\rbrace.\nonumber\\
	\vert \varphi^{+}\rangle = \frac{1}{\sqrt{2}}\Big\lbrace \vert 0_{A}\rangle \vert 1_{B}\rangle + \vert 1_{A}\rangle \vert 0_{B}\rangle\Big\rbrace.\nonumber\\
	\vert \varphi^{-}\rangle = \frac{1}{\sqrt{2}}\Big\lbrace \vert 0_{A}\rangle \vert 0_{B}\rangle - \vert 1_{A}\rangle \vert 1_{B}\rangle\Big\rbrace.
	\end{eqnarray}\\
	The states of (\ref{bellstates}) are known as Bell states \cite{bell1988speakable} in the name of John Bell. An elaborate discussion on Bell states will follow in later sections. Bell states are also called maximally entangled pure states and play significant role in understanding the main content of this book.\\\\
	\textbf{Problem $1.38$:} The Hadamard operator on one qubit may be written as $\frac{1}{\sqrt{2}}\lbrace (\vert 0\rangle + \vert 1\rangle)\langle 0\vert + (\vert 0\rangle - \vert 1\rangle)\langle 1\vert \rbrace$. This is the outer product representation of the operator which can easily be shown to be same as equation (\ref{P5}).5 Find explicit matrix representation of $H^{\otimes 2}$. [Hints: Remember the fact that $\vert i\rangle\langle i\vert \otimes \vert j\rangle\langle j\vert = \vert ij\rangle\langle ij\vert$.] A quick glance tells us that $n-$ qubits Hadamard operator, denoted by $H^{\otimes n}$, is defined as \cite{nielsen2010quantum}
	\begin{eqnarray}
	\label{hadamard}
	H^{\otimes n} = \frac{1}{\sqrt{2^{n}}}\sum_{x=0}^{1}\sum_{y=0}^{1}(-1)^{x.y}\vert x\rangle\langle y\vert.
	\end{eqnarray}\\
\subsection{Density operator:}
Quantum mechanics can be formalized using a very useful mathematical tool known as \textit{density operator} approach or sometimes, is also called \textit{density matrix} approach.\footnote{Density operator approach and state vector approach are both used to describe quantum mechanics.} Being a convenient way, this density operator methodology is often used when the state of the quantum system is not completely known. Suppose a quantum system is in one of number of states $\vert \psi_{i}\rangle$ where $i \in \mathcal{I}$, (the notation $\mathcal{I}$ here stands for index set), with respective probabilities $p_{i}$ , then the collection $\lbrace p_{i}, \vert \psi_{i}\rangle\rbrace$ is called an \textbf{ensemble of pure states}. The density operator of such a system is then defined as
\begin{eqnarray}
\label{densityoperator1}
\rho = \sum_{i}p_{i}\vert \psi_{i}\rangle\langle \psi_{i}\vert.
\end{eqnarray}
Let us try to understand this density operator formulation with an example.\\\\
We know $\mathbb{C}^{2}$ denotes the two dimensional Hilbert space whose basis is given by $\lbrace \vert 0\rangle, \vert 1\rangle\rbrace$. We prepare, say, 1000 systems, where each member of the system can be in one of the two state vectors, either $\vert \psi_{1}\rangle = \alpha_{1}\vert 0\rangle + \beta_{1}\vert 1\rangle$ or $\vert \psi_{2}\rangle = \alpha_{2}\vert 0\rangle + \beta_{2}\vert 1\rangle$, (these state vectors are assumed to be normalized). We suppose that $500$ of these systems are in the state $\vert \psi_{1}\rangle$ and remaining $500$ is in the state $\vert \psi_{2}\rangle$. Then if we randomly select a member of the ensemble, the probability that it is found in state $\vert \psi_{1}\rangle$ is given by $\frac{500}{1000} = \frac{1}{2}$ and the probability that it is found in state $\vert \psi_{2}\rangle$ is $1-\frac{1}{2}$ i.e. $\frac{1}{2}$. It is very interesting to note here that both Born's rules of probability (let me call them quantum probability laws) and Kolmogorov rules (classical probability axioms) play a vital role in this density matrix mechanism.\footnote{We shall use the terms, density operator and density matrix interchangeably}. 
\begin{itemize}
\item With respect to single quantum system Born's rule gives us the probability (that the square of the modulus of the probability amplitudes) of obtaining a given measurement result.($|\alpha_{i}|^{2}$ and $|\beta_{i}|^{2}$, $i=1,2$ are such probabilities).
\item With respect to the ensemble level, while if we draw a member of the ensemble, there is a certain probability that the system is prepared in one state vector or another. At the ensemble level, however, the use of probability is acting in a classical way. This reflects that, at the ensemble level, we have a simple statistical mixture which is equivalent to saying that we have incomplete information about the quantum system. (In the above example $\frac{1}{2}$ is such a probability).
\end{itemize}
Thus to describe a system like where there are classical probabilities of finding each member of the system in different states, one needs to calculate usual quantities, such as the expectation values of operators and the probabilities of obtaining different measurement results. In case of statistical mixture of states, the calculated quantities are weighted by the probabilities of finding different states. This is what is called the \textit{density operator machinery}\cite{steeb1998quantum}.\\\\
Using equation (\ref{densityoperator1}) and considering our example above, if we try to give a density operator formalism to the quantum system prepared in a statistical mixture of $\vert \psi_{1}\rangle$ and $\vert \psi_{2}\rangle$, then we get
\begin{eqnarray}
\label{densityoperator2}
\rho &=& \sum_{i=1}^{2}\frac{1}{2}\vert \psi_{i}\rangle\langle \psi_{i}\vert\nonumber\\
&=& \left(%
\begin{array}{cc}
	\frac{\vert \alpha_{1}\vert^{2}+\vert \alpha_{2}\vert^{2}}{2} & \frac{\alpha_{1}\beta_{1}^{*}+\alpha_{2}\beta_{2}^{*}}{2}\\
	\frac{\alpha_{1}^{*}\beta_{1}+\alpha_{2}^{*}\beta_{2}}{2} & \frac{\vert \beta_{1}\vert^{2}+\vert \beta_{2}\vert^{2}}{2}\\
\end{array}%
\right).
\end{eqnarray}\\
However, if the quantum system is found to be in a definite known state $\vert \psi_{1}\rangle$ (or  $\vert \psi_{2}\rangle$) then density matrix of the quantum system is given as
\begin{eqnarray}
\label{densityoperator3}
\rho_{1} &=& \vert \psi_{1}\rangle\langle \psi_{1}\vert\nonumber\\
&=& \left(%
\begin{array}{cc}
	\vert \alpha_{1}\vert^{2} & \alpha_{1}\beta_{1}^{*}\\
	\alpha_{1}^{*}\beta_{1} & \vert \beta_{1}\vert^{2}\\
\end{array}%
\right), ~~~
\end{eqnarray}
\begin{eqnarray}
\label{densityoperator4}
\rho_{2} &=& \vert \psi_{2}\rangle\langle \psi_{2}\vert\nonumber\\
&=& \left(%
\begin{array}{cc}
	\vert \alpha_{2}\vert^{2} & \alpha_{2}\beta_{2}^{*}\\
	\alpha_{2}^{*}\beta_{2} & \vert \beta_{2}\vert^{2}\\
\end{array}%
\right).
\end{eqnarray}
When a system is in a definite state like (\ref{densityoperator3}) or (\ref{densityoperator4}) then the quantum system is said to be in a \textbf{pure state}\footnote{The term \textit{pure state} is often used in reference to a state vector $\vert\psi\rangle$, to distinguish it from a density operator $\rho$.} whereas the equation (\ref{densityoperator2}) and (equation (\ref{densityoperator1}) in general) are called \textit{ensemble of pure states}.\\\\
More precisely, if we consider a system that is in some known state $\vert \psi\rangle$, defined in some Hilbert space having orthonormal basis vectors $\lbrace \vert u_{i}\rangle\rbrace_{i=1}^{n}$, then $\vert \psi\rangle  = c_{1}\vert u_{1}\rangle + c_{2}\vert u_{2}\rangle +  \cdots +c_{n}\vert u_{n}\rangle$, ($c_{i}$'s are scalars from complex field). The probability of finding the system in some state $\vert u_{i}\rangle$ upon measurement is given by $|c_{i}|^{2}$, by Born's rule. When a system is in a definite state like this, we say that the system is in a \textit{pure state}.\\\\
In this respect one particular point is worth mentioned. Suppose there be two pure states $\vert 0\rangle$ and $\vert 1\rangle$ and a quantum system is prepared in state $\vert 0\rangle$ with probability $\frac{2}{3}$ and in state $\vert 1\rangle$ with probability $\frac{1}{3}$ so that the density matrix is given by\\ 
\begin{eqnarray}
\label{2by3}
\rho = \frac{2}{3}\vert 0\rangle\langle 0\vert + \frac{1}{3}\vert 1\rangle\langle 1\vert.
\end{eqnarray}\\
We can interpret this as the ensemble $\lbrace \frac{2}{3}, \vert 0\rangle, \frac{1}{3}, \vert 1\rangle\rbrace$ giving rise to the density matrix $\rho$. Now again we consider two states $\vert \alpha\rangle$ and $\vert \beta\rangle$ which are defined as
\begin{eqnarray}
\label{alphabeta}
\vert \alpha\rangle &=& \sqrt{\frac{2}{3}}\vert 0\rangle + \sqrt{\frac{1}{3}}\vert 1\rangle\nonumber\\
\vert \beta \rangle &=&  \sqrt{\frac{2}{3}}\vert 0\rangle - \sqrt{\frac{1}{3}}\vert 1\rangle
\end{eqnarray} 
Suppose a quantum system is now prepared in the state $\vert \alpha\rangle$ with probability $\frac{1}{2}$ and in the state $\vert 1\rangle$ with probability $\frac{1}{2}$. The density matrix of such a system is then given as
\begin{eqnarray}
\label{alphabeta1}
\varrho &=& \frac{1}{2}\vert \alpha\rangle\langle \alpha\vert + \frac{1}{2}\vert \beta\rangle\langle \beta\vert\nonumber\\
&=& \frac{2}{3}\vert 0\rangle\langle 0\vert + \frac{1}{3}\vert 1\rangle\langle 1\vert.
\end{eqnarray}\\
We see that the ensemble $\lbrace \frac{1}{2}, \vert \alpha\rangle, \frac{1}{2}, \vert\beta\rangle\rbrace$ gives rise to the density matrix $\varrho$ as shown in equation (\ref{alphabeta1}). A careful observation of the equations (\ref{2by3}) and (\ref{alphabeta1}), however, discloses that the density matrices $\rho$ and $\varrho$ have the same form. This indicates that two different ensembles of quantum states may give rise to the same density matrix.\\\\
\textbf{Problem $1.39$:} For a two dimensional Hilbert space $\mathbb{C}^{2}$ we consider the set $\lbrace \vert \phi^{+}\rangle, \vert \phi^{-}\rangle, \vert \varphi^{+}\rangle, \vert \varphi^{-}\rangle\rbrace$ as Bell basis where the states $\vert \phi^{+}\rangle,\: \vert \phi^{-}\rangle,\: \vert \varphi^{+}\rangle,\: \vert \varphi^{-}\rangle$ have already been defined in equation (\ref{bellstates}). Verify that Bell basis forms an orthonormal basis for a two qubit state space. Now suppose a quantum system is prepared in any one of the Bell states, then what will be the density matrix representation for the quantum system? (Do it for each of the four cases). If, however, a quantum system is prepared in ensemble of four Bell states (or in other words in statistical mixtures of four Bell states) what will be the density operator representation of such a quantum system? [You may consider the probabilities $p_{\phi^{+}},\: p_{\phi^{-}},\:p_{\varphi^{+}},\:p_{\varphi^{-}}$  for each of the four Bell states $\vert \phi^{+}\rangle,\: \vert \phi^{-}\rangle,\: \vert \varphi^{+}\rangle,\: \vert \varphi^{-}\rangle$ respectively.]\\\\
\textit{Hints:} [Second Part: Construct $p_{\phi^{+}}\vert \phi^{+}\rangle\langle \phi^{+}\vert + p_{\phi^{-}}\vert \phi^{-}\rangle\langle \phi^{-}\vert + p_{\varphi^{+}}\vert \varphi^{+}\rangle\langle \varphi^{+}\vert + p_{\varphi^{-}}\vert \varphi^{-}\rangle\langle \varphi^{-}\vert$ in the matrix form.]\\\\
Careful observations of the above density matrices reveal two characterizations of the density operators which have been stated in the following as a statement of a theorem.\\\\
\textbf{Theorem:} An operator $\rho$ is the density operator associated to some ensemble $\lbrace p_{i},\vert \psi_{i}\rangle\rbrace$ if and only if it satisfies the conditions: (a) Trace of $\rho$ is equal to one [this condition is known as \textit{trace condition}] \footnote{The condition that trace of a density operator is equal to $1$ is equivalent to the total probability of the ensemble being equal to $1$.}, (b) $\rho$ is a positive operator [the condition is known as \textit{positivity condition}]\footnote{If $\lambda_{i}$'s are the eigenvalues of a positive operator then, $\lambda_{i}>0$.}.\\\\
\textit{Proof:} \textbf{Necessity:} Let $\rho$ be the density operator which is given by $\rho = \sum_{i} p_{i}\vert \psi_{i}\rangle\langle \psi_{i}\vert$ for a given ensemble $\lbrace p_{i},\vert \psi_{i}\rangle\rbrace$.\\\\ (a) Then
\begin{eqnarray}
\label{charactden1}
Tr(\rho) &=& \sum_{i}p_{i}Tr(\vert \psi_{i}\rangle\langle \psi_{i}\vert)\nonumber\\ &=& \sum_{i} p_{i}\langle \psi_{i}\vert \psi_{i}\rangle = \sum_{i}p_{i} = 1.
\end{eqnarray}\\\\ 
(b) Let $\vert \varphi\rangle$ be an arbitrary vector in the state space. Then we have 
\begin{eqnarray}
\label{charactden2}
\langle \varphi \vert \rho \vert \varphi\rangle &=& \langle \varphi \vert \sum_{i} p_{i} \vert \psi_{i}\rangle\langle \psi_{i}\vert \varphi\rangle\nonumber\\ &=&  \sum_{i} p_{i}  \langle \varphi \vert \psi_{i}\rangle\langle \psi_{i}\vert \varphi\rangle\nonumber\\ &=& \sum_{i} p_{i}  \langle \varphi \vert \psi_{i}\rangle\langle \varphi \vert \psi_{i} \rangle^{*} = \sum_{i} p_{i} \vert \langle \varphi \vert \psi_{i} \rangle \vert^{2} \geq 0.
\end{eqnarray}\\\\
\textbf{Sufficiency:} Conversely let $\rho$ be any operator which satisfies the conditions (a) and (b) as given in the statement. Now, since $\rho$ is positive operator so it has a spectral decomposition
\begin{eqnarray}
\label{charactden3}
\rho = \sum_{k}\lambda_{k}\vert k\rangle\langle k\vert,
\end{eqnarray}
where the vectors $\vert k\rangle$ are orthogonal, and $\lambda_{k}$ are real, non-negative eigenvalues of $\rho$. Thus from the trace condition (a), we get $\sum_{k} \lambda_{k} = 1$. Hence a system in state $\vert k\rangle$ with probability $\lambda_{k}$ will have density operator $\rho$. Hence, the ensemble $\lbrace \lambda_{k}, \vert k\rangle\rbrace$ is an ensemble of states giving rise to the density operator $\rho$.    ~~~~~~~~~~~~~~~~~~~~~~~~~~~~~~~~~~~~~~~~~~~~~~~~~~~~~~~~~~~~~~~~~~~~~~~~~~~~~~~~~~~~~~~~~~~~~~~ \textbf{Q. E. D}. \\\\
We have already known that a positive operator is necessarily hermitian (see section $1.2.8$) . This means that if $A$ is a positive operator then $A$ is hermitian. But of course an operator which is not hermitian is definitely not positive\footnote{A statement $P$ implies a statement $Q$. Then negation of statement $Q$ also implies negation of $P$.}.  For example if we suppose a matrix $H_{1}=\left(%
\begin{array}{cc}
	\frac{1}{4} & \frac{1-i}{4}\\\\
	\frac{1-i}{4} & \frac{3}{4}\\\\
\end{array}%
\right)$ then we can see that $H_{1}^{\dagger}=\left(%
\begin{array}{cc}
	\frac{1}{4} & \frac{1+i}{4}\\\\
	\frac{1+i}{4} & \frac{3}{4}\\\\
\end{array}%
\right)\ne H_{1}$ implying the fact that $H_{1}$ is not hermitian and so is not positive. Although $Tr(H_{1})=1$, which is the first criteria for a matrix to be a density matrix, the second criteria that $H_{1}$ must also be a positive operator, fails. Consequently it implies that $H_{1}$ cannot be density matrix.\cite{steeb1998quantum}\\\\
So far we have known that a quantum system whose state $\vert \psi\rangle$ is known exactly is said to be in a \textit{pure state}. In this case the density operator $\rho$ can simply be expressed as $\rho = \vert \psi\rangle\langle \psi\vert$. Otherwise $\rho$ is in a \textbf{mixed state}, that is, in a state which is a mixture of different pure states in the ensemble for $\rho$. For a pure state $\rho$ we have $Tr(\rho^{2})=1$. This happens only when $\rho^{2}=\rho$. On the other hand if $\rho$ is a mixed state then $Tr(\rho^{2})<1$.\\\\
In the context of mixed state one can study the situation where not all of the $N$ systems of the ensemble are in the same state that is $N_{i}$ systems are in the state $\vert \psi_{i}\rangle$ respectively, such that $\sum_{i}N_{i}=N$. The probability $p_{i}$ to find an individual system of the ensemble described by the state $\vert \psi_{i}\rangle$ is given by $p_{i}=\frac{N_{i}}{N}$ where $\sum_{i}p_{i}=1$. We can thus write down the mixed state as a convex sum of pure state density matrices as 
\begin{eqnarray}
\label{mixedstate1}
\rho_{mix} = \sum_{i}p_{i}\rho_{i}^{pure} = \sum_{i}p_{i}\vert \psi_{i}\rangle\langle \psi_{i}\vert.
\end{eqnarray} 
The expectation value is however given by 
\begin{eqnarray}
\label{mixedstate2}
\langle A\rangle_{\rho_{mix}} = Tr(\rho_{mix}A).
\end{eqnarray}
Here we can express the expectation value of mixed state as a convex sum of expectation values of it's constituent pure states. Then from eqs.(\ref{mixedstate1})-(\ref{mixedstate2}) (and also taking references from eq.(\ref{exp})) we get 
\begin{eqnarray}
\label{mixedstate3}
\langle A\rangle_{\rho_{mix}} = \sum_{i}p_{i}\langle \psi_{i}\vert A\vert \psi_{i}\rangle.
\end{eqnarray}
\textit{Proof:}
\begin{eqnarray}
\label{rho2<1a}
Tr(\rho_{mix}A) &=& Tr\Bigg[\sum_{i}p_{i}\vert \psi_{i}\rangle\langle \psi_{i}\vert A\Bigg]\nonumber\\
&=& \sum_{j}\sum_{i}p_{i}\langle \psi_{j}\vert \psi_{i}\rangle\langle \psi_{i}\vert A\vert \psi_{j}\rangle\nonumber\\
&=& \sum_{j}\sum_{i}p_{i}\langle \psi_{i}\vert A\vert \psi_{j}\rangle\langle \psi_{j}\vert \psi_{i}\rangle\nonumber\\
&=& \sum_{i}p_{i}\langle \psi_{i}\vert A\sum_{j}\vert \psi_{j}\rangle\langle \psi_{j}\vert \psi_{i}\rangle \nonumber\\
&=& \sum_{i}p_{i}\langle \psi_{i}\vert A\vert \psi_{i}\rangle.
\end{eqnarray}
\\\\
Again $\rho_{mix}^{2}=\sum_{i}\sum_{j}p_{i}p_{j}\vert \psi_{i}\rangle\langle \psi_{i}\vert \psi_{j}\rangle\langle \psi_{j}\vert = \sum_{i}p_{i}^{2}\vert \psi_{i}\rangle\langle \psi_{i}\vert \ne \rho_{mix}$. Without any loss of generality we have assumed here that $\vert \psi_{i}\rangle$ and $\vert \psi_{j}\rangle$ are orthonormal. Now we shall calculate the quantity $Tr(\rho^{2})$ for mixed state $\rho_{mix}$.
\begin{eqnarray}
\label{mixedstate4}
Tr(\rho_{mix}^{2}) &=& \sum_{k}\langle \psi_{k}\vert \sum_{i} \sum_{j}p_{i}p_{j}\vert \psi_{i}\rangle\langle \psi_{i}\vert \psi_{j}\rangle\langle \psi_{j}\vert \psi_{k}\rangle \nonumber\\
&=& \sum_{i}\sum_{j}p_{i}p_{j}\langle \psi_{i}\vert \psi_{j}\rangle\langle \psi_{j}\vert \sum_{k}\vert \psi_{k}\rangle\langle \psi_{k}\vert\psi_{i}\rangle \nonumber\\
&=& \sum_{i}\sum_{j}p_{i}p_{j}\vert \langle \psi_{i}\vert \psi_{j}\rangle\vert^{2}\nonumber\\
&=& \sum_{i}p_{i}^{2} < \sum_{i}p_{i} = 1, ~~~~~~~~ 0 \leq p_{i} \leq 1,~~p_{i}^{2}\leq p_{i}.
\end{eqnarray}
This proves that for any mixed state $\rho$ we have $Tr(\rho^{2})<1$.~~~~~~\textbf{Q.E.D}\\\\ We conclude that the trace of $\rho^{2}$ is a good measure for the mixedness of a density matrix, since it is equal to $1$ for pure states and strictly smaller than $1$ for mixed states. For a \textbf{maximally mixed state}, we have for a given dimension $d$ of the system
\begin{eqnarray}
\label{mixedstate5}
Tr(\rho_{mix}^{2}) = \frac{1}{d}>0.
\end{eqnarray}
\\\\
\textbf{Problem $1.40$:} Suppose we want to prepare a quantum system which can either be in state $\vert \phi^{+}\rangle$ or in state $\vert \phi^{-}\rangle$. Also assume that the system being in these two states are equally probable. What will be the density matrix representation of the quantum system. On the other hand if the system can be found in state $\vert \phi^{+}\rangle$ only (say), what will then be the density matrix representation of the system? Try to give a physical interpretation of it.\\\\
\textit{Solution:} We assume that whenever we try to prepare a quantum system then this system is going to be in either the state $\vert \phi^{+}\rangle = \frac{\vert 0\rangle\vert 0\rangle + \vert 1\rangle \vert 1\rangle}{\sqrt{2}}$ or in $\vert \phi^{-}\rangle=\frac{\vert 0\rangle \vert 0\rangle - \vert 1\rangle \vert 1\rangle}{\sqrt{2}}$. And if we assume that system behaves in an unbiased fashion so as to get these states with equal probability, we consequently can assume an ensemble $\lbrace \frac{1}{2}, \vert \phi^{+}\rangle, \frac{1}{2}, \vert \phi^{-}\rangle\rbrace$. Since Bell states are pure entangled states so this ensemble that we have obtained is an ensemble of pure states. The density operator associated with the states therefore is given by
\begin{eqnarray}
\label{problem38a}
\rho &=& \frac{1}{2}\vert \phi^{+}\rangle\langle \phi^{+}\vert + \frac{1}{2} \vert \phi^{-}\rangle\langle \phi^{-}\vert \nonumber\\
&=& \frac{\vert 00\rangle\langle 00\vert + \vert 11\rangle\langle 11\vert}{2}\nonumber\\
&=& \frac{\vert 0\rangle\langle 0\vert \otimes \vert 0\rangle\langle 0\vert + \vert 1\rangle\langle 1\vert \otimes \vert 1\rangle\langle 1\vert}{2}\nonumber\\
&=& \left(%
\begin{array}{cccc}
	\frac{1}{2} & 0 & 0 & 0\\
    0 &  0 & 0 & 0\\
    0 &  0 & 0 & 0\\
    0 &  0 & 0 & \frac{1}{2}\\
\end{array}%
\right).
\end{eqnarray}
This implies that 
\begin{eqnarray}
\label{problem38b}
\rho^{2} &=&  \left(%
\begin{array}{cccc}
	\frac{1}{4} & 0 & 0 & 0\\
    0 &  0 & 0 & 0\\
    0 &  0 & 0 & 0\\
    0 &  0 & 0 & \frac{1}{4}\\
\end{array}%
\right).
\end{eqnarray}
Hence $Tr(\rho^{2})=\frac{1}{4}+\frac{1}{4}=\frac{1}{2}<1$. Therefore the system (which is an ensemble of pure states) is actually in a mixed state. On the other hand if the system can only be in either of the Bell states (say only in state $\vert \phi^{+}\rangle$) then a similar calculation will reveal that $Tr(\rho^{2})=1$ that is the system is in pure state.\\\\ 
In section $1.3.6$ we stated the postulates of quantum mechanics. The postulates $1$, $2$, $4$ and $5$ can also be restated in terms of density operator approach.
\begin{itemize}
\item \textit{Postulate $1$:} Associated to any isolated physical system is a finite dimensional complex vector space with inner product (that is, a Hilbert space) known as the state space of the system. The system is completely described by its density operator, which is a positive operator $\rho$ with trace one, acting on the state space of the system. If a quantum system is in the state $\rho_{i}$ with probability $p_{i}$, then the density operator for the system is $\sum_{i}p_{i}\rho_{i}$ \cite{nielsen2010quantum}.
\item \textit{Postulate $2$:} The evolution of a closed quantum system is described by a unitary transformation. That is, the state $\rho$ of the system at time $t_{1}$ is related to the state $\rho^{\prime}$ of the system at time $t_{2}$ by a unitary operator $U$ which depends only on the times $t_{1}$ and $t_{2}$, and $\rho^{\prime} = U\:\rho\:U^{\dagger}$ \cite{nielsen2010quantum}.
\item \textit{Postulate $4$:} Quantum measurements are described by a collection $\lbrace M_{m}\rbrace$ of measurement operators. These are operators acting on the state space of the system being measured. The index $m$ refers to the measurement outcomes that
may occur in the experiment. If the state of the quantum system is $\rho$ immediately before the measurement then the probability that result $m$ occurs is given by $p(m) = Tr(M_{m}^{\dagger}M_{m}\rho)$, and the state of the system after the measurement is     $\frac{M_{m}\rho\:M_{m}^{\dagger}}{Tr(M_{m}^{\dagger}M_{m}\rho)}$. The measurement operators satisfy the completeness equation $\sum_{m}M_{m}^{\dagger}M_{m} = I$ \cite{nielsen2010quantum}.
\item The state space of a composite physical system is the tensor product of the state spaces of the component physical systems. Moreover, if we have systems numbered $1$ through $n$, and system number $i$ is prepared in the state $\rho_{i}$, then the joint state of the total system is $\rho_{1}\otimes \rho_{2}\otimes \cdots \otimes \rho_{n}$ \cite{nielsen2010quantum}.
\end{itemize}
We know from postulate $4$ of section $1.3.6$ that if we have a quantum system in the state $\vert \psi\rangle$, then the probability of measurement outcome $m$ when measurement operators $M_{m}$ act on the state $\vert \psi\rangle$ is given by $p(m) = \langle \psi\vert M_{m}^{\dagger}M_{m}\vert \psi\rangle$. Let the operator $M_{m}^{\dagger}M_{m}$ be denoted by $P_{vm}$. Then from completeness relation (\ref{measurement3}) we can immediately say $\sum_{m}P_{vm} = I$ and $p(m) = \langle \psi\vert P_{vm}\vert \psi\rangle$. The operators $P_{vm}$ are \textbf{positive operator valued measure} (abbreviated as POVM). The collection $\lbrace P_{vm}\rbrace$ is sufficient to determine the probabilities of the different measurement outcomes.
\subsection{Reduced Density operator:}
The deepest application of the density operator is as a descriptive tool for sub-systems of a composite quantum system. Such a description is provided by the \textbf{reduced density operator}. For two physical systems $A$ and $B$, whose combined state is described by a density operator (say) $\rho^{AB}$, the \textit{reduced density operator} for system $A$ (or system $B$) is defined by
\begin{eqnarray}
\label{reduceddensityoperator}
\rho^{A}=Tr_{B}(\rho^{AB}),\:\: \rho^{B}=Tr_{A}(\rho^{AB}).
\end{eqnarray}
$Tr_{B}$ (or $Tr_{A}$) is a map of operators known as the \textbf{partial trace} over the system $B$ (or $A$). The reduced density operators $\rho^{A}$ and $\rho^{B}$ can be explicitly written as
\begin{eqnarray}
\label{reddensop}
\rho^{A} &=& \sum_{j=1}^{n_{B}}(I_{A}\otimes \langle \phi_{j}\vert)\rho^{AB}(I_{A}\otimes \vert \phi_{j}\rangle)\nonumber\\
\rho^{B} &=& \sum_{k=1}^{n_{A}}(\langle \psi_{k}\vert \otimes I_{B})\rho^{AB}(\vert \psi_{k}\rangle \otimes I_{B}).
\end{eqnarray}
Here $I_{A}$ and $I_{B}$ are the identity operators in $H_{A}$ and $H_{B}$. Here $H_{A}$ and $H_{B}$ are the Hilbert spaces corresponding to the systems $A$ and $B$ respectively and $H_{AB}$ is the Hilbert space of the joint quantum mechanical system $AB$. Also $\vert \phi_{j}\rangle$, ($j=1,\:2,\:\cdots ,n_{B}$) is an orthonormal basis in $H_{B}$. Similarly $\vert \psi_{k}\rangle$, ($k=1,\:2,\:\cdots ,n_{A}$) is an orthonormal basis in $H_{A}$. Also $n_{A}$ and $n_{B}$ are the dimensions of Hilbert spaces $H_{A}$ and $H_{B}$ respectively.\\\\
However, if a state $\rho^{AB}$ is \textbf{separable} then we can always express $\rho^{AB}$ as the tensor product of the sub-systems $\rho_{A}$ and $\rho_{B}$\footnote{Remember that the notation $\rho_{A}$ stands for state of the subsystems $A$ while $\rho^{A}$ stands for reduced density operator of the subsystem $A$.}. Mathematically,
\begin{eqnarray}
\label{seperablereduced1}
\rho^{AB} = \rho_{A} \otimes \rho_{B}.
\end{eqnarray}
Using eqs.(\ref{reddensop}) one immediately obtains 
\begin{eqnarray}
\rho^{A} &=& \sum_{j=1}^{n_{B}}(I_{A}\otimes \langle \phi_{j}\vert)(\rho_{A}\otimes \rho_{B})(I_{A}\otimes \vert \phi_{j}\rangle)\nonumber\\
&=& \sum_{j=1}^{n_{B}}(I_{A}\otimes \langle \phi_{j}\vert)(\rho_{A} I_{A} \otimes \rho_{B}\vert \phi_{j}\rangle)\nonumber\\
&=& \sum_{j=1}^{n_{B}}(I_{A}\otimes \langle \phi_{j}\vert)(\rho_{A} \otimes \rho_{B}\vert \phi_{j}\rangle)\nonumber\\
&=& \sum_{j=1}^{n_{B}}(I_{A}\rho_{A} \otimes \langle \phi_{j}\vert\rho_{B}\vert \phi_{j}\rangle)\nonumber\\
&=& \sum_{j=1}^{n_{B}}(\rho_{A} \otimes \langle \phi_{j}\vert\rho_{B}\vert \phi_{j}\rangle)\nonumber\\
&=& \rho_{A} \otimes \sum_{j=1}^{n_{B}}\langle \phi_{j}\vert\rho_{B}\vert \phi_{j}\rangle\nonumber\\
&=& \rho_{A}, ~~~~~~ \sum_{j=1}^{n_{B}}\langle \phi_{j}\vert\rho_{B}\vert \phi_{j}\rangle =1.
\end{eqnarray}
Here $\sum_{j=1}^{n_{B}}\langle \phi_{j}\vert\rho_{B}\vert \phi_{j}\rangle$ is the sum of the diagonal elements of the density matrix $\rho_{B}$. Similarly one can also show that $\rho^{B}= \rho_{B}$ when $\rho^{AB}$ is separable. [In the above calculation property (\ref{property4b}) has also been used.\\\\
To understand the mechanism of partial trace we consider one of the Bell states from the equation (\ref{bellstates}) and let it be $\vert \varphi^{-}\rangle = \frac{1}{\sqrt{2}}(\vert 01\rangle_{AB}-\vert 10\rangle_{AB})$. Here the suffix $AB$ denotes that it is a two qubit system as well as the first qubit corresponds to system $A$ whereas the second qubit is with system $B$. The density operator corresponding to the state vector $\vert \varphi^{-}\rangle$ is\footnote{Here I have used the notation $\vert ab\rangle\langle ba\vert$ for $\vert a\rangle\langle b\vert \otimes \vert b\rangle\langle a\vert$.}
\begin{eqnarray}
\label{partialtrace1}
\rho^{AB} &=& \vert \varphi^{-}\rangle_{AB}\langle \varphi^{-} \vert = \frac{1}{\sqrt{2}}\Big\lbrace \vert 0_{A}1_{B}\rangle-\vert 1_{A}0_{B}\rangle\Big\rbrace \frac{1}{\sqrt{2}}\Big\lbrace \langle 0_{A}1_{B}\vert - \langle 1_{A}0_{B}\vert\Big\rbrace \nonumber\\
&=& \frac{1}{2}\Big\lbrace \vert 0_{A}1_{B}\rangle\langle 0_{A}1_{B}\vert - \vert 0_{A}1_{B}\rangle\langle 1_{A}0_{B}\vert - \vert 1_{A}0_{B}\rangle\langle 0_{A}1_{B}\vert + \vert 1_{A}0_{B}\rangle\langle 1_{A}0_{B}\vert\Big\rbrace\nonumber\\
&=& \left(%
	\begin{array}{cccc}
		0 & 0 & 0 & 0\\\\
		0 & \frac{1}{2} & -\frac{1}{2} & 0\\\\
	    0 & -\frac{1}{2} & \frac{1}{2} & 0\\\\
		0 &   0 & 0 & 0
	\end{array}%
	\right).
\end{eqnarray}\\
Now if we want to find the density matrix corresponding to the quantum mechanical subsystem $A$ of the composite system $AB$ we proceed as follows.
\begin{eqnarray}
\label{partialtrace2}
\rho^{A} &=& \frac{1}{2}\:Tr_{B}\Big\lbrace \vert 0_{A}1_{B}\rangle\langle 0_{A}1_{B}\vert - \vert 0_{A}1_{B}\rangle\langle 1_{A}0_{B}\vert - \vert 1_{A}0_{B}\rangle\langle 0_{A}1_{B}\vert + \vert 1_{A}0_{B}\rangle\langle 1_{A}0_{B}\vert\big\rbrace\nonumber\\
&=& \frac{1}{2} \Big\lbrace\: \vert 0_{A}\rangle\langle 0_{A}\vert \langle 1_{B}\vert 1_{B}\rangle - \vert 0_{A}\rangle\langle 1_{A}\vert \langle 1_{B}\vert 0_{B}\rangle - \vert 1_{A}\rangle\langle 0_{A}\vert \langle 0_{B}\vert 1_{B}\rangle +\vert 1_{A}\rangle\langle 1_{A}\vert \langle 0_{B}\vert 0_{B}\rangle\Big\rbrace\nonumber\\
&=& \frac{1}{2}\Big\lbrace \vert 0_{A}\rangle\langle 0_{A}\vert + \vert 1_{A}\rangle\langle 1_{A}\vert\Big\rbrace\nonumber\\
&=& \left(%
	\begin{array}{cc}
		\frac{1}{2} & 0 \\\\
		0 & \frac{1}{2}
	\end{array}%
	\right).
\end{eqnarray}
In a similar manner one can also find the density matrix of the quantum mechanical subsystem $B$ of the system $AB$ which is given by,

\begin{eqnarray}
\label{partialtrace2}
\rho^{B} &=& \frac{1}{2}\:Tr_{A}\Big\lbrace \vert 0_{A}1_{B}\rangle\langle 0_{A}1_{B}\vert - \vert 0_{A}1_{B}\rangle\langle 1_{A}0_{B}\vert - \vert 1_{A}0_{B}\rangle\langle 0_{A}1_{B}\vert + \vert 1_{A}0_{B}\rangle\langle 1_{A}0_{B}\vert \Big\rbrace\nonumber\\
&=& \frac{1}{2} \Big\lbrace\: \langle 0_{A}\vert 0_{A}\rangle\vert 1_{B}\rangle\langle 1_{B}\vert - \langle 0_{A}\vert 1_{A}\rangle\vert 1_{B}\rangle\langle 0_{B}\vert -\langle 1_{A}\vert 0_{A}\rangle\vert 0_{B}\rangle\langle 1_{B}\vert + \langle 1_{A}\vert 1_{A}\rangle\vert 0_{B}\rangle\langle 0_{B}\vert\Big\rbrace\nonumber\\
&=& \frac{1}{2}\Big\lbrace \vert 1_{A}\rangle\langle 0_{A}\vert + \vert 0_{A}\rangle\langle 1_{A}\vert\Big\rbrace\nonumber\\
&=& \left(%
	\begin{array}{cc}
		\frac{1}{2} & 0 \\\\
		0 & \frac{1}{2}
	\end{array}%
	\right).
\end{eqnarray}
\\\\
\textbf{Problem $1.41$:} Consider the Bell states $\vert \phi^{+}\rangle$, $\vert \phi^{-}\rangle$ and $\vert \varphi^{+}\rangle$ from the equation (\ref{bellstates}). These are two qubit entangled quantum mechanical systems also. Find the density matrix of subsystems $A$ and $B$ of the joint quantum mechanical system described by Bell states.\\\\
Pure states and mixed states are of contrasting nature. When a system is in a mixed state it means that we do not have maximal knowledge about that system. However it is not true for pure states. Above discussion on partial trace and the reduced density operator of Bell states reveal that the subsystems of the two qubit joint quantum mechanical system are all of the form $\frac{I}{2}$. Consequently we also notice that $Tr\lbrace(\frac{I}{2})^{2}\rbrace = Tr\left(%
	\begin{array}{cc}
		\frac{1}{2} & 0 \\\\
		0 & \frac{1}{2}
	\end{array}%
	\right)^{2}=\frac{1}{2}<\:1$. Hence the state $\frac{I}{2}$ is a mixed state. It is very interesting to note that although the Bell states are two qubit pure entangled states, the subsystems are mixed. This implies that we can have information of the joint system but the knowledge about the individual subsystems are not completely known. This is a remarkable characteristic trait of \textit{quantum entanglement}\cite{steeb1998quantum}.\\\\
\textbf{Problem $1.42$:} Suppose a composite of systems $A$ and $B$ is in the state $\vert a\rangle\vert b\rangle$ or $\vert ab\rangle$, where $\vert a\rangle$ is a pure state of system $A$, and $\vert b\rangle$ is a pure state of system $B$. Show that the reduced density operator of system $A$ alone is a pure state.\\\\
\textit{Solution:} We know that a quantum system whose state $\vert \psi\rangle$ is known exactly is said to be in a pure state. In this case the density operator is simply $\rho = \vert \psi\rangle\langle \psi\vert$. The given state $\vert a\rangle$ is a pure state and so the density operator is $\rho^{A} = \vert a\rangle\langle a\vert$ and the state $\vert b\rangle$ is also a pure state (as given) whose density matrix is given by $\rho^{B} = \vert b\rangle\langle b\vert$. The joint system is $\vert a\rangle \vert b\rangle$ and its density matrix is then represented as $\rho^{AB}=\vert a\rangle\vert b\rangle\langle a\vert \langle b\vert$. Then the reduced density matrix of the system $A$ is $Tr_{B}(\rho^{AB})$ which is calculated as $Tr_{B}(\vert ab\rangle\langle ab\vert)= \vert a\rangle\langle a\vert \langle b\vert b\rangle = \vert a\rangle\langle a\vert =\rho^{A}$. This proves the result.~~~~~~~~\textbf{Q.E.D}\\\\
At this point we would like to mention an interesting state which has significant applications in quantum information processing protocol. In later chapters we will elaborately discuss about the implications of such states. The state is constructed in a special way. Let us discuss this and we will also get to see how reduced density operator tool is applied in the construction of such a state.\\\\
Let us consider two tripartite (or $3$ party ) pure entangled state, one of which is famously known as $GHZ$ state\cite{greenberger1989bell}, which is defined as
\begin{eqnarray}
\label{ghz}
\vert GHZ\rangle_{ABC} = \frac{1}{\sqrt{2}}\Big(\vert 0_{A}0_{B}0_{C}\rangle + \vert 1_{A}1_{B}1_{C}\rangle\Big),
\end{eqnarray}
while another tripartite pure entangled state, known as $W-$ state \cite{dur2000three} is defined as below
\begin{eqnarray}
\label{w}
\vert W\rangle_{ABC} = \frac{1}{\sqrt{3}}\Big(\vert 1_{A}0_{B}0_{C}\rangle + \vert 0_{A}1_{B}0_{C}\rangle + \vert 0_{A}0_{B}1_{C}\rangle\Big).
\end{eqnarray}
The above states (\ref{ghz}) and (\ref{w}) are both pure maximally entangled but their characteristic features are different which will be discussed later. Also both of these states are useful in quantum information protocols. Using eq.(\ref{densityoperator1}) the density operators for the $GHZ$ and $W$  states are calculated. Suppose $\rho^{ABC}_{G}$ be the density operator for the $GHZ$ state and $\rho^{ABC}_{W}$ is the density operator for $W$ state. Then we get
\begin{eqnarray}
\label{ghz1}
\rho^{ABC}_{G} = \vert GHZ\rangle_{ABC}\langle GHZ\vert \nonumber\\
 = \frac{1}{2}\Big\lbrace \vert 0_{A}0_{B}0_{C}\rangle\langle 0_{A}0_{B}0_{C}\vert + \vert 0_{A}0_{B}0_{C}\rangle\langle 1_{A}1_{B}1_{C}\vert \nonumber\\ + \vert 1_{A}1_{B}1_{C}\rangle\langle 0_{A}0_{B}0_{C}\vert + \vert 1_{A}1_{B}1_{C}\rangle\langle 1_{A}1_{B}1_{C}\vert \Big\rbrace,
\end{eqnarray}
and 
\begin{eqnarray}
\label{w1}
\rho^{ABC}_{W} = \vert W\rangle_{ABC}\langle W\vert \nonumber\\
= \frac{1}{3} \lbrace \vert 1_{A}0_{B}0_{C}\rangle\langle 1_{A}0_{B}0_{C}\vert + \vert 1_{A}0_{B}0_{C}\rangle\langle 0_{A}1_{B}0_{C}\vert + \nonumber\\ \vert 1_{A}0_{B}0_{C}\rangle\langle 0_{A}0_{B}1_{C}\vert + \nonumber\\
+ \vert 0_{A}1_{B}0_{C}\rangle\langle 1_{A}0_{B}0_{C}\vert + \vert 0_{A}1_{B}0_{C}\rangle\langle 0_{A}1_{B}0_{C}\vert + \vert 0_{A}1_{B}0_{C}\rangle\langle 0_{A}0_{B}1_{C}\vert +\nonumber\\  \vert 0_{A}0_{B}1_{C}\rangle\langle 1_{A}0_{B}0_{C}\vert + \vert 0_{A}0_{B}1_{C}\rangle\langle 0_{A}1_{B}0_{C}\vert + \vert 0_{A}0_{B}1_{C}\rangle\langle 0_{A}0_{B}1_{C}\vert \rbrace.
\end{eqnarray}
Now if one finds the reduced density operators of the states $\rho^{ABC}_{G}$ and $\rho^{ABC}_{W}$ by tracing out party $C$ from the systems using eq.(\ref{partialtrace1}) then such states become mixed states which have the following forms, $\rho^{AB}_{G}$ is the reduced density operator corresponding to $GHZ$ state and $\rho^{AB}_{W}$ is the reduced density operator for the $W$ state. Thus we get
\begin{eqnarray}
\label{ghz2}
\rho^{AB}_{G}=  \left(%
\begin{array}{cccc}
	\frac{1}{2} & 0 & 0 & 0\\
    0 &  0 & 0 & 0\\
    0 &  0 & 0 & 0\\
    0 &  0 & 0 & \frac{1}{2}\\
\end{array}%
\right),
\end{eqnarray}
and 
\begin{eqnarray}
\label{w2}
\rho^{AB}_{W} &=&  \left(%
\begin{array}{cccc}
	\frac{1}{3} & 0 & 0 & 0\\
    0 &  \frac{1}{3} & \frac{1}{3} & 0\\
    0 &  \frac{1}{3} & \frac{1}{3} & 0\\
    0 &  0 & 0 & 0\\
\end{array}%
\right).
\end{eqnarray}
The author of this book, along with his other research collaborators had taken a convex combination of the states $\rho^{AB}_{G}$ and $\rho^{AB}_{W}$ i.e. $p\: \rho^{AB}_{G} + (1-p) \:\rho^{AB}_{W},  0 \leq p \leq 1$ \cite{adhikari2008teleportation}. We had shown that this state is a special class of \textit{non-maximally entangled mixed state} (NMEMS) with many interesting features and having its significant role in \textit{Quantum Teleportation}. Let us denote this state using the density operator form $\rho^{AB}_{nmems}$ and it is given as
\begin{eqnarray}
\label{roynmems}
\rho^{AB}_{nmems}=  \left(%
\begin{array}{cccc}
	\frac{p+2}{6} & 0 & 0 & 0\\
    0 &  \frac{1-p}{3} & \frac{1-p}{3} & 0\\
    0 &  \frac{1-p}{3} & \frac{1-p}{3} & 0\\
    0 &  0 & 0 & \frac{p}{2}\\
\end{array}%
\right).
\end{eqnarray}
Entanglement properties of this class of states will be discussed later.
\subsection{Schmidt decomposition, purification and entanglement:}
Two additional tools which are of great importance in quantum information processing science are the \textit{Schmidt decomposition} and \textit{purification}.\\\\
\textbf{Schmidt decomposition:} It is known that if two systems $A$ and $B$ are defined in the Hilbert spaces $H_{A}$ and $H_{B}$ respectively, then their composite system $AB$ is defined in the Hilbert space $H_{AB}=H_{A}\otimes H_{B}$. Any bipartite (or two party) pure state $\vert \psi\rangle_{AB}$ for the composite quantum system $AB$ is expressed as 
\begin{eqnarray}
\label{schmidtdecompose1}
\vert \psi\rangle_{AB} = \sum_{i,\:j}c_{ij}\vert i_{A}\rangle\vert j_{B}\rangle,
\end{eqnarray}
where $\lbrace \vert i_{A}\rangle\rbrace$ and $\lbrace \vert j_{B}\rangle\rbrace$ are the orthonormal bases in $H_{A}$ and $H_{B}$ respectively. According to linear algebra, for every state $\vert \psi\rangle_{AB}$ there exist bases $\lbrace \vert u_{i}\rangle_{A}\rbrace$ and $\lbrace\vert v_{i}\rangle_{B}\rbrace$ \footnote{Remember that the bases $\lbrace\vert u_{i}\rangle_{A}\rbrace$ and $\lbrace\vert v_{i}\rangle_{B}\rbrace$ are also orthonormal bases.} such that
\begin{eqnarray}
\label{schmidtdecompose2}
\vert \psi\rangle_{AB} = \sum_{i = 1}^{n} \sqrt{\tilde{c_{i}}}\vert u_{i}\rangle_{A}\otimes \vert v_{i}\rangle_{B},
\end{eqnarray}
where $n = \min (dim(H_{A}),\: dim(H_{B}))$ and $\sum_{i = 1}^{n}\tilde{c_{i}} = 1$.\footnote{$dim$ stands for dimension which is defined as the total number of elements present in the basis of the Hilbert space.} $\sqrt{\tilde{c_{i}}}$'s are known as the \textbf{Schmidt coefficients} which are non-negative. These \textit{Schmidt coefficients} are the square roots of the eigenvalues of the two partial traces of the density matrix $\rho^{AB}=\vert \psi\rangle_{AB}\langle \psi\vert$ of the composite quantum system $AB$. The \textbf{partial traces} (over the system $A$ and over the system $B$) of the density matrix $\rho^{AB}$ however are given by
\begin{eqnarray}
\label{schmidtdecompose3}
\rho^{A} &=& Tr_{B}(\rho^{AB}) = \sum_{i}^{n}\tilde{c_{i}}\vert u_{i}\rangle_{A}\langle u_{i}\vert\nonumber\\
\rho^{B} &=& Tr_{A}(\rho^{AB}) = \sum_{i}^{n}\tilde{c_{i}}\vert v_{i}\rangle_{B}\langle v_{i}\vert.
\end{eqnarray}\\
Moreover the bases $\lbrace \vert u_{i}\rangle_{A}\rbrace$ and  $\lbrace \vert v_{i}\rangle_{B}\rbrace$  are called \textbf{Schmidt bases} for systems $A$ and $B$. However, the number of non-zero values $\tilde{c_{i}}$ is called \textbf{Schmidt Number} or sometimes is also called \textbf{Schmidt Rank}.
\\\\
A simple example would be to consider the Bell state $\vert \phi^{+}\rangle$ of equation (\ref{bellstates}). The state is a two qubit quantum mechanical system where the first qubit is with system $A$ and the second qubit is with system $B$. The systems $A$ and $B$ both have orthonormal basis $\lbrace \vert 0\rangle, \vert 1\rangle\rbrace$. Therefore the state $\vert \phi^{+}\rangle$ can be expressed as
\begin{eqnarray}
\label{schmidtdecompose4}
\vert \phi^{+}\rangle_{AB} = \frac{1}{\sqrt{2}}\vert 0_{A}\rangle\vert 0_{B}\rangle + \frac{1}{\sqrt{2}}\vert 1_{A}\rangle\vert 1_{B}\rangle.
\end{eqnarray}
The Schmidt coefficients, here, are $\frac{1}{\sqrt{2}}$. So all the Bell states are already in their Schmidt decomposition form. The number of Schmidt coefficients for Bell state $\vert \phi^{+}\rangle$ is 2 (as two $\frac{1}{\sqrt{2}}$'s are there).
It is to be noted that if we consider the states like $\vert 0_{A}\rangle \vert 0_{B}\rangle$ or $\vert 1_{A}\rangle \vert 1_{B}\rangle$, we find that all the expansion coefficients are real and there is no cross-term so the state is already in Schmidt decomposed form. For both the systems, the Schmidt basis is the computational basis. The Schmidt number for either of these states is $1$.
\\\\
Schmidt decomposition is useful for the separability characterization of Pure states in bipartite system. The following points  need to be remembered.
\begin{itemize}
\item The state $\vert \psi\rangle_{AB}$ is separable if and only if there is only one non-zero Schmidt coefficient $\tilde{c_{i}} = 1$, $\tilde{c_{j}} = 0$ $\forall$ $j \ne i$.
\item If more than one Schmidt coefficients are non-zero, then the state is entangled.
\item If all the Schmidt coefficients are non-zero and equal, then the state is said to be \textit{maximally entangled}. For example $\vert \phi^{+}\rangle$ is a maximally entangled state.
\end{itemize}
At this stage we state a very important theorem which goes as follows.\\\\
\textbf{Theorem:} A state $\vert \psi\rangle_{AB}$ of a composite system $AB$ (or in other words the bipartite two qubit system) is a \textit{product state} if and only if it has Schmidt number $1$. If the Schmidt number is greater than $1$ for a bipartite state, the state is entangled.\\\\
The statement of the theorem immediately hints at the notion that the four Bell states are entangled while the states $\vert 0\rangle \vert 0\rangle$ and $\vert 1\rangle \vert 1\rangle$ are product states. Let us try to be a little bit more elaborate in discussing Schmidt decomposition. It is because representation in Schmidt decomposition is important in the sense that it witnesses the signature of entanglement in bipartite system. If we consider two Hilbert spaces $H_{A}$ and $H_{B}$ with dimensions $m$ and $n$ corresponding to two systems $A$ and $B$, then we know that the joint quantum mechanical system $AB$ will be defined in the Hilbert space $H_{A} \otimes H_{B}$ whose dimension is $mn$. Therefore if $\lbrace \vert i_{A}\rangle\rbrace$ and $\lbrace \vert j_{B}\rangle\rbrace$ are the orthonormal bases of the systems $A$ and $B$ then $\lbrace \vert i_{A}\rangle\otimes \vert j_{B}\rangle\rbrace$ is the orthonormal basis of the joint system $AB$ and it has already been observed that any bipartite state can then be expressed as shown in eq. (\ref{schmidtdecompose1}). Now if we compare the eqs. (\ref{schmidtdecompose1}) and (\ref{schmidtdecompose2}) we see that there lie differences between the two expressions. In eq. (\ref{schmidtdecompose1}) the coefficients are all complex numbers and all the cross terms are present so that the summation is on the two indices $i$ and $j$ whereas in eq. (\ref{schmidtdecompose2}) the coefficients are all real and non-negative and the cross terms are absent there so that the summation is only on a single index $i$. It is also to be remembered that the maximum number of terms in eq. (\ref{schmidtdecompose2}) is minimum of the two dimensions $m$ and $n$. This is easily understood by the examples considered above. We see that in the Bell state $\vert \phi^{+}\rangle$ the maximum number of terms in its Schmidt decomposed form is $2$ (which is $\min\lbrace 2, 2\rbrace$, $dim(H_{A})=dim(H_{B})=2$)  while the maximum number of terms in states like $\vert 0\rangle\vert 0\rangle$ or $\vert 1\rangle \vert 1\rangle$ are 1.\\\\ 
It is not always easy to find Schmidt decomposition using inspection only. A clear prescription, however, to obtain Schmidt bases and Schmidt coefficients for an arbitrary state is as follows. It is seen from eq. (\ref{schmidtdecompose3}) the reduced density operators are diagonal in the Schmidt bases. Also as we know that since density operators are normal operators, then by spectral decomposition theorem such operators are diagonalizable in their own eigenbases \footnote{Remember that a linear operator defined on a finite dimensional vector space is diagonalizable if and only if there exists an ordered basis of that vector space which contains the eigenvectors of the transformation as elements \cite{insel2003linear}.}. Consequently the reduced density operators will also be diagonal in its own eigenbasis. One may utilize these eigenbases as Schmidt bases of the subsystem and inserting that in the given state one may find out the Schmidt bases of the other subsystem \cite{pathak2013elements}.\\

An algorithmic approach that can be followed to obtain Schmidt decomposed form of the states is described below.
\begin{itemize}
\item From the composite system $AB$ we first trace out the second subsystem i.e. $B$ using partial trace operation from equations (\ref{reduceddensityoperator}) and (\ref{reddensop}). In this manner we obtain the reduced density operator for the first subsystem $A$ which is denoted by $\rho^{A}$.\\
\item We then compute the eigenvalues and eigenvectors of the reduced density operator obtained in the previous step. The square roots of the eigenvalues are Schmidt coefficients and the set of eigenvectors is the Schmidt basis of the first subsystem.\\
\item We then rewrite the given state in a way so that the first subsystem is in the Schmidt basis states obtained in the previous step. This will automatically reveal the Schmidt basis of the other subsystem.
\end{itemize}
We shall try to understand the above algorithm by considering the following problem.\\\\
\textbf{Problem $42$:} Consider a composite system consisting of two qubits, which is given as $\vert \psi\rangle_{AB} = \frac{1}{2}(\vert 0_{A}\rangle \vert 0_{B}\rangle + \vert 0_{A}\rangle \vert 1_{B}\rangle + \vert 1_{A}\rangle \vert 0_{B}\rangle + \vert 1_{A}\rangle \vert 1_{B}\rangle)$. Find the Schmidt decomposition of this state \cite{nielsen2010quantum}.\\\\
\textit{Solution:} \begin{itemize}
\item First of all we shall construct the density matrix (or operator) for the given state $\vert \psi\rangle_{AB}$. This is calculated as 
\begin{eqnarray}
\label{problem41a}
\rho^{AB} = \vert \psi\rangle_{AB}\langle \psi\vert\nonumber\\ 
= \frac{1}{4}\Big\lbrace \vert 0_{A}\rangle \vert 0_{B}\rangle\langle 0_{A}0_{B}\vert + \vert 0_{A}0_{B}\rangle\langle 0_{A}1_{B}\vert + \vert 0_{A}0_{B}\rangle\langle 1_{A}0_{B}\vert + \nonumber\\
\vert 0_{A}0_{B}\rangle\langle 1_{A}1_{B}\vert  + 
 \vert 0_{A}1_{B}\rangle\langle 0_{A}0_{B}\vert + \vert 0_{A}1_{B}\rangle\langle 0_{A}1_{B}\vert \nonumber\\ + \vert 0_{A}1_{B}\rangle\langle 1_{A}0_{B}\vert  + \nonumber\\ \vert 0_{A}1_{B}\rangle\langle 1_{A}1_{B}\vert + \vert 1_{A}0_{B}\rangle\langle 0_{A}0_{B}\vert + \vert 1_{A}0_{B}\rangle\langle 0_{A}1_{B}\vert +  \vert 1_{A}0_{B}\rangle\langle 1_{A}0_{B}\vert + \nonumber\\  \vert 1_{A}0_{B}\rangle\langle 1_{A}1_{B}\vert + \vert 1_{A}1_{B}\rangle\langle 0_{A}0_{B}\vert + \vert 1_{A}1_{B}\rangle\langle 0_{A}1_{B}\vert + \vert 1_{A}1_{B}\rangle\langle 1_{A}0_{B}\vert + \vert 1_{A}1_{B}\rangle\langle 1_{A}1_{B}\vert\Big\rbrace \nonumber\\
\end{eqnarray}
Tracing out system $B$ we get the following reduced density operator for system $A$, denoted by $\rho^{A}$ and is given by
\begin{eqnarray}
\rho^{A} &=& \frac{1}{2}\Big\lbrace \vert 0_{A}\rangle\langle 0_{A}\vert  + \vert 0_{A}\rangle\langle 1_{A}\vert + \vert 1_{A}\rangle\langle 0_{A}\vert +  \vert 1_{A}\rangle\langle 1_{A}\vert\Big\rbrace \nonumber\\
&=& \frac{1}{2}\left(%
	\begin{array}{cc}
		1 & 1 \\\\
		1 & 1
	\end{array}%
	\right).
\end{eqnarray}
\item Next we calculate the eigenvalues and eigenvectors of the above reduced density operator $\rho^{A}$\footnote{One can take the help of the Maple software in this case also otherwise one must go through the usual mathematical approach to calculate the eigenvalues and eigenvectors defined in the article $1.2.2$.}. The eigenvalues are $0$ and $1$ and the corresponding eigenvectors are respectively $\left(%
	\begin{array}{c}
		-1 \\\\
		1
	\end{array}%
	\right)$ and $\left(%
	\begin{array}{c}
		1 \\\\
		1
	\end{array}%
	\right)$. It is to be noted that the square roots of the eigenvalues, which are $0$ and $1$ in this case, are Schmidt coefficients  .(Recall from the definition that Schmidt coefficients are non-negative !)
	\item Since we need Schmidt basis, we normalize the eigenvectors by using the formula defined in eq. (\ref{norm2}) and is given by 
	$\frac{\left(%
	\begin{array}{c}
		-1 \\\\
		1
	\end{array}%
	\right)}{\|\left(%
	\begin{array}{cc}
		-1 & 1
	\end{array}%
	\right)^{T}\|}$ and $\frac{\left(%
	\begin{array}{c}
		1 \\\\
		1
	\end{array}%
	\right)}{\|\left(%
	\begin{array}{cc}
		1 & 1
	\end{array}%
	\right)^{T}\|}$ which in turn equals $\frac{1}{\sqrt{2}}\left(%
	\begin{array}{c}
		-1 \\\\
		1
	\end{array}%
	\right)$ and $\frac{1}{\sqrt{2}}\left(%
	\begin{array}{c}
		1 \\\\
		1
	\end{array}%
	\right)$. The set containing these two normalized vectors is the Schmidt basis for the subsystem $A$. The explanation for normalizing eigenvectors to get Schmidt basis is due to the following theorem.\\\\
	\textit{Theorem:} Let $T$ be a linear operator on a finite-dimensional complex inner product space (i.e. Hilbert space)
 $V$. Then $T$ is normal if and only if there exist an orthonormal basis for $V$ consisting of eigenvectors of $T$ \cite{insel2003linear}.\\\\
 Try to recall that the density operators are normal operators.
 \item After we have calculated the Schmidt coefficients and Schmidt basis of the first system we write original state $\vert \psi\rangle_{AB}$ in terms of them as
 \begin{eqnarray}
 \label{problem41b}
 \vert \psi\rangle_{AB} &=& \frac{1}{2}\lbrace \vert 0_{A}0_{B}\rangle + \vert 0_{A}1_{B}\rangle + \vert 1_{A}0_{B}\rangle + \vert 1_{A}1_{B}\rangle\rbrace \nonumber\\ &=& 0\times \left(%
	\begin{array}{c}
		-\frac{1}{\sqrt{2}} \\\\
		\frac{1}{\sqrt{2}}
	\end{array}%
	\right)_{A}\otimes \left(%
	\begin{array}{c}
		x_{1} \\\\
		y_{1}
	\end{array}%
	\right)_{B} + 1\times \left(%
	\begin{array}{c}
		\frac{1}{\sqrt{2}} \\\\
		\frac{1}{\sqrt{2}}
	\end{array}%
	\right)_{A}\otimes \left(%
	\begin{array}{c}
		x_{2} \\\\
		y_{2}
	\end{array}%
	\right)_{B}\nonumber\\ &=& \frac{1}{\sqrt{2}}\left(%
	\begin{array}{c}
		x_{2} \\\\
		y_{2}\\\\
		x_{2} \\\\
		y_{2}	
	\end{array}%
	\right)_{AB}
 \end{eqnarray}
 At this point, we can recall that the ket vectors $\vert 0\rangle = \left(%
	\begin{array}{c}
		1 \\\\
		0
	\end{array}%
	\right)$ and $\vert 1\rangle = \left(%
	\begin{array}{c}
		0 \\\\
		1
	\end{array}%
	\right)$, so that taking the tensor products of these vectors on the left-hand side of eq. (\ref{problem41b}) (such as $\vert 0\rangle\vert 0\rangle, \vert 0\rangle \vert 1\rangle, \vert 1\rangle \vert 0\rangle, \vert 1\rangle \vert 1\rangle$) and equating this to the right hand side of the same equation we ultimately get $x_{2} = \frac{1}{\sqrt{2}}$ and $y_{2} = \frac{1}{\sqrt{2}}$. Thus from eq. (\ref{problem41b}) we get $\vert \psi\rangle_{AB} = \frac{1}{\sqrt{2}}\times\left(%
	\begin{array}{c}
		1 \\\\
		1
	\end{array}%
	\right)_{A}\otimes \left(%
	\begin{array}{c}
		1 \\\\
		1
	\end{array}%
	\right)_{B}$ which is the required Schmidt decomposed form of the given state. 
	\item To find the Schmidt basis of the subsystem $B$ we need to find the vector $\left(%
	\begin{array}{c}
		x_{1}\\\\
		y_{1}
	\end{array}%
	\right)$ and since the Schmidt basis is orthonormal, this vector must be orthogonal to the vector $\left(%
	\begin{array}{c}
		x_{2}\\\\
		y_{2}
	\end{array}%
	\right)$ This implies that $\left(%
	\begin{array}{cc}
		x_{1} & y_{1}\\\\
	\end{array}%
	\right)_{B} \left(%
	\begin{array}{c}
		x_{2}\\\\
		y_{2}
	\end{array}%
	\right)_{B} = \frac{1}{\sqrt{2}}\left(%
	\begin{array}{cc}
		x_{1} & y_{1}\\\\
	\end{array}%
	\right)_{B}\left(%
	\begin{array}{c}
		1\\\\
		1
	\end{array}%
	\right)_{B} = 0$. Solving this we get $y_{1} = -x_{1}$  so that the required vector is $\frac{1}{\sqrt{2}}\left(%
	\begin{array}{c}
		1\\\\
		-1
	\end{array}%
	\right)_{B}$. Therefore the Schmidt basis of the subsystem $B$ contains two vectors $\frac{1}{\sqrt{2}}\left(%
	\begin{array}{c}
		1\\\\
		-1
	\end{array}%
	\right)_{B}$  and $\frac{1}{\sqrt{2}}\left(%
	\begin{array}{c}
		1\\\\
		1
	\end{array}%
	\right)_{B}$.
	\item If we recall problem $2$ of this chapter we find that $\frac{1}{\sqrt{2}}\left(%
	\begin{array}{c}
		1\\\\
		1
	\end{array}%
	\right)=\vert a\rangle$ and $\frac{1}{\sqrt{2}}\left(%
	\begin{array}{c}
		1\\\\
		-1
	\end{array}%
	\right)=\vert b\rangle$ so that the Schmidt basis of subsystem $A$ is given as $\lbrace \vert a_{A}\rangle,-\vert b_{A}\rangle\rbrace$ while the Schmidt basis of the subsystem $B$ is $\lbrace \vert a_{B}\rangle,\vert b_{B}\rangle\rbrace$.
	\item Also it is important to note down that the state $\vert \psi_{AB}\rangle$ has only one non-zero Schmidt coefficient and hence the Schmidt number (or Schmidt rank) of the state is $1$. Consequently the state $\vert \psi\rangle_{AB}$ is separable. A quick glance at the state however reveals that the state can always be expressed as $\frac{1}{2}\Big(\vert 0_{A}\rangle+\vert 1_{A}\rangle)\otimes (\vert 0_{B}\rangle + \vert 1_{B}\rangle\Big)$.
 \end{itemize}
\textbf{Problem $1.43$:} Consider a composite system consisting of two qubits, which is given as $\vert \psi\rangle_{AB} = \frac{1}{\sqrt{3}}(\vert 00\rangle_{AB} + \vert 01\rangle_{AB} + \vert 10\rangle_{AB})$. Find the Schmidt decomposition of this state \cite{nielsen2010quantum,pathak2013elements}.\\\\
\noindent \textbf{Purification:} When we have a joint quantum mechanical system $AB$, we know that we always can trace out one subsystem, say, $A$ to get the information about the other subsystem, say $B$ and vice-versa. The reduced system is however mixed in nature. Now suppose we have a reduced system whose reduced density operator is given by $\rho^{A}$ and which is mixed. We can always think of a dummy system $D$ where $AD$ denotes the joint system and $\vert AD\rangle$ is the pure state such that if we trace out the dummy system $D$ from the joint system $AD$ by using partial trace we will get back $\rho^{A}$, i.e. $\rho^{A} = Tr_{D}(\vert A\rangle \langle A\vert \otimes \vert D \rangle \langle D\vert)$. This process is called \textit{purification}. One, however, should keep in mind that the dummy system $D$ does not have any physical significance. Since $\rho^{A}$ is a kind of density operator so it is necessarily Hermitian and hence using spectral decomposition one can express $\rho^{A}$ as $\rho^{A}=\sum_{i}p_{i}\vert i_{A}\rangle\langle i_{A}\vert$, where $\lbrace \vert i_{A}\rangle\rbrace$ is the orthonormal basis of the system $A$. Let us consider one such system, say,  $\rho^{A}=\frac{1}{3}\vert 0\rangle\langle 0\vert + \frac{2}{3}\vert 1\rangle\langle 1\vert$. To purify the given state $\rho^{A}$ a dummy system $D$ is introduced which has the same state space as system $A$. The orthonormal basis of the dummy system is $\lbrace \vert i_{D}\rangle\rbrace$. Thus pure state of the combined system is defined by
\begin{eqnarray}
\label{purification}
\vert A\rangle \vert D\rangle = \sum_{i}\sqrt{p_{i}}\vert i_{A}\rangle\vert i_{D}\rangle.
\end{eqnarray}
We can now easily see that 
\begin{eqnarray}
\label{purification1}
Tr_{D}(\vert A\rangle\langle A\vert \otimes \vert D\rangle\langle D\vert) &=& \sum_{ij}\sqrt{p_{i}p_{j}}\vert i_{A}\rangle\langle j_{A}\vert Tr(\vert i_{D}\rangle\langle j_{D}\vert)\nonumber\\
&=& \sum_{ij}\sqrt{p_{i}p_{j}}\vert i_{A}\rangle\langle j_{A}\vert \delta_{ij}\nonumber\\
&=& \sum_{i}p_{i}\vert i_{A}\rangle\langle i_{A}\vert \nonumber\\
&=& \rho^{A}.
\end{eqnarray}
Here $\vert A\rangle\langle A\vert \otimes \vert D\rangle\langle D\vert = \rho^{AD}$.
In the example that we have considered above, the pure state using (\ref{purification}), however, will be $\vert A\rangle \vert D\rangle = \sqrt{\frac{1}{3}}\vert 0\rangle \vert 0\rangle+ \sqrt{\frac{2}{3}}\vert 1\rangle \vert 1\rangle$. Therefore the density matrix $\rho^{AD} = \frac{1}{3}\vert 0\rangle\langle0\vert \otimes \vert 0\rangle\langle 0\vert  + \frac{\sqrt{2}}{3}\vert 0\rangle\langle0\vert \otimes \vert 1\rangle\langle 1\vert + \frac{\sqrt{2}}{3}\vert 1\rangle\langle1\vert \otimes \vert 0\rangle\langle 0\vert + \frac{2}{3}\vert 1\rangle\langle1\vert \otimes \vert 1\rangle\langle 1\vert = \left(%
\begin{array}{cccc}
	\frac{1}{3} & 0 & 0 & \frac{\sqrt{2}}{3}\\
    0 &  0 & 0 & 0\\
    0 &  0 & 0 & 0\\
   \frac{\sqrt{2}}{3} &  0 & 0 & \frac{2}{3}\\
\end{array}%
\right)$. Consequently $(\rho^{AD})^{2} = \left(%
\begin{array}{cccc}
	\frac{1}{3} & 0 & 0 & \frac{\sqrt{2}}{3}\\
    0 &  0 & 0 & 0\\
    0 &  0 & 0 & 0\\
   \frac{\sqrt{2}}{3} &  0 & 0 & \frac{2}{3}\\
\end{array}%
\right)$ and so $Tr((\rho^{AD})^{2})=1$. Thus $\rho_{AD}$ is a pure state.\\\\
\noindent \textbf{Entanglement:} The more intrinsic quantum mechanical sense in which quantum states can embody vastly more information than its classical counterparts is due to the non-classical feature of \textit{quantum entanglement} or simply \textit{entanglement}. The phenomenon of entanglement remains predominant in most aspects of quantum information theory.\\\\
\noindent Let us consider a system consisting of two sub-systems where each sub-system is associated with a Hilbert space. Let $H_{A}$ and $H_{B}$ denote these two Hilbert spaces with respect to the two subsystems $A$ and $B$ respectively. Let $\lbrace \vert i_{A}\rangle\rbrace$ and $\lbrace \vert j_{B}\rangle\rbrace$ ($i, j= 1,2,3,\cdots$) represent two complete orthonormal bases for $H_{A}$ and $H_{B}$ respectively. The two subsystems taken together is associated with the Hilbert space $H_{A}\otimes H_{B}$, spanned by the states $\vert i_{A}\rangle\otimes \vert j_{B}\rangle$.  A linear combination of the basis states $\vert i_{A}\rangle \otimes \vert j_{B}\rangle$ is a state of the composite system $AB$. We know the pure state $\vert \psi\rangle_{AB}$ of the system can be written as
\begin{eqnarray}
\label{entangledstate1}
\vert \psi\rangle_{AB} = \sum_{i,\: j} c_{ij}\vert i_{A} \rangle\otimes \vert j_{B}\rangle,
\end{eqnarray}
where $c_{ij}$'s are the complex coefficients satisfying the normalization condition $\sum_{i,\: j}\vert c_{ij}\vert^{2} = 1$. If $\vert \psi\rangle_{AB}$ factors into a normalized state $\vert \psi\rangle_{A} = \sum_{i}^{dim(H_{A})}c_{i}\vert i_{A}\rangle$ in $H_{A}$ and a normalized state $\vert \psi\rangle_{B} = \sum_{j}^{dim(H_{B})}c_{j}\vert j_{B}\rangle$\footnote{Here $dim(H_{A})$ and $dim(H_{B})$ are known as the dimensions of the two respective Hilbert spaces.} in $H_{B}$ i.e. $\vert \psi\rangle_{AB} = \vert \psi\rangle_{A}\otimes \vert \psi\rangle_{B}$, then the state $\vert \psi\rangle_{AB}$ is called a \textbf{separable state} or \textbf{product state}. If a state belonging to the Hilbert space $H_{A}\otimes H_{B}$ is not a product state, then such a state is called \textbf{entangled state}. \\\\
If $\vert \psi\rangle_{AB}$ represents a pure state of a composite system consisting of two Hilbert spaces $H_{A}$ and $H_{B}$ for the individual systems $A$ and $B$, then $\vert \psi\rangle_{AB}$ can always be written in the Schmidt form as
\begin{eqnarray}
\label{entangledstate2}
\vert \psi\rangle_{AB} = \sum_{i}^{k\leq \min\lbrace dim H_{A}, dim H_{B}\rbrace}\sqrt{\lambda_{i}}\vert i_{A}\rangle\otimes \vert i_{B}\rangle,
\end{eqnarray}
where $\lbrace \vert i_{A}\rangle\rbrace$ and $\lbrace \vert i_{B}\rangle\rbrace$ are orthonormal bases of systems $A$ and $B$ respectively with the conditions $\lambda_{i}\geq 0$, $\sum_{i}\lambda_{i} = 1$, $\lambda_{i}$'s being the \textit{Schmidt Coefficients}. If two or more Schmidt coefficients are non-zero, then the state $\vert \psi\rangle_{AB}$ is referred to as \textit{Pure Entangled State}. The eq.(\ref{entangledstate2}) is similar to that of eq.(\ref{schmidtdecompose2}).\\\\
In the next unit, we discuss two fundamental arena in which quantum entanglement plays a significant role in the domain of quantum information processing.
\section{Unit - V}\label{sec:unit5}
\subsection{Bell non-locality:}
After spending a considerable amount of time discussing the mathematical preliminaries, it is now time to delve deeper into the domains of quantum information science and to see how  Quantum Mechanics (QM) was saved from sinking into the  abyss. In the words of Feynman, \textit{uncertainty principle protects quantum mechanics}\cite{feynman1979feynman}. But Albert Einstein was a firm disbeliever of QM. He, along with his students Boris Podolsky and Nathan Rosen, in the year $1935$, wrote a seminal paper where they raised two questions on the existence of quantum mechanical theory\cite{einstein1935can}. Their thought process was based on two entities viz. (a) Realism and (b) Locality\footnote{Realism is sometimes called Reality while Locality is also known as Localism.}. Let us try to understand first what Einstein meant by these terms actually. Here we set a simplectic classical example from pedagogical perspective. When we go out for a vacation, lights of our flat remain switched off until we come back and turn them on. During our absence, the furniture of our flat stay firmly in their respective positions (if not externally disturbed). In other words we can say that the physical observables, which in this case are positions of the furniture, retain their pre-existing values irrespective of whether one observes them or not. This property of the physical object is what is known as the physical \textit{reality}. Thus according to Einstein, \textit{if, without in any way disturbing a system, we can predict with certainty (i.e. with unit probability) the value of a physical quantity, then there exists an element of physical reality corresponding to this physical quantity}\cite{einstein1935can}. Let us now discuss another facet which was the barrier in truly understanding the quantum mechanical behaviour. Suppose you and your friend are playing a game where both of you toss with classical dice, which, each one of you have in your respective possessions. But also at the same time both of you are spatially separated from one another\footnote{Spatial separation means you and your friends are located at two places which are some distant apart.}. We also assume that both of you are tossing your respective dice simultaneously. From classical probability theory it is obvious that the outcomes of your toss are independent of what your friend gets in his respective toss. We may say that each of your outcomes are localized. Classical physics demands this \textit{local-realism} from its physical objects. This is not, however, true with quantum mechanical systems\footnote{Although \textit{locality} argument is tagged with Einstein always, it was conjectured way before Einstein\cite{peres2002quantum}.}. We will now try to understand why Bell's inequality is so intricately connected to Quantum Mechanics (QM). It is to be remembered, however, that Bell's inequality is not actually talking about QM, rather \textit{it is a general proof, independent of any specific physical theory, that there is an upper limit to the correlation of distant events, if one just assumes the validity of the principle of local causes}\cite{peres2002quantum}. Let us consider the following figure to understand the argument of Bell\footnote{The structure of the diagram is in parity with that depicted in article by Alain Aspect\cite{aspect2002bell}}.\\\\
\begin{figure}[h!]
		\begin{center}
			\includegraphics[width=9.5cm]{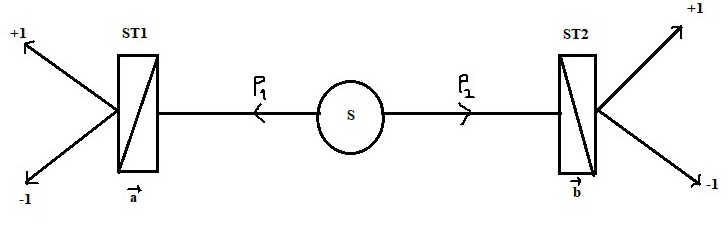}
		\end{center}
		\label{f2}
	\end{figure}
Suppose a source S produces a pair of spin-one half particles\footnote{A spin of $\frac{1}{2}$ means that the particle must be rotated by two full turns (through $720$ degree) before it has the same configuration as when it started.} prepared in a singlet state. The two particles are then sent towards two opposite directions where on one side the observer $A$ is located with his Stern-Garlach magnet $S1$ to measure some observable of particle $P1$ and on the other side the observer $B$ is located with his Stern-Garlach magnet $S2$ to measure the same observable of particle $P2$. Observables are actually the physical properties that can be observed or measured, (like one may be interested in measuring the spin of a particle or polarization of a particle etc). According to Bohm and Aharaonov, the wave function of a molecule of total spin zero consisting of two atoms, each of spin one-half, can be thought of in a singlet form given by,
\begin{eqnarray}
\label{bohmsinglet}
\vert \psi\rangle =  \frac{1}{\sqrt{2}}\Big\lbrace \vert \psi_{+}(1)\rangle\vert \psi_{-}(2)\rangle-\vert \psi_{-}(1)\rangle \vert \psi_{+}(2)\rangle\Big\rbrace,
\end{eqnarray}
where, $\vert \psi_{+}(1)\rangle$ refers to the wave function of the atomic state in which particle $1$ has spin $+\frac{\hbar}{2}$, $\hbar$ is the Planck's constant and so on\cite{bohm1957discussion}. By denoting $\vert \psi_{+}\rangle$ by $\vert 0\rangle$ and $\vert\psi_{-}\rangle$ by $\vert 1\rangle$  the singlet state is re-written in the form (for reference see eq.(\ref{bellstates})),
\begin{eqnarray}
\label{singletstate}
\vert \psi\rangle = \frac{1}{\sqrt{2}}\Big\lbrace \vert 0_{A}\rangle\vert 1_{B}\rangle -\vert 1_{A}\rangle\vert 0_{B}\rangle\Big\rbrace,
\end{eqnarray}
which is similar to the state $\vert \varphi^{-}\rangle$ defined in (\ref{bellstates}). Here the first qubit is with party $A$ and the second qubit is with party $B$. Now suppose that the two spatially separated magnets $S1$ and $S2$ measure the spin components of these two particles $P1$ and $P2$ respectively. Let the spin of particle $P1$ in general be denoted by $\vec{\sigma_{A}}$ and that of particle $P2$ be denoted by $\vec{\sigma_{B}}$. Now if $\vec{a}$ is the three dimensional unit vector in the direction of $\vec{\sigma_{A}}$ (orientation of magnet $S1$ be considered in some direction) and $\vec{b}$ is another three dimensional unit vector in the direction of $\vec{\sigma_{B}}$ (orientation of magnet $S2$ be considered in some direction), then we can define observables in the locations of party $A$ and $B$ as\cite{nielsen2010quantum}
\begin{eqnarray}
\label{observables}
\vec{a}\cdot \vec{\sigma} &=& a_{1}\sigma_{1}+a_{2}\sigma_{2}+a_{3}\sigma_{3}\nonumber\\
\vec{b}\cdot \vec{\sigma} &=& b_{1}\sigma_{1}+b_{2}\sigma_{2}+b_{3}\sigma_{3},
\end{eqnarray}
where $\sigma_{i}$'s are Pauli spin matrices. Measurements of these observables $\vec{a}\cdot \vec{\sigma}$ and $\vec{b}\cdot \vec{\sigma}$ are sometimes referred to as a \textit{measurement of spin along $\vec{a}$ and $\vec{b}$ axes respectively}\cite{nielsen2010quantum}. From Bohm's perspective, we can say that since the total spin of the molecule is zero, then immediately one can conclude under the assumption that parties $A$ and $B$, when cease to interact with one another, the outcomes of the measurements of the similar spin components on the parts of the two parties $A$ and $B$ will have values opposite to one another, i.e. if the measurement made by party $A$ yields $+1$, then the measurement made by party $B$ will yield $-1$ and vice-versa\cite{bohm1957discussion}.\\\\
\textbf{Problem $1.5.1(1)$:} Suppose $\vec{v}$ is any three dimensional unit vector. We can define an observable as $\vec{v}\cdot\vec{\sigma} = v_{1}\sigma_{1}+v_{2}\sigma_{2}+v_{3}\sigma_{3}$. Prove that $\vec{v}\cdot\vec{\sigma}$ has eigenvalues $\pm 1$. Here remember that the $\sigma_{i}$'s are Pauli spin matrices. Measurement of the observable $\vec{v}\cdot\vec{\sigma}$ is known as the measurement of spin along the $\vec{v}$ axis.\\\\
At this point let's deviate a little from the quantum world and come back to the classical world where two players Alice and Bob are tossing coins simultaneously, situating at their respective places, that are remote from one another. We assume that Alice can't communicate anything to Bob and vice-versa. The outcomes of their experiments can be tabulated below.
\begin{center}
		\begin{tabular}{|c|c|}
			\hline Alice's result & Bob's result\\
			\hline H & H\\
			\hline H & T\\
			\hline T & H\\
			\hline T & T\\
			\hline
		\end{tabular}
	\end{center}\
	
\noindent Although the system of Alice and Bob are independent, it may sometimes happen that their outcomes match, if the experiments are repeated many a times. But it is not that we are not aware of the possibilities of Bob's outcome and in no way measurement on Alice's end affects that of Bob's. It is only a matter of time when they both get similar results on their end. In other words the possibilities of their outcomes are predetermined.\\\\
Likewise argument of Bell further claims that in experiment described above by Bohm and Aharaonov\cite{aharonov2008quantum} one can predict in advance the result of measuring any chosen component of  $\vec{\sigma_{B}}$, by previously measuring the same component of $\vec{\sigma_{A}}$ and it follows that the outcome of any such measurement must actually be pre-determined and this pre-determination implies the possibility of a more complete specification of the state. Bell further argues that if such complete specification be affected by means of parameters $\lambda$\footnote{The parameter $\lambda$ may be a single variable or a set, may be discrete or continuous. They are also called set of hidden variables or hidden variables are parametrized by $\lambda$.}, the result measured by $A$ of the observable $\vec{\sigma_{a}}\cdot \vec{a}$ is then determined by $\vec{a}$ and $\lambda$ whereas  the result measured by $B$ of the observable $\vec{\sigma_{b}}\cdot \vec{b}$ is then determined by $\vec{b}$ and $\lambda$. Also we have 
\begin{eqnarray}
\label{parameterlambda}
A(\vec{a},\lambda) = \pm 1, ~~~~~ B(\vec{b},\lambda) = \pm 1.
\end{eqnarray}
If $\rho(\lambda)$ denotes the probability distribution of parameter $\lambda$ then $\rho(\lambda)\geq 0$ and the expectation value of the product of the two components $\vec{\sigma_{A}}\cdot\vec{a}$ and $\vec{\sigma_{B}}\cdot\vec{b}$ is 
\begin{eqnarray}
\label{probdistparameter}
E(\vec{a},\vec{b}) = \int d\lambda \rho(\lambda)A(\vec{a},\lambda)B(\vec{b},\lambda).
\end{eqnarray}
\textbf{Problem $1.5.1(2)$:} Consider the singlet state $\vert \varphi^{-}\rangle = \frac{1}{\sqrt{2}}\lbrace \vert 01\rangle -  \vert 10\rangle\rbrace$. With the help of eq.(\ref{observables}) calculate the expectation value of the operator $\vec{a}\cdot \vec{\sigma} \otimes \vec{b}\cdot \vec{\sigma}$ with respect to the given singlet state. The expectation value is defined as $\langle \varphi^{-}\vert \vec{a}\cdot \vec{\sigma} \otimes \vec{b}\cdot \vec{\sigma} \vert \varphi^{-}\rangle$\footnote{The calculation of $\langle \varphi^{-}\vert \vec{\sigma} \cdot \vec{a}  \otimes \vec{\sigma} \cdot \vec{b}\vert \varphi^{-}\rangle$ will yield the same result.}.\\\\
Let the expectation value calculated in the above problem $1.5.1(2)$ be  denoted by $\langle \vec{\sigma} \cdot \vec{a}  \otimes \vec{\sigma} \cdot \vec{b}\rangle$. This expectation value is the quantum mechanical expectation value for the singlet state. John Bell in his famous paper showed that under the assumption of existence of parameter $\lambda$, $E(\vec{a},\vec{b})$ from eq. (\ref{probdistparameter}) is not same as the quantum mechanical expectation value $\langle \vec{\sigma} \cdot \vec{a}  \otimes \vec{\sigma} \cdot \vec{b}\rangle$\cite{bell1964einstein}. The vital assumption based on which Bell put forward above argument was that of locality i.e. the result obtained by party B does not depend on the setting $\vec{a}$, of the magnet for particle $1$ received by party A and vice versa. We shall now show the value of the expectation of the term given by eq.(\ref{probdistparameter}). For this we shall stick to Clauser-Horne-Shimony-Holt (CHSH) version of the proof where they generalized Bell's theorem\cite{clauser1969proposed}.\\\\
Since the possible values of the observables measured by party A and B are $-1$, $0$ and $1$, then we have
\begin{eqnarray}
\label{condition1}
\vert A(\vec{a},\lambda)\vert \leq 1, ~~~~~ \vert B(\vec{b},\lambda)\vert \leq 1.
\end{eqnarray}
Consider $a$, $a^{\prime}$, $b$ and $b^{\prime}$ are alternative settings for the Stern-Garlach detectors $S1$ and $S2$ on the sides of party $A$ and party $B$. We now try to find the expression $E(\vec{a},\vec{b})-E(\vec{a},\vec{b^{\prime}})$ which is
\begin{eqnarray}
\label{condition2}
E(\vec{a},\vec{b})-E(\vec{a},\vec{b^{\prime}}) &=& \int d\lambda \rho(\lambda)A(\vec{a},\lambda)B(\vec{b},\lambda) - \int d\lambda \rho(\lambda)A(\vec{a},\lambda)B(\vec{b^{\prime}},\lambda)\nonumber\\
&=& \int d\lambda \rho(\lambda)\lbrace A(\vec{a},\lambda)B(\vec{b},\lambda) - A(\vec{a},\lambda)B(\vec{b^{\prime}},\lambda)\rbrace 
\end{eqnarray}
Adding extra term ($\pm A(\vec{a},\lambda)B(\vec{b},\lambda)A(\vec{a^{\prime}},\lambda)B(\vec{b^{\prime}},\lambda)\mp A(\vec{a},\lambda)B(\vec{b},\lambda)A(\vec{a^{\prime}},\lambda)B(\vec{b^{\prime}},\lambda)$) in the integrand and rearranging the terms we get from eq. (\ref{condition2}) 
\begin{eqnarray}
\label{condition2}
\vert E(\vec{a},\vec{b})-E(\vec{a},\vec{b^{\prime}}) \vert \leq  \Bigg\vert \int d\lambda \rho(\lambda) A(\vec{a},\lambda)B(\vec{b},\lambda)\lbrace 1 \pm A(\vec{a^{\prime},\lambda})B(\vec{b^{\prime}}, \lambda)\rbrace\Bigg\vert + \nonumber\\\Bigg \vert \int d\lambda \rho(\lambda)  A(\vec{a},\lambda)B(\vec{b^{\prime}},\lambda)\lbrace 1 \pm A(\vec{a^{\prime}},\lambda)B(\vec{b},\lambda)\rbrace\Bigg \vert \nonumber\\
= \int \vert d\lambda \rho(\lambda)\vert \vert A(\vec{a},\lambda)B(\vec{b},\lambda)\vert \vert  1 \pm A(\vec{a^{\prime},\lambda})B(\vec{b^{\prime}}, \lambda)\vert  +\nonumber\\
\int \vert d\lambda \rho(\lambda)\vert  \vert  A(\vec{a},\lambda)B(\vec{b^{\prime}},\lambda)\vert \vert  1 \pm A(\vec{a^{\prime}},\lambda)B(\vec{b},\lambda)\vert , \nonumber\\
\end{eqnarray}
since $\vert \vec{p}-\vec{q}\vert \leq \vert \vec{p}\vert + \vert \vec{q}\vert$ (from triangle inequality). Remember also, that the terms $[1+A(\vec{a^{\prime}},\lambda)B(\vec{b^{\prime}},\lambda)]\rho(\lambda)$ and $[1+A(\vec{a^{\prime}},\lambda)B(\vec{b},\lambda)]\rho(\lambda)$ are both non-negative. Using eq.(\ref{condition1}) we can see that the last expression of eq.(\ref{condition2}) is less than or equal to $\int d\lambda \rho(\lambda)[1 \pm A(\vec{a^{\prime}},\lambda)B(\vec{b^{\prime}},\lambda)] + \int d\lambda \rho(\lambda)[1 \pm A(\vec{a^{\prime}},\lambda)B(\vec{b},\lambda)]$ and also since $\int d\lambda \rho(\lambda) = 1$, hence
\begin{eqnarray}
\label{condition3}
\vert E(\vec{a},\vec{b})-E(\vec{a},\vec{b^{\prime}})\vert \leq 2 \pm \Bigg[\int d\lambda \rho(\lambda) A(a^{\prime},\lambda)B(b^{\prime},\lambda) +  \int d\lambda \rho(\lambda) A(a^{\prime},\lambda)B(b,\lambda)\Bigg].\nonumber\\
\end{eqnarray}
This implies that
\begin{eqnarray}
\label{condition4}
\vert E(\vec{a},\vec{b})-E(\vec{a},\vec{b^{\prime}})\vert \leq 2 \pm [E(\vec{a^{\prime}},\vec{b^{\prime}})-E(\vec{a^{\prime}},\vec{b})].
\end{eqnarray}
Therefore we get\footnote{If $\vert X\vert \leq 2+ Y$ and also at the same time $\vert X\vert \leq 2- Y$, then it is obvious that $\vert X\vert \leq 2 - \vert Y\vert$.}
\begin{eqnarray}
\label{condition4}
\vert E(\vec{a},\vec{b})-E(\vec{a},\vec{b^{\prime}})\vert \leq 2 \pm \vert E(\vec{a^{\prime}},\vec{b^{\prime}})-E(\vec{a^{\prime}},\vec{b})\vert.
\end{eqnarray}
In this way we get 
\begin{eqnarray}
\label{bellchsh}
\vert E(\vec{a},\vec{b})-E(\vec{a},\vec{b^{\prime}}) + E(\vec{a^{\prime}},\vec{b^{\prime}})-E(\vec{a^{\prime}},\vec{b})\vert  \leq 
\vert E(\vec{a},\vec{b})-E(\vec{a},\vec{b^{\prime}})\vert  + \vert E(\vec{a^{\prime}},\vec{b^{\prime}})-E(\vec{a^{\prime}},\vec{b})\vert \leq 2.\nonumber\\
\end{eqnarray}
The above eq.(\ref{bellchsh}) is one of the several forms of Bell inequalities, defined for some specific settings of apparatus on the parts of parties $A$ and $B$. The author of the book suggests readers to go through another beautiful proof of Bell's inequality in reference \cite{nielsen2010quantum}. The result derived in eq. (\ref{bellchsh}) is sometimes also known that \textit{CHSH inequality}. Bell was not actually talking about QM, rather he was suggesting that any physical system for its complete specification supplemented by local hidden variables, must satisfy inequality defined in eq. (\ref{bellchsh}) or the form alike. However, Einstein, Podolsky and Rosen (EPR) used entanglement features to ascribe values to physical quantities prior to measurement and showed the quantum mechanical description of physical system is not a complete theory\cite{einstein1935can}, rather a paradoxical situation. In 1964, Bell accepted the EPR conclusion as a working hypothesis and formalized the EPR deterministic world idea in terms of local hidden variable model (LHVM), where he assumed the \textit{realism} and \textit{locality} as working hypothesis. Also he assumed that the setting of local apparatus are independent of the hidden variables which determine local results. With these, Bell proved that the probabilities for the outcomes obtained when suitably measuring some entangled quantum state violate Bell inequality. Hence entanglement is that resource of quantum formalism which makes impossible to simulate the quantum correlations within any classical formalism.\\\\
Alain Aspect, was born in $15^{th}$ June $1947$, Agen, France, (the year India got Independence from colonial British rule). He won Nobel prize in Physics in the year 2022 and shared this prize with two other distinguished scientists viz. John Clauser (the first of CHSH, i.e. Clauser-Horne-Shimony-Holt). They receive the prize for their pioneering experiments with entangled photons, establishing the violation of Bell inequalities and pioneering quantum information science\footnote{https://www.nobelprize.org/prizes/physics/2022/aspect/facts/.}. In the paper titled $``$Experimental Test of Bell's Inequalities Using Time-Varying Analyzers"  (year $1982$), Alain with his team members showed correlations of linear polarizations of pair of photons. He measured these correlations with time-varying analyzers. The results are in good agreement with quantum mechanical predictions but violate Bell's inequalities by $5$ standard deviations.
\begin{figure}[h!]
		\begin{center}
			\includegraphics[width=5.5cm]{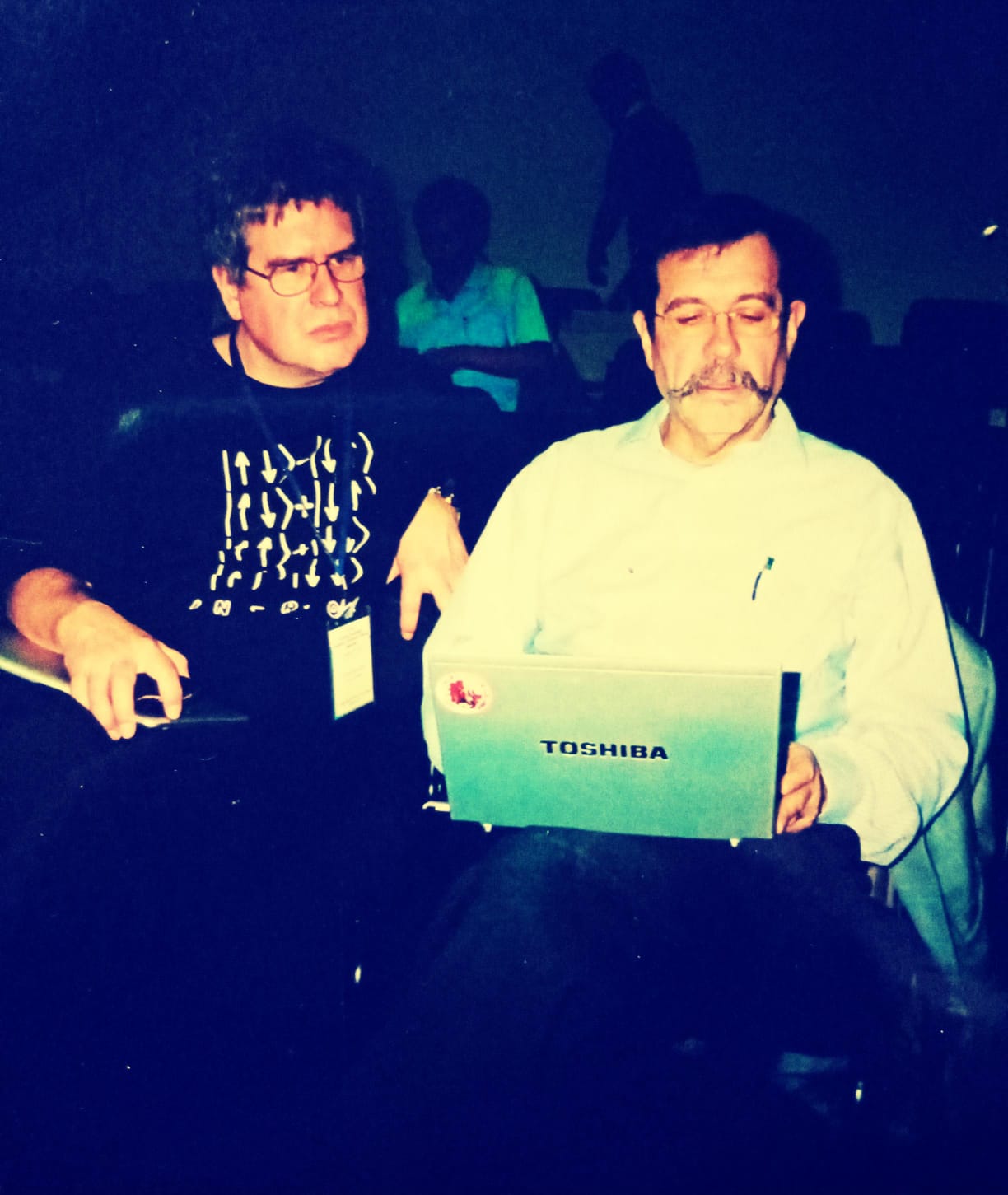}
		\end{center}
		\label{aspect}
        \caption{Photo clicked by the author in 2011 at an International Conference celebrating 75 years of quantum entanglement, organized by S.N.Bose National Centre for Basic Science and Bose Institute, Kolkata, India. Left: Samuel Braunstein and Gilles Brassard (pioneer in Quantum Cryptography). Right: Gilles Brassard with Alain Aspect (2022 Nobel Laureate in Physics).}
	\end{figure}
\chapter{Quantum Teleportation: What is it?}
\label{ch:qt}
To understand the technicalities involved in Quantum Teleportation, I shall first discuss about another important aspect of modern quantum mechanics which is known as \textit{Quantum no-cloning theorem}. Cloning simply means copying something. just as we like to make photocopies of the pages, (scientifically can be considered as classical states). So the basic question is, given a quantum state, can we make a single exact copy of it? The answer is, $``$No"! Let us first understand this from pedagogical point of view. This view was once articulated to me by one of my research collaborators. I was new to modern quantum mechanics then. In the classical scenario, when we want to copy a page, we insert the page inside the photocopying machine. We put a blank page on the tray and then start the machine. Most modern photocopiers use a technology called xerography, a dry process that uses electrostatic charges on a light-sensitive photoreceptor to first attract and then transfer toner particles (a powder) onto paper in the form of an image. The toner is then fused onto the paper using heat, pressure, or a combination of both. Copiers can also use other technologies, such as inkjet, but xerography is standard for office copying. Now my collaborator's imagination was that if we consider a quantum photocopying machine, which is, say, capable of copying a qubit, then the nature of the machine would be such that the output that would come out of the machine would get entangled with the original qubit in such a way that perfect copy couldn't be materialized. In due process, some of the information would get lost, when one would try to disentangle the photocopied qubit from the original qubit. (According to him, classically one may think of that scenario where the photocopy is coming out of the xerox machine but being glued to the original page.) The schematic diagram of my collaborator's version has been shown below.
\begin{figure}[h!]
		\begin{center}
			\includegraphics[width=5.5cm]{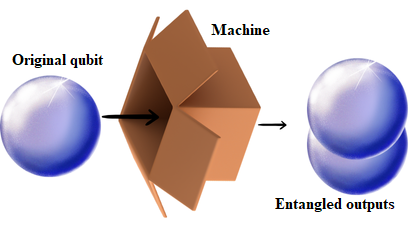}
		\end{center}
		\label{imagine}
        \caption{The figure represents a visualization of the fact the single quantum qubit cannot be perfectly cloned.}
	\end{figure}
When I could not understand what he was talking about, the collaborator suggested me to read an article by Wootters and Zurek, published in nature in the year 1982\cite{wootters1982single}. Below I briefly describe what I found.
\subsection{Quantum no-cloning theorem:}
Suppose we consider a hypothetical quantum cloning machine that can copy the state of a photon or an electron. If such a machine indeed exists then it is supposed to copy the state $\vert 0\rangle$ and make a copy $\vert 0\rangle\vert0\rangle$ while the state $\vert 1\rangle$ when passed through that quantum cloning machine, is expected to produce the state $\vert 1\rangle\vert 1\rangle$. In analogy with classical cloning machine, it is quite obvious case. When a classical state is given input to the cloning machine then in the output we get the copy of the classical state as well as the input state is also retrieved. Let us now talk about the quantum superposition of the states $\vert 0\rangle$ and $\vert 1\rangle$ which is given by $\vert s\rangle_{original} = \alpha\vert 0\rangle + \beta\vert 1\rangle$. If such a superposition state is passed through the quantum cloning machine, then from the linearity of the quantum mechanics the output state must be of the form $\vert e\rangle_{copy} = \alpha\vert 0\rangle \vert 0\rangle + \beta\vert 1\rangle \vert 1\rangle$. But in actual case, as one wants the original state and the copy both from the output section of the quantum cloning machine, the output state will be of the form $\vert s\rangle_{original}\vert s\rangle_{copy}$. A mere calculation reveals that such an output is of the type $\vert s\rangle_{original}\vert s\rangle_{copy} = (\alpha\vert 0\rangle + \beta \vert 1\rangle)_{original}(\alpha\vert 0\rangle + \beta \vert 1\rangle)_{copy} = \vert \alpha\vert^{2}\vert 0\rangle_{original}\vert 0\rangle_{copy} + \alpha\beta^*\vert 0\rangle_{original}\vert 1\rangle_{copy}+\beta\alpha^*\vert 1\rangle_{original}\vert 0\rangle_{copy}+\vert \beta\vert^{2} \vert 1\rangle_{original}\vert 1\rangle_{copy}$ which is definitely not same as $\vert e\rangle_{copy}$. This proves that the arbitrary single qubit quantum state cannot be copied perfectly. This is known as \textbf{no cloning theorem}\cite{wootters1982single}.
\subsection{Quantum Teleportation: Bennett Scheme}
One of the fundamental protocols, for sending information in quantum information science, was conceived by Charles H. Bennett and his collaborators in the year $1993$. The protocol was named as Quantum Teleportation (QT), but the idea is different from its classical counterpart (as perceived in the science fiction movie $``$\textit{The Fly}"). The basic idea of QT involves two parties Alice, who will play the role of sender and another party is Bob, playing the role of receiver of the information that Alice intends to communicate to him. Alice will possess a quantum system prepared in a state $\vert \Phi\rangle$. The state is unknown to Alice. She wants to send sufficient information to Bob, so that Bob can make an accurate copy of $\vert \Phi\rangle$. Now one can ask, what if Alice clones the state $\vert \Phi\rangle$ and sends it to Bob. This will be prohibited due to \textit{quantum no-cloning theorem}, discussed in the previous section, and which states that a single quantum system cannot be cloned. Hence Alice will devise another strategy to fulfil her requirements. In the words of J.L.Park, a $``$\textit{spin exchange measurement}" will take place in QT. The prosaic idea in QT is as follows. Alice will divide the complete information encoded in the system $\vert \Phi\rangle$ into two parts, viz. (a) classical and (b) non-classical. These parts are sent to Bob through two different channels. Having received these information from Alice, Bob can construct an accurate replica of $\vert \Phi\rangle$. In due process, the original state $\vert \Phi\rangle$ is destroyed at Alice's end, as it must obey no-cloning theorem. The classical part of the protocol puts restriction on instantaneous information transfer from Alice to Bob and thus doesn't defy physical laws. I shall now describe the Bennett QT protocol in a concise manner below. For the sake of clarity and as I had used the term spin exchange, I therefore, designate the two bits as $\vert \uparrow\rangle$ (spin-up state) and  $\vert \downarrow\rangle$ (spin-down state) (where $\vert \uparrow\rangle$ corresponds to $\vert 0\rangle$ and $\vert \downarrow\rangle$ corresponds to $\vert 1\rangle$). One can also represent the two bits as $``$vertical" and $``$horizontal" polarization of photons. The unknown system that contains complete information and which Alice holds will thus be represented by the qubit
\begin{eqnarray}
\label{teleportunknownstate}
\vert \phi_1\rangle = \alpha \vert \uparrow_{1}\rangle + \beta \vert \downarrow_{1}\rangle,
\end{eqnarray}
with the condition $\vert \alpha\vert^2 + \vert \beta\vert^2 = 1$. However the entangled channel that Alice and Bob will share is represented by
\begin{eqnarray}
\label{entangledalicebob}
\vert \varphi^-_{23}\rangle = \frac{1}{\sqrt{2}} \Big(\vert \uparrow_{2}\rangle \vert \downarrow_{3}\rangle - \vert \downarrow_{2} \rangle \vert \uparrow_{3}\rangle\Big).
\end{eqnarray}
The subscript $1$ in eq.(\ref{teleportunknownstate}) and $2$ in eq.(\ref{entangledalicebob}) denote the qubits are with Alice and $3$ in eq.(\ref{entangledalicebob}) denotes that the qubit is with Bob. The channel (\ref{entangledalicebob}) is known as \textit{singlet state}, with which Bennett ran His protocol. This state is one of the four Bell states that was defined in eq.(\ref{bellstates}). (We can represent all the Bell states of eq.(\ref{bellstates}) with $\vert \uparrow\rangle$ and $\vert \downarrow\rangle$). Now the complete state of three particles before Alice's measurement is thus
\begin{eqnarray}
\label{entangledaliceunbob1}
\vert \Phi_{123}\rangle &=& \nonumber\\ \vert \phi_1\rangle \otimes \vert \varphi^-_{23}\rangle &=&
\frac{1}{\sqrt{2}}\Big[\alpha \vert \uparrow_1\rangle\vert \uparrow_2\rangle\vert \downarrow_3\rangle - \alpha \vert \uparrow_1\rangle\vert \downarrow_2\rangle\vert \uparrow_3\rangle  + \beta \vert \downarrow_1\rangle\vert \uparrow_2\rangle\vert \downarrow_3\rangle - \beta \vert \downarrow_1\rangle\vert \downarrow_2\rangle\vert \uparrow_3\rangle\Big].\nonumber\\
\end{eqnarray}
Now from eq.(\ref{bellstates}), a little calculation will lead us to the following.
\begin{eqnarray}
\label{bellconvertcompbas}
\vert \uparrow \rangle \vert \uparrow\rangle &=& \frac{1}{\sqrt{2}}\Big[\vert \phi^+\rangle + \vert \phi^-\rangle\Big],\nonumber\\
\vert \downarrow \rangle \vert \downarrow\rangle &=& \frac{1}{\sqrt{2}}\Big[\vert \phi^+\rangle - \vert \phi^-\rangle\Big],\nonumber\\
\vert \uparrow \rangle \vert \downarrow\rangle &=& \frac{1}{\sqrt{2}}\Big[\vert \varphi^+\rangle + \vert \varphi^-\rangle\Big],\nonumber\\
\vert \downarrow \rangle \vert \uparrow\rangle &=& \frac{1}{\sqrt{2}}\Big[\vert \varphi^+\rangle - \vert \varphi^-\rangle\Big].
\end{eqnarray}
Using eqs.(\ref{bellstates}) and (\ref{bellconvertcompbas}) in (\ref{entangledaliceunbob1}) we get
\begin{eqnarray}
\label{entangledaliceunbob2}
\vert \Phi_{123}\rangle = \frac{1}{2}\Big[\vert \phi^{-}_{12}\rangle(-a\vert \uparrow_3\rangle - b\vert \downarrow_3\rangle) + \vert \phi^{+}_{12}\rangle(-a\vert \uparrow_3\rangle + b\vert \downarrow_3\rangle) +\nonumber\\  \vert \varphi^-_{12}\rangle(a\vert \downarrow_3\rangle + b\vert \uparrow_3\rangle) + \vert \varphi^+_{12}\rangle(a\vert \downarrow_3\rangle - b\vert \uparrow_3\rangle)\Big].
\end{eqnarray}
It follows that, regardless of the unknown state $\vert \phi_1\rangle$, the four measurement outcomes are equally likely, each occurring with probability $\frac{1}{4}$. After Alice's Bell State (BS) joint measurement, particle $3$ will be projected into one of the four pure states superposed in eq.(\ref{entangledaliceunbob2}), according to the measurement outcome, viz. $-a\vert \uparrow_3\rangle - b\vert \downarrow_3\rangle$, $-a\vert \uparrow_3\rangle + b\vert \downarrow_3\rangle$, $a\vert \downarrow_3\rangle + b\vert \uparrow_3\rangle$ and $a\vert \downarrow_3\rangle - b\vert \uparrow_3\rangle$ respectively. Each of the pure states is related in a simple way to the original state $\vert \phi_1\rangle$ which Alice sought to teleport to Bob. Now it is known that the Pauli spin operators are unitary operators. 
\begin{table}[h!]
\begin{center}
\caption{Teleportation of an unknown qubit from Alice to Bob}
\label{table1}
\begin{tabular}{|c|c|c|}
\hline
Alice's  outcome & $\vert q_3\rangle$ & Bob applies\\
\hline
$\vert \phi^{-}_{12}\rangle$ & $-a\vert \uparrow_3\rangle - b\vert \downarrow_3\rangle$ & $I\vert q_3\rangle$ \\
\hline
 $\vert \phi^{+}_{12}\rangle$ & $-a\vert \uparrow_3\rangle + b\vert \downarrow_3\rangle$ & $-(\vert \uparrow\rangle\langle \uparrow\vert - \vert \downarrow\rangle\langle \downarrow\vert)\vert q_3\rangle $\\
\hline
 $\vert \varphi^-_{12}\rangle$ & $a\vert \downarrow_3\rangle + b\vert \uparrow_3\rangle$ & $(\vert \uparrow\rangle\langle \downarrow\vert + \vert \downarrow\rangle\langle \uparrow\vert)\vert q_3\rangle $  \\
\hline
$\vert \varphi^+_{12}\rangle$ & $a\vert \downarrow_3\rangle - b\vert \uparrow_3\rangle$ & $-(\vert \uparrow\rangle\langle \downarrow\vert - \vert \downarrow\rangle\langle \uparrow\vert)\vert q_3\rangle $ \\
\hline
\end{tabular}
\end{center}
\end{table}
The column $3$ of table \ref{table1} shows what Bob needs to apply on $\vert q_3\rangle$ of column $2$ to get a perfect replica of $\vert \phi_1\rangle$ while column $1$ shows that Alice is left with states $\vert \phi^{\pm}\rangle$ and $\vert \varphi^{\pm}\rangle$, without any trace of original state $\vert \phi_1\rangle$. It is to be noted that Alice sends her measurement outcomes to Bob via a classical channel (which may be throug email or simple telephone call). This protocol is famously known as \textit{Bennett's protocol} of teleporting an unknown quantum state\cite{bennett1993teleporting}. Any of the four Bell states can be used as quantum channel for teleporting the unknown qubit under Bennett's protocol.
\subsection{Quantum Teleportation: First experimental proposal}
Physicists derive protocols and experimentalists implement them in laboratory. The same was true for QT, when Dirk Bowmeester et al publihsed their paper in Nature, in the year 1997. The paper was titled $``$Experimental Quantum Teleportation"\footnote{In the paper published in Nature, it is written Dik Bowmeester, but the author of this book, found conflict in the first name of Bowmeester as Dirk in Wiki source.}. We already know that teleportation involves reconstruction of the state of an unknown qubit at some distant location, on the basis of a few measurements performed on the part of sender and receiver. A few questions may arise from experimental perspective. (a) How precisely can this be a true copy of the original qubit? (b) What if these parts and pieces are electrons, atoms and molecules? (c) What happens to their individual quantum properties, which as per Heisenberg's uncertainty principle cannot be measured with arbitrary precision. These questions have been answered by Dirk Bowmeester and his team\cite{bouwmeester1997experimental}. 
The transfer of the state of an unknown qubit from Alice to Bob can happen over arbitrary distances though experimentally it was shown then (in the $90$'s) that quantum entanglement survived over the distances of the order of 10 kilometers\cite{tittel1998experimental}. But for the teleportation scheme it is not necessary for Alice to know where actually Bob is. The two most challenging tasks for experimental realization of teleportation demands production and measurement of entangled states. Just as Bennett et al modeled teleportation with the singlet state of eq.(\ref{entangledalicebob}), so did Bowmeester et al. They produced the entangled photons 2 and 3 by parametric down-conversion. In this technique, inside a nonlinear crystal, an incoming pump photon can decay spontaneously into two photons which,
in the case of type II parametric down-conversion, are in the state given by eq.(\ref{entangledalicebob}). To achieve projection of photons $1$ and $2$ into a Bell state (see eq.(\ref{bellstates}), the photons needed to be indistinguishable; and if they are incident one from each side, the question was how could it happen that they emerged still one on each side? The answer lied in the fact that, this could happen if the photons are either both reflected or both transmitted. Quantum Mechanics (QM) suggests that if the amplitudes for the above two possibilities are superimposed, the question raised above can be answered. Now when both photons are reflected, unitarity of the system imposes additional minus sign, thus the two process cancel each other and this is always true for symmetric states. However, for antisymmetric states, the two possibilities obtain another relative minus sign, and in this case they constructively interfere. Hence Bowmeester and group for projecting photons $1$ and $2$ onto the anti-symmetric state $\vert \varphi^{-}_{12}\rangle$, placed detectors in each of the outputs of the beam-splitter and thereby registered simultaneous detections (coincidence). To make sure that photons 1 and 2 could not be distinguished by their arrival times, they were generated using a pulsed pump beam and sent through narrow-bandwidth filters producing a coherence time much longer than the pump pulse length. In the experiment, the pump pulses had a duration of $200 \:fs$ at a repetition rate of 76 MHz. Observing the down-converted photons at a wavelength of $788 \:nm$ and a bandwidth of $4 nm$ resulted in a coherence time of $520 \:fs$\footnote{$fs$ denotes femto-second and $1 \:fs = 1\times 10^{-15}$ of a second. Also $nm$ denotes nano-meter and $1\: nm = 1 \times 10^-9$ $m$.}. It should be mentioned that, because photon $1$ was also produced as part of an entangled pair, its partner could serve to indicate that it was emitted. Below I show the schematic of the experimental design that Bowmeester et al proposed exactly in the form as it was there in the paper\cite{bouwmeester1997experimental}.
\begin{figure}[h!]
		\begin{center}
			\includegraphics[width=5.5cm]{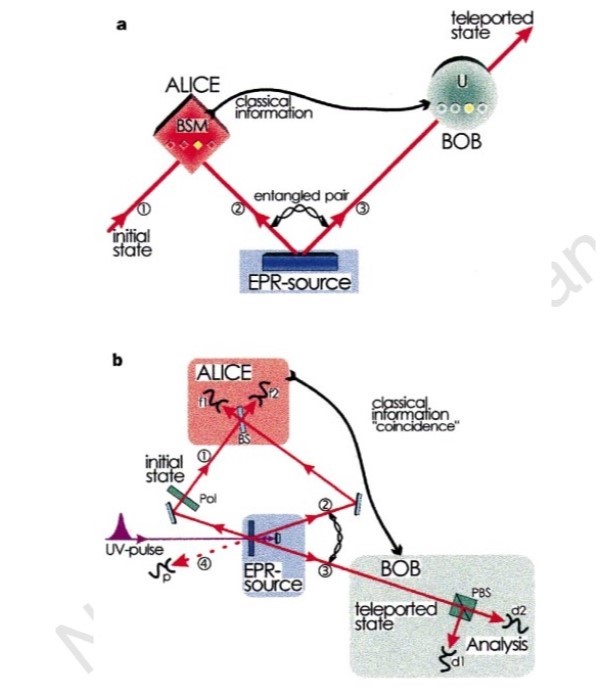}
		\end{center}
		\label{bowmeester1}
        \caption{The figure is the schematic that Bouwmeester et al showed in their original paper.}
	\end{figure}
Bennett et al teleportation protocol works for a complete basis, a set of known states into which any other states can be decomposed. Now a basis for polarization states has just two components, horizontal ($\vert H\rangle$) and vertical ($\vert V\rangle$). According to Bowmeester et al, these two polarization directions are preferred directions in their experiment. Hence, they chose as the basis for teleportation the two states linearly polarized at $-45^0$ and $+45^0$ which are superposition of $\vert H\rangle$ and $\vert V\rangle$ polarizations. They also demonstrated teleportation for circular polarization. In the Fig.\ref{bowmeester1}, $f1$ and $f2$ are the pair of detectors on Alice's side while $d1$ and $d2$ are the detectors at Bob's side. First, photon $1$ was polarized at $+45^0$. Teleportation would work when photons $1$ and $2$ were detected in the state $\vert \varphi^{-}_{12}$, which would occur in $\frac{1}{4}$ of all possible cases. The state $\vert \varphi^{-}_{12}\rangle$ was identified by recording a coincidence between two detectors, $f1$ and $f2$, placed behind the beam splitter (BS). If a $f1f2$ type coincidence, between detectors $f1$ and $f2$, is detected, photon $3$, should then also be polarized at $+45^0$, which is analyzed by passing photon $3$ through a polarizing beam splitter selecting $+45^0$ and $-45^0$ polarization. To demonstrate teleportation, only detector $d2$ at the $+45^0$ output of the polarizing beam splitter should click (that is, register a detection) once detectors $f1$ and $f2$ click. Detector $d1$ at the $-45^0$ output of the polarizing beam splitter should not detect a photon. Therefore, recording a three-fold coincidence $d2f1f2$ ($+45^0$ analysis) together with the absence of a
three-fold coincidence $d1f1f2$ ($-45^0$ analysis) is a proof that the polarization of photon $1$ has been teleported to photon $3$ in Bowmeester et al set up. Secondly, to rule out any classical explanation for the experimental results, they had produced further confirmation that their procedure worked by additional experiments. In these experiments they teleported photons linearly polarized at $0^0$ and at $90^0$, and also teleported circularly polarized photons. Dirk (Dik) is currently involved in researches in the fields of  solid-state cavity quantum electrodynamics, knotted states of light, micro-optomechanical systems,and DNA-templated optical emitters. He was awarded the 2014 Spinoza Prize\footnote{The Spinoza Prize (Dutch: Spinozapremie) is an annual award of 1.5 million euro prize money, to be spent on new research given by the Dutch Research Council (NWO).}, the highest scientific award in the Netherlands. This is also to be noted that, when Bowmeester worked on experimental quantum teleportation, one of his collaborators was an Austrian Physicist named Anton Zeilinger, the $2022$ Nobel Laureate in Phsyics, for his experiments with entangled photons, establishing the violation of Bell inequalities and pioneering quantum information science, along with Alain Aspect and John Caluser.\\\\ 
In contemporary time of Bowmeester et al experiment, another group reported on a quantum optical experimental implementation of teleportation of unknown pure quantum states. Boschi et al exhibited results for the teleportation of a linearly polarized state and of an elliptically polarized state\cite{boschi1998experimental}. However, the Boschi group was not happy with the published version of the experimental realization of quantum teleportation proposed by Bowmeester et al. In an article published in $``$Physics World", Francesco De Martini wrote and I am quoting \textit{$``$The article on quantum teleportation (January p5), which said that this phenomenon had been demonstrated for the first time by a group from the University of Innsbruck, is not quite correct. The first successful quantum teleportation experiment was in fact performed by our group in Rome. Our paper, "Experimental realisation of teleporting an unknown pure quantum state via dual classical and Einstein-Podolsky-Rosen channels" (D Boschi et al. 1998 Phys. Rev. Iett. 80 1121), was submitted in July of last year and published last month. The teleportation paper by the Innsbruck group, led by Anton Zeilinger, was submitted on 16 October of last year and published in December (Bouwmeester et al. 1997 Nature390 575)."}\cite{de1998teleportation}.\\\\
In the year $2004$, a very interesting paper got published in Nature\cite{ursin2004quantum}, the paper was titled $``$Quantum teleportation across the Danube". In the book $``$Dance of the Photons"\cite{zeilinger2010dance}, Anton, very beautifully and lucidly described his Danube experiment on teleportation. In this experiment, an $800$-meter-long optical fiber wire was installed in a public sewer system underneath the Danube River in Vienna Austria, and it was exposed to temperature changes and other environmental influences. Alice must perform a joint Bell state measurement (BSM) on photon $1$, the input photon, and photon $2$, her part of the entangled photon pair (photons $2$ and $3$). Photon $3$, Bob's receiver photon, will contain all of the information on the input photon $1$, except for a phase rotation that depends on the state that Alice observed. This experiment implemented an active feed-forward system that sends Alice's measurement results via a classical microwave channel with a fast electro-optical modulator in order to exactly replicate Alice's input photon. The teleportation fidelity obtained from the linear polarization state at $+45^0$ varied between $0.84$ and $0.90$, which is well above the classical fidelity limit of $0.66$\footnote{The word fidelity will be discussed in the later chapters.}. 
\begin{figure}[h!]
		\begin{center}
			\includegraphics[width=7.5cm]{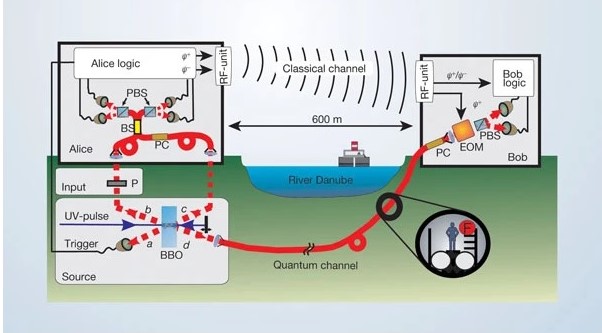}
		\end{center}
		\label{danube1}
        \caption{The figure shows the set up of teleportation experiment across Danube river by Zeilinger and his team.}
	\end{figure}\\\\
The schematic of Danube experiment is shown below in Fig.$2.4$.
Another research paper is worth noting here, though I personally think that the details of it will be a little harder for under graduate students to understand until they start a full scale serious research works on quantum information processing. So I skip detailing the work and limit myself to deliver only a glimpse of it. In the year $1995$, two physicists named S. L. Braunstein and A. Mann proposed a scheme for measuring an optical version of the Bell operator, using a generalization of the H.O.Mandel interferometer. The calculation was on discrete mode and their study consequently showed to be sufficient to allow teleportation of the state of polarization of a photon with conditional efficiency approaching \cite{braunstein1995measurement}.
\begin{figure}[h!]
		\begin{center}
			\includegraphics[width=4.2cm]{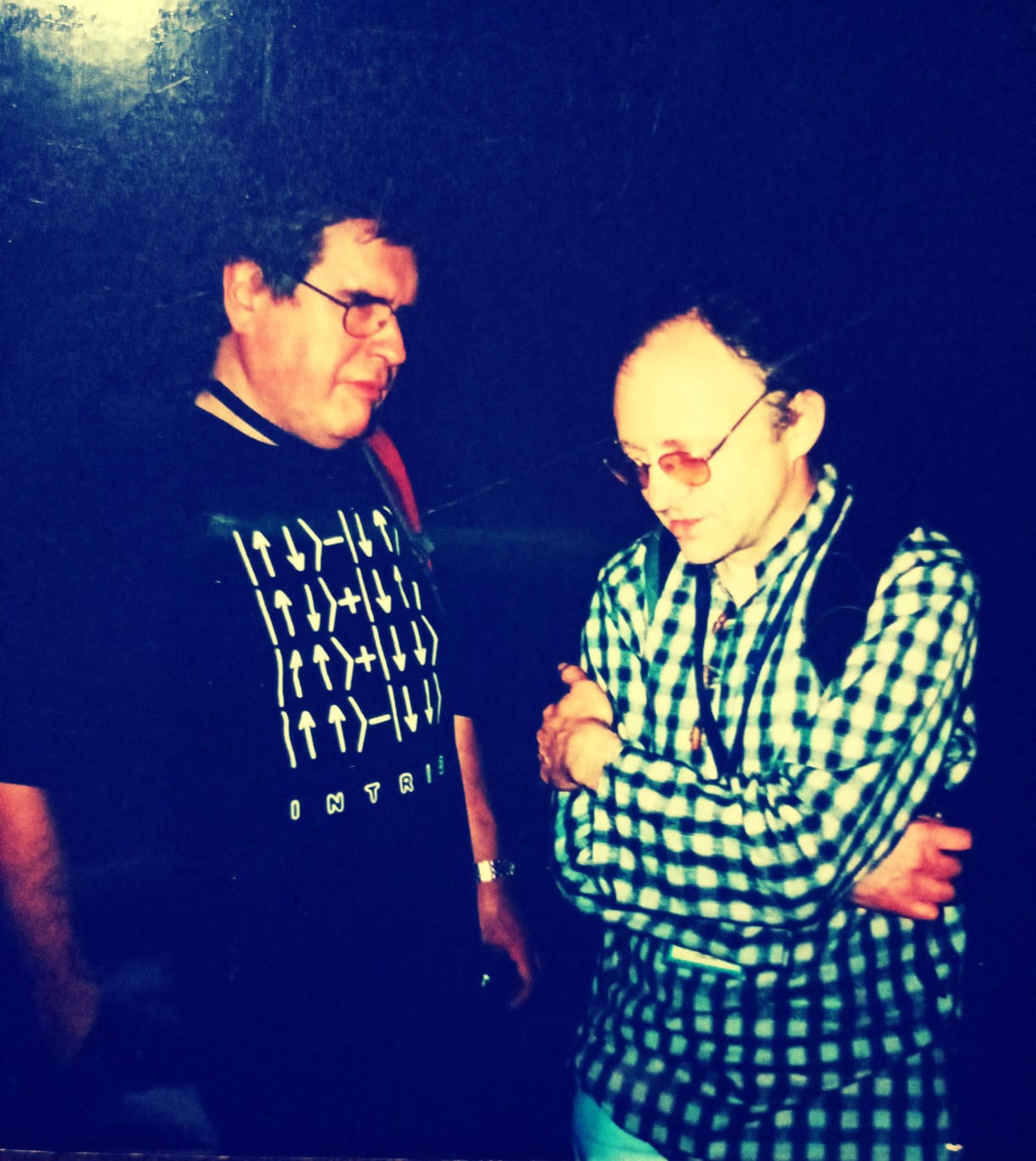}
		\end{center}
		\label{danube1}
        \caption{Gilles Brassard and Samuel L. Braunstein in a serious discussion mode. The photo has been clicked by the author himself in the year 2011.}
	\end{figure}\\\\
Before concluding this chapter I shall briefly describe another important type of teleportation protocol conceived by two eminent Indian Scientists, showing that in this rat-race of researches on quantum teleportation, Indians are also not falling behind. Prof. Arun Kumar Pati is a prominent Indian physicist notable for his researches in quantum information, quantum computation and Foundations of quantum mechanics. He has made pioneering contributions in the area of quantum information. Professor Pati is the recipient of the India Physics Association Award for Young Physicist of the Year ($2000$) and the Indian Physical Society Awards for Young Scientists ($1996$). Professor Pankaj Agrawal received his PhD from the University of Washington, Seattle. After spending more than $25$ years at the Institute of Physics, Bhubaneswar, he joined CQuERE\footnote{CQuERE: Centre for Quantum Engineering, Research and Education}. He has broad interest in quantum information, including quantum correlations, quantum nonlocality, quantum communication, and quantum cryptography. Agrawal and Pati, in the year $2002$, considered a generalized quantum teleportation protocol for an unknown qubit using non-maximally entangled state as a shared resource. All the four Bell states defined in eq.(\ref{bellstates}) are maximally entangled pure states. Agrawal and Pati investigated the teleportation of an unknown state with unit fidelity but less than unit probability when two parties share a non-maximally entangled state. They termed it as \textit{probabilistic teleportation}\cite{agrawal2002probabilistic}.

\subsection{Probabilistic Quantum Teleportation: Agrawal \& Pati approach}
The pure non-maximally entangled state that Alice and Bob would share as quantum channel for teleportation, as proposed by Agrawal and Pati is defined as
\begin{eqnarray}
    \label{agrapatinonmax}
    \vert \varphi_{nm}\rangle_{23} = \frac{1}{\sqrt{1 + \vert n\vert^2}}\Big(\vert \uparrow_2\rangle\vert \uparrow_3\rangle + n\vert \downarrow_2\rangle\vert \downarrow_3\rangle\Big),
\end{eqnarray}
where $n$ is a known complex number and also qubits $2$ and $3$ are with Alice and Bob respectively. It is a known fact that because of the existence
of Schmidt decomposition any two qubit entangled state $\phi \in \mathcal{H}^2\otimes \mathcal{H}^2$ i.e.
\begin{eqnarray}
    \label{gentwoqubitstate}
    \alpha\vert\uparrow\rangle\vert \uparrow\rangle + \beta\vert\uparrow\rangle\vert \downarrow\rangle + \gamma\vert\downarrow\rangle\vert \uparrow\rangle + \delta\vert\downarrow\rangle\vert \downarrow\rangle,
\end{eqnarray}
can be written as a superposition of two basis vectors and here $\mathcal{H}^2$ is the two-dimensional Hilbert space\cite{pati2000existence}. In general given an arbitrary two-qubit state Eq.(\ref{gentwoqubitstate}), the computational basis $\vert \uparrow\rangle$ and $\vert \downarrow\rangle$ need not be the Schmidt basis, but Agrawal and Pati assumed that Alice and Bob knew the Schmidt basis and coefficients. Then Eq.(\ref{agrapatinonmax}) is the most general non-maximally entangled state up to local unitary transformations relating Schmidt basis and computational basis states. Now the state Alice wished to teleport to Bob was $\vert \psi\rangle_{1} = a\vert \uparrow_1\rangle + b\vert \downarrow_1\rangle$, with $\vert a\vert^2 + \vert b\vert^2 = 1$. In order to send the unknown qubit $\vert \psi\rangle_{1}$ to Bob, Alice would make a joint measurement on the two qubits, viz. $2$, that comes out of the entangled state (\ref{agrapatinonmax}) and held by Alice and the other qubit is $1$ which is the unknown qubit that Alice wished to send. The difference between the Bennett's teleportation protocol and that put forward by Agrawal and Pati lied in the fact that, if Alice performed a measurement in the Bell-basis states (see Eq.(\ref{bellstates}), then $\vert \psi\rangle_{1}$ could not be teleported faithfully i.e. with unit fidelity and unit probability. However, if the measurement was taken with respect to a non-maximally entangled basis, it would be then possible for Alice to send the unknown qubit to Bob with unit fidelity, thought not with unit probability. Therefore, Agrawal and Pati called it probabilistic teleportation. They constructed the mutually orthogonal basis vectors as
\begin{eqnarray}
    \label{agrawalpatibasis}
  \vert \varphi^+_{l}\rangle &=& \frac{1}{\sqrt{1 + \vert l\vert}^2}\Big(\vert \uparrow\rangle\vert\uparrow\rangle + \vert \downarrow\rangle\vert\downarrow\rangle\Big),\nonumber\\
  \vert \varphi^-_{l}\rangle &=& \frac{1}{\sqrt{1 + \vert l\vert}^2}\Big(l^*\vert \uparrow\rangle\vert\uparrow\rangle - \vert \downarrow\rangle\vert\downarrow\rangle\Big),\nonumber\\
  \vert \varphi^+_{p}\rangle &=& \frac{1}{\sqrt{1 + \vert p\vert}^2}\Big(\vert \uparrow\rangle\vert\downarrow\rangle + p\vert \downarrow\rangle\vert\uparrow\rangle\Big),\nonumber\\
  \vert \varphi^+_{p}\rangle &=& \frac{1}{\sqrt{1 + \vert p\vert}^2}\Big(p^*\vert \uparrow\rangle\vert\downarrow\rangle - \vert \downarrow\rangle\vert\uparrow\rangle\Big).
\end{eqnarray}
Here $l$ and $p$ are complex numbers in general. When $l = p = 0$, this basis reduces to the computational basis which is not entangled. For $l = p = 1$, it reduces to the Bell basis which is maximally entangled. Therefore this set interpolates between untangled and maximally entangled set of basis vectors. Using Eq.(\ref{agrawalpatibasis}), the computational basis vectors $\vert \uparrow\rangle\vert \uparrow\rangle, \: \vert \uparrow\rangle \vert \downarrow\rangle,\: \vert \downarrow\rangle\vert \uparrow\rangle,\: \vert \downarrow\rangle\vert \downarrow\rangle$ can be written as follows.
\begin{eqnarray}
    \label{agrapatibasischange}
    \vert \uparrow\rangle\vert \uparrow\rangle &=& \frac{1}{1+\vert l\vert^2}\Big(\vert \varphi_{l}^+\rangle + l \vert \varphi_{l}^-\rangle\Big),\nonumber\\
     \vert \uparrow\rangle\vert \downarrow\rangle &=& \frac{1}{1+\vert p\vert^2}\Big(\vert \varphi_{p}^+\rangle + p \vert \varphi_{p}^-\rangle\Big),\nonumber\\
     \vert \downarrow\rangle\vert \uparrow\rangle &=& \frac{1}{1+\vert p\vert^2}\Big(p^*\vert \varphi_{p}^+\rangle -  \vert \varphi_{p}^-\rangle\Big),\nonumber\\
      \vert \downarrow\rangle\vert \downarrow\rangle &=& \frac{1}{1+\vert l\vert^2}\Big(l^*\vert \varphi_{l}^+\rangle -  \vert \varphi_{l}^-\rangle\Big).\nonumber\\
\end{eqnarray}
Using the non-maximally entangled basis states given in Eq.(\ref{agrawalpatibasis}), one can rewrite the combined state of the input (the qubit to be sent) and resource state (the entangled channel between Alice and Bob) as
\begin{eqnarray}
    \label{clubbedstateagrapati}
\vert \psi\rangle_{1}\vert \varphi_{nm}\rangle_{23} = N\Big[\vert \varphi^+_{l}\rangle_{12}(La\vert \uparrow_3\rangle + Lnbl^*\vert \downarrow_3\rangle) + \vert \varphi^-_{l}\rangle_{12}(Lla\vert \uparrow_3\rangle - Lnb\vert \downarrow_3\rangle) \nonumber\\+ \vert \varphi^+_{p}\rangle_{12}(Ppb\vert \uparrow_3\rangle + Pan\vert \downarrow_3\rangle)  +\varphi^-_{p}\rangle_{12}(-Pb\vert \uparrow_3\rangle + Panp\vert \downarrow_3\rangle)\Big],\nonumber\\
\end{eqnarray}
where, $L=\frac{1}{\sqrt{1+\vert l\vert^2}}$, $N=\frac{1}{\sqrt{1+\vert n\vert^2}}$ and $P=\frac{1}{\sqrt{1+\vert p\vert^2}}$ are real numbers. Given the value of $n$, Agrawal and Pati agreed upon various choices of the parameters $l$ and $p$, to make the unknown qubit teleported to Bob with unit fidelity and non-zero probability. Three cases, they have considered.
\begin{itemize}
    \item \textit{Standard teleportation protocol:} If one chooses $n=l=\frac{1}{l^*}=p=\frac{1}{p^*}$, where, $l,\:n,\: p$ can
be pure phases, i.e., complex numbers of unit modulus, then, faithful teleportation is possible with unit fidelity and unit probability. This is classic teleportation\cite{bennett1993teleporting}.
\item \textit{Probabilistic teleportation protocol:} If one makes the choice such as $l=n=p^*$ or $l=n=\frac{1}{p^*}$ or $l^* = \frac{1}{n} = p$ or $l^* = \frac{1}{n} = \frac{1}{p^*}$, then for any of these choices, reliable (i.e., with unit fidelity) teleportation is possible for only two out of four possible results of the measurement. For
example, in the case of first choice, when the outcome is $\vert \varphi^{-}_{l=n}$, then the state at Bob's hand will be $a\vert \uparrow\rangle - b\vert \downarrow\rangle$ and when the outcome is $\vert \varphi^{+}_{l=n}$, then the state at Bob's hand is $b \vert \uparrow\rangle + a\vert \downarrow\rangle$. Therefore, when Alice sends two classical bits to Bob he will apply $\sigma_{z}$ in the former and $\sigma_{x}$ in the later case to recover the unknown state with unit fidelity. The total probability of this successful teleportation will be given by $P_{succ} = \frac{2\vert n\vert^2}{(1+\vert n\vert^2)^2}$.
\item \textit{No teleportation:} If the values of $p$ and $l$ are not related with that of $n$, then teleportation is not possible with unit fidelity.
\end{itemize}
\chapter{Quantum Teleportation: How did $3$ qubits come into play?}
\label{ch:qtm}
Daniel M. Greenberger, Michael A. Horne and Anton Zeilinger, in the years $1989$ and $1990$, back to back published two milestone papers in the field of Quantum Mechanics (QM); the papers were titled as (a) $``$ Going beyond Bell's theorem"\cite{greenberger1989going} and (b) $``$Bell's theorem without inequalities"\cite{greenberger1990bell}. The essence of these two papers are as follows. Bell proved that one couldn't, in general, reproduce the results of quantum theory with a classical deterministic local model. Greenberger, Horne and Zeilinger ($GHZ$), in their first paper, considered a slightly more complicated model than Bell's and made the premise for making declaration that, for the appropriate $3-$ particle (or even $4-$ particle) system, quantum theory did hold experimentally and could not be classically duplicated. This first paper was a conference paper and set the outline of $GHZ$'s more elaborate experimental paper published in the subsequent year of $1990$. The second paper\cite{greenberger1990bell} (which Abner Shimony also co-authored) seems, however, very mysterious to me, in the sense that, the authors of the paper while making comments on their experimental arguments, kept on writing $GHZ$ instead of $GHSZ$. Anyways, $GHZ$ demonstrated that the premises of Einstein-Podolsky-Rosen (EPR) paper were inconsistent when applied to quantum systems consisting of at least three partcles. I shall leave behind the detailed interesting discussion put forward by $GHZ$ (any interested reader can go through the contents of the paper\cite{greenberger1990bell}) for the limitations of pages as well as the motivation of writing this book. After publication of the above two papers, scientific community started calling the state proposed by Greenberger-Horne-Zeilinger as $GHZ$ state (or may be the premise had been set in their paper itself). 

The state is defined as
\begin{eqnarray}
    \label{ghz1}
    \vert GHZ\rangle = \frac{1}{\sqrt{2}}\Big[\vert H\rangle \vert H\rangle \vert H\rangle + \vert V\rangle \vert V\rangle \vert V\rangle\Big],
\end{eqnarray}
or more poularly as
\begin{eqnarray}
    \label{ghz2}
    \vert GHZ\rangle = \frac{1}{\sqrt{2}}\Big[\vert 0\rangle \vert 0\rangle \vert 0\rangle + \vert 1\rangle \vert 1\rangle \vert 1\rangle\Big].
\end{eqnarray}
Here $\vert H\rangle$ and $\vert V\rangle$ denote respectively the horizontal and vertical polarization states.\\\\
Now as teleportation had already been successfully performed, first with pen and paper, and then was demonstrated experimentally, the immediate next question to be asked was that, what would happen, when one introduced more qubits into the picture? What new scheme could be developed using a three-particle entangled state? Anders Karlsson and Mohamed Bourennane, from Department of Electronics, Laboratory of Photonics and Microwave Engineering, Royal Institute of Technology (KTH), Electrum 229, 164 40 Kista, Sweden, in the year $1998$ came up with the answer in their paper titled, $``$ Quantum teleportation using three-particle entanglement"\cite{karlsson1998quantum} (they only considered polarization entanglement case though).

\subsection{QT: Karlsson-Bourennane Scheme} 
As before Alice ($A$) holds an unknown qubit, which is denoted by
\begin{eqnarray}
    \label{karlssonqubit}
    \vert \phi\rangle_{1} = a \vert H_{1}\rangle + b\vert V_{1}\rangle.
\end{eqnarray}
To perform teleportation, Alice now shares $3-$ qubit $GHZ$ state as quantum channel for teleporting her unknown state, and she shares it with two other parties Bob ($B$) and Cliff ($C$). The $GHZ$ state has already been defined in Eq.(\ref{ghz1}), yet we re-write it once again as
\begin{eqnarray}
    \label{karlssonghz}
    \vert \Psi_{GHZ}\rangle_{234} =  \frac{1}{\sqrt{2}}\Big[\vert H_{2}\rangle\vert H_{3}\rangle\vert H_{4}\rangle +\vert V_{2}\rangle\vert V_{3}\rangle\vert V_{4}\rangle\Big].
\end{eqnarray}
Here the particles $1$ and $2$ are kept by Alice whereas the particles $3$ and $4$ are held by Bob and Cliff respectively and that the parties Bob and Cliff are remotely separated. The situation is like, as shown below.
\begin{figure}[h!]
		\begin{center}
			\includegraphics[width=6.5cm]{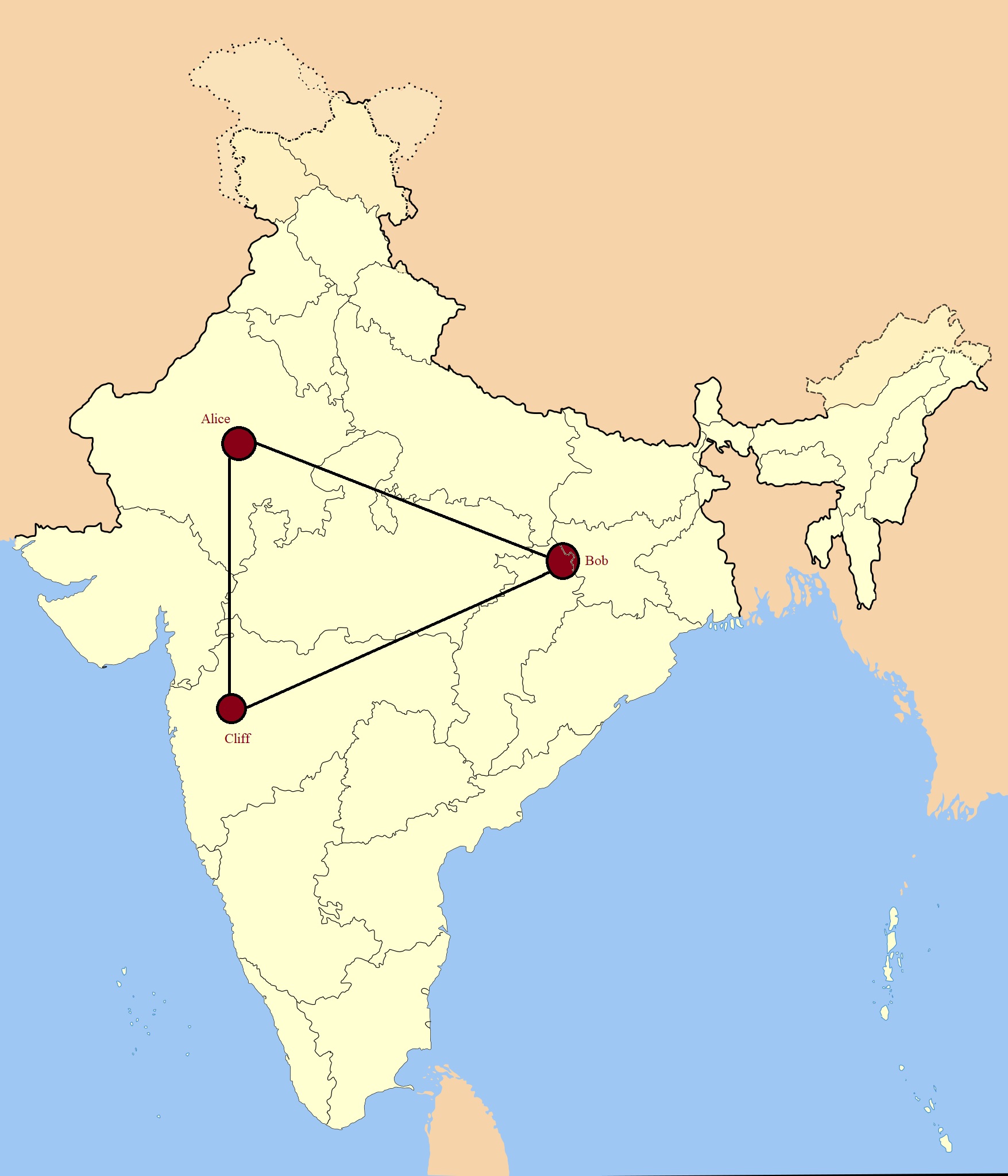}
		\end{center}
		\label{danube1}
        \caption{The figure shows how the qubits are located spatially throughout different parts of India and are still entangled to one another.}
	\end{figure}\\\\
At this stage, I shall discuss the Karlsson-Bourennane scheme in my own style and not in the way it was depicted in the original paper.\\\\
Now Alice, as before, clubs her qubits with the shared entangled state given in Eq.(\ref{karlssonghz}) and subsequently the following is obtained after a little manipulation.
\begin{eqnarray}
\label{karl1}
    \vert \Phi\rangle_{1234} = \frac{1}{2}\Big[\vert \phi^+\rangle_{12}(a\vert H_3\rangle\vert H_4\rangle + b \vert V_3\rangle\vert V_4\rangle) + \vert \phi^-\rangle_{12}(a\vert H_3\rangle\vert H_4\rangle - b \vert V_3\rangle\vert V_4\rangle)   \nonumber\\ \vert \varphi^+\rangle_{12}(a\vert V_3\rangle\vert V_4\rangle + b \vert H_3\rangle\vert H_4\rangle + \vert \varphi^-\rangle_{12}(a\vert V_3\rangle\vert V_4\rangle - b \vert H_3\rangle\vert H_4\rangle) \Big]\nonumber\\ 
\end{eqnarray}
We can re-express Eq.(\ref{karl1}) as 
\begin{eqnarray}
\label{karl2}
    \vert \Phi\rangle_{1234} = \frac{1}{2}\Big[\vert \phi^+\rangle_{12}(a\vert H_3\rangle\otimes \vert H_4\rangle + b\vert V_3 \rangle \otimes \vert V_4\rangle) + \vert \phi^-\rangle_{12}(a\vert H_3\rangle\otimes b\vert H_4\rangle - \vert V_3 \rangle \otimes \vert V_4\rangle)   \nonumber\\ \vert \varphi^+\rangle_{12}(a\vert V_3\rangle\otimes \vert V_4\rangle + b\vert H_3 \rangle \otimes \vert H_4\rangle)  + \vert \varphi^-\rangle_{12}(a\vert V_3\rangle\otimes \vert V_4\rangle - b\vert H_3 \rangle \otimes \vert H_4\rangle) \Big]\nonumber\\ 
\end{eqnarray}
 We know that the qubit $4$ is with Cliff. So now Cliff expresses his qubit as linear superposition of two other qubits (may be diagonal qubits). Mathematically there is a change of basis at Cliff's hand. He chooses diagonal polarizations as $\lbrace \vert \nearrow\rangle, \; \vert \nwarrow\rangle\rbrace$ and expresses his qubits as
 \begin{eqnarray}
     \label{cliffchanges}
     \vert H_4\rangle &=& \frac{1}{\sqrt{2}} \Big[\vert \nearrow_4\rangle + \vert \nwarrow_4\rangle\Big],\nonumber\\
     \vert V_4\rangle &=& \frac{1}{\sqrt{2}} \Big[\vert \nearrow_4\rangle - \vert \nwarrow_4\rangle\Big].
 \end{eqnarray}
 Using Eq.(\ref{cliffchanges}) in Eq.(\ref{karl2}), we get the following.
 \begin{eqnarray}
     \label{withcliff'sbasis}
  \vert \Phi\rangle_{1234} =  \frac{1}{2\sqrt{2}}\Big[\vert \phi^+\rangle_{12}\lbrace (a\vert H_3\rangle + b\vert V_3\rangle)\otimes \vert \nearrow_4\rangle + (a\vert H_3  - b\vert v_3\rangle) \otimes \vert \nwarrow_4\rangle\rbrace  +\nonumber\\ \vert \phi^-\rangle_{12}\lbrace (a\vert H_3\rangle - b\vert V_3\rangle)\otimes \vert \nearrow_4\rangle + (a\vert H_3  + b\vert v_3\rangle) \otimes \vert \nwarrow_4\rangle\rbrace\nonumber\\\ + \vert \varphi^+\rangle_{12}\lbrace (a\vert V_3\rangle + b\vert H_3\rangle)\otimes \vert \nearrow_4\rangle + (-a\vert V_3  + b\vert H_3\rangle) \otimes \vert \nwarrow_4\rangle\rbrace \nonumber\\ + \vert \varphi^-\rangle_{12}\lbrace (a\vert V_3\rangle - b\vert H_3\rangle)\otimes \vert \nearrow_4\rangle + (-a\vert V_3  - b\vert H_3\rangle) \otimes \vert \nwarrow_4\rangle\rbrace\Big].\nonumber\\
 \end{eqnarray}
 Now for every joint von-Neumann measurement that Alice makes on her qubits, Cliff has two possibilities. As for example, if Alice's measurement result reveals $\phi^+$, Cliff is left with $(a\vert H_3\rangle + b\vert V_3\rangle)\otimes \vert \nearrow_4\rangle + (a\vert H_3  - b\vert v_3\rangle) \otimes \vert \nwarrow_4\rangle$. (Here all the measurements that Alice makes are with respect to Bell basis as per Eq.(\ref{bellstates}). Cliff, after receiving telephone call from Alice, then makes measurements with respect to his diagonal basis elements and sends an email to Bob declaring his measurement outcome. If the out come is say $\vert \nearrow\rangle$, then Bob does nothing and is left with the unknown qubit recreated at his side while if the measurment outcome of Cliff is $\vert \nwarrow\rangle$, then Bob applies $Z-$ gate to retreive the original unknown qubit possessed initially by Alice. Thus, with Cliff as mediator, Bob Alice teleports her unknown qubit to Bob. Such a protocol is also sometimes called \textbf{controlled teleportation}. Below we summarize the measurement outcomes of Alice and Cliff and how Bob retrieves the qubit.
 \begin{table}[h!]
\begin{center}
\caption{Controlled teleportation with $\vert GHZ\rangle$ state}
\label{table1}
\begin{tabular}{|c|c|c|}
\hline
Alice's  outcome & Cliff's outcome & Bob applies\\
\hline
$\vert \phi^+\rangle_{12}$ & $\vert \nearrow_{4} \rangle$  &  $\vert \phi\rangle_3$ \\
\hline
$\vert \phi^+\rangle_{12}$& $\vert \nwarrow_{4} \rangle$  & $Z\vert \phi\rangle_3$ \\
\hline
$\vert \phi^-\rangle_{12}$& $\vert \nearrow_4 \rangle$  & Z$\vert \phi\rangle_3$ \\
\hline
$\vert \phi^-\rangle_{12}$& $\vert \nwarrow_4 \rangle$  & $\vert \phi\rangle_3$ \\
\hline
$\vert \varphi^+\rangle_{12}$& $\vert \nearrow_4 \rangle$  & $X\vert \phi\rangle_3$\\
\hline
$\vert \varphi^+\rangle_{12}$& $\vert \nwarrow_4 \rangle$  & $-iY\vert \phi\rangle_3$\\
\hline
$\vert \varphi^-\rangle_{12}$&  $\vert \nearrow_4 \rangle$ & $iY\vert \phi\rangle_3$\\
\hline
$\vert \varphi^-\rangle_{12}$& $\vert \nwarrow_4 \rangle$  & $-X\vert \phi\rangle_3$\\
\hline
\hline
\end{tabular}
\end{center}
\end{table}
The third column of above table is representing what Bob needs to do to after receiving the classical message from Cliff. Sometimes Bob does nothing and at another point of time he applies appropriate unitary gate to retrieve the unknown qubit $\vert \phi\rangle$, shown in Eq.(\ref{karlssonqubit}).\\\\
In tripartite system there is another important class of states known as $W-$ state. When I was Ph.D student at Satyendra Nath Bose National Center for Basic Science (SNBNCBS), Kolkata (India) and first read one of the articles on $W$ state, I mis-interpreted that the state was due to Physicist Werner and his name's first letter was being used to designate the state. My supervisor at SNBNCBS center, Prof. Archan S. Majumdar rectified it and informed me of the fact that the state was indeed proposed by the physicist named Wolfgang D$\Ddot{u}r$ and hence is named after him. These class of states can be of many types. But the type I am interested here, is a prototypical $W$ state and is defined as\cite{dr2000three}
\begin{eqnarray}
\label{wstate}
\vert W\rangle = \frac{1}{\sqrt{3}}\Big[\vert 1\rangle\vert 0\rangle\vert 0\rangle + \vert 0\rangle\vert 1\rangle\vert 0\rangle + \vert 0\rangle\vert 0\rangle\vert 1\rangle\Big]
\end{eqnarray} 

Along with D$\Ddot{u}$r, G. Vidal and J. I. Cirac, in their paper titled, \textit{Three qubits can be entangled in two inequivalent ways}, showed that the $W-$ class of states are inequivalent to that of $GHZ$ class of states in the sense that one class of states cannot be converted into the other under stochastic local operations and classical communications (SLOCC).\\\\
After D$\Ddot{u}$r et al proposed $W$ state, physicists started examining the viability of the states from the perspective of quantum information processing. Along this line, in the year 2002, Bao-Sen Shi and Akhihisa Tomita of Imai Quantum Computation and Information Project, NEC Tsukuba Laboratories, Japan published a paper in Elsevier where they claimed that they had successfully teleported an unknown state from Alice to Bob via prototypical $W$ state of E.(\ref{wstate})\cite{shi2002teleportation}. Immediately after the publication of this paper, Jaewoo Joo and Young-Jai Park of Department of Physics, Sogang University, Seoul, South Korea, identified an error in Bao-Sen Shi's calculation and subsequently published their comment in Elsevier\cite{joo2002comment}. And then as usual Bao-Sen Shi and Akihisa Tomita, wrote a reply to the comment by Jaewoo Joo and Young-Jai and published in Elsevier itself\cite{shi2002reply}. I am personally not convinced with Bao-Sen Shi's paper and so I shall skip their proposed teleportation protocol with $W$ state. Rather I shall divert our focus on another important paper which, I think is more authentic in this regard. We again shift our attention to the works of prominent Indian physicists Prof. Arun Kr. Pati and Prof. Pankaj Agrawal. Agrawal and Pati, in their paper, published in Physical Review A (Year: 2006), claimed that the prototype $W$ state of Eq.(\ref{wstate}) may not be suitable for perfect teleportation. The author of this book also tried and verified their claim by himself. On the other hand, Agrawal and Pati worked with non-prototype $W$ state and performed successful teleportation\cite{agrawal2006perfect}. The non-protoype $W$ state is defined as
\begin{eqnarray}
    \label{nonprototype1}
    \vert W_{n}\rangle_{123} = \frac{1}{\sqrt{2+2n}}\Big(\vert 1_1\rangle\vert 0_2\rangle \vert 0_3\rangle +\sqrt{n}e^{i\gamma}\vert 0_1\rangle\vert 1_2\rangle \vert 0_3\rangle + \sqrt{n+1}e^{i\delta}\vert 0_1\rangle\vert 0_2\rangle \vert 1_3\rangle\Big).\nonumber\\
\end{eqnarray}
Here suffices $1,\:2,\: 3$ are representing the qubits held by the three parties (say, Alice, Bob and Cliff) who will share the state among them, $n$ is a real number and $\gamma$ and $\delta$ are phases. In particular, if one takes $n = 1$, one gets a special non-prototype $W$ state as
\begin{eqnarray}
    \label{nonprototype2}
    \vert W_{1}\rangle_{123} = \frac{1}{2}\Big(\vert 1_1\rangle\vert 0_2\rangle \vert 0_3\rangle + \vert 0_1\rangle\vert 1_2\rangle \vert 0_3\rangle + \sqrt{2}\vert 0_1\rangle\vert 0_2\rangle \vert 1_3\rangle\Big).\nonumber\\
\end{eqnarray}
    
\subsection{QT: Non-prototype $W$ state as channel}
In their teleportation protocol, Agrawal and Pati considered the non-prototype $W$ state (Eq.(\ref{nonprototype2}) as quantum channel for performing teleportation as prototype $W$ state is not useful as a quantum resource for the usual
teleportation protocol\cite{gorbachev2003can}. But here they assumed that the qubits $1$ and $2$ will be retained by Alice and $3$ will be with Bob. They have used the notations $\vert 0\rangle$ and $\vert 1\rangle$ as spin-up and spin-down state in their paper. The unknown state that Alice wants to teleport to Bob is $\vert \psi\rangle_{a} =\alpha\vert 0_{a} + \vert 1_a\rangle$, $|\alpha|^2 + |\beta|^2 = 1$ and suffix $a$ denotes again that the qubit is in possession of Alice. To
examine the possibility of teleportation, Agrawal and Pati combined the input state and  the entangled state and rewrote the expression as follows.
\begin{eqnarray}
    \label{clubbednonprotopati}
    \vert \psi\rangle_{a}\vert W_1\rangle_{123} = \frac{1}{2}\Big(\alpha\vert 0_{a}\rangle + \vert 1_a\rangle\Big)\Big(\vert 1_1\rangle\vert 0_2\rangle \vert 0_3\rangle + \vert 0_1\rangle\vert 1_2\rangle \vert 0_3\rangle + \sqrt{2}\vert 0_1\rangle\vert 0_2\rangle \vert 1_3\rangle\Big)\nonumber\\
= \frac{1}{2}\Big[\vert \eta_1^{+}\rangle_{a12}(\alpha \vert 0_3\rangle + \beta \vert 1_3\rangle) + \vert \eta_1^{-}\rangle_{a12}(\alpha \vert 0_3\rangle - \beta \vert 1_3\rangle)  + \vert \xi_1^{+}\rangle_{a12}(\beta \vert 0_3\rangle + \alpha \vert 1_3\rangle)\nonumber\\ + \vert \xi_1^{-}\rangle_{a12}(\beta \vert 0_3\rangle - \alpha \vert 1_3\rangle) \Big].\nonumber\\
\end{eqnarray}
Here $\lbrace \vert \eta_1^{\pm}\rangle,\:\vert \xi_1^{\pm}\rangle\rbrace$ are a set of orthogonal states in the $W$ state category given by
\begin{eqnarray}
    \label{basisnonprototype}
    \vert \eta_1^{\pm}\rangle &=& \frac{1}{2}\Big(\vert 0\rangle\vert 1\rangle \vert 0\rangle + \vert 0\rangle\vert 0\rangle \vert 1\rangle \pm \sqrt{2}\vert 1\rangle\vert 0\rangle \vert 0\rangle\Big)\nonumber\\
    \vert \xi_1^{\pm}\rangle &=&  \frac{1}{2}\Big(\vert 1\rangle\vert 1\rangle \vert 0\rangle + \vert 1\rangle\vert 0\rangle \vert 1\rangle \pm \sqrt{2}\vert 0\rangle\vert 0\rangle \vert 0\rangle\Big).
\end{eqnarray}
Alice can now make a joint von-Neumann measurement in a basis that includes the states $\lbrace \vert\eta^{\pm}_1\rangle, \vert\xi^{\pm}_1\rangle\rbrace$ on the combined system of
three particles $a12$. She then sends the result of her measurements using two classical bits to Bob who can apply one of the unitary transformations $\lbrace \mathcal{I},\: \sigma_1,\: i\sigma_2, \: \sigma_3\rbrace$ to convert the state of his particle $3$ to that of particle $a$. For example, if the
outcome is $\vert \xi^+_1\rangle_{a12}$, then Bob has to apply $\sigma_1$ to get the desired state $\vert \psi\rangle_3$. This completes the teleportation protocol using the $W$ state. In this way Agrawal and Pati have shown that there exists a class of $W$ states which are useful for quantum teleportation. The
scheme presented here works analogously to the original protocol. The only difference is that Alice needs to carry out a three-qubit von-Neumann projection instead of a Bell-state measurement.
\subsection{QT: Tripartite partially entangled states as channels}
It is now clear from above discussions that quantum entanglement is efficiently used as a key resource in QT for sending quantum as well as classical information from a sender to a receiver. For communication tasks including only two parties, Bell states play a pivotal role in transferring the information from one location to another arbitrary location, which was original Bennett's protocol. Quantum information transfer protocols with more than two parties allow for controlled quantum communication where the controller controls or assists the successful information transfer between the sender and the receiver. I have already discussed it with respect to Karlsson and Bourennane scheme. In general, for an optimal and successful information transfer, the shared quantum channel between two or more than two parties is considered to be a maximally entangled resource, such as $\vert GHZ\rangle$ state or non-prototypical $\vert W\rangle$ state. However, in real experimental setups, it is always a challenge to obtain a multiqubit maximally entangled resource\cite{bennett1996concentrating}. Therefore, it is important to identify multiqubit entangled systems which are partially entangled but can be efficiently used as a resource in quantum information processing such as in QT, with optimal success. I shall now address this issue by discussing an important research work which dealt with this scenario where a set of three-qubit partially entangled states were used as resource for controlled quantum teleportation. I co-authored in that research work\cite{kumar2013optimal}. Although the paper dealt with many other relevant issues with respect to quantum information processing, I shall not go into those details, and will rather concentrate on the discussion about QT with partially entangled states very briefly. The discussion of partially entangled state is very important for complete understanding of tripartite scanario.\\\\
Let us first consider the following figure. 
\begin{figure}[h!]
		\begin{center}
			\includegraphics[width=8.5cm]{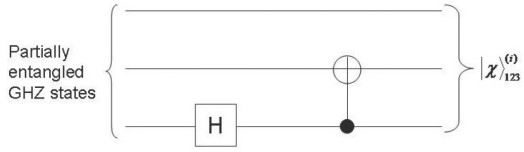}
		\end{center}
		\label{pes}
        \caption{The figure denotes a quantum circuit to prepare the partially entangled states $\vert \chi\rangle_{123}$.}
	\end{figure}
The three-qubit partially entangled set of states that we characterized in that paper can be represented as
\begin{eqnarray}
    \label{pes1}
    \vert \chi\rangle_{123}^{(1)} &=& \frac{1}{\sqrt{2}}\Big[\sin(\theta)\vert 000\rangle_{123} + \sin(\theta)\vert 011\rangle_{123} + \cos(\theta)\vert 110\rangle_{123} - \cos(\theta)\vert 101\rangle_{123}\Big]\nonumber\\
    \vert \chi\rangle_{123}^{(2)} &=& \frac{1}{\sqrt{2}}\Big[\sin(\theta)\vert 000\rangle_{123} - \sin(\theta)\vert 011\rangle_{123} + \cos(\theta)\vert 110\rangle_{123} + \cos(\theta)\vert 101\rangle_{123}\Big].\nonumber\\
\end{eqnarray}
The sets in Eq.(\ref{pes1}) are orthonormal entangled set of states $\vert\chi\rangle_{123}^{(i)}$, where $i = 1,\: 2$. The input in the circuit shown in Fig.$3.5$ is the generalized $GHZ$ ($GGHZ$) states. For example, if the input in the circuit is
\begin{eqnarray}
    \label{gghz}
    \vert \psi\rangle_{GHZ}^{(1)} = \sin(\theta)\vert 000\rangle_{123} +  \cos(\theta)\vert 111\rangle_{123}.
\end{eqnarray}
Then applying a Hadamard operation on the third qubit will result in
\begin{eqnarray}
    \label{constgghz}
    \vert \psi^{\prime}\rangle^{(1)} = \frac{1}{\sqrt{2}}\Big[\sin(\theta)\vert 000\rangle_{123} + \sin(\theta)\vert 001\rangle_{123} + \cos(\theta)\vert 110\rangle_{123} - \cos(\theta)\vert 111\rangle_{123}\Big].\nonumber\\
\end{eqnarray}
$\vert \psi^{\prime}\rangle^{(1)}$ can then be transformed to the three-qubit partially entangled state $\vert \chi^1\rangle_{123}$ by performing a CNOT operation on qubits $2$ and $3$ with qubit $3$ as control and qubit $2$ as target. Similarly, one can obtain other orthogonal states of the set $\vert \chi\rangle_{123}^{(i)}$, depending on the input given to the circuit. Thus, the partially entangled set represented in Eq. (\ref{pes1}) can be obtained starting from the GGHZ states by performing one local and one nonlocal operation. This additional nonlocal CNOT operation leads to an efficient and optimally
controlled quantum communication when the $\vert \chi\rangle_{123}^{(i)}$ states are used as a quantum channel compared to other partially entangled sates.
\\\\
Suppose, Alice wants to communicate an unknown quantum state $\vert \psi\rangle_1 = a\vert 0_1\rangle + b\vert 1_1\rangle$ and Bob is the receiver. Alice shares a partially entangled state as of type shown in Eq.(\ref{pes1}) and suppose it is $\vert \chi\rangle_{234}^{(1)}$; here $0<\theta<\frac{\pi}{4}$, with Bob and a third party Cliff. It is assumed that qubit $2$ is with Cliff, qubit $3$ is retained by Alice and the qubit $4$ is held by Bob. In fact, Alice and Bob can have any of qubits $3$ or $4$ in the shared entangled resource but qubit $2$ is fixed for Cliff. The joint state of four qubits including the arbitrary qubit $\vert \psi\rangle_1$ and the shared quantum channel $\vert \chi\rangle_{234}^{(1)}$ can be written as 
\begin{eqnarray}
    \label{jointpes1}
    \vert \psi\rangle_{1234} = \vert \psi\rangle_1 \otimes \vert \chi\rangle_{234}^{(1)}.
\end{eqnarray}
A joint measurement by Alice in the Bell basis on qubits $1$ and $3$ will project the joint state of Cliff and Bob's particles onto one of four possible states with equal probability of $\frac{1}{4}$. For example, if Alice’s measurement outcome is $\vert \phi^+\rangle_{13}$, then
the joint state of rest of the qubits can be given as
\begin{eqnarray}
    \label{jointstatepes11}
    \vert \phi\rangle_{24} = a\sin(\theta)\vert 0_{2}0_{4}\rangle - a\cos(\theta)\vert 1_{2}1_{4}\rangle + b\sin(\theta)\vert 0_{2}1_{4}\rangle + b\cos(\theta)\vert 1_{2}0_{4}\rangle.\nonumber\\
\end{eqnarray}
In order to teleport the unknown state to Bob successfully, Cliff measures his qubit $2$ in the new basis
\begin{eqnarray}
    \label{newbasispes}
    \vert x^+\rangle_2 &=& \cos(\alpha)\vert 0_2\rangle + \sin(\alpha)\vert 1_2\rangle\nonumber\\
    \vert x^-\rangle_2 &=& \sin(\alpha)\vert 0_2\rangle - \cos(\alpha)\vert 1_2\rangle.
\end{eqnarray}
The joint state of qubits $2$ and $4$ can be re-expressed in terms of Cliff’s measurement basis $\lbrace \vert x^+\rangle_2,\: \vert x^-\rangle_2\rbrace$ as
\begin{eqnarray}
    \label{pess}
    \vert \phi\rangle_{24} = \vert x^+\rangle_2 [\lbrace a\sin(\theta)\cos(\alpha) + b\cos(\theta)\sin(\alpha)\rbrace \vert 0_4\rangle \nonumber\\
    + \lbrace b\sin(\theta)\cos(\alpha) - a\cos(\theta)\sin(\alpha)\rbrace \vert 1_4\rangle ]\nonumber\\ +\vert x^-\rangle_2 [\lbrace  a\sin(\theta)\sin(\alpha) -b\cos(\theta)\cos(\alpha)\rbrace \vert 0_4\rangle \nonumber\\
    + \lbrace a\cos(\theta)\cos(\alpha) + b\sin(\theta)\sin(\alpha)\rbrace \vert 1_4\rangle].
\end{eqnarray}
Thus, for Cliff’s measurement in the basis $\lbrace \vert x^+\rangle_2,\: \vert x^-\rangle_2\rbrace$ Bob's qubit will be projected onto one of the states
\begin{eqnarray}
    \label{cliffpes1}
    \vert \psi\rangle_1 = \lbrace a\sin(\theta)\cos(\alpha) + b\cos(\theta)\sin(\alpha)\rbrace \vert 0_4\rangle \nonumber\\
    + \lbrace b\sin(\theta)\cos(\alpha) - a\cos(\theta)\sin(\alpha)\rbrace \vert 1_4\rangle,
\end{eqnarray}
and 
\begin{eqnarray}
    \label{cliffpes1}
    \vert \psi\rangle_2 = \lbrace  a\sin(\theta)\sin(\alpha) -b\cos(\theta)\cos(\alpha)\rbrace \vert 0_4\rangle \nonumber\\
    + \lbrace a\cos(\theta)\cos(\alpha) + b\sin(\theta)\sin(\alpha)\rbrace \vert 1_4\rangle.
\end{eqnarray}
Let us assume that Cliff's measurement outcome is $\vert x^+\rangle_2$; then in order to recover the teleported state successfully Bob
needs to perform a single-qubit unitary transformation.
\begin{eqnarray}
\label{unitpes}
U_4 = \left(%
	\begin{array}{cc}
		\sin(\theta) & -\cos(\theta)\\
		\cos(\theta) & \sin(\theta)\\
	\end{array}%
	\right)
    \end{eqnarray}
which will evolve the state of qubit $4$ as
\begin{eqnarray}
    \label{uniarypes1}
    U_4\vert \psi\rangle_1 = \frac{1}{\sqrt{N}}\Big[\lbrace a\sin^2(\theta)\cos(\alpha)+b\sin(\theta)\cos(\theta)\lbrace\sin(\alpha)-\cos(\alpha)\rbrace + a\cos^2(\theta)\sin(\alpha)\rbrace\vert 0_4\rangle + \nonumber\\ \lbrace b\cos^2(\theta)\sin(\alpha)+a\sin(\theta)\cos(\theta)\lbrace\cos(\alpha)-\sin(\alpha)\rbrace + b\sin^2(\theta)\cos(\alpha)\rbrace \vert 1_4\rangle\Big].\nonumber\\
\end{eqnarray}
where $N_1$ is the normalization constant given by 
\begin{eqnarray}
    \label{n1norma}
    N_1 = [a\sin^2(\theta)\cos(\alpha) + b\sin(\theta)\cos(\theta)\lbrace \sin(\alpha)-\cos(\alpha)\rbrace + a\cos^2(\theta)\sin(\alpha)]^2\nonumber\\
    +[b\cos^2(\theta)\sin(\alpha) + a\sin(\theta)\cos(\theta)\lbrace \cos(\alpha) - \sin(\alpha)\rbrace + b\sin^2(\theta)\cos(\alpha)]^2.\nonumber\\
\end{eqnarray}
The square of the overlap between the input state $\vert \psi\rangle_1$ and $U_4\vert \psi\rangle_1$ is
\begin{eqnarray}
\label{fidpes1}
    \langle \psi_1\vert U_4 \vert \psi_1\rangle^2 = \frac{1}{N_1}[\sin^2(\theta)\cos(\alpha) + \cos^2(\theta)\sin(\alpha)]^2.
\end{eqnarray}
Similarly, for Cliff's measurement outcome $\vert x^-\rangle_2$ , the required unitary transformation is
\begin{eqnarray}
    V_4 = \left(%
	\begin{array}{cc}
		\sin(\theta) & \cos(\theta)\\
		-\cos(\theta) & \sin(\theta)\\
	\end{array}%
	\right)
\end{eqnarray}
and the overlap with the input state $\vert \psi\rangle_1$ is
        \begin{eqnarray}
            \label{fidpes2}
   \langle \psi_1\vert U_4^{\prime} \vert \psi_1\rangle^2 = \frac{1}{N_2}\Big[\sin^2(\theta)\sin(\alpha) + \cos^2(\theta)\cos(\alpha)\Big]^2. 
        \end{eqnarray}
Here, 
\begin{eqnarray}
    \label{n1norma}
    N_2 = [a\sin^2(\theta)\sin(\alpha) + b\sin(\theta)\cos(\theta)\lbrace \sin(\alpha)-\cos(\alpha)\rbrace + a\cos^2(\theta)\sin(\alpha)]^2\nonumber\\
    +[b\cos^2(\theta)\sin(\alpha) + a\sin(\theta)\cos(\theta)\lbrace \cos(\alpha) - \sin(\alpha)\rbrace + b\sin^2(\theta)\cos(\alpha)]^2.\nonumber\\
\end{eqnarray}
Then, the average fidelity can be expressed as
\begin{eqnarray}
\label{fidpesss}
    F_d = \sin^4(\theta) + \cos^4(\theta) + 4\sin^2(\theta)\cos^2(\theta)\sin(\alpha)\cos(\alpha).
\end{eqnarray}
At this stage, I must confess that, I have not yet defined the term fidelity , which will rather be discussed in the next chapter. Anyways, for $\alpha = \frac{\pi}{4}$, the average fidelity would be unity for any shared resource state. Moreover, the state of Bob's qubit will
be the same as the state of Alice's qubit after performing the unitary transformation $U_4$ and teleportation will be successful. However, for all other $\alpha$, ($0< \alpha < \frac{\pi}{4}$) average fidelity, Eq.(\ref{fidpesss}), depends on the controller’s measurement basis,
characterized by $\alpha$, and Bob can always recover the original state communicated by Alice in a similar fashion as discussed
above. Thus, controlled teleportation is always successful irrespective of the degree of entanglement between the three
qubits, which may provide flexibility to the experimental setups by releasing the constraint of using a maximally entangled shared resource for faithful teleportation. Therefore, for a certain choice of Cliff's measurement basis, one can always obtain unit probability and fidelity of the teleportation process using our state $\vert \chi\rangle_{234}^{(i)}$ (where $i=1,\:2)$ as an entangled resource. \\\\
There is another kind of tripartite partially entangled states known as \textit{Maximal Sliced States}, which are defined as
\begin{eqnarray}
    \label{pesms}
    \vert MS\rangle_{123} = \frac{1}{\sqrt{2}}\Big[\vert 000\rangle_{123} + c\vert 111\rangle_{123} + d\vert 011\rangle_{123}\Big],
\end{eqnarray}
where $c$ and $d$ are assumed to be real numbers with the condition that $c^2 + d^2 = 1$. When $c=0$, the resulting state is a product state of the first qubit with a maximally entangled Bell pair of qubits $2$ and $3$. When $c=1$, the state becomes the maximally entangled $GHZ$ state and for all other values of $c$, the three qubits are partially entangled. The MS states have been shown to have interesting entanglement and nonlocality properties due to their inherent symmetries. These type of partially entangled states have also shown to be useful for perfect and deterministic controlled teleportation\cite{gao2008optimal,li2014control}.
\\\\
I shall now deviate our focus on another aspect of quantum information processing, where it is important to learn how to measure the entanglement of a quantum state and whether violation of the quantum state has anything to do with teleportation fidelity of that quantum state. I would also like discuss about what teleportation fidelity means. It is important to know about these concepts.
\chapter{Measures: Entanglement and Teleportation fidelity}
\label{ch:qtm}
To begin with, I shall make a disclaimer here. This book is not a review article while it gives a brief overview of teleportation protocols. And only those protocols so far have been discussed which have extreme historical importance. For teleportation, one needs a quantum channel which will be shared between two parties or among multi-parties. The entanglement of these channels need to be tested first and one has to make sure that the channels will be capable of teleporting unknown qubit from sender to receiver with high fidelity. I shall discuss first the entanglement measures which are commonly used for two-party or bipartite systems.\\\\
It is already a known fact that entanglement cannot be created using local operations and classical communications (LOCC). Mintert et.al showed that the quantities that do not increase under LOCC operations \cite{mintert2005measures}, can be used to quantify entanglement. Any scalar valued function that satisfies this criterion is called \textit{entanglement monotone}.
Entanglement monotones that satisfy certain additional axioms are called \textit{entanglement measures} and is generally denoted by $E$. Such potential axioms can be listed below. For any mixed state $\varrho$,
\begin{itemize}
\item $E(\varrho)$ vanishes exactly for separable states.
\item The entanglement of several copies of a state adds up to $n$ times the entanglement of a single copy. Symbolically, $E(\varrho^{\otimes\:n})=n\:E(\varrho)$.
\item The entanglement of two states ($\varrho$ and $\varrho^{/}$) is not larger than the sum of the entanglement of both individual states. Symbolically, $E(\varrho\otimes\varrho^{/})\:\leq E(\varrho) + E(\varrho^{/})$.
\item Entanglement measure $E$ satisfies the convexity property, i.e. $E(\lambda\:\varrho + (1-\lambda)\:\varrho^{/})\:\leq\:\lambda\:E(\varrho) + (1-\lambda)\:E(\varrho^{/})$, where, $0\:\leq\lambda\:\leq\: 1$.
\end{itemize}
I shall discuss not many entanglement measures but a few which will be relevant to the objective of my book. 
\subsection{Entanglement of formation:}
\textit{Entanglement of formation} \cite{wootters1998entanglement} quantifies the amount of entanglement necessary to create the entangled state. It is defined as
\begin{eqnarray}
E_{F}(\rho) = \min_{\lbrace p_{i},\psi_{i}\rbrace}\:\sum_{i}\:p_{i}\:E(\vert \psi_{i}\rangle \langle \psi_{i}\vert)
\label{eef1},
\end{eqnarray}
where the minimum is taken over those probabilities $\lbrace p_{i}\rbrace$ and pure states $\lbrace \psi_{i}\rbrace$ that, when taken together, reproduce the density matrix $\rho=\sum_{i}\:p_{i}\:\vert \psi_{i}\rangle\langle \psi_{i}\vert$. $E(\vert \psi_{i}\rangle \langle \psi_{i}\vert)$ is the entropy of entanglement.
For two qubit systems, $E_{f}$ can be expressed explicitly as 
\begin{eqnarray}
E_{f}(\rho) = h\:\left\{\:\frac{1+\sqrt{1-C^{2}(\rho)}}{2}\:\right\},
\label{eef2}
\end{eqnarray}
where, $h(x)=\:-x\:\log_{2}\:x-(1-x)\:\log_{2}\:(1-x)$ is Shannon's entropy function and $C(\rho)$ is the \textit{concurrence} of the state $\rho$. The quantity $C^{2}(\rho)=\tau$ is sometimes called \textbf{squared concurrence} \cite{coffman2000distributed}. The entanglement of formation $E_{F}$ is a strictly monotonic function of $C^2$, the maximum of $C^2$ corresponds to the maximum of $E_{f}$. Hence $`$concurrence squared' can also be considered as direct measure of entanglement. For a maximally entangled pure state $C^2=1$ while for an unentangled state $C^2=0$.
\subsection{Concurrence:}
\textit{Concurrence} is a non-negative real number. For a bipartite $2\otimes2$ or $2\otimes3$ mixed state $\rho$, it is defined in \cite{wootters1998entanglement,hill1997entanglement}  as\\
\begin{eqnarray}
C(\rho)=\max\,(0, \sqrt\la_1-\sqrt\la_2-\sqrt\la_3-\sqrt\la_4),
\label{concurrence}
\end{eqnarray} \\
where the $\la_{i}$'s are the  eigenvalues of $\rho\,\tilde{\rho}$ in decreasing order. The spin flipped density matrix $\tilde{\rho}$ is defined as $\uu$ . Since $E_{f}$ is a monotonic function of $C$ and $C$ ranges from zero to one (i.e. for unentangled to maximal entanglement), so the concurrence $C$ is also a measure of entanglement. Remember $C$ lies between zero and 1 (both inclusive).\\\\
\textbf{Problem:} Consider any of the Bell states defined in Eq.(\ref{bellstates}). Evaluate concurrence $C$ for each. Also calculate entanglement of formation ($E_f$) for each.\\\\
\textbf{Problem:} Consider $GHZ$ and prototype $W$ state [see Eqs.(\ref{ghz1}) and (\ref{wstate})]. Now form a mixture $p\vert GHZ\rangle\langle GHZ\vert + (1-p)\vert W\rangle\langle W\vert$. Trace out, say party $C$ (i.e. the third party sharing one qubit). You will get bipartite mixed state. Let us denote this by $p\rho^{ghz}_{AB}+(1-p)\rho^{w}_{AB}$. Calculate the concurrence $C$ of that state.
\subsection{Relative entropy of entanglement:}
The measure of \textit{relative entropy of entanglement} \cite{audenaert2005continuity,miranowicz2004comparative,plenio1998teleportation} is based on distinguishability and geometric distance. The idea is basically to compare a given quantum state $\sigma$ of a pair of particles with separable states. The \textit{relative entropy of entanglement} of a given state $\sigma$ is denoted by $E_{re}\:(\sigma)$ and is defined as\\
\begin{eqnarray}
E_{re}\:(\sigma) = \min_{\rho \in M}\: D(\sigma \parallel \rho).
\label{re}
\end{eqnarray}\\
Here, $M$ denotes the set of all separable states and $D$ can be any function that describes a measure of separation between two density operators. A particular form of the function $D$ is the relative entropy which is defined as $S(\sigma \parallel \rho)=\:Tr\:(\sigma\:\ln\:\sigma -\sigma\:\ln\:\rho)$.
\subsection{Negativity:}
The concept of \textit{negativity} of a state is closely related to the well-known \textbf{Peres-Horodecki criterion} or \textbf{positive partial transposition} (PPT) criteria for the separability of a state \cite{horodecki1997inseparable,peres1996separability}.\\\\
Peres-Horodecki criterion states that a necessary and sufficient condition for the state $\rho$ of two spins to be inseparable is that at least one of the eigenvalues of the partially transposed operator, defined as $\rho^{T_{2}}_{m\mu\:,\:n\nu} = \rho_{m\nu\:,\:n\mu}$, is negative. This is equivalent to the condition that at least one of the following two determinants of eq.(\ref{peres1}) is negative.\\
\begin{eqnarray}
W_{3}=\left|%
\begin{array}{ccc}
\rho_{00\:,\:00}&\rho_{01\:,\:00}& \rho_{00\:,\:10}\\
\rho_{00\:,\:01}&\rho_{01\:,\:01}&\rho_{00\:,\:11}\\
\rho_{10\:,\:00}&\rho_{11\:,\:00}&\rho_{10\:,\:10}\\
\end{array}%
\right|~~~~ \mbox{and}~~~~
W_{4}=\left|%
\begin{array}{cccc}
\rho_{00\:,\:00}&\rho_{01\:,\:00}& \rho_{00\:,\:10}&\rho_{01\:,\:10}\\
\rho_{00\:,\:01}&\rho_{01\:,\:01}&\rho_{00\:,\:11}&\rho_{01\:,\:11}\\
\rho_{10\:,\:00}&\rho_{11\:,\:00}&\rho_{10\:,\:10}&\rho_{11\:,\:10}\\
\rho_{10\:,\:01}&\rho_{11\:,\:01}&\rho_{10\:,\:11}&\rho_{11\:,\:11}\\
\end{array}%
\right|, \nonumber\\\nonumber\\
\label{peres1}
\end{eqnarray}\\
and the determinant of eq. (\ref{peres2}) is non-negative.\\
\begin{eqnarray}
W_{2}=\left|%
\begin{array}{cc}
\rho_{00\:,\:00}&\rho_{01\:,\:00}\\
\rho_{00\:,\:01}&\rho_{01\:,\:01}\\
\end{array}%
\right|
\label{peres2}.
\end{eqnarray}
If a state is separable (i.e. not entangled), then the partial transpose of its density matrix is again a valid state i.e. it is positive semi-definite \footnote{A linear self adjoint map $\Lambda: \Re(H_{B})\rightarrow \Re(H_{C})$ is called positive semi-definite if for all $\rho \in \Re(H_{B})$, (where $H_{B}, \: H_{C}$ are the Hilbert spaces and $\Re(H_{i})_{i\:=B,\:C}$ are the set of linear operators acting on $H_{i}$), with $\rho\:\geq\: 0\Rightarrow \Lambda(\rho)\geq 0$. The map $\Lambda$ is Positive definite if $\Lambda(\rho)> 0$.}. It also turns out that  the partial transpose of a non-separable state may have one or more negative eigenvalues \cite{wei2003maximal}.\\\\
The \textit{negativity}, however, of a state \cite{miranowicz2004comparative} indicates the extent to which a state violates the positive partial transpose (separability) criterion. The negativity of the state $\rho$ is defined as follows:\\
\begin{eqnarray}
N^{\rho} = 2\:\mbox{max}\:(0\:,\:-\lambda_{neg}),
\label{negativity1}
\end{eqnarray}
where $\lambda_{neg}$ is the sum of the negative eigenvalues of $\rho^{T_{B}}$. In $C^{2}\otimes C^{2}$ systems, it can be shown that the partial transpose of the density matrix can have at most one negative eigenvalue. It was proved later that negativity is an entanglement monotone and hence a good measure of entanglement \cite{wei2003maximal}. For mixed states, Eisert and Plenio \cite{eisert1999comparison} conjectured that \textit{negativity never exceeds concurrence} and the conjecture was proved later by Audenaert \textit{et. al} in \cite{audenaert2000negativity}\\\\
For higher dimensions, the negativity can be generalized as \cite{lee2003convex}\\
\begin{eqnarray}
\mathscr{N}(\rho)=\frac{\parallel \rho^{T_{B}} \parallel - 1}{d-1},
\label{negativity2}
\end{eqnarray}
where, $\rho^{T_B}$ is the partial transpose of a state $\rho$ in the quantum system $d\otimes d^{\prime}$ ($d\leq d^{\prime}$) quantum system and $\|\cdot\|$ is the smaller of the dimensions of the bipartite system. Also see that in the above formula, Eq.(\ref{negativity2}) a norm function has been used. Remember, $\|X\| = Tr| X | = Tr \sqrt{X^{\dagger}X}$ is the \textit{trace norm} or the \textit{sum of the singular values} of the operator $X$ ($Tr$ represents Trace, which has been defined in chapter $1$). An alternative and equivalent definition is the absolute sum of the negative eigenvalues of $\rho^{T_{B}}$ is
\beq
\label{negativity3}
\mathscr{N}(\rho) = \sum_i\frac{|\lambda_i| - \lambda_i}{2},
\eeq
where $\lambda_i$s are all of the eigenvalues.\\\\
However, although the positivity of the partial transpose is a necessary and sufficient condition for non-distillability in $2\otimes n$ quantum system, there exist entangled states with PPT in any bipartite system except in $2\otimes 2$ and $2\otimes 3$ quantum systems, that is, there exist entangled states whose negativity are not positive. Such states are known as \textbf{PPT bound entangled states}, which can be useful in a quasi-distillation process, called the activation of bound entanglement. Nevertheless, the negativity cannot distinguish the PPT bound entanglement from separability. Hence, it is not sufficient for the negativity to be a good measure of entanglement even in $2\otimes n$ quantum system.\\\\
Now a days, multipartite entanglement or multipartite entangled systems are at the scientific focal point. We call a system to be multipartite if the number of parties sharing the qubits is more than $3$. Hence we generally begin with the study of tripartite systems. I have already discussed to inequivalent class of tripartite states viz. $GHZ$ and $W$ class in the previous chapter. Consequently the quantification of entanglement present in such tripartite systems is also important. I shall discuss some other special characteristics of these two inequivalent class of tripartite states later. However, here I bring into picture a measure of entanglement which was originally proposed by Coffman et al in the year $2000$ and over the time has become one of the most cited papers in quantum information science community. William "Bill" Kent Wootters is an American theoretical physicist, and one of the founders of the field of quantum information theory. In a 1982 joint paper with Wojciech H. Zurek, Wootters proved the no-cloning theorem (which has already been discussedin the chapter $2$.). Valerie Coffman and Joydip Kundu, under the supervision of Willam published a paper named \textit{distributed entanglement}, where they proposed a formula for quantifying \textbf{genuine tripartite entanglement} and the measure is known as \textbf{Tangle}\cite{coffman2000distributed}.

Let the tripartite system be denoted by $\vert \xi_{ABC}\rangle$ where, $A,\:B,\: C$ denote the three parties Alice, Bob and Charlie and the state is represented by
\beq
\label{coffman1}
\vert \xi_{ABC}\rangle =  \sum_{ijk} a_{ijk} \vert ijk\rangle.
\eeq
The quantity $\tau_{ABC}$ is called \textit{Tangle} where,
\beq
\label{coffman2}
\tau_{ABC} =  C^2_{A(BC)} - C^2_{AB} - C^2_{AC}.
\eeq
Here $C^2$ is \textit{concurrence squared}\footnote{In this book I shall not discuss the concept of monogamy.} . There is another computational form of formula given in Eq.(\ref{coffman2}) which is
\beq
\label{coffman3}
\tau_{ABC} =  4 \vert d_1 - 2d_2  + 4d_3\vert, 
\eeq
where
\beq
\label{coffman4}
d_1 = a^2_{000}a^2_{111} + a^2_{001}a^2_{110} + a^2_{010}a^2_{101} + a^2_{100}a^2_{011}\nonumber\\
d_2 = a_{000}a_{111}a_{011}a_{100} + a_{000}a_{111}a_{101}a_{010} + a_{000}a_{111}a_{110}a_{001}+\nonumber\\
a_{011}a_{100}a_{101}a_{010} + a_{011}a_{100}a_{110}a_{001} + a_{101}a_{010}a_{110}a_{001}\nonumber\\
d_3 = a_{000}a_{110}a_{101}a_{011} + a_{111}a_{001}a_{010}a_{100}.
\eeq
If we consider, for example, the general $W$ class of states such as $\vert \phi\rangle_{ABC} = \alpha \vert 100\rangle + \beta 010\rangle + \gamma 100\rangle$ (with normalization condition satisfied), then using the Eqs.(\ref{coffman1})-(\ref{coffman4}), it can easily be shown that, $C_{AB} = 2\vert \alpha\beta \vert$, $C_{AC} = 2 \vert \alpha \gamma\vert$ and $C_{A(BC)} = 2\vert \alpha\vert \sqrt{\vert \beta\vert^2 + \vert \gamma\vert^2}$. Thus, the tangle of state $\phi$ is found to be $\tau(\vert \phi\rangle_{ABC}) = 0$. Now if we consider $\alpha = \beta = \gamma = \frac{1}{\sqrt{3}}$, i.e. we are considering the prototype $W$ state (see Eq.(\ref{wstate})), we can immediately find that the genuinely tripartite entanglement found in the prototype $W$ state is zero. Then we are not amazed by this fact as prototype $W$ state is not suitable as quantum channel for teleportation. Now the interesting part is when we consider non-prototype $W$ state as defined in Eq.(\ref{nonprototype2}), it is again found that $\tau(\vert W_1\rangle) = 0$, whereas Agrawal and Pati had successfully shown that this non-prototype $W$ state is a viable candidate as quantum teleportation channel.
Again if we consider the generalized version of $GHZ$ state which is $\vert G\rangle_{ABC} = \alpha\vert 000\rangle + \beta \vert 111\rangle$ (with normalization condition satisfied), then again using Eqs.(\ref{coffman1})-(\ref{coffman4}), it can be shown that $\tau(\vert G\rangle_{ABC}) = 4 \vert \alpha\vert^2\vert \beta\vert^2$. Now, if one takes $\alpha = \beta = \frac{1}{\sqrt{2}}$, then $\tau(\vert G\rangle_{ABC}) = 1$. From our previous chapter we have seen that the $GHZ$ state is a potential candidate as quantum channel for teleportation. What have we learned from the above observations? We have seen an interesting fact and we can immediately conclude that genuine tripartite entanglement of the tripartite state is not entirely responsible for the state's viability as quantum teleportation channel.\\\\
\textbf{Problem:} Evaluate the genuine tripartite entanglement present in (a) tripartite partially entangled state of Eq.(\ref{pes1}) and in (b) maximal sliced state of Eq.(\ref{pesms}). \\\\
Now I shall discuss another important aspect of quantum information science which is known as \textit{fidelity}. The main purpose of calculating fidelity is to find the distance between the input state ($\rho^{in}$) and the output state ($\rho^{out}$). With respect to the quantum cloning machine one calculates the fidelity between the state that is put inside cloning machine and the output state that comes out of the cloning machine. Likewise, when Alice wants to teleport an unknown qubit to Bob, one can also find the fidelity between the state that is intended to be sent and the state that Bob retrieves. There are many fidelity measures which have been discussed thoroughly by Prof. Michael Nielsen and Prof. Issac Chuang\cite{nielsen2010quantum}. I shall discuss only about teleportation fidelity here which is necessary for the purpose of writing this book.
\subsection{Discussion on teleportation fidelity:}
We know that the protocol of quantum teleportation (QT), is a manifestation of quantum non-locality. In QT, a pair of quantum systems in state $\mathscr{D}$ is distributed between two partners, Alice ($A$) and Bob ($B$), where $\mathscr{D}$ is a pure or mixed two-spin $\frac{1}{2}$ state. The aim of teleportation is to transfer the spin state of that test particle which Alice possess, as faithfully as possible, to Bob using only the common pair of particles in the state $\mathscr{D}$ (i.e. quantum channel) and two classical bits (classical channel). In the teleportation protocol; $\vert \psi_{A}\rangle$ and $\rho_{B}$ denote the (normalized) state of the test particle prepared on Alice's side (pure state) and transferred to Bob's end (possibly a mixed state). The successful teleportation relies on three facts viz. (a) the state $\rho_B$ may depend on $\vert \psi_A\rangle$, (b) on the state $\mathscr{D}$ and on the strategy used by Alice and Bob. The \textbf{fidelity} $\mathcal{F}(\mathscr{D})$ is the mean distance between $\vert \psi\rangle_{A}$ and $\rho_B$ corresponding to the optimal strategy, is given by
\beq
\label{fid1}
\mathcal{F}(\mathscr{D}) =\max_{all\:strategies}[M\lbrace \langle \psi_A \vert \rho_B \vert \psi_A\rangle\rbrace],
\eeq
where the mean $M$ is taken over all possible states $\vert \psi_A\rangle$. Bennett et al in their seminal paper, worked with singlet state ($\frac{1}{\sqrt{2}}(\vert 01\rangle + \vert 10\rangle)$) as quantum channel for teleportation. They proved that if $\mathscr{D}$ is the singlet state or any maximally entangled pure state, then
\beq
\label{fidsinglet}
\mathcal{F}(\mathscr{D}) = 1.
\eeq
On the other hand if $\mathscr{D}$ is the product state, Alice can only measure $\vert \psi_a\rangle$ long an arbitrary  but given
direction $z$ and tells Bob the measurement result, (which can be spin-up or spin-down), and Bob prepares $\rho_B$ in the spin-up or spin-down state. In this manner, only one bit of classical information is transferred. The corresponding fidelity is 
\beq
\label{fidsinglet2}
\mathcal{F}(\mathscr{D})(product\:state) = \frac{2}{3}.
\eeq
Sandu Popescu is a Romanian born British physicist who did foundational work on quantum mechanics and quantum information. He is one of the co-editors and co-authors of the first text book on quantum information and computation\cite{lo1998introduction}. Popescu in his paper\cite{popescu1995bell} concluded that \textbf{a state $\mathscr{D}$ with fidelity larger than $\frac{2}{3}$ has $``$nonclassical aspects" (i.e. something that is not explainable by local hidden variables)}. This claim by Popescu raised a question as to what would be the fidelity for teleportation of a hypothetical theory with local hidden variables? Nicholas Gisin is a Swiss physicist and professor at the University of Geneva. He also worked on the foundations of quantum mechanics, quantum information and quantum communication principles. He co-founded ID-Quantique, a company that provides quantum based technologies. To answer the question raised by Popescu, Gisin, in his paper\cite{gisin1996nonlocality} showed that 
\beq
\mathcal{F}(local\:hidden\:variables) =  \mathcal{F}(lhv) &=& \frac{1}{2} + \frac{\sqrt{3}}{2}\frac{\tan^{-1}(\sqrt{2})}{\pi}\nonumber\\
\mathcal{F}(lhv) &=& 0.87\: (approx.)
\eeq
Gisin showed fidelity of a state $\mathscr{D}$ revealed non-classical aspects (i.e. is incompatible with local hidden variables)  only if $\mathcal{F}(\mathscr{D})>0.87$. Two messages were clear from Gisin's paper. 
\begin{itemize}
    \item If the teleportation fidelity $\mathcal{F}(\mathscr{D})>0.87$, then the state $\mathscr{D}$ is non-local in the sense of incompatibility with local hidden variables. If, however, $\mathscr{D}$ is local, Alice would not be able to make use of quantum channel for the purpose of teleportation.
    \item The teleportation fidelity compatible with local hidden variables differs from the teleportation fidelity compatible with local quantum states. Hence teleportation reveals two different aspects of non-locality, one related to classical local variables, with upper bound of $0.87$, while the other related to local quantum states, with upper bound of $\frac{2}{3}$.

\end{itemize}
The efficiency of a quantum channel used for teleportation is measured in terms of its average teleportation fidelity given by \cite{horodecki1996information}
\begin{eqnarray}
f_{opt}^T(\rho_{\phi}) = \int_S d M(\phi) \sum_k p_k Tr (\rho_k \rho_{\phi}),
\label{telfid0}
\end{eqnarray}
where $\rho_{\phi}$ is the input pure state and $\rho_k$ is the output state provided the outcome $k$ is obtained by Alice. The quantity
$Tr (\rho_k \rho_{\phi})$ which is a measure of how the resulting state is similar to the input one, is averaged over the probabilities of outcomes $p_k$, and then over all possible input states ($M$ denotes the uniform distribution on the Bloch sphere $S$). It has been shown \cite{horodecki1996information} that if a state is useful for standard teleportation, the optimal teleportation fidelity can be expressed as
\begin{eqnarray}
f_{opt}^{T}(\rho)=\frac{1}{2}\Big[1+\frac{\mathcal{N}(\rho)}{3}\Big],
\label{tel.fid.N}
\end{eqnarray}
where $\mathcal{N}(\rho)=\sum_{i=1}^{3}\sqrt{u_{i}}$ and $u_{i}$'s are the
eigenvalues of the matrix $T^{\dagger}T$. The elements of the
matrix $T$ are given by
\begin{eqnarray}
t_{nm}=Tr(\rho~\sigma_{n}\bigotimes\sigma_{m}),
\label{tmatrix}
\end{eqnarray}
 where $\sigma_{i}$'s
denote the Pauli spin matrices. Now, in terms of the quantity $\mathcal{N}(\rho)$, a general result\cite{horodecki1999general} holds that any mixed spin-$\frac{1}{2}$ state is useful for (standard) teleportation if and only if
\begin{eqnarray}
N(\rho)>1.
\label{teleportcond}
\end{eqnarray}
In the next chapter I shall reflect upon the concept of teleportation fidelity from the perspective of Bell inequality violation.
\chapter{Bell Non-locality and Teleportation Fidelity:}
\label{ch:bntf}
As I teach Mathematics in my college, let me begin this chapter by making you learn a bit of $``$Mathematical Logic". I think this is the best way to initiate the discussion of this chapter. Let $\mathcal{P}$ and $\mathcal{Q}$ be two statements\footnote{In Mathematical Logic, there is a difference between sentence and statement. Statement is something which can be judged as $``$True" or $``$False".}, such that $\mathcal{P}$ implies $\mathcal{Q}$, symbolically, $\mathcal{P} \Rightarrow \mathcal{Q}$. Then negation of $\mathcal{Q}$ implies negation of $\mathcal{P}$. Symbolically $\neg \mathcal{Q}\Rightarrow \neg \mathcal{P}$. Let me take a simple example. Suppose $\mathcal{P}:=$ $``$ I work in Techno Main Salt Lake (TMSL) engineering college" and $\mathcal{Q}:=$ $``$I live in Kolkata". Now here, $\mathcal{P}$ implies $\mathcal{Q}$ as, if I work in TMSL then it directly implies that I reside in Kolkata. Consequently, if I don't live in Kolkata then I am not working in TMSL which is represented by $\neg \mathcal{Q}\Rightarrow \neg \mathcal{P}$.\\\\
Horodecki et.al \cite{horodecki1996teleportation} had shown \textbf{that every mixed two spin-$\frac{1}{2}$ state which violates any generalized Bell-CHSH inequality is useful for teleportation}. If we denote by $\mathcal{P}$ the statement that $``$ $\rho$ violates Bell-CHSH inequality" and by $\mathcal{Q}$ we mean $``$ the state $\rho$ is useful for teleportation", then by Horodecki's paper we have $``$ $\mathcal{P}$ $\Rightarrow$ $\mathcal{Q}$". By mathematical logic one can immediately conclude that $``$ $\neg \mathcal{Q}$ $\Rightarrow$ $\neg \mathcal{P}$". This means that the state which is not useful for teleportation will satisfy Bell-CHSH inequality. But one cannot immediately conclude that $``$ $\mathcal{Q}$ $\Rightarrow$ $\mathcal{P}$", i.e. there may exist states which satisfy Bell-CHSH inequality, but still can be used as teleportation channel. One such example is \textbf{Werner state}\cite{werner1989quantum}. And it is to be remebered that this feature can only be observed in case mixed states. I shall discuss about several other such states in details in the next chapter. Here I shall focus on the mathematics of Bell-violation as proposed by Horodecki et. al.
Horordecki et. al\cite{horodecki1996teleportation} have proved that the necessary and sufficient condition for violating the Bell-CHSH inequality involves a real valued function 
\beq
\label{mrho}
M(\rho) = \max_{i>j}(u_i + u_j),
\eeq
where $u_i$'s are the eigenvalues of the matrix $T^{\dagger}T$ (how to construct the elements of $T$ has already been discussed in the previous chapter). Consequently the inequality $M(\rho) \leq 1$  is equivalent to the Bell-CHSH one. It is to be noted that for any state which violates the Bell-CHSH inequality, we have $M(\rho)>1$. In this chapter we shall discuss about Werner state from the perspectives of Bell inequality violation and the state's teleportation fidelity. To understand this we define another quantifier which is often used in quantum information and the quantifier is known as \textbf{maximal singlet fraction}. The maximal singlet fraction is defined for a general state $\rho$ as\cite{bose2000mixedness}
\beq
\label{singlet fraction}
\mathscr{F}(\rho) = \max\:\langle \Psi\vert \rho\vert \Psi\rangle,
\eeq
where the maximum is taken over all maximally entangled states $\vert \Psi\rangle$. Let us now review the Werner state as a resource for teleportation. Though the Werner state can be represented in various ways, in the present work we express it in terms of the maximal singlet fraction, which is\cite{lee2000entanglement}
\beq
\label{wernerstate}
\rho_{wr}  =  \frac{1 - \mathscr{F}_{wr}}{3}\:I_4 + \frac{4\:\mathscr{F}_{wr} - 1 }{3}\vert \varphi^{-}\rangle\langle \varphi^{-}\vert,
\eeq
where $\vert \varphi^-\rangle$ is the singlet state as defined in Eq.(\ref{bellstates}) and $\mathscr{F}_{wr}$ is the maximal singlet fraction corresponding to the Werner state. The concurrence of $\rho_{wr}$ is calculated using Eq.(\ref{concurrence}) and is found to be
\begin{eqnarray}
\textit{C}(\rho_{wr})&=& max
\{0,2\mathscr{F}_{wr}-1\},\nonumber\\ &=&\left\{\begin{array}{cccc} 0
& & & 0\leq \mathscr{F}_{wr} \leq \frac{1}{2}\\
2\mathscr{F}_{wr}-1 & & & \frac{1}{2}< \mathscr{F}_{wr} \leq 1
\end{array}
\right. \label{werner-con.}
\end{eqnarray}
When the Werner state is used as a quantum channel for teleportation, the average optimal teleportation fidelity is calculated using Eq.(\ref{telfid0}) and is given by\cite{badziag2000local}
\begin{eqnarray}
f_{opt}^{T}(\rho_{wr})= \frac{2\mathscr{F}_{wr}+1}{3},~~~~~ \frac{1}{2}< \mathscr{F}_{wr}
\leq 1\label{werner.tel.fid.}
\end{eqnarray}
Similarly, the relation between the teleportation fidelity and the concurrence of the Werner state is given by
\begin{eqnarray}
f_{opt}^{T}(\rho_{wr})=\frac{2+C(\rho_{wr})}{3}.
\label{fidconc}
\end{eqnarray}
For the teleportation fidelity to be greater than the classical limit of $\frac{2}{3}$, (i.e. by considering $0.68<f_{opt}^{T}(\rho_{wr})<1$, the concurrence $C(\rho_{wr})$ must vary between $[0.04,\:\:1]$ and in this range $f_{opt}^{T}(\rho_{wr})$ is directly proportional to  $C(\rho_{wr})$. This implies, that if $C(\rho_{wr})$ increases, so is $f_{opt}^{T}(\rho_{wr})$.\\\\
We now review the status of the violation of the Bell-CHSH inequality by the Werner state.
Using Eq.(\ref{tmatrix}) the eigenvalues of the matrix $T_{wr}^{\dagger}T_{wr}$  are given by
$u_{1}=u_{2}=u_{3}=\frac{(4\mathscr{F}_{wr}-1)^2}{9}$, where $(T_{wr})_{nm}=Tr(\rho_{wr}\sigma_{n}\otimes\sigma_{m})$ denotes the
elements of the matrix $T_{wr}$. The Werner state violates the Bell-CHSH inequality iff
$M(\rho_{wr})>1$, where $M(\rho_{wr})$ is given by
\begin{eqnarray}
M(\rho_{wr})=  2\frac{(4\mathscr{F}_{wr}-1)^2}{9}
\label{werner-m-rho}
\end{eqnarray}
Using Eq.(\ref{werner-con.}) it follows that the Werner state satisfies the Bell-CHSH inequality although
it is entangled when the maximal singlet fraction $\mathscr{F}_{wr}$ lies within the range
\begin{eqnarray}
\frac{1}{2}\le \mathscr{F}_{wr}\leq \frac{3+\sqrt{2}}{4\sqrt{2}}
\label{wernerBell-CHSH2}
\end{eqnarray}
The optimal teleportation fidelity in terms of $M(\rho)$ is given by
\begin{eqnarray}
f_{opt}^T(\rho_{wr}) = \frac{\sqrt{\frac{M(\rho_{wr})}{2}} +1}{2}
\label{werner.fid.bell}
\end{eqnarray}
Moreover, from Eqs.(\ref{werner.tel.fid.}) and (\ref{werner.fid.bell}) it follows that the Werner state can be used as a quantum teleportation channel (average optimal fidelity exceeding $2/3$) even without violating the Bell-CHSH inequality in the above domain.\\\\
Thus we see that there exists mixed state of Werner class which satisfies Bell-CHSH inequality but is still useful as channel for quantum teleportation. There are many other important class of mixed states which follow the same characteristics. I shall discuss about them in the next chapter. However, before concluding this chapter, let me discuss about how we quantum mixedness of the states. We can use \textbf{Linear Entropy} for this quantification.\\\\
A quantum state is more often studied as an isolated system.  This is not exactly true and quantum states are in contact with an external environment which influences these states externally.  The environment causes a degradation of the quantumness of these states and causes them to decohere and become classical.  The quantumness of a state is  maximal when it is pure and once it interacts with the environment, it usually loses its quantumness and consequently it appears mixed.  A completely  classical state is the maximal mixed state.  When the state changes from being pure to becoming mixed, there is an entropy introduced which is quantified using the linear entropy.  For an arbitrary $d$-dimensional quantum mixed  state $\rho$, the mixedness is defined using the normalized linear entropy  $S_L(\rho)$ and is defined as\cite{peters2004mixed}
\begin{eqnarray}
\label{linearentropy1}
S_L(\rho)=\frac{d}{d-1}(1-{\rm Tr}(\rho^{2})).
\end{eqnarray}
Here the quantity ${\rm Tr} (\rho^{2})$  describes the purity of the quantum system.  For a two-qubit system, the value of $L$ ranges from $0$ to $1$. The entropy $S_L(\rho)= 0$ for any pure state, and the maximum value $S_L(\rho) = 1$ is attained for the maximally mixed state $\frac{I}{4}$.
In terms of the linear entropy $S_L$, it can easily be shown that, for the Werner state
\beq
\label{slwerner}
f_{opt}^T(\rho_{wr}) =  \frac{1 - \sqrt{1-S_L(\rho_{wr})}}{2};\:\: 0\leq S_L < \frac{8}{9}.
\eeq
If we plot the teleportation fidelity of the Werner state against the parameters $S_L$ (linear entropy) and $C$ (concurrence) (varying them from $0.04$ to $0.89$) we get the following pattern
\begin{figure}[h!]
		\begin{center}
			\includegraphics[width=7.5cm]{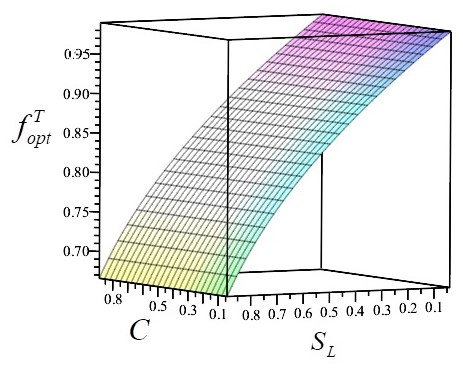}
		\end{center}
		\label{werner}
        \caption{The figure shows the change of teleportation fidelity of the Werner state against its concurrence and mixedness.}
	\end{figure}
From the figure, it is clear that the teleportation fidelity $f^T_{opt}$ of the Werner state varies from $0.68$ to $1$, in the common range $[0.04,\:0.89]$ where concurrence and mixedness vary.
\chapter{Mixed States and Teleportation:}
\label{ch:mixedstatesmaximalnonmaximal}
In the previous chapters I have already defined pure and mixed states. Pure states may be maximally entangled or non-maximally entangled. All the Bell states of Eq.(\ref{bellstates}) are maximally entangled pure states. Say for example the singlet state $\frac{1}{\sqrt{2}}\Big(\vert 0\rangle\vert 1\rangle - \vert 1\rangle\vert 0\rangle\Big)$ with which Bennett et al worked is one of the examples of pure maximally entangled state. If, however, we consider the state $\alpha \vert 0\rangle\vert 1\rangle - \beta \vert 1\rangle\vert 0\rangle$, with the normalization condition $\vert \alpha\vert^2 + \vert \beta \vert^2 = 1$, such states are non-maximally entangled pure states. Likewise, mixed states can also be classified into two categories, viz. (a) Maximally entangled mixed state (MEMS) and (b) Non-maximally entangled mixed state (NMEMS). \textbf{Those states that achieve the greatest possible entanglement for a given mixedness are known as MEMS, otherwise they are NMEMS.} I shall now give a few examples of maximally entangled mixed states. 
\subsection{MEMS:}
The notion of maximally entangled mixed state in bipartite system was first proposed by Satoshi Ishizaka and Tohya Hiroshima in the year $2000$\cite{ishizaka2000maximally}. They proposed a class of mixed states in bipartite $2\otimes 2$ systems (or two qubit system), where the states in this class showed a property of having a maximum amount of entanglement in the sense that the entanglement of formation (and even negativity) of these states cannot be increased by any (local or non-local) unitary transformation. They proved this property in the case of states with rank less than $4$ and also confirmed the results numerically by generating $10000$ density matrices.

The states, that Ishizaka and Hiroshima proposed, were those obtained by applying any local unitary transformation to the states of the form
\beq
\label{ishihiromems1}
\mathcal{M}_1 = p_1\vert \varphi^-\rangle\langle \varphi^-\vert  + p_2\vert 00\rangle\langle 00\vert + p_3 \vert \varphi^+\rangle\langle \varphi^+\vert + p_4 \vert 11\rangle\langle 11\vert,
\eeq
where $\vert \varphi^{\pm}\rangle = \frac{1}{\sqrt{2}}\Big(\vert 01\rangle \pm \vert 10\rangle\Big)$ are the Bell states, and $\vert 00\rangle$ and $\vert 11\rangle$ are product states orthogonal to $\vert \varphi^{\pm}\rangle$. Here $p_i$s are the eigenvalues of $\mathcal{M}_1$ in decreasing order (i.e. $p_1\geq p_2\geq p_3\geq p_4$), and $p_1 + p_2 + p_3 + p_4 = 1$. These included states such as
\beq
\label{ishihiromems2}
\mathcal{M}_2 = p_1\vert \phi^-\rangle\langle \phi^-\vert  + p_2\vert 01\rangle\langle 01\vert + p_3 \vert \phi^+\rangle\langle \phi^+\vert + p_4 \vert 10\rangle\langle 10\vert,
\eeq
where $\vert \phi^{\pm}\rangle = \frac{1}{\sqrt{2}}\Big(\vert 00\rangle \pm \vert 11\rangle\Big)$ are also Bell states, and include those that are obtained by exchanging $\vert \varphi^-\rangle \leftrightarrow \vert \varphi^+\rangle$, $\vert 00\rangle \leftrightarrow \vert 11\rangle$, in Eq.(\ref{ishihiromems1}), or $\vert \phi^-\rangle \leftrightarrow \vert \phi^+\rangle$, $\vert 01\rangle \leftrightarrow \vert 10\rangle$ in Eq.(\ref{ishihiromems2}). One of the examples of MEMS is Werner state as the entanglement of formation of this state cannot be increased by any unitary transformation. I have already discussed in the previous chapter that Werner state is a viable test bed to check quantum teleportation using the state as quantum channel. In this respect, one point worth notable, is that on exceeding a certain degree of mixedness (as quantum by the von Neumann entropy)\footnote{I have skipped discussion on the measure of von Neumann entropy. The readers are referred to the book written by Nielsen and Chuang.}, entangled states become useless for teleportation\cite{bose2000mixedness}.\\\\
There is another class MEMS which were proposed by T-C. Wei et.al. They showed\cite{wei2003maximal}
showed that there exist a class of states that have significantly
greater degree of entanglement for a given linear entropy than the
Werner state. In this section we will investigate whether the
class of states introduced by Munro {\it et al.} could be used as
a teleportation channel. We begin with the
analytical form of the MEMS given by
\begin{eqnarray}
\label{wei}
\rho_{MEMS}= \left(\begin{matrix} h(\textit{C}) & 0 & 0 &
\frac{\textit{C}}{2}\cr 0 & 1-2h(\textit{C})& 0 &0 \cr 0  & 0 &
0&0 \cr \frac{\textit{C}}{2} &0 & 0 & h(\textit{C})
\end{matrix}\right) \label{T16}
\end{eqnarray}
where
\begin{eqnarray}
h(\textit{C})=\left\{\begin{array}{cccc}
\textit{C}/2 & & &\textit{C}\geq \frac{2}{3}\\
1/3 & & & \textit{C}< \frac{2}{3}
\end{array}
\right. \label{h(C)}
\end{eqnarray}
with $\textit{C}$ denoting the concurrence of $\rho_{MEMS}$. The form of the linear entropy (using Eq.(\ref{linearentropy1}) of this class is given by
\begin{eqnarray}
S_{L}(\rho_{MEMS})=\left\{
\begin{array}{cccc}
\frac{8}{3}(\textit{C}-\textit{C}^{2}), & & &\textit{C}\geq 2/3.\\
\frac{2}{3}(\frac{4}{3}-\textit{C}^{2}), & & & \textit{C}< 2/3.
\end{array}
\right.
\end{eqnarray}
However, the optimal teleportation fidelity of $\rho_{MEMS}$ in terms of concurrence and mixedness are found to be
\begin{eqnarray}
f_{opt}^{T}(\rho_{MEMS})=\left\{
\begin{array}{cccc}
\frac{2\textit{C}+1}{3} & & &\textit{C}\geq 2/3\\
\frac{5+3\textit{C}}{9} & & & \textit{C}< 2/3
\end{array}
\left. \right. \right\} \label{Tel.fid.(C)}
\end{eqnarray}
and
\begin{eqnarray}
f_{opt}^{T}(\rho_{MEMS})=\left\{
\begin{array}{cccc}
\frac{2}{3}+\frac{\sqrt{2-3S_{L}}}{3\sqrt{2}} & & & 0\leq S_{L}\leq \frac{16}{27}\\
\frac{5}{9}+\frac{\sqrt{8-9S_{L}}}{3\sqrt{6}} & & & \frac{16}{27}<
S_{L} \leq \frac{8}{9}
\end{array}
\right\} \end{eqnarray}
These Wei-class MEMS can further be analyzed for Bell-inequality violation. It has been shown in\cite{adhikari2008teleportation} that $\mathcal{M}(\rho_{MEMS})>1$ when $\frac{\sqrt{153}-8}{18}<\:C\:< \frac{2}{3}$. This implies, in this range $\rho_{MEMS}$ violates the Bell-CHSH inequality. On the contrary, when $\frac{1}{3}<\: C\: \leq \frac{\sqrt{153}-3}{18}$ and hence the state $\rho_{MEMS}$ the Bell-CHSH inequality although it is entangled. Also it has been observed that, Wei-class of MEMS, defined in Eq.(\ref{wei}), needs a much higher degree of entanglement to violate the Bell-CHSH inequality compared to the Werner class of MEMS.
\subsection{NMEMS:}
In the year $2009$, I joined as Ph.D Scholar in Satyendra Nath Bose National Centre of Basic Science (SNBCBS) under the supervision of Prof. Archan Subhra Majumdar, an eminent physicist in the Department of Astrophysics and High Energy Physics. His primary interest was Relativity and Cosmology and he also did considerable work on Quantum information Processing. He is a senior professor in SNBCBS now. Dr. Satyabrata Adhikari was post-doctoral student under Majumdar and was working in the field of Quantum Information Science then. He is now an Assistant Professor in Delhi Technology University. I had known Dr. Adhikari since $2006$ when he was writing his Ph.D thesis and Dr. Adhikari was the person who introduced me to the domain of Quantum Information and because of him I came in contact with Prof. Majumdar of SNBCBS. SNBCBS is one of the quantum hubs in the east zone of India.  I joined SNBCBS at the time when the Institute was engaged in the joint venture with Bose Institute, Kolkata (India) to prepare for upcoming international seminar which would be held in Kolkata on the occasion of celebrating $75$ years of quantum entanglement. We started working on a project then which was based on the study of teleportation fidelity of a class of states which are not MEMS. Dr. Adhikari proposed the idea and it was excellent. Later in this venture two other physicists joined viz. Dr. Biplab Ghosh (who is now the Head of the Department of Physics of Vivekananda college for Women) and Prof. Nilkantha Nayak (who was Associate Professor Head Optical Physics Group in SNBCBS then). Together we defined a class of Non Maximally Entangled Mixed States (NMEMS) and published our work in Quantum Information and Computation (Rinton Press)\cite{adhikari2008teleportation}. For the sake of our discussion let me name that NMEMS as $AMRGN$ state and let me denote this by $\rho_{amrgn}$.

To maintain the sanctity of the research journal in which our paper was published, I shall give a very brief overview of our findings in this book and readers are requested to study the original paper to get the complete essence.
I, in chapter $3$, already defined two inequivalent class of tripartite states, viz. $GHZ$ and $W$ class. These two classes are way different from one another in the sense that (a) one class of state cannot be converted into the other under stochastic local operations and classical communications (SLOCC), (b) the genuine tripartite entanglement (measured by tangle) of $GHZ$ class is zero, while that of $W$ class is one and (c) $GHZ$ class has no entanglement in its bipartite level whereas $W$ class retains bipartite entanglement. By the last point, we mean that, if we remove any party from $GHZ$ class, the resultant state has no entanglement left in it while this is not true for $W$ class. In $W$ class, if we remove any party (or trace out any party), the resultant state is still left with some entanglement in it, which we can measure by concurrence. In our paper, we exploited these properties that the $GHZ$ state and the $W$ state are two qubit separable and inseparable states, respectively, when a qubit is lost from the corresponding three qubit states. By constructing
this type of a non-maximally entangled mixed state, our aim was to show that it can be used as a better teleportation channel than another class of NMEMS known as \textit{Werner derivative}\cite{hiroshima2000local}. The construction of our proposed NMEMS is as follows.
The two-qubit state described by the density matrix $\rho_{amrgn}$ can be explicitly written as
\begin{eqnarray}
\rho_{amrgn}=p\rho^{G}_{12}+(1-p)\rho^{W}_{12},~~~~0\leq p \leq 1
\label{T1}
\end{eqnarray}
The matrix representation of the density matrix $\rho_{amrgn}$ in the computational basis is given by
\begin{eqnarray}
\label{T2}
\rho_{amrgn} = \left(%
	\begin{array}{cccc}
		\frac{p+2}{6} & 0 & 0 & 0\\
		0 & \frac{1-p}{3} & \frac{1-p}{3} & 0\\
		0 & \frac{1-p}{3} & \frac{1-p}{3} & 0\\
        0 & 0 & 0 & \frac{p}{2}\\
	\end{array}%
	\right). 
\end{eqnarray}
Since the state described by the density matrix (\ref{T2}) is of the form
\begin{eqnarray}
\label{T3}
\sigma =  \left(%
	\begin{array}{cccc}
		a & 0 & 0 & 0\\
		0 & b & c & 0\\
		0 & c^{*} & d & 0\\
        0 & 0 & 0 & e\\
	\end{array}%
    \right).
\end{eqnarray}
its amount of entanglement \cite{bruss2003entanglement} is given by
\begin{eqnarray}
C(\rho_{new}) && =C(\sigma)=2\:\max\:(|c|-\sqrt{ae},0) \nonumber\\
&& = 2\:\:\max\Big[\Big(\frac{1-p}{3}-\sqrt{\frac{p(p+2)}{12}}\Big),0\Big] \label{T4}
\end{eqnarray}
Therefore, $\rho_{new}$ is entangled only if
$\frac{1-p}{3}-\sqrt{\frac{p(p+2)}{12}}>0$, i.e., when
$0\leq p<0.292$.
Note that in the limiting case of $p=0$ the state $\rho_{new}$ reduces to
\begin{eqnarray}
\rho_{12}^{W} =
\frac{1}{3}|00\rangle\langle00|+\frac{2}{3}|\psi^{+}\rangle\langle\psi^{+}|
\label{mat.p=0}
\end{eqnarray}
where $|\psi^+\rangle = (|01\rangle + |10\rangle)/\sqrt{2}$.
The state
$\rho_{12}^{W}$ is maximally entangled since it can be put into Ishizaka and Hiroshima's \cite{ishizaka2000maximally} proposed class of MEMS. The concurrence of this state is $\frac{2}{3}$. The average optimal teleportation fidelity of our constructed NMEMS is given by
\begin{eqnarray}
f^{T}_{opt}(\rho_{amrgn})= \frac{7-4p}{9},~~~~0\leq p <\frac{1}{4}
 \label{T6}
\end{eqnarray}
and it follows that
\begin{eqnarray}
\frac{2}{3}<f^{T}_{opt}(\rho_{amrgn})\leq \frac{7}{9}
 \label{T7(a)}
\end{eqnarray}
We note here an interesting fact that the state $\rho_{amrgn}$ cannot be used as an efficient teleportation channel when $0.25<p<0.292$ although the state is entangled there. We noted here an interesting fact that the state $\rho_{amrgn}$ cannot be used as an
efficient teleportation channel when $0.25< p < 0.292$ although the state is entangled there.  Also it was observed that the constructed state $\rho_{new}$ does not violate the Bell-CHSH inequality although it is entangled for $0\leq p\:< 0.292$. Thus we showed that in the domain of NMEMS states there is our proposed NMEMS which in some range of the state's parameter $p$ although satisfies Bell-inequality is a viable candidate as quantum channel for teleportation. 
I stop here and encourage interested readers to study the original article published in\cite{adhikari2008teleportation}.
\chapter{Some other notable works on QT:}
\label{notableqt}
Teleportation is a process which, since its inception, has gone thus a long way, and all over the world many physicists contributed in this journey in a theoretical as well as an experimental way. If I were writing a review article on this topic I would have included many  references on teleportation works. And for this I am extremely sorry for not considering many such notable works in this brief overview. However, while scanning through the literature, some of the works caught my attention which I would like to mention in this chapter. Quantum teleportation allows for sending quantum information encoded in a qubit (a spin$\frac{1}{2}$ particle or any quantum two-level system) from one place to another without any material transfer of the particle itself. The two parties involved in this process initially share a maximally entangled state. The protocol is carried out using only local measurements (Bell measurement) and classical communication. We also know that in the original protocol by Bennett et al. they used singlet state as channel. However, any Bell states can serve this purpose. Bell states are orthogonal entangled states. In other words, Alice teleports her unknown qubit to Bob using a basis of orthogonal pure entangled states that she shares with Bob. Satyabrata Adhikari, Archan S. Majumdar, Dipankar Home, Alok K. Pan and Pankaj Joshi, devised a protocol of quantum teleportation using non-orthogonal entangled channels and published their article in the year $2012$\cite{adhikari2012quantum}.

Prof. Dipankar Home is an emeritus professor and globally recognized scientist in the field of quantum information and foundation. He is currently associated with Bose Institute, Kolkata and did seminal works on quantum contextuality. I personally consider him as a living legend in the field of quantum mechanics. He has authored/coauthored many good books which I would like to mention here. $``$Conceptual Foundations of Quantum Physics – An Overview from Modern Perspectives" is a research level book, which he authored (the forward was written by Sir Anthony Leggett) whose title is $``$Einstein’s Struggles with Quantum Theory: A Reappraisal". Then Prof. Home coauthored a book with Andrew Whitaker (the forward was written by legendary Nobel Laureate Sir Roger Penrose). He also coauthored a popular level book with Prof. Partho Ghose, titled, $``$Riddles in your Teacup – Fun with Everyday Scientific Puzzles" whose forward was written by Paul Davies \cite{home2013conceptual,home2007einstein,ghose2020riddles}\footnote{For futher information one can visit his website https://dipankarhome.com/}.

Prof. A. K. Pan is currently associated to Department of Physics at Indian Institute of Technology Hyderabad, Telengana (India) and Prof. Pankaj Joshi is an Indian astrophysicist and cosmologist whose research is mainly focused on areas of gravitational collapse and spacetime singulariy. Currently, he is a Distinguished Professor of Physics, and founding director of the International Center for Space and Cosmology at Ahmedabad University, Gujarat, India.
\subsection{QT: Using non-orthogonal channels}
To implement teleportation, one ideally needs maximally entangled two-qubit states, i.e. singlet states. But in a typical experiment it is very difficult to prepare singlet states because the preparation is never perfect. As a result, one may generally have to deal with non-maximally entangled states or mixed states. So it becomes necessary to generalize the idea of using the maximally entangled states as quantum channels to the case of non-maximally entangled or noisy channels between two distant partners. To switch over from a
maximally entangled state to a non-maximally entangled state or a mixed state, one has to pay a price in terms of the loss of teleportation fidelity. This was the primary idea, on which Adhikari et. al started working on\cite{adhikari2012quantum}. They considered a bipartite entangled state of the form
\beq
\label{nonorthentan}
\vert \Psi\rangle_{AB} = \mu\vert \alpha_A\rangle\vert \beta_B\rangle + \nu\vert \nu_A\rangle\vert \delta_B\rangle,
\eeq
where, the state vectors $\vert \alpha_A\rangle$ and $\vert \nu_A\rangle$ are with Alice and $\vert \beta_B\rangle$ and $\vert \delta_B\rangle$ are the qubits held by Bob. Here $\lbrace \vert \alpha_A\rangle, \vert \nu_A\rangle\rbrace$ and $\lbrace \vert \beta_B\rangle, \vert \delta_B\rangle\rbrace$ represent linearly independent non-orthogonal basis that span a $2D$ subspace of each Hilbert space $H_A$ and $H_B$ to which Alice and Bob belong to. Since quantum state $\vert \alpha\rangle$ and $\vert \beta\rangle$ are non-orthogonal, we have $\langle \alpha\vert \beta\rangle \neq  0$ and $\langle \alpha\vert \beta\rangle$ is complex number. Then one can take $\langle \alpha\vert \beta\rangle = r\:e^{i\:\theta}$, where $r$ and $\theta$ are real parameters representing respectively the modulus and argument of the complex number. Let us choose the normalized non-orthogonal basis vectors $\vert \alpha\rangle$ and $\vert \beta\rangle$ to be
\beq
\label{nonortholet}
\vert \alpha\rangle = \left(%
		\begin{array}{c}
			0\\
			1
		\end{array}%
		\right),\:\:\: \left(%
		\begin{array}{c}
			r\:e^{i\:\theta}\\
			N_\beta
		\end{array}%
		\right),
\eeq
where $N = \sqrt{1-r^2}$. Using Gram-Schmidt orthonormalization process\footnote{See Chapter $1$.}, no-orthogonal basis basis vectors $\vert \alpha\rangle$ and $\vert \beta\rangle$ can be transformed into orthogonal basis vectors $\vert 0\rangle$ and $\vert 1\rangle$ as,
\beq
\label{orthobasis}
\vert 0\rangle = \vert \alpha\rangle;\:\:\: \vert 1\rangle = N_{G}(\vert \beta\rangle - \langle \alpha\vert \beta\rangle \vert \alpha\rangle),
\eeq
where $N_G = \frac{1}{N_{\beta}} = \frac{1}{\sqrt{1-r^2}}$. The state $\vert 1\rangle$ contains implicitly information on non-orthogonality of the original non-orthogonal system. Hence one can proceed with quantum teleportation with the states like $\vert 0\rangle$ and $\vert 1\rangle$. \\\\
Now a bipartite entangled state in the non-orthogonal basis can be written as
\beq
\label{nonorthostaten1}
\vert \Phi\rangle_{AB} = N_1[\vert \alpha_A\rangle \vert \beta_B\rangle + \vert \beta_A\rangle \vert \alpha_B\rangle],
\eeq
where $N_1 = \frac{1}{\sqrt{2(1+r^2)}}$ is the normalization constant. In terms of the orthogonal basis vectors $\vert 0\rangle$ and $\vert 1\rangle$, Eq.(\ref{nonorthostaten1}) can be re-expressed as
\beq
\label{nonorthostaten2}
\vert \Phi\rangle_{AB} = \frac{N_1}{N_G}\Big[\vert 0_A\rangle \vert 1_B\rangle + \vert 1_A\rangle \vert 0_B\rangle\Big] + 2N_1\langle \alpha\vert\beta\rangle \vert 0\rangle\vert 0\rangle.
\eeq
In general, the concurrence for an arbitrary two-qubit pure state $a\vert 00\rangle + b\vert 01\rangle + c\vert 10\rangle + d\vert 11\rangle$ is given by $2\vert ad - bc\vert$\cite{miranowicz2004ordering}. Therefore, the amount of entanglement contained in the bipartite entangled state given in Eq.(\ref{nonorthostaten2}) can be quantified by the measure concurrence and that is given by
\beq
\label{connonortho}
C(\vert \Phi\rangle_{AB} ) = \frac{1-r^2}{1 + r^2},\:\: 0\leq r \leq 1.
\eeq
The parameter $r$ is a measure of the non-orthogonality. From expression Eq.(\ref{connonortho}), it is seen that $C$ is a decreasing function of $r$ and hence as the amount of non-orthogonality increases, the amount of entanglement in a bipartite system decreases and
it goes to zero when the non-orthogonal parameter $r$ tends to $1$. The maximum amount of entanglement is achieved when $r = 0$, i.e. when the basis state vectors are orthogonal to each other. 
Adhikari et al then formulated their teleportation scheme as follows. There are two parties Alice ($A$) and Bob ($B$) where Alice would play the role of sender and Bob, the receiver. The input state can be prepared by a third party Cliff, who after preparation can send the state to Alice and thus Alice will have no idea what the state is. It was also assumed that during this transmission of state from Cliff to Alice there was no distortion observed on the state. Let the unknown state be
\beq
\label{unknownstatefromcliff}
\vert s_a\rangle = x\:\vert 0_a\rangle + y\: \vert 1_a\rangle;\:\: |x|^2+|y|^2 = 1.
\eeq
Once the single qubit state of Eq.(\ref{unknownstatefromcliff}) is combined with shared entangled state of Eq.(\ref{nonorthostaten2}), that result in a three-qubit system, as a tensor product of a single qubit $B$ and the Bell-basis involvling the two qubits $a$ and $A$, as
\beq
\label{unknownclubbecliff}
\vert \chi\rangle_{aAB} = \vert s_a\rangle \otimes \vert \phi\rangle_{AB}\nonumber\\
=\frac{1}{\sqrt{2}}\Big[ \vert \phi^+\rangle_{aA} (P_+\vert 0_B\rangle + M^{\prime}\vert 1_{B}\rangle) +  \vert \phi^-\rangle_{aA} (P_-\vert 0_B\rangle + M^{\prime}\vert 1_{B}\rangle) + \nonumber\\
\vert \varphi^+\rangle_{aA}(Q_+\vert 0_B\rangle + M^{\prime\prime}\vert 1_B\rangle) +  \vert \varphi^-\rangle_{aA}(Q_-\vert 0_B\rangle  -M^{\prime\prime}\vert 1_B\rangle)\Big],
\eeq
where $M^{\prime} = \frac{x\:N_1}{N_G}$, $M^{\prime\prime} = \frac{y\:N_1}{N_G}$, $P_{\pm} = N_1\Big(2\:x\:r\;e^{i\theta} \pm \frac{y}{N_G}\Big)$ and $Q_{\pm} = N_1\Big(\pm 2\:y\:r\;e^{i\theta} \pm \frac{x}{N_G}\Big)$ while $\vert \phi^{\pm}\rangle$ and $\vert \varphi^{\pm}\rangle$ are the Bell states from Eq.(\ref{bellstates}). Since the qubits $a$ and $A$ are with Alice, she makes a Bell state measurement on her qubits and then sends the
measurement result to Bob at the expense of two classical bits. On the basis of those measurement outcomes viz. $\vert \phi^+\rangle$, $\vert \phi^-\rangle$, $\vert \varphi^+\rangle$ or $\vert \varphi^-\rangle$, Bob performs a suitable unitary operation on his qubit $B$ as follows:
\begin{itemize}
    \item If the measurement result is $\vert \phi^+\rangle$, Bob operates $X$ operator on his qubit.
    \item If the measurement result is $\vert \phi^-\rangle$, Bob operates $Y$ operator on his qubit.
    \item If the measurement result is $\vert \varphi^+\rangle$, Bob operates $I$ (the identity) operator on his qubit, which means he needs to do nothing.
    \item If the measurement result is $\vert \varphi^-\rangle$, Bob operates $Z$ operator on his qubit.
\end{itemize}
All these operators that Bob applies are Pauli operators, which have already been discussed in Chapter $1$.
Another eminent Indian quantum physicist that I would like to mention about is Prof. Anirban Pathak of Jaypee Institute of Technology, (JIIT), Noida, India. He did his Ph.D. from Visva-Bharati, Santiniketan, India. Subsequently, he was a post-doctoral fellow at the Freie University, Berlin. He joined JIIT, Noida in 2002. At present, he is actively involved in teaching and research related to several aspects of quantum optics and quantum information with a focus on secure quantum communication and secure multiparty quantum computation. He is one of the editors of the journal $``$Quantum Information in India" which is affiliated $``$ Quantum Information Processing (Springer)". He has authored many books, one of which I would like to specifically mention is $``$Elements of Quantum Computation and Quantum Communication"\cite{pathak2013elements}. This book helped me a lot during my Ph.D days. This book has superb introduction in Indian Stylized way. Among his many notable works, one that is relevant to my book is his work on QT where he proposed a scheme of bidirectional QT with $5$ qubit states. I shall briefly describe it here.

\subsection{Bidirectional Controlled QT (BCQT): With $5$ qubit state} 
The 5-qubit quantum states that are useful for BCQT may be described as (I have kept the notations as it was used in the paper)\cite{shukla2013bidirectional}
\begin{eqnarray}
\label{bcqtstate}
\vert \psi\rangle_{12345} = \frac{1}{\sqrt{2}}\Big(\vert \psi_1\rangle_{A_1B_1} \vert \psi_2\rangle_{A_2B_2}\vert a\rangle_{C_1} \pm  \vert \psi_3\rangle_{A_1B_1} \vert \psi_4\rangle_{A_2B_2}\vert b\rangle_{C_1}\Big).
\end{eqnarray}
Here, single qubit states $\vert a\rangle$ and $\vert b\rangle$ satisfy the condition $\langle a\vert b\rangle = \delta_{a,b}$ and $\vert \psi_i\rangle$ belongs to the set of Bell states $\lbrace \vert \phi^+\rangle, \vert \phi^-\rangle, \vert \varphi^+\rangle, \vert \varphi^-\rangle\rbrace$. In Eq.(\ref{bcqtstate}), $\vert \psi_1\rangle \neq \vert \psi_3\rangle$ and $\vert \psi_2\rangle \neq \vert \psi_4\rangle$ while the subscripts $A,\:B,\:C$ indicate the qubits of Alice, Bob and Cliff respectively. Thus $\vert \psi_i\rangle$s which constitute Eq.(\ref{bcqtstate}) are two of the four Bell states. Also the condition 
\begin{eqnarray}
\label{bcqtcond}
  \vert \psi_1\rangle \neq \vert \psi_3\rangle ;\:\:\: \vert \psi_2\rangle \neq \vert \psi_4\rangle,   
\end{eqnarray}
however, ensures that Cliff's qubit is \textit{appropriately entangled} with remaining four qubits. By appropriately entangled Pathak et al. mean that unless Cliff measures his qubit in $\lbrace \vert a\rangle, \vert b\rangle\rbrace$  basis and discloses the outcome. Alice and Bob are unaware of the entangled (Bell) states they share and consequently the receiver does not know upon the receipt of the measurement outcome of the sender which unitary operation is to be applied.\\\\
In case $\vert \psi_1\rangle = \vert \psi_3\rangle$  ($\vert \psi_2\rangle = \vert \psi_4\rangle$) is allowed then the qubits $1$ and $2$ ($3$ or $4$) are separable from the remaining qubits and consequently Charlie has no control over the teleportation done using those two qubits. Now when the state (\ref{bcqtstate}) satisfies the condition (\ref{bcqtcond}) then on the disclosure of the outcome of
Cliff’s measurement on $\lbrace \vert a\rangle, \vert b\rangle\rbrace$ basis, Alice and Bob know with certainty which two
Bell states they share and consequently they can use the conventional teleportation scheme to teleport unknown quantum states. Without the knowledge of the initial Bell states shared by Alice and Bob, the receiver cannot decide the operation to be implemented by him/her. As the condition in Eq.(\ref{bcqtcond}) ensures that without the disclosure of Cliff the receiver and the sender do not know the entangled state shared by them so Cliff has a control over the bidirectional teleportation scheme. \\\\
I think, I must stop here and encourage the readers to go through this wonderful work proposed by Pathak et al.\cite{shukla2013bidirectional}\\\\
I have already discussed what maximally and non-maximally entangled mixed states are in the previous chapters. There is a very good work done by K.G. Paulson and Prasanta K. Panigrahi on Teleportation devised under the umbrella of these mixed states\cite{paulson2019tripartite}, which I shall describe in brief now. Prasanta K. Panigrahi is one of the eminent quantum physicists of eastern zone of India. At the time of doing this work he was a professor of Indian Institute of Science Education and Research (IISER), Kolkata. Once we met and he said "I love those who do wonderful Mathematics". Let me discuss his work on teleportation now.
       
\subsection{Controlled QT: With NMEMS states} Paulsson and  Panigrahi\cite{paulson2019tripartite} explored the use of three-qubit mixed states as quantum channels for controlled teleportation of single qubits. They investigated how various channel properties influence the achievement of maximum teleportation fidelity in the controlled setting. Surprisingly, their findings revealed that a specific class of non-maximally entangled mixed $X$ states ($X-$ NMEMS) could attain optimal controlled quantum teleportation fidelity for a given level of multipartite entanglement and mixedness. In contrast, maximally entangled mixed X states (X-MEMS) failed to do so. This result challenged the conventional understanding from standard quantum teleportation of single qubits, where maximal entanglement is often considered crucial. Furthermore, they demonstrated that even biseparable X-NMEMS could serve as valuable resources for achieving high controlled quantum teleportation fidelity within a certain range of mixedness, suggesting relaxed requirements for quantum channels in controlled teleportation. The class of three qubit $X-$ MEMS as a convex sum of maximally entangled pure $GHZ$ and separable states is given as\footnote{MEMS stands for Maximally Entangled Mixed States and NMEMS stands for Non-Maximally Entangled Mixed States.}
\begin{eqnarray}
    \label{rhonmemsx}
    \rho(X)_{MEMS} = p_1\vert GHZ_1^+\rangle\langle GHZ_1^+\vert + p_2\vert 001\rangle\langle 001\vert + p_3 \vert 010\rangle\langle 010\vert + \nonumber\\  p_4\vert 011\rangle\langle 011\vert + p_5\vert GHZ_1^-\rangle\langle GHZ_1^-\vert + p_6 \vert 100\rangle\langle 100\vert + p_7\vert 101\rangle\langle 101\vert + p_8\vert 110\rangle\langle 110\vert\nonumber\\
\end{eqnarray}
where $p_1\geq p_2 \geq p_3 \geq p_4 \geq p_5 \geq p_6 \geq p_7 \geq p_8 \geq 0$ are eigenvalues of density matrix $\rho(X)_{MEMS}$ and $\sum_{i=1}^8 p_i = 1$. Here,
\beq
\label{ghzs}
\vert GHZ_{1}^{\pm}\rangle &=& \frac{1}{\sqrt{2}}\Big[\vert 000\rangle \pm \vert 111\rangle\Big]\nonumber\\
\vert GHZ_{2}^{\pm}\rangle &=& \frac{1}{\sqrt{2}}\Big[\vert 001\rangle \pm \vert 110\rangle\Big]\nonumber\\
\vert GHZ_{3}^{\pm}\rangle &=& \frac{1}{\sqrt{2}}\Big[\vert 010\rangle \pm \vert 101\rangle\Big]\nonumber\\
\vert GHZ_{4}^{\pm}\rangle &=& \frac{1}{\sqrt{2}}\Big[\vert 011\rangle \pm \vert 100\rangle\Big].
\eeq
Now the three qubit symmetric mixed states $X$ states are defined with diagonal elements denoted by $a_1,\:a_2,\:a_3,\:a_4,\:b_1,\:b_2,\:b_3,\:b_4 \geq 1$ and anti diagonal elements by $z_1,\:z_2,\:z_3,\:z_4,\:z_1^*,\:z_2^*,\:z_3^*,\:z_4^*$. The \textbf{Genuine Multipartite Entanglement (GME)}\cite{yu2005evolution} of a three qubit $X$ state is given by
\beq
\label{gme}
C_{GME} = 2\: \max \lbrace 0, |z_j| - w_j\rbrace,
\eeq
where, $\sum_{i}(a_i + b_i) = 1$ and $w_j = \sum_{k\neq j}\sqrt{a_k\:b_k}$. The positivity criterion of the X matrix is satisfied with the condition $|z_i| \leq \sqrt{a_i\:b_i}$. The tripartite $X-$ states are entangled for $0< C_{GME} \leq 1$ and $C_{GME}$ is zero for bi-separable states.\\\\
Panigrahi et al. showed that the given density matrix $\rho(X)_{MEMS}$ possessed a maximum value of \textit{Genuine Multipartite Entanglement (GME)} for a given spectrum of eigenvalues. The $GME$ of the state (\ref{rhonmemsx}) using (\ref{gme}) had been calculated as 
\begin{eqnarray}
    \label{gmerhonmemes}
    C^{*}[ \rho(X)_{MEMS}] = \max \lbrace 0, p_1 - p_5 - 2[\sqrt{p_2p_8} + \sqrt{p_3p_7}+\sqrt{p_4p_6}]\rbrace.
\end{eqnarray}
The $GME$ of a given state $\rho(X)$ is equal to $C^*(\rho(X))$ then the state $\rho(X)$ belonged to the class of $\rho(X)_{MEMS}$.
In Controlled Quantum Teleportation (CQT) another measure plays a significant role, the measure is called \textit{Control Power} (or CP in short). The CP is a quantity which is used to define the authority of the  controller in CQT and is the difference of conditioned and non-conditioned fidelity. The influence of the control qubit in the CQT process is quantified by estimating CP and is defined as
\beq
\label{cp}
CP(\rho_X) = F_{CQT}(\rho_{X}) -  F_{NC}(\rho_{X}),
\eeq
where $F_{CQT}(\rho_{X})$ and $F_{NC}(\rho_{X})$ are conditioned and non-conditioned fidelity. The two conditions $F_{CQT}(\rho_{X})>\frac{2}{3}$ and $F_{NC}(\rho_{X})\leq \frac{2}{3}$ should be satisfied by tripartite quantum channels to ensure the active participation of the controller in the controlled quantum teleportation process. Mixedness of quantum states is also an important parameter that influences fidelity of controlled quantum. teleportation.\\\\
I am not going to the details of their work, rather in brief Panigrahi et al. made an in-depth analysis of tripartite rank-dependent $X$ states as resources for controlled quantum teleportation (CQT) has illuminated several key aspects of multipartite quantum systems and their potential for efficient quantum information processing. This investigation delved into the performance of these states under varying degrees of entanglement and mixedness, revealing nuanced behaviors that challenge conventional assumptions. An in-depth analysis of tripartite rank-dependent $X$ states as resources for controlled quantum teleportation (CQT) has illuminated several key aspects of multipartite quantum systems and their potential for efficient quantum information processing. Specifically, the study demonstrated that for a given level of multipartite entanglement and mixedness, a specific class of non-maximally entangled mixed $X$ states ($X$-NMEMS) consistently achieved optimal fidelity in controlled quantum teleportation. This finding is significant because it contradicts the intuition that maximal entanglement is always necessary for optimal performance. Instead, it suggests that a careful selection of non-maximal entanglement can also yield superior results in practical quantum information protocols  Furthermore, the investigation revealed a contrasting behavior for tripartite maximally entangled mixed states ($X$-MEMS). These states, despite their maximal entanglement, were found to be incapable of attaining optimal teleportation fidelity. This observation underscores the complexity of multipartite entanglement and highlights that the degree of entanglement alone is not a sufficient predictor of a state's utility in quantum communication tasks. \\\\
Importantly, the study confirmed that these results hold true across different measures of multipartite entanglement. This consistency across various entanglement measures strengthens the validity and generality of the findings, indicating that the observed behaviors are intrinsic to the structure of the $X$ states and not artifacts of a specific entanglement metric. In summary, this research provides valuable insights into the performance of tripartite rank-dependent $X$ states for controlled quantum teleportation. It challenges the conventional view of maximal entanglement as the sole determinant of optimal performance and demonstrates the potential of non-maximally entangled and even biseparable states for efficient quantum information processing. These findings have significant implications for the development of practical quantum communication protocols and the design of robust quantum information systems.The study further explored the potential of biseparable $X$-NMEMS as resources for CQT.
 Biseparable states, which possess a specific form of partial entanglement, are often considered to be less powerful than fully entangled states. However, the analysis showed that certain biseparable $X$-NMEMS could achieve high controlled quantum teleportation fidelity, particularly in highly mixed scenarios. This finding is crucial for practical applications, as highly mixed states are more prevalent in real-world quantum system as a result of unavoidable interactions with the environment. Thus, the ability to utilize biseparable states for CQT offers a pathway to robust quantum communication in noisy environments. \\\\
 At this stage, I can't resist mentioning one of our papers published recently in a prestigious journal in which we worked in the direction of analysing the teleportation fidelities of certain mixed states which belong to the family of $X$ states\cite{bhattacharjee2024non}. In this work, I was assisted by my collaborators, Dr. Anushree Bhattacharjee of Techno Main Salt Lake (Engg. Colg.), Prof. Md Manirul Ali of Chennai Institute of Technology and Dr. Biplab Ghosh of Vivekananda College of Women, Kolkata\footnote{Dr.Biplab Ghosh has already been introduced in sec: $6.0.2$.}. 

In this paper, we considered first, two different types of tripartite pure states, one of which is $\vert W\overline{W}\rangle$ and the other one is $\vert Star\rangle$. The state $\vert \overline{W}\rangle$ is the spin-flipped version of the $\vert W\rangle$ state and $\vert W\overline{W}\rangle$ is an equal linear superposition of standard $\vert W\rangle$ state and its spin-flipped version $\vert \overline{W}\rangle$. Below I show the mathematical abstraction of the said state.
\beq
\label{wwbarstate}
\vert W\overline{W}\rangle = \frac{1}{\sqrt{2}}\Big(\vert W\rangle + \vert \overline{W}\rangle\Big),
\eeq
where if $\vert W\rangle = \frac{1}{\sqrt{3}}\Big(\vert 001\rangle + \vert 010\rangle + \vert 100\rangle\Big)$ then $\vert \overline{W}\rangle = \frac{1}{\sqrt{3}}\Big(\vert 110\rangle + \vert 101\rangle + \vert 011\rangle\Big)$. On the other hand pure tripartite $\vert Star\rangle$ is defined as
\beq
\label{starstate}
\vert Star\rangle = \frac{1}{2}\Big(\vert 000\rangle + \vert 100\rangle + \vert 101\rangle + \vert 111\rangle\Big).
\eeq
The $\vert W\overline{W}\rangle$ state is significant as different types of correlations can be found in its bipartite and tripartite levels. This makes $\vert W\overline{W}\rangle$ state robust under local decoherence. The $\vert Star\rangle$ state has correlations spread in it in an asymmetric way in contrast to the  $\vert W\overline{W}\rangle$ state, which is symmetric for all qubits. Moreover, in $\vert Star\rangle$ state there is a central qubit and two peripheral qubits. If the central qubit is traced out, the remaining qubits are left in a separable state\cite{cao2020fragility}. The characteristics of these two states i.e. $\vert W\overline{W}\rangle$ and $\vert Star\rangle$ can be well understood by the following pictorial representations.
\begin{figure}[h!]
		\begin{center}
			\includegraphics[width=7.6cm]{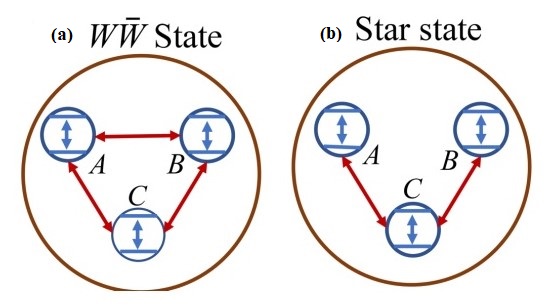}
		\end{center}
		\label{werner}
        \caption{(a) $W\overline{W}$ state (b) $Star$ state (photo courtesy: Author).}
        \end{figure}
Then we constructed two distinct classes of bipartite mixed states, by taking convex combination of a three qubit pure tripartite states (like $\vert W\overline{W}\rangle$ and $\vert Star\rangle$ and any one of the four Bell states (see Eq.(\ref{bellstates})) and thereby removing any of the three qubits by taking partial trace. The resultant class of states are written below. 
\begin{eqnarray}
    \label{mix1}
    \rho^{1} &=& p\rho^{W} + (1-p) \rho^{\phi^{\pm}},\nonumber\\
    \rho^{2} &=& \hat{p}\rho^{W} + (1-\hat{p}) \rho^{\varphi^{\pm}}; \: 0\leq p,\hat{p}\leq 1\nonumber\\
    \rho^{3} &=& q\varrho^{\overline{W}} + (1-q) \varrho^{\phi^{\pm}},\nonumber\\
    \rho^{4} &=& \hat{q}\varrho^{\overline {W}} + (1-\hat{q}) \varrho^{\varphi^{\pm}},  0\leq q,\hat{q}\leq 1.
\end{eqnarray}
In the states depicted in the Eq.(\ref{mix1}), the qubits held by third party in $\vert W\rangle$ and $\vert \overline{W}\rangle$ have been traced out. The states have been named as $Class-I$ states.  Again, we considered
\begin{eqnarray}
    \label{mix2}
     \rho^{5} &=& r\rho^{W\overline{W}} + (1-r) \rho^{\phi^{\pm}},\nonumber\\
    \rho^{6} &=& \hat{r}\rho^{W\overline{W}} + (1-\hat{r}) \rho^{\varphi^{\pm}}; \: 0\leq r,\hat{r}\leq 1,
\end{eqnarray}
and 
\begin{eqnarray}
    \label{mix3}
     \tau^{1} &=& s\varrho^{star} + (1-s) \varrho^{\phi^{\pm}},\nonumber\\
    \tau^{2} &=& \hat{s}\varrho^{star} + (1-\hat{s}) \varrho^{\varphi^{\pm}}; \: 0\leq s,\hat{s}\leq 1.
\end{eqnarray}
In the Eq.(\ref{mix2}) third qubit is traced out from $\vert W\overline{W}\rangle$ and in the Eq.(\ref{mix3}), one of the peripheral qubits have been traced out from $\vert Star\rangle$. It is to be noted that except $\vert Star\rangle$, removing any party from the rest of $\vert W\rangle$, $\vert \overline{W}\rangle$ and $\vert W\overline{W}\rangle$ won't change the final conclusion drawn on the constructed states. The states described in Eqs.(\ref{mix1}),(\ref{mix2}) and (\ref{mix3}) are non-maximally entangled mixed states (NMEMS). Nevertheless, the states (\ref{mix1}) and (\ref{mix2}) are $X-$ NMEMS whereas the states (\ref{mix2}) and (\ref{mix3}) are not of $X$ types and we called it Non-$X$ NMEMS. We showed that the teleportation fidelities of all these bipartite  mixed states of $X$ and Non-$X$ types exceed the teleportation fidelities of maximally entangled mixed state (MEMS) such as $\rho^{W} = \frac{1}{3}\vert 00\rangle\langle 00\vert + \frac{2}{3}\vert \phi^+\rangle\langle \phi^+\vert$ and $\rho^{\overline{W}} = \frac{1}{3}\vert 11\rangle\langle 11\vert + \frac{2}{3}\vert \phi^+\rangle\langle \phi^+\vert$. Also it was shown that states $\tau^1$ and $\tau^2$ (of Eqs.(\ref{mix3})), which are Non-$X$ states are not only useful as quantum teleportation channels but also are better performers than the $X$ type NMEMS. Moreover, the teleportation fidelities of these two states exceed that of Werner state (which is a type of MEMS). Myself and my collaborators Dr. Anushree Pandey along with Dr. Abhijit Mandal later did another interesting work on QT\cite{bhattacharjee2025teleportation}. Eylee Jung et al. once conjectured that with respect to, the Groverian measure of quantum entanglement which involves a quantity like $P_{max}$, $P_{max}=\frac{1}{2}$  is a necessary and sufficient condition for the perfect two-party teleportation\cite{jung2007perfect}. Myself, Bhattacharjee and Mandal had shown that this was not the case. With a counter example we showed that there might exist states whose $P_{max}$ is less than half but can still be used as quantum channel for teleportation.

Let me discuss another aspect of quantum teleportation. The key ingredient for performing many information processing tasks is provided by quantum entanglement and hence detection of this trait in quantum states is also very significant. The \textit{entanglement witness} is the signature operator for experimental detection of entanglement. This tool can also be used to detect whether an unknown entangled state could be used as a resource for performing information processing tasks such as teleportation. We can call such as operator as \textit{teleportation witness operator}. In this Dr. Nirman Ganguly et al. made a considerable progress in the year $2011$. Dr. Nirman Ganguly is a dear friend of mine and we started our research career with quantum information togther under the supervision of Prof. Archan S. Majundar of Satyendra Nath Bose National Center for Basic Science (SNBNCBS). Dr. Ganguly is currently holding the post of associate professor in Birla Institute of Technology, Pilani, Hyderabad.

The ability of entangled states to act as a resource for teleportation is linked to a property of the fully entangled fraction. Nirman et al. showed that the set of states with their fully entangled fraction bounded by a threshold value required for performing teleportation is both convex and compact. This feature enables the existence of Hermitian witness operators, the measurement of which could distinguish unknown states useful for performing teleportation\cite{ganguly2011entanglement}. After that, in the year $2012$, Ganguly et al. contributed another interesting paper where they provided systematic method to construct teleportation witnesses from entanglement witnesses corresponding to general qudit systems. The witnesses so constructed are shown to be optimal for qubit and qutrit systems and therefore detect the largest set of states useful for teleportation within a given class\cite{adhikari2012construction}.
\subsection{QT: With four particle cluster states}
My book will probably remain incomplete if I don't write anything about contributions of our neighbor country, China, in the field of quantum information science. China has progressed a lot in the theoretical as well as experimental works in this field, and because of the page constraint, it is not possible to include all the contributions. Rather, I shall focus on only one work which was by Li Song-Song et al.\cite{song2008controlled}. They showed a controlled teleportation scheme with four particle cluster states as a quantum channel. The $n$ partcle entangled \textbf{cluster state} is expressed as
\begin{eqnarray}
    \label{nparticleculster}
    \vert Cls\rangle_n = \frac{1}{2^{\frac{n}{2}}}\otimes_{a=1}^n\Big(\vert 0_a\rangle\sigma_z^{(a+1)} + \vert 1_a\rangle\Big).
\end{eqnarray}
Here by the notation $\otimes_{a=1}^n\vert 0_a\rangle$ we mean $\vert 0_1\rangle \otimes \vert 0_2\rangle \otimes \cdots \otimes \vert 0_n\rangle$. This state was first introduced by H.J. Briegel and R. Raussendorf in the year $2001$\cite{briegel2001persistent}. It has also been pointed out that when $n =2$ (or $n = 3$), the state $\vert Cls\rangle_n$ of Eq.(\ref{clusterstate}) and Bell state of Eq.(\ref{bellstates}) (or $GHZ$ state of Eq.(\ref{ghz})) can be transformed into each other by local operations and classical communications (LOCC). But when $n>3$ the state $\vert Cls\rangle_n$ and the $n-$ particle $GHZ$ state i.e. $\vert GHZ\rangle_{n}$ i.e. $\frac{1}{\sqrt{2}}\Big[\vert 0_1\rangle\vert 0_2\rangle\cdots\vert 0_n\rangle + \vert 1_1\rangle\vert 1_2\rangle\cdots\vert 1_n\rangle\Big]$ cannot be transformed into each other by LOCC.\\\\
For $n=4$, the four particle cluster state is defined as 
\begin{eqnarray}
    \label{clusterstate}
    \vert Cls\rangle_{abcd} = \frac{1}{2}\Big[\vert 0_a\rangle\vert 0_b\rangle\vert 0_c\rangle\vert 0_d\rangle + \vert 0_a\rangle\vert 0_b\rangle\vert 1_c\rangle \vert 1_d\rangle + \vert 1_a\rangle \vert 1_b\rangle +  \vert 0_c\rangle \vert 0_d\rangle - \vert 1_a\rangle \vert 1_b\rangle \vert 1_c\rangle \vert 1_d\rangle\Big]\nonumber\\.
\end{eqnarray}
The cluster state cluster state to have particular characteristics when $n>3$\cite{song2008controlled}. For example, this cluster state of Eq.(\ref{clusterstate}) has the properties of both the $GHZ$ and $W$ class of states. The state is persistent with respect to its property of entanglement which means it is harder to destroy the entanglement of the state by local operations (which is the case for $GHZ$ state, as we know, when one party is removed or traced out from the $GHZ$ state, it loses its entanglement). The creation of a cluster state requires only the Ising-type interactions between two-state particles in a lattice configuration, which can be easily found in various condensed matter systems with proper spin-spin coupling on lattices\cite{briegel2001persistent}. Now with this state shown in Eq.(\ref{clusterstate}), Song-Song et al, devised controlled teleportation strategy to teleport a single qubit from sender Alice to the receiver Cliff with Bob playing the intermediary role as controller of the scheme.\\\\
The sender Alice should first prepare a four particle cluster state and then send qubit $2$ to supervisor (or controller) Bob and qubit $4$ to receiver Cliff, respectively. There is a prevention policy, that Song-Song et al. implemented, so that Bob wouldn't be able to steal the information, which I am not going to write in details. Suppose the single-qubit state Alice wants to teleport to Cliff was originally carried by qubit $A$, 
\begin{eqnarray}
    \label{songqubit}
    \vert \psi_A\rangle = \alpha\vert 0_A\rangle + \beta \vert 1_A\rangle,
\end{eqnarray}
where $|\alpha|^2 + |\beta|^2 = 1$. The state of the composite quantum system can be expressed as $\vert \Psi\rangle_{Aabcd} = \vert \psi_A\rangle \otimes  \vert Cls\rangle_{abcd}$. Now Alice has three qubits $A$, $a$ and $c$, she performs a measurement on the three qubits under the basis $\lbrace \vert 0\rangle\vert 0\rangle\vert 0\rangle \pm \vert 1\rangle\vert 1\rangle\vert 1\rangle,\:\vert 0\rangle\vert 0\rangle \vert 1\rangle \pm \vert 1\rangle \vert 1\rangle \vert 0\rangle, \: \vert 0\rangle\vert 1\rangle\vert 0\rangle \pm \vert 1\rangle\vert 0\rangle\vert 1\rangle, \: \vert 1\rangle\vert 0\rangle\vert 0\rangle \pm \vert 0\rangle\vert 1\rangle\vert 1\rangle \rbrace$, one gets,
\begin{eqnarray}
    \label{songsongclubbed}
    \vert \Psi\rangle_{Aabcd} =\frac{1}{4}\Big\lbrace (\vert 000\rangle + \vert 111\rangle)_{Aac}(\alpha\vert 00\rangle - \beta \vert 11\rangle_{bd}) + \nonumber\\  (\vert 000\rangle - \vert 111\rangle)_{Aac}(\alpha\vert 00\rangle - \beta \vert 11\rangle_{bd}) + \nonumber\\ (\vert 001\rangle + \vert 110\rangle)_{Aac}(\alpha\vert 01\rangle + \beta \vert 10\rangle_{bd})_{bd} +\nonumber\\+ 
    (\vert 001\rangle - \vert 110\rangle)_{Aac}(\alpha\vert 01\rangle - \beta \vert 10\rangle_{bd}) +\nonumber\\ 
    (\vert 010\rangle + \vert 101\rangle)_{Aac}(\alpha\vert 10\rangle + \beta \vert 01\rangle_{bd}) +\nonumber\\
    (\vert 010\rangle - \vert 101\rangle)_{Aac}(\alpha\vert 10\rangle - \beta \vert 01\rangle_{bd}) + \nonumber\\
    (\vert 100\rangle + \vert 011\rangle)_{Aac}(-\alpha\vert 11\rangle - \beta \vert 00\rangle_{bd}) + \nonumber\\
    (\vert 100\rangle - \vert 011\rangle)_{Aac}(\alpha\vert 11\rangle + \beta \vert 00\rangle_{bd})\rbrace\Big\rbrace.
\end{eqnarray}
After the measurement, Alice broadcasts her result, so that the state of qubits $b$ and $d$ can be transformed, by Bob and Cliff to a common form
\begin{eqnarray}
    \label{songsong24}
    \vert \psi\rangle_{bd} = (\alpha\vert 00\rangle + \beta \vert 11\rangle)_{bd}.
\end{eqnarray}
The complete set of messages that Alice broadcasts is $\frac{1}{\sqrt{2}}\Big(\vert 000\rangle \pm \vert 111\rangle\Big);\: \frac{1}{\sqrt{2}}\Big(\vert 001\rangle \pm \vert 110\rangle\Big); \: \frac{1}{\sqrt{2}}\Big(\vert 010\rangle \pm \vert 101\rangle\Big); \: \frac{1}{\sqrt{2}}\Big(\vert 100\rangle \pm \vert 011\rangle\Big)$ and consequently the unitary transformation that Bob and Cliff apply are respectively $\lbrace I_b\otimes \sigma^Z_{d};\: I_b\otimes I_d\rbrace$, $\lbrace I_b\otimes \sigma^X_{d};\: I_b\otimes i\sigma^Y_{d}\rbrace$, $\lbrace \sigma_b^{X}\otimes I_d;\: \sigma^X_{b}\otimes \sigma^Z_{b}\rbrace$ and $\lbrace \sigma^{X}_{b}\otimes (-i\sigma^Y_{d}), \sigma_{b}^X\otimes \sigma_d^{X}\rbrace$, which would subsequently generate Eq. (\ref{songsong24}). The state $ \vert \psi\rangle_{bd}$ of Eq. (\ref{songsong24}) is an entangled state from which Cliff cannot get anything useful without cooperation from Bob. If Bob, indeed, would like to help Cliff, he should measure just the qubit $b$ on the basis $\Big\lbrace \frac{1}{\sqrt{2}}(\vert 0\rangle + \vert 1\rangle)_{b},\: \frac{1}{\sqrt{2}}(\vert 0\rangle - \vert 1\rangle)_{b}\Big\rbrace$ and transfer the outcome of his measurement to Cliff via a classical channel. Thereafter, the Eq.(\ref{songsong24}) can be re-expressed as
\begin{eqnarray}
    \label{songsong241}
    \vert \psi\rangle_{bd} = \frac{1}{\sqrt{2}}\Big(\vert 0\rangle + \vert 1\rangle\Big)_b\otimes \frac{1}{\sqrt{2}}\Big(\alpha \vert 0\rangle + \beta 1\rangle\Big)_d +\nonumber\\ \frac{1}{\sqrt{2}}\Big(\vert 0\rangle - \vert 1\rangle\Big)_b\otimes \frac{1}{\sqrt{2}}\Big(\alpha \vert 0\rangle - \beta 1\rangle\Big)_d.
\end{eqnarray}
As soon as Charlie is informed of Bob’s result, he can perform an appropriate unitary transformation on qubit $d$ to obtain the original single-qubit state shown in Eq.(\ref{songqubit}). According to the possible outcomes $\frac{1}{\sqrt{2}}\Big(\vert 0\rangle + \vert 1\rangle\Big)_b$ and $\frac{1}{\sqrt{2}}\Big(\vert 0\rangle - \vert 1\rangle\Big)_d$, the corresponding transformation are $I_d = \vert 0\rangle\langle 0\vert + \vert 1\rangle\langle 1\vert$ and $\sigma_d^Z = \vert 0\rangle\langle 0\vert - \vert 1\rangle\langle 1\vert$ respectively.\\\\
In the above, the notations $\sigma^i$s (where, $i = X, Y, Z, iY$ and $I$ is the identity operator) denote the Pauli operators and when we write $b$ and/or $d$ in the suffix, it means the operators are being used by Bob and/or Cliff. In their paper Song et al. generated another protocol with permuting qubits and distributing them among the parties involved, but I am skipping this part (which will be a mere repetition of the Song protocol). The readers are advised to go through the original paper. Song song Li et al. did another work on teleportation where a new scheme of controlled teleportation was proposed\cite{nie2009non}. In that scheme, a four-particle cluster state is shared by a sender, a controller and a receiver. The sender first performs a Bell-basis measurement on the qubits at hand, and the controller performs measurements under a non-maximally entangled Bellbasis after he knows the sender’s measurement result. Then the receiver introduces an auxiliary qubit and performs some appropriate unitary transformations on his qubits. Quantum teleportation is realized after the receiver performs a local measurement on the auxiliary qubit and an appropriate unitary transformation on his qubit.
\subsection{QT: What else?}
Since its inception, many more works have been published on QT. It is not possible for me to include each one of these works here. But a few more notable works are worth mentioning. A quantum teleportation scheme based on energy-time entangled EPR-pairs was presented, and its application to the teleportation of a multi-mode single-photon wave packet was explored by S.N. Molotkov\cite{molotkov1998quantum}. Bound entanglement, a form of noisy entanglement that cannot be distilled into singlets, is generally unsuitable for direct quantum communication. However, Horodecki et al. demonstrated that bound entanglement could be \textit{pumped} into a single pair of freely entangled particles, enabling teleportation with a fidelity unattainable without its assistance\cite{horodecki1999bound}. A quantum \textit{telecloning} process combining quantum teleportation and optimal quantum cloning from one input to $M$ outputs was proposed by Murao et al.\cite{murao1999quantum}. The scheme relied on the establishment of particular multi-particle entangled states, which functioned as multi-user quantum information channels. The entanglement structure of these states was analyzed and shown to be crucial for this type of information processing. In chemistry, \textit{Chiral molecules}, characterized by their non-superimposable mirror-image forms (enantiomers), can, in principle, exist in quantum superpositions of their left- and right-handed states. The work done by Christopher S. Maierle et al. explored a novel approach to transferring the complex amplitudes defining these superposition states to the polarization degrees of freedom of a single photon, thereby enabling their measurement. They proposed two distinct experimental schemes for this purpose. The first scheme theoretically achieved perfect teleportation of the chiral superposition amplitudes onto the photon's polarization state, providing a complete and lossless transfer of information. The second scheme, while achieving state-dependent teleportation, still allowed for a complete determination of the amplitudes, albeit with some inherent limitations in fidelity. Both methods offerrd a pathway to extract comprehensive information about the chiral superposition, enabling detailed analysis of these quantum states\cite{maierle1998teleport}. Quantum systems possess information processing capabilities exceeding those of classical devices, as exemplified by quantum teleportation. This process enables the transfer of a quantum state between locations without physical transport. While partial quantum teleportation has been demonstrated over macroscopic distances using optical systems, complete recovery of the original state—the final stage of teleportation—has remained elusive. In this respect, Nielsen et al reported the experimental realization of full quantum teleportation over interatomic distances using liquid-state nuclear magnetic resonance. Employing natural phase decoherence of carbon nuclei in trichloroethylene molecules, they successfully teleported the quantum state of a carbon nucleus to a hydrogen nucleus. This demonstration established a potential subroutine for larger quantum computations and a pathway for quantum communication applications\cite{nielsen1998complete}. Quantum teleportation with \textit{squeezed vaccuum state} was worked upon by G. J. Milburn and Samuel L. Braunstein\cite{milburn1999quantum}. Quantum teleportation was also analyzed for states of dynamical variables with continuous spectra, in contrast to previous work with discrete (spin) variables. The entanglement fidelity of the scheme was computed, including the roles of finite quantum correlation and non-ideal detection efficiency. A protocol was presented for teleporting the wave function of a single mode of the electromagnetic field with high fidelity using squeezed-state entanglement and current experimental capability, by  P. van Loock, Samuel L. Braunstein and H. J. Kimble\cite{braunstein1998teleportation,van2000broadband}. E. Solano and his team studied a method for the implementation of a reliable teleportation protocol (theoretically, 100\% of success) of internal states in trapped ions. The generation of the quantum channel (any of four Bell states) might be done respecting technical limitations on individual addressing and without claiming the Lamb-Dicke regime. An adequate Bell analyzer, that transformed unitarily the Bell basis into a completely disentangled one, was considered. Probable sources of error and fidelity estimations of the teleportation process were studied too\cite{solano2001reliable}. Hai-Woong Lee in his paper, in the year $2001$, demonstrated a setup for the direct, total teleportation of a single-particle entangled state. His scheme utilized a parametric down conversion source to generate entangled photon pairs and a dual Bell measurement system comprising two beam splitters and four detectors. This allowed for the direct transfer of an entangled state between two pairs of beams with a 50\% success probability. The scheme could be readily extended for the total teleportation of multiparticle entangled states\cite{lee2001total}. Quantum teleportation can be directly derived from a standard classical circuit for qubit state exchange using only controlled-NOT gates. This transparent classical origin allows for a straightforward generalization of teleportation to d-level systems. This interesting work was done by N. David Mermin\cite{mermin2001classical}. Sibasish Ghosh et al. explored the teleportation of bipartite pure entanglement through two identical noisy quantum channels modeled as mixtures of Bell states. They identified a specific mixed-state channel that preserved the entanglement of all pure entangled states within a known bi-orthogonal product basis during the teleportation process. In contrast, the Werner state channel did not exhibit these properties. They also examined the connection between these noisy channels and the violation of Bell's inequality and the quantum two-entropy inequality\cite{ghosh2002entanglement}. Experimental demonstration of continuous-variable quantum teleportation utilizing squeezed-state entanglement were performed. Thus T. C. Zhang et al. explicitly calculated the teleportation fidelity for the real system, accounting for imperfections like propagation losses, detection inefficiencies, and phase fluctuations. The measured fidelity for input coherent states was $F=0.61 \pm 0.02$, which, when corrected for output detection efficiency, yielded $0.62$. This was lower than the projected fidelity of $0.69$ based on independent measurements of entanglement and efficiencies. They detailed the teleportation protocol in their work and discussed the discrepancy between experimental results and theory, as well as the limitations of our current setup\cite{zhang2003quantum}. Stephen D. Bartlett1 and William J. Munro demonstrated that a universal gate set for optical quantum computation could be quantum teleported using EPR entanglement, homodyne detection, and conditional linear optics and squeezing. This approach enabled fault-tolerant quantum computation in both qubit and continuous-variable optical systems. They also discussed the teleportation of nondeterministic nonlinear gates used in linear optics quantum computation\cite{bartlett2003quantum}. Ye Yeo used quantum teleportation to probe the thermally entangled state of a three-qubit Heisenberg XX ring. His analysis revealed entanglement features beyond pairwise thermal concurrence. Specifically, two mixtures of distinct pairs of W states with identical concurrence exhibited significantly different average teleportation fidelities\cite{yeo2003studying}. Paul M. Alsing and G. J. Milburn's work described quantum teleportation between an inertial Alice and a uniformly accelerating Bob. They found that Davies-Unruh radiation in Bob's frame reduces the teleportation fidelity. Since teleportation relies on entanglement, their results indicated that quantum entanglement degraded in non-inertial frames\cite{alsing2003teleportation}. Jaeyoon Cho and Hai-Woong Lee proposed a quantum teleportation scheme for single atoms trapped in cavities, utilizing adiabatic passage and polarization measurement. They demonstrated that this method can teleport an atomic internal state between two cavities with a success probability of $\frac{1}{2}$ and a fidelity of $1$. Notably, the scheme was robust against significant imperfections like Lamb-Dicke condition violation, weak atom-cavity coupling, spontaneous emission, and detection inefficiency\cite{cho2004quantum}. H. de Riedmatten et al. reported a long-distance quantum teleportation experiment using a fiber-delayed Bell state measurement (BSM). Two $2$ kilometer optical fibers connected the qubit source and the entanglement source to the BSM. The teleported qubits were then analyzed after an additional $2.2$ kilometer fiber in a separate laboratory $55$ meter away. Time-bin qubits at $1310$ nano-meter were teleported onto $1550$ nano-meter photons with a fidelity of 77\%, surpassing the classical limit. This demonstrated the first elementary quantum relay over significant distances, paving the way for extended-range quantum communication and key distribution\cite{de2004long}. F. de Pasquale et al. proposed a quantum teleportation protocol utilizing a double quantum dot array. The unknown qubit was encoded in a tunnel-coupled pair of quantum dots with a single excess electron. They demonstrated the creation of a maximally entangled state via adiabatic increase of Coulomb repulsion between dot pairs. This entanglement facilitated teleportation through adiabatic coupling with the unknown qubit. A sudden separation of Bob's qubit enabled a time evolution of Alice's, effectively performing a modified Bell measurement. Long-distance transmission could be achieved using an entangled chain of $N$ coupled double quantum dots, with increasing robustness against phonon-induced decoherence for larger $N$\cite{de2004teleportation}.A. Dantan et al. proposed a protocol for high-fidelity quantum state teleportation of a macroscopic atomic ensemble using two quantum-correlated ensembles. they detailed the preparation of these correlated ensembles via quasi-perfect light-atom quantum state transfer. their teleportation scheme relied on optical joint measurements of the atomic ensembles and magnetic feedback reconstruction\cite{dantan2005teleportation}. G. Gordon and G. Rigolin presented a generalized telecloning (GTC) protocol where the quantum channel was non-optimally entangled and they studied how the fidelity of the telecloned states depended on the entanglement of the channel. They showed that one could increase the fidelity of the telecloned states, achieving the optimal value in some situations, by properly choosing the measurement basis at Alice's, albeit turning the protocol to a probabilistic one. They also showed how one could convert the GTC protocol to the teleportation protocol via proper unitary operations\cite{gordon2007generalized}. Hidehiro Yonezawa et al. presented a sequence of two quantum teleportations of optical coherent states, achieved by cascading two high-fidelity continuous-variable teleporters. In their experiment, the individual teleportation fidelities were evaluated as $F_1  = 0.70 \pm  0.02$ and $F_2 = 0.75 \pm  0.02$, while the fidelity between the input and the sequentially teleported states was determined as $0.57 \pm 0.02$. This still exceeded the optimal fidelity of one half for classical teleportation of arbitrary coherent states and almost attained the value of the first (un-sequential) QT experiment with optical coherent states\cite{yonezawa2007sequential}. Hua Wei proposed a scheme for decoherence-free quantum state transfer and teleportation in a quantum network, utilizing qubits encoded in atomic internal states within cavities and mediated by cavity-assisted single-photon pulse interactions between spatially separated nodes. They detailed the implementation of a logic-qubit Hadamard gate and a two-logic-qubit conditional gate, and discussed the experimental feasibility of our approach\cite{wei2007transfer}. Guoming Wang and Mingsheng Ying investigated perfect quantum teleportation from multiple senders to a single receiver using a pre-shared stabilizer state. This state was distributed among senders as partitioned subsystems. They presented two sufficient conditions for a stabilizer state to achieve a nonzero teleportation capacity for a given partition. Corresponding teleportation protocols were provided. Notably, they found that even mixed stabilizer states could enable perfect many-to-one teleportation. Their work offered a stabilizer formalism perspective on standard teleportation and a technique for analyzing multipartite entangled state teleportation capabilities\cite{wang2008perfect}. My one time collaborators Dr. Satyabrata Adhikari and Dr. Nirman Ganguly along with two other co-authors did a brilliant work on quantum teleportation. In their work, they investigated the potential of employing the two-qubit output of the Buzek–Hillery quantum copying machine (not necessarily a universal quantum cloner) as a teleportation channel. Their analysis revealed that the output state exhibited entanglement and violated the Bell-CHSH inequality within a specific range of the machine parameter $\xi$, while remaining entangled but not violating the inequality in another range of that parameter. They found that for certain $\xi$ values, this mixed two-qubit state could function as a teleportation channel. This application of the Buzek–Hillery cloning machine's output as a teleportation channel enhanced its significance\cite{adhikari2008quantum}. A scheme for the teleportation of a beam of light including its temporal fluctuations was proposed by Changsuk Noh et al. Expressions for the teleported degrees of first- and second-order optical coherence were presented. Teleportation of an antibunched photon stream illustrated their proposal\cite{noh2009quantum}. Also, The usefulness of the genuinely entangled six-qubit state that had been introduced by Borras et al., was investigated by Prasanta K. Panigrahi and his team for the quantum teleportation of an arbitrary three-qubit state and for quantum state sharing (QSTS) of an arbitrary two-qubit state\cite{choudhury2009quantum,borras2007multiqubit}. In an interesting work by Indranil Chakrabarty, Nirman Ganguly and Binayak S. Choudhury, the efficacy of the entangled output of Pati-Braunstein deletion machine as a teleportation channel, was analyzed. The possibility of such output violating the Bell’s inequality was investigated. Interestingly they found that for all values of the input parameter $\alpha$ the state did not violate the Bell’s inequality but when used as a teleportation channel could give a fidelity higher than the classical optimum value which was $\frac{2}{3}$\cite{chakrabarty2011deletion}. Masahiro Hotta analyzed a quantum energy teleportation protocol that transferred energy from one end of a linear ion crystal to the other using local operations and classical communication (LOCC), achieving a speed exceeding the crystal's phonon velocity. The protocol involved briefly and strongly coupling a probe qubit to ground-state phonon fluctuations at the initial edge, followed by a projective measurement to extract information. This measurement excited phonons, requiring external energy input. The measurement outcome was classically transmitted to the opposite edge. Despite the excited phonons not physically traversing the crystal at the speed of information transfer, the received information enabled rapid energy extraction from the ions at the destination. Critically, the protocol avoided exciting intermediate ions, ensuring energy transfer without heat generation in the channel\cite{hotta2009quantum}. Efficiently transferring quantum information between disparate physical systems is crucial for quantum technologies. A particularly demanding task is the exchange between discrete and continuous variable encodings, especially within harmonic oscillator systems. Alexander E. Ulanov et al. demonstrated a protocol for teleporting a continuous-variable optical qubit, encoded in low-amplitude coherent states, onto a discrete-variable, single-rail qubit (a vacuum/single-photon superposition) using a hybrid entangled resource. Testing on a one-dimensional subspace of the input qubit space, they achieved an average teleportation fidelity of $0.83 \pm 0.04$, mapping onto the equator of the Bloch sphere. This work paved the way for broader integration of discrete- and continuous-variable quantum information processing within optical circuits\cite{ulanov2017quantum}.\\\\
Having charted the significant milestones in the development of quantum teleportation, from its theoretical inception to its increasingly sophisticated experimental realizations, it becomes evident that this once-futuristic concept is now poised to transcend the laboratory. The remarkable progress detailed in this chapter lays a crucial foundation, and as we turn our attention to the next, I will explore how these advancements are actively contributing to and shaping the burgeoning landscape of quantum technologies, promising a revolutionary leap in computation, communication, and sensing.
\chapter{QT: In Quantum 2.0 Era}
\label{ch:qt2.0}
Quantum 2.0 signifies a graceful and ambitious ascent in the landscape of quantum technology, artfully moving beyond the foundational revelations of Quantum 1.0. It embodies a deliberate and nuanced pursuit of engineering and deploying intricate quantum systems, mastering the delicate interplay of superposition and the profound interconnectedness of entanglement within increasingly refined architectures. This meticulously cultivated control unlocks the gateway to transformative applications poised to reshape the very contours of our world.
India's National Quantum Mission (NQM) emerges as a visionary and strategically vital undertaking, resonating harmoniously with the aspirations of Quantum 2.0. It marks a resolute and purposeful stride towards realizing the profound potential of quantum technologies for the nation's progress and sovereignty. By nurturing indigenous expertise in research and development across the pivotal domains of quantum computing, secure communication, advanced sensing, and novel materials, the NQM endeavors to cultivate a dynamic and self-sustaining ecosystem of quantum innovation. This ambitious mission aspires to position India as a leading architect in the unfolding quantum epoch, translating the inherent elegance of Quantum 2.0 into tangible solutions that address societal imperatives, bolster national security, and drive economic prosperity. The NQM, therefore, acts as the indispensable catalyst, channeling the theoretical brilliance of quantum mechanics into the tangible fabric of India's technological and economic future, paving the way for a quantum-enabled tomorrow.
Within this vibrant context, the research endeavor in quantum teleportation has also experienced a significant upswing. Recent years have witnessed a surge in experimental advancements in this fascinating domain, and India is actively contributing to this global race. I shall first delineate the overall progress achieved in experimental quantum teleportation in brief, followed by a concise overview of India's notable contributions in this exciting field. One of the early and highly influential papers in this field is "Ground-to-satellite quantum teleportation," published in Nature in 2017. This study was carried out by a group of 25 scientists with J-W. Pan as the principal investigator\cite{ren2017ground}. Quantum teleportation allows the reliable transfer of unknown quantum states over long distances, crucial for future quantum networks and computing. Previous terrestrial experiments were limited to approximately $100$ kilometers due to photon loss. This work by Pan et al. reported a significant breakthrough which ensured successful teleportation of single-photon qubits from the ground to a satellite over $1400$ kilometer using optimized techniques to overcome atmospheric challenges. Achieving a high fidelity, this ground-to-space link was a vital step towards realizing a global quantum internet. An emerging scientific discipline, machine learning has found a compelling counterpart in the burgeoning field of quantum machine learning. In their 2020 publication, Julius Wallnofer et al. demonstrated the efficacy of machine learning in identifying pivotal quantum protocols such as teleportation, entanglement purification, and the quantum repeater. These schemes are fundamental to the advancement of long-distance quantum communication, and their discovery has significantly shaped the landscape of quantum information processing. Significantly, the utility of learning agents extends beyond the mere reproduction of established protocols; the same methodology enables the discovery of enhanced solutions for persistent long-distance communication challenges, particularly in asymmetric scenarios characterized by nonuniform channel noise and varying segment distances. Their insightful findings stemmed from the application of projective simulation, a sophisticated learning agent model that elegantly integrates reinforcement learning and decision-making within a physically inspired framework. This learning agent was provided with a universal gate set, and the desired objective was precisely defined through a reward scheme. From a technical perspective, the learning agent adeptly navigates stochastic environments and their inherent reactions. Their methodology incorporated a concept akin to hierarchical skill acquisition, wherein solutions to discrete subproblems were learned and subsequently reused within the overarching scheme. This innovative approach holds particular significance in the ongoing development of robust long-distance communication protocols and promises to pave the way for the seamless integration of machine learning in the design and implementation of increasingly sophisticated quantum networks\cite{wallnofer2020machine}. The research by Hung Do et al. explored a flexible quantum teleportation method using a satellite to connect diverse quantum devices, whether they encoded information in discrete variable (DV) or continuous variables (CV). The study focused on CV teleportation via a lossy satellite link and, unlike previous work on hybrid entanglement swapping, accounted for transmission loss. The researchers developed a general framework for CV teleportation through lossy channels and analyzed the teleportation of DV states encoded in photon number. Their findings indicated that under certain conditions, teleported DV states exhibit higher entanglement quality compared to direct satellite-sourced DV entanglement. The implications for establishing a global, ultra-secure communication network between various quantum devices with satellite connections were discussed, specifically illustrating the impact of this teleportation method on quantum key distribution rates\cite{do2019hybrid}. Using fiber-coupled devices, including state-of-the-art low-noise superconducting nanowire single-photon detectors and off-the-shelf optics, Raju Valivarthi et al achieved conditional quantum teleportation of time-bin qubits at the telecommunication wavelength of $1536.5$ nanometer. They measured teleportation fidelities of  $\geq 90\%$ that were consistent with an analytical model of their system, which included realistic imperfections. To demonstrate the compatibility of their setup with deployed quantum networks, they teleport qubits over $22$ kilometer of single-mode fiber while transmitting qubits over an additional $22$ kilometer of fiber. Their systems, which were compatible with emerging solid-state quantum devices, provided a realistic foundation for a high-fidelity quantum Internet with practical devices\cite{valivarthi2020teleportation}. Niccolo Fiaschi et al. demonstrated quantum teleportation of a polarization-encoded optical input state onto the joint state of a pair of nanomechanical resonators. Their protocol also allowed to store and retrieved an arbitrary qubit state onto a dual-rail encoded optomechanical quantum memory. This work showed the full functionality of a single quantum repeater node and presented a key milestone towards applications of optomechanical systems as quantum network nodes\cite{fiaschi2021optomechanical}.  N. Sehati, M. K. Tavassoly and M. Ghasemi successfully teleported the unknown quantum state which is a superposition of squeezed vacuum state and squeezed one-photon state using the beam splitter in the absence of dissipation. In the continuation, they tried to implement the same teleportation protocol, however, in the presence of dissipation effects. To do this task, they used proper entangled channel to reach to perfect teleportation under the influence of decoherence. Also, they considered another superposition of two squeezed vacuum states with separation in phase by and teleported it with a different appropriate entangled channel\cite{sehati2019teleportation}. Hossein Rangani Jahromi addressed the idea of remote sensing in quantum teleportation with topological qubits against the background of robust noise. He investigated the enhancement of quantum teleportation through non-local characteristics of the topological qubits to achieve near-perfect quantum teleportation even with mixed quantum states\cite{jahromi2022remote}. Cheng-Qian Xu and D. L. Zhou tackled the challenge of understanding information flow in quantum networks built with anyons. While anyons are promising for topological quantum computation, how they communicate quantum information isn't well-established. The researchers proposed a new, topologically protected quantum teleportation protocol specifically designed for the Ising anyon model. Importantly, their protocol demonstrated that an unknown quantum state encoded in any number of Ising anyons could be successfully teleported from Alice to Bob. This work was significant because it extended the concept of standard quantum teleportation (used for distinguishable particles) to the realm of Ising anyons. This generalization could deepen our understanding of quantum entanglement in anyonic systems and how it could be used as a resource\cite{xu2022quantum}. In the tapestry of modern wireless communication, microwave technology stands as a central thread, weaving together mobile networks and local area connections. Beyond its conventional applications, the microwave spectrum harbors inherent advantages for open-air transmission, boasting minimal absorption losses and efficient energy utilization; moreover, it resonates as the natural frequency for cutting-edge superconducting quantum technologies. At the heart of secure quantum communication lies the distribution of entanglement across distant parties. Consequently, a profound understanding of its limitations within realistic open-air environments, particularly in the relatively uncharted microwave domain, is paramount in its evolution towards a mainstream quantum technology. Tasio Gonzalez-Raya et al. computed the fidelity of a continuous-variable quantum teleportation protocol using open-air-distributed entanglement as a resource\cite{gonzalez2022open}. The nascent challenge of teleporting energy to distant sites beckons at the forefront of quantum information science and technology. The creation of a methodology to channel localized energy within laboratory settings to remote outposts promises to unlock intricate energy flows within quantum networks. Through the lens of quantum information engineering, Kazuki Ikeda introduced a novel approach for the distribution of local energy to a multitude of remote nodes, leveraging the elegant framework of hyperbolic geometry. Given their exponential node scalability, hyperbolic networks emerge as particularly well-suited architectures for energy allocation across expansive quantum networks. To achieve the ambitious goal of long-range Quantum Energy Teleportation (QET), the author proposed a synergistic hybrid method, intertwining quantum state teleportation with QET protocols. By transmitting local quantum information via teleportation and subsequently enacting conditional operations upon it, QET can, in theory, transcend the limitations of geographical separation. This proposed method offers a fresh perspective, potentially illuminating novel applications for future large-scale quantum networks and fostering the integration of quantum physics within the realm of information engineering\cite{ikeda2024long}.\\\\
The landscape of quantum teleportation has been significantly enriched by the profound contributions of Indian scientists, their research deeply rooted in the intricate realms of quantum information, entanglement, and allied technologies. Pioneering works by esteemed figures stand as notable milestones in this journey: Prof. Dipankar Home of Bose Institute, Kolkata; Prof. Guruprasad Kar of the Indian Statistical Institute, Kolkata; Prof. Arun Kumar Pati of Synergy Quantum India, Research and Innovation Park, Indian Institute of Technology Delhi, New Delhi; Prof. Pankaj Agrawal of the Center for Quantum Engineering, Research and Education (CQuERE), TCG Centres for Research and Education in Science and Technology (TCG CREST), Kolkata; Prof. Archan S. Majumdar of S. N. Bose National Centre for Basic Sciences; Prof. Prasanta K. Panigrahi of the Department of Physical Sciences, Indian Institute of Science Education and Research (IISER) Kolkata; Prof. Anirban Pathak of Jaypee Institute of Technology Noida; Profs. Ujjawal Sen and Aditi Sen of the Harish-Chandra Research Institute, A CI of Homi Bhabha National Institute, Jhansi; Prof. Manirul Ali of Chennai Institute of Technology; and Profs. Surajit Sen and Tushar Kanti Dey of the Centre of Advanced Studies and Innovation Lab, Assam, amongst many others, have indelibly shaped this vibrant field.
\chapter{Motivating Engineers of 21$^{st}$ century}
\label{ch:qt21st}
The 21st century unfolds as a remarkable epoch, poised to witness the flourishing of two pivotal domains of engineering and science: the expansive realm of Artificial Intelligence (AI) and the enigmatic potential of Quantum Computing (QC). Though their foundational seeds were sown in the fertile intellectual landscape of the 20th century, we now stand at the cusp of experiencing the profound fruits of that early cultivation. It is pertinent to note the hierarchical relationship within AI, where Machine Learning (ML) forms a significant subset, and Deep Learning (DL) resides at its core – akin to three concentric circles, each nested within the other. Alongside these advancements, Data Science has also burgeoned into a field of considerable influence. Yet, a discernible divergence exists in the perception and intuitive grasp of these domains among students. The future of AI often feels tangible and readily visualizable, while the intricacies of Quantum Computing remain more elusive, residing in a realm less readily accessed by conventional understanding.\\\\
Intriguingly, the nascent field of Quantum Machine Learning emerges as a compelling intersection of these two powerful disciplines. One fundamental distinction lies in their underlying principles: AI, in its current dominant paradigms, rests upon the bedrock of classical physics, whereas Quantum Computing draws its profound capabilities from the counter-intuitive principles of quantum mechanics.
This reliance on quantum physics introduces a unique challenge, as a significant portion of the populace may not yet fully embrace the often-astonishing tenets of quantum science. As the profound words of Niels Bohr, a luminary in atomic structure and quantum theory, remind us: “Anyone who is not shocked by quantum theory has not understood it.” This very shock, this necessary leap of imagination, is intrinsic to comprehending a field that primarily grapples with the intricacies of the microscopic world. However, contemporary scientific endeavors are diligently striving to bridge the chasm between the quantum realm and our macroscopic experience. In this context, two noteworthy papers warrant particular attention, (a) and (b) \cite{mal2016quantum,das2024mass}.\\

\subsection*{Concluding Remarks: A Message to Future Engineers}

As we arrive at the end of this brief journey through Quantum Information Processing and Quantum Teleportation, it is worthwhile to pause and reflect on how far science has travelled. A little more than a century ago, the ideas of quantum mechanics appeared strange, counter-intuitive, and even unacceptable to many of the greatest scientists of the time. Today, those very ideas form the foundation of technologies that are gradually reshaping our world. Quantum teleportation began as a theoretical curiosity. The original proposal by Bennett and his collaborators in 1993 demonstrated that an unknown quantum state could be transferred from one location to another without physically moving the particle carrying that state. What initially appeared to belong to the realm of science fiction has now become an active area of research with experimental demonstrations over optical fibres, free-space channels, and even satellite-based platforms. Throughout this book, we have seen how the mathematics of Hilbert spaces, linear operators, density matrices, entanglement, Bell non-locality, and teleportation fidelity collectively contribute to our understanding of quantum communication. We have also witnessed how the notion of entanglement evolved from a philosophical puzzle into a valuable resource for information processing. Yet, the story is far from complete. Many of the technologies that future generations will routinely use are still under development. Quantum computers capable of solving large-scale practical problems, quantum internet infrastructures connecting distant quantum processors, ultra-secure communication systems, and distributed quantum sensing networks remain grand engineering challenges. The solutions to these challenges will not come solely from physicists. They will require mathematicians, computer scientists, electronics engineers, communication engineers, material scientists, data scientists, and artificial intelligence researchers working together. The engineers of the twenty-first century therefore stand at a unique crossroads. Unlike previous generations who witnessed the rise of classical computing, the present generation has the opportunity to participate in the birth of an entirely new technological paradigm. Quantum technologies are no longer confined to academic laboratories. Governments, industries, startups, and research organizations around the world are investing heavily in quantum computing, quantum communication, quantum cryptography, and quantum sensing. For students, the message is simple: do not be intimidated by quantum mechanics. Every expert in the field began with confusion. The mathematical formalism may appear difficult at first, but persistence gradually transforms complexity into clarity. The purpose of education is not merely to learn what is already known; it is to acquire the ability to discover what is still unknown.\\

\noindent I encourage readers to view this book not as an endpoint but as a starting point. Explore advanced topics such as quantum algorithms, quantum error correction, quantum machine learning, quantum networks, quantum foundations, and quantum thermodynamics. Read original research articles. Learn to use modern quantum programming frameworks. Participate in research projects. Most importantly, cultivate curiosity. The future of quantum science will be written by those who are willing to ask questions that seem impossible today. Perhaps among the readers of this book is a future scientist who will discover a new teleportation protocol, design a scalable quantum network, develop fault-tolerant quantum computers, or uncover deeper principles governing the quantum world. The next breakthrough may emerge from a classroom, a small laboratory, or even from a student's notebook. The quantum revolution is still unfolding. Its most exciting chapters have not yet been written.

\begin{center}
\rule{5cm}{0.4pt}\, $\ast$ \,\rule{5cm}{0.4pt}
\end{center}
\printindex

\printbibliography
\end{document}